\documentclass[aps,rmp,twocolumn,showpacs,10pt,superscriptaddress,preprintnumbers,longbibliography]{revtex4-1}
\usepackage{epsfig,amssymb,amsmath,psfrag,epstopdf,color}
\pdfoutput=1
\usepackage{graphicx}
\definecolor{darkblue}{rgb}{0.1,0.1,.7}
\usepackage[colorlinks, linkcolor=darkblue, citecolor=darkblue, urlcolor=darkblue, linktocpage]{hyperref} 
\usepackage{subcaption}
\usepackage{ragged2e}
\DeclareCaptionJustification{justified}{\justifying}
 \usepackage[normalem]{ulem}
\usepackage{paralist}
\usepackage{hyperref}
\allowdisplaybreaks

\def\beq{\begin{equation}}
\def\eeq{\end{equation}}
\def\bsp#1\esp{\begin{split}#1\end{split}}
\newcommand{\be}{\begin{equation}}
\newcommand{\ee}{\end{equation}}
\newcommand{\bea}{\begin{eqnarray}}
\newcommand{\eea}{\end{eqnarray}}
\newcommand{\eqn}[1]{eq.~\eqref{#1}}

\def\Fig#1{Fig.~{\ref{#1}}}

\def\eqn#1{eq.~(\ref{#1})}

\def\Sec#1{Section~\ref{#1}}

\def\cN{{\mathcal N}}
\def\cO{{\mathcal O}}

\def\to{\rightarrow}

\newcommand{\comment}[1]{}

\newcommand{\cE}{\mathcal{E}}

\def\tr{\mathop{\rm tr}\nolimits}

\newcommand{\img}{{\rm i}}

\newcommand{\nn}{\nonumber}
\newcommand{\df}{\mathrm{d}}

\newcommand{\nt}{\notag\\} 

\newcommand{\gcusp}{\Gamma^{\mathrm{cusp}}}
\newcommand{\gammacusp}{\Gamma_\text{cusp}^q}

\newcommand{\ga}{\gamma}

\newcommand{\vev}[1]{\langle #1 \rangle}

\usepackage{slashed}

\newcommand{\mae}[3]{\langle#1\rvert#2\rvert#3\rangle}

\begin{document}

\title{Energy Correlators: A Journey From Theory to Experiment}


\author{Ian Moult}
\email{ian.moult@yale.edu}
\affiliation{Department of Physics, Yale University, New Haven, CT 06511}

\author{Hua Xing Zhu}
\email{zhuhx@pku.edu.cn}
\affiliation{School of Physics, Peking University, Beijing, 100871, China}
\affiliation{Center for High Energy Physics, Peking University, Beijing 100871, China}


\begin{abstract}
Collider experiments offer a unique opportunity to explore the Standard Model (SM), and to search for new physics, new interactions, and new principles of nature. 
The theoretical abstraction of a collider, namely the study of correlations in asymptotic fluxes, provides one of the most basic examples of an observable in quantum field theory (QFT) and quantum gravity.
Energy flux is described in QFT by energy flow operators, a particular example of light-ray operators.
In addition to their central role in the theoretical description of collider physics, energy flow operators play an important role in diverse areas of formal QFT and gravity, providing a connection between real world collider phenomenology,  and the deep underlying principles of QFT.
 
Recently it has become possible to measure correlation functions of energy flow operators in a wide variety of collider experiments, providing an exciting new connection between collider physics and formal theory.
In this review, we provide a survey of recent progress in our understanding of energy operators and their correlators, highlighting their importance in both formal theory and collider phenomenology, and in particular, their great potential for bridging these areas to provide new ways to understand the real world. 
We intend this article as a resource for both formal theorists interested in understanding how light-ray operators are being applied in particle and nuclear physics, as well as for experimentalists interested in the theoretical motivation for these observables.
Most importantly, we aim to stimulate further interaction between the formal, phenomenological and experimental communities through the common lens of energy correlators.
\end{abstract}

\maketitle

\tableofcontents

\section{Introduction}\label{sec:intro}

After a particle collision, the underlying microscopic physics gets imprinted into detailed correlations in macroscopic fluxes, much in analogy to how our cosmic history is imprinted into correlations in the Cosmic Microwave Background.  Understanding how to map these correlations to properties of the underlying microscopic description is key to addressing a wide variety of questions in particle and nuclear physics.

In quantum field theory (QFT), energy flux in colliders is described by energy flow operators. In addition to their central role in the theoretical description of collider physics, energy flow operators play an important role in diverse areas of formal QFT and gravity, ranging from providing constraints on renormalization group flows, to understanding the emergence of causality in the AdS/CFT correspondence. Energy flow operators provide the connection between real world collider phenomenology,  and the deep underlying principles of QFT.

\begin{figure}
\includegraphics[width=0.955\linewidth]{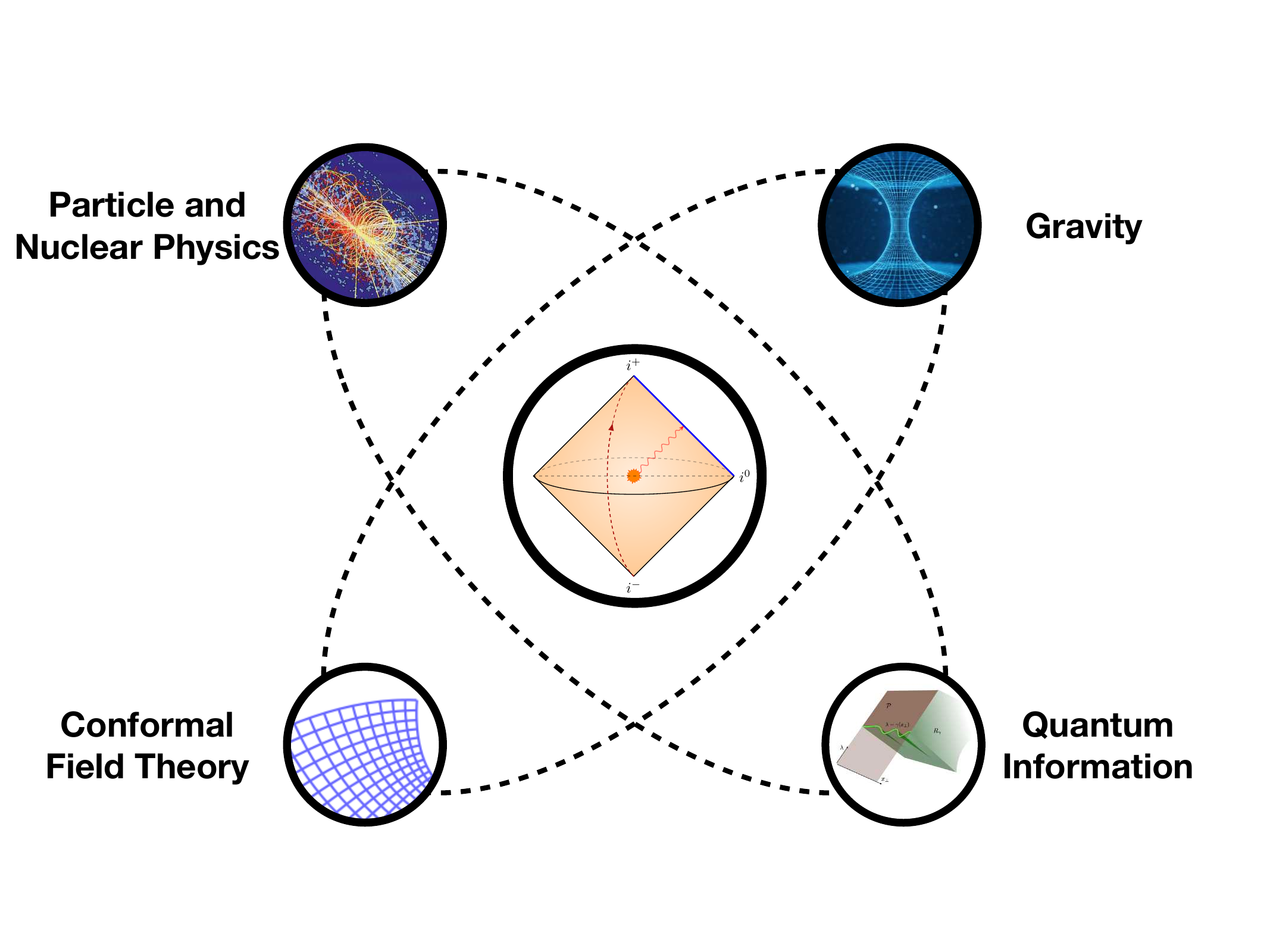}
\caption{Energy flow operators play an important role in diverse areas of formal QFT and gravity, connecting the deep underlying principles of QFT with real world experiments.
}
\label{fig:fig3}
\end{figure}

The study of correlation functions of energy flow operators has recently seen a rejuvenated interest, giving rise to a lively interplay between formal QFT, phenomenology, and experiment, see \Fig{fig:fig3}. On the phenomenological side, this has led to new theoretical techniques for performing calculations in quantum chromodynamics (QCD) with numerous applications in particle and nuclear physics, and has motivated measurements at a wide variety of collider experiments, providing new opportunities for precision Standard Model measurements, and new insights into the dynamics of QCD in extreme conditions. 

In this article, we will provide an overview of this emerging field, emphasizing the connections between different areas of study. Due to the breadth of topics covered, our goal will not be to present detailed derivations of results, but rather to emphasize the underlying physics of energy operators and their correlators, highlighting why they are of interest to different areas of the theoretical community, and how they are used in real world experiment.

\subsection{50 Years of Energy Flux}\label{sec:history}

The study of energy correlator observables has a remarkable history that is intertwined with the history of the field theory of QCD itself. It is a testament to their fundamental nature that they are still avidly being studied 50 years after their introduction. Although the goal of this article is not to provide a comprehensive history of QCD (to which we refer the reader to \cite{Gross:2022hyw}), we find it important to highlight certain key aspects in the history of energy correlators. This also helps to place into context a number of recent developments in the study of energy correlators.

Immediately after the discovery of asymptotic freedom \cite{Gross:1974cs,Gross:1973ju,Gross:1973id,Politzer:1973fx}, and the formulation of QCD as the theory of the strong interactions, it was apparent that one could use perturbation theory to begin to understand the distribution of energy in collider physics experiments, and that detailed structure of this distribution was a manifestation of the interactions of asymptotically free quarks and gluons at short distances. 

The first attempt made in this direction was by Appelquist and Georgi \cite{Appelquist:1973uz} and Zee \cite{Zee:1973sr}, who showed that the total cross section for $e^+e^-$ to hadrons could be systematically computed in perturbation theory at large $Q^2$, an example of the general Kinoshita-Lee-Nauenberg theorem \cite{Kinoshita:1962ur,Lee:1964is}. The total cross-section can be written as the two-point correlator of the electromagnetic current
\begin{align}\label{eq:twopoint}
\sigma &=\int \df^4x\, e^{\img Q\cdot x} \langle0| J(x) J(0)|0 \rangle\,,
\end{align}
which depends on the scale $Q^2$ (Here and throughout we suppress indices for simplicity of presentation). In a conformal field theory (CFT), the two-point function is fixed by the dimension of the operators up to a numerical coefficient. For the case of $J^\mu=\bar \psi \gamma^\mu \psi$, this leads to the well known $\sigma (Q^2) \sim 1/Q^2$ scaling of the total cross-section. In QCD the two-point function has a more complicated structure due to the presence of numerous hadronic resonances, but can be well described using perturbation theory after smearing \cite{Poggio:1975af}. The total cross-section provides a first, and in our eyes extremely beautiful, example of relating a macroscopic measurement in colliders to a fundamental correlator, in this case the two-point correlator, in QCD.

\begin{figure}
\includegraphics[width=0.555\linewidth]{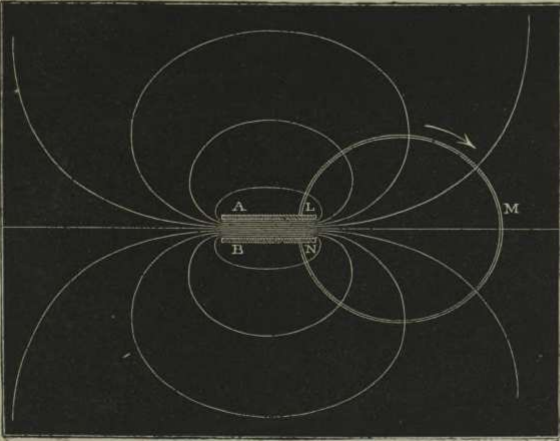}
\includegraphics[width=0.555\linewidth]{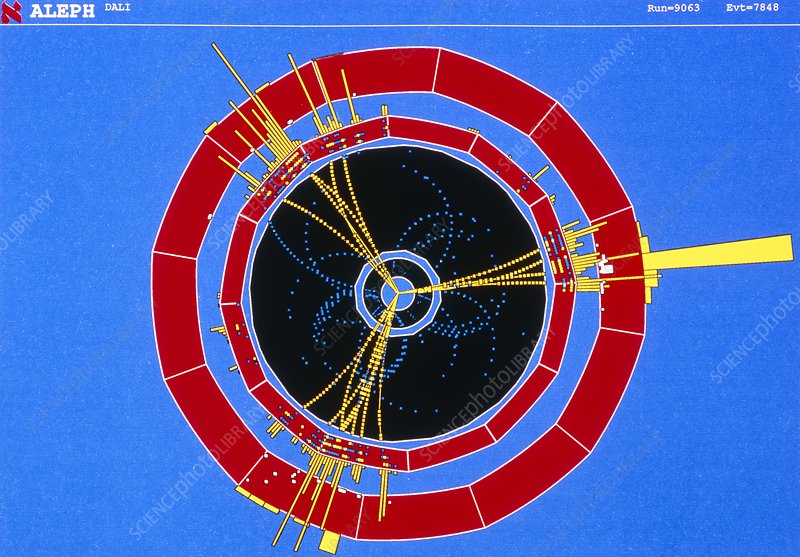}
\caption{The evolution of the study of energy flux in gauge theories. Top: The slow discharge of a capacitor in electromagnetism, as studied in Poynting's original work \cite{poynting}. Bottom: The rapid neutralization of a quark-antiquark color dipole in Yang-Mills theory. 
}
\label{fig:poynting}
\end{figure}

However, this is not entirely satisfying. We would really like to be able to ``see" the states that connect the two currents, much like we do when we look at a collider event display. In the study of condensed matter systems it is clear how to proceed: we can simply measure higher point functions, for example those obtain through the insertion of the stress tensor, $\langle J T J \rangle$. In gauge theories, this problem also has a remarkable history going back to the original work of Poynting \cite{poynting}, who considered the discharge of a capacitor through a high resistance wire, allowing him to follow the flow of energy. As illustrated in \Fig{fig:poynting}, $e^+e^-$ collider experiments are conceptually similar to Poynting's original setup, in that they study the rapid neutralization of a quark anti-quark dipole. While the Poynting vector extends to classical Yang-Mills (see e.g. \cite{Coleman:1977ps} for the case of plane waves), since QCD confines with a time scale of $t\sim 1/\Lambda_{\text{QCD}}$, we can neither measure higher point correlation functions of local operators, nor follow the flow of energy in the Yang-Mills fields. Instead, we must learn to characterize asymptotic fluxes of confined hadrons. The study of such observables in Yang-Mills theories is a giant conceptual leap, both in understanding that they can be calculated in perturbation theory, since asymptotic measurements are ultimately sensitive to the infrared structure of the theory, and in formulating the precise nature of the observables to study.

Generalizing, \eqn{eq:twopoint}, it is natural to view the two-point function as the norm of a state, and to define a three-point function, which is the expectation value of a particular operator in this state. Motivated by what a physical detector does in an experiment, namely measure the expectation value of energy flux in a particular direction, Sterman introduced the energy flux operator\footnote{Throughout this review we will interchangeable use the language ``energy flux operator", ``energy flow operator", ``energy operator", ``Average Null Energy (ANE) operator" and ``Average Null Energy Condition (ANEC) operator. } \cite{Sterman:1975xv}, via its 
action on on-shell particle states
\begin{align}\label{eq:ANEC_particle}
\mathcal{E}(\hat n) |X \rangle = \sum\limits_{k\in X} k^0 \delta^{2}(\Omega_{n}-\Omega_{k}) |X\rangle\,.
\end{align}
where $\Omega_n$ and $\Omega_k$ are the angular positions of the detector and a particle with momentum $k$. In his own words, ``energy flow became the focus of calculability", and this has been the foundation of our study of colliders for the subsequent 50 years.
As written, this particular definition is only valid in a theory with asymptotic particles, as is the case in QCD, but we will later see that it can be formulated in terms of the stress tensor of the underlying theory, generalizing it to any theory with a local stress tensor. 

Armed with the energy flux operator, it is now natural to consider its correlation functions in various states. From an experimental perspective, these correlation functions characterize the distribution of energy flux, and played an important role in establishing QCD as the theory of the strong interaction. The simplest of all collider observables is  the three-point function, also referred to as the one-point energy correlator, as first introduced by Sterman 
\begin{align}
\text{EC}(\hat n) &=
\frac{1}{\sigma Q} \int \df^4x\, e^{\img Q\cdot x} \langle0| J(x) \cE(\hat n) J(0)|0 \rangle 
\,.
\end{align}
This observable characterizes the direction of energy flux in an event. Sterman \cite{Sterman:1975xv} further argued that this one-point function is infrared and collinear safe, enabling its systematic calculation in perturbation theory, a property of crucial importance in later developments.

\begin{figure}
\includegraphics[width=0.955\linewidth]{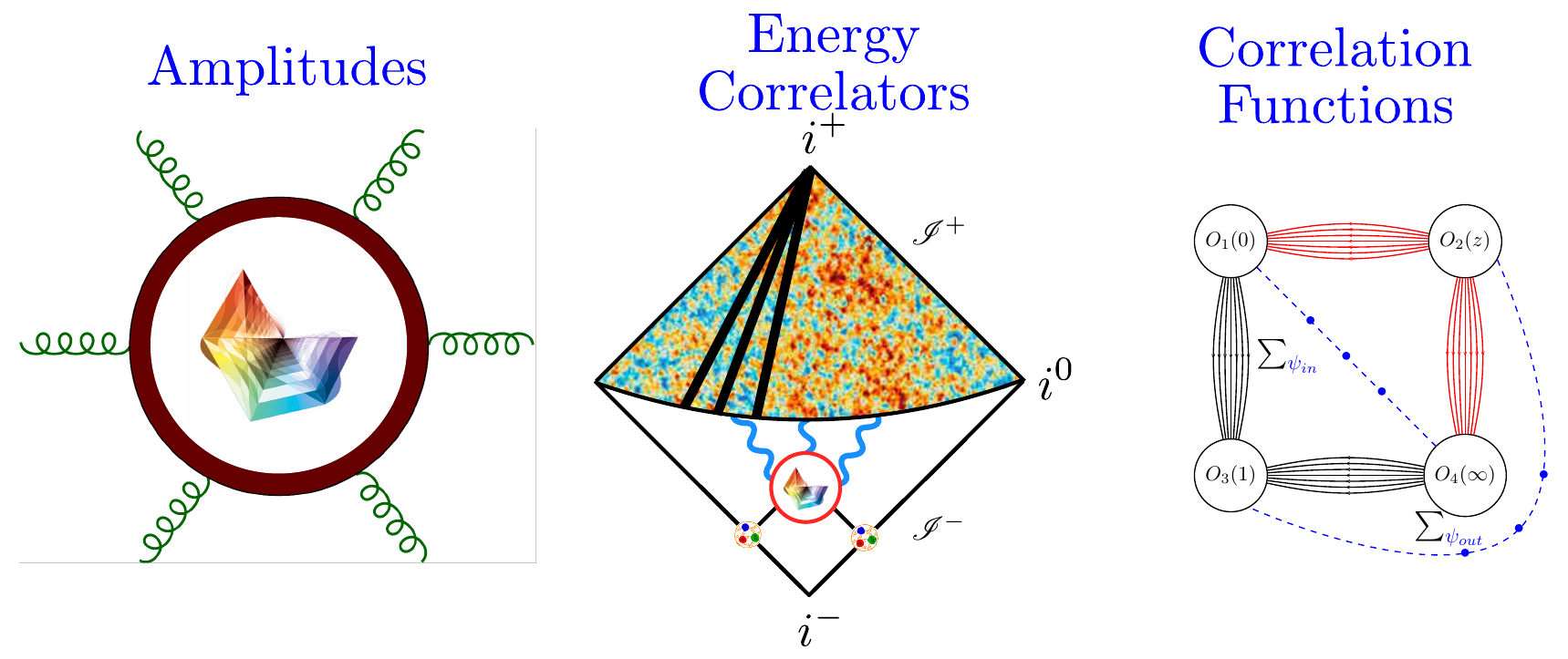}
\caption{Three interesting observables in QFT: amplitudes, energy correlators, and correlation functions of local operators. Energy correlators combine properties of both amplitudes and correlation functions, providing a well defined observable directly measuredable in experiments.
}
\label{fig:fig2}
\end{figure}

The next seminal step was the introduction of an operator definition of the energy flux operator, relating it to the stress tensor of the underlying theory
\cite{Sveshnikov:1995vi,Tkachov:1995kk,Korchemsky:1999kt}
\begin{align}\label{eq:ANEC_op}
\mathcal{E}(\hat n) = \lim_{r\to \infty}  \int\limits_0^\infty \df t\, r^2 n_i T_{0i}(t,r \hat n)\,.
\end{align}
In the case where we have asymptotic particle states, it is easy to verify that this definition acts as in Eq.~\ref{eq:ANEC_particle}. However, this definition applies more generally to any theory that has a stress tensor.

This energy flux operator is a specific instance of  the Average Null Energy (ANE) operator
\begin{align}
\mathcal{E}_u=\int du T_{uu}\,,
\end{align}
with $u$ a null-coordinate. In the case relevant for collider physics, the operator is placed at infinity. This operator takes its name from the condition it satisfies
\begin{align}
\langle \Psi | \mathcal{E}_u | \Psi \rangle \geq 0\,,
\end{align}
referred to as the ``average null energy condition" (ANEC). While the ANE operator and the ANEC played an important role in general relativity, where it is used as inputs to focussing theorems \cite{Borde:1987qr, Tipler:1978zz}, the chronology conjecture \cite{Hawking:1991nk}, topological censorship \cite{Friedman:1993ty}, and related causality theorems \cite{Tipler:1976bi,Gao:2000ga}, it is unclear that this connection was realized at this stage. We will return to the interesting implications of this statement shortly.

Providing a theoretical definition of an idealized calorimeter cell introduces a sharp link between the observables measured in collider physics, and correlation functions of local operators in a QFT \cite{Hofman:2008ar}. Indeed, given the correlation function of local operators $\langle \mathcal{O}^\dagger (x_1) T(x_2) \mathcal{O}(x_3) \rangle$, with $\mathcal{O}$ a local operator, and $T$ the stress tensor, one can apply the limiting procedure in Eq.~\ref{eq:ANEC_op} to obtain the corresponding collider observable. We will often be interested in the case where the local operator is a conserved current, and we will denote the corresponding correlation function of local operators with the shorthand $\langle J T J\rangle$. Correlation functions of detector operators, therefore provide the crucial link between the measurements of asymptotic energy flux at macroscopic scales, and the correlation functions $\langle J T J\rangle$  of the underlying microscopic QFT. It is interesting to note that although such an observable could have been defined abstractly within the study of formal QFT, it was motivated by collider experiments. The fact that it has since proven to be a useful observable for understanding properties of QFT highlights the interplay between theory and experiment.

We can therefore understand energy correlators as an interesting intermediate observable between correlation functions of local operators and scattering amplitudes, as shown in \Fig{fig:fig2}. It is interesting to again emphasize the distinction. Correlation functions of local operators are non-perturbatively well defined observables, and in perturbation theory they are free of IR divergences.  On the other hand, scattering amplitudes generically have infrared divergences in perturbation theory, and are usually squared and integrated over phase space to compute a physical observable. Energy correlator observables, or more general correlation functions of detector operators, are physical observables that can be directly measured in collider experiment. For appropriately chosen detector operators, such as the energy flow operator, they are non-perturbatively well defined, and free of infrared divergences.

We can also form detector operators from other operators in the theory, for example conserved currents. However, depending on the details of the operator and the theory, the limiting procedure used to define a detector operator may not exist. In general, understanding the space of detector operators in a generic theory is an open problem.

\begin{figure}
\hspace{-0.6cm}\includegraphics[width=0.955\linewidth]{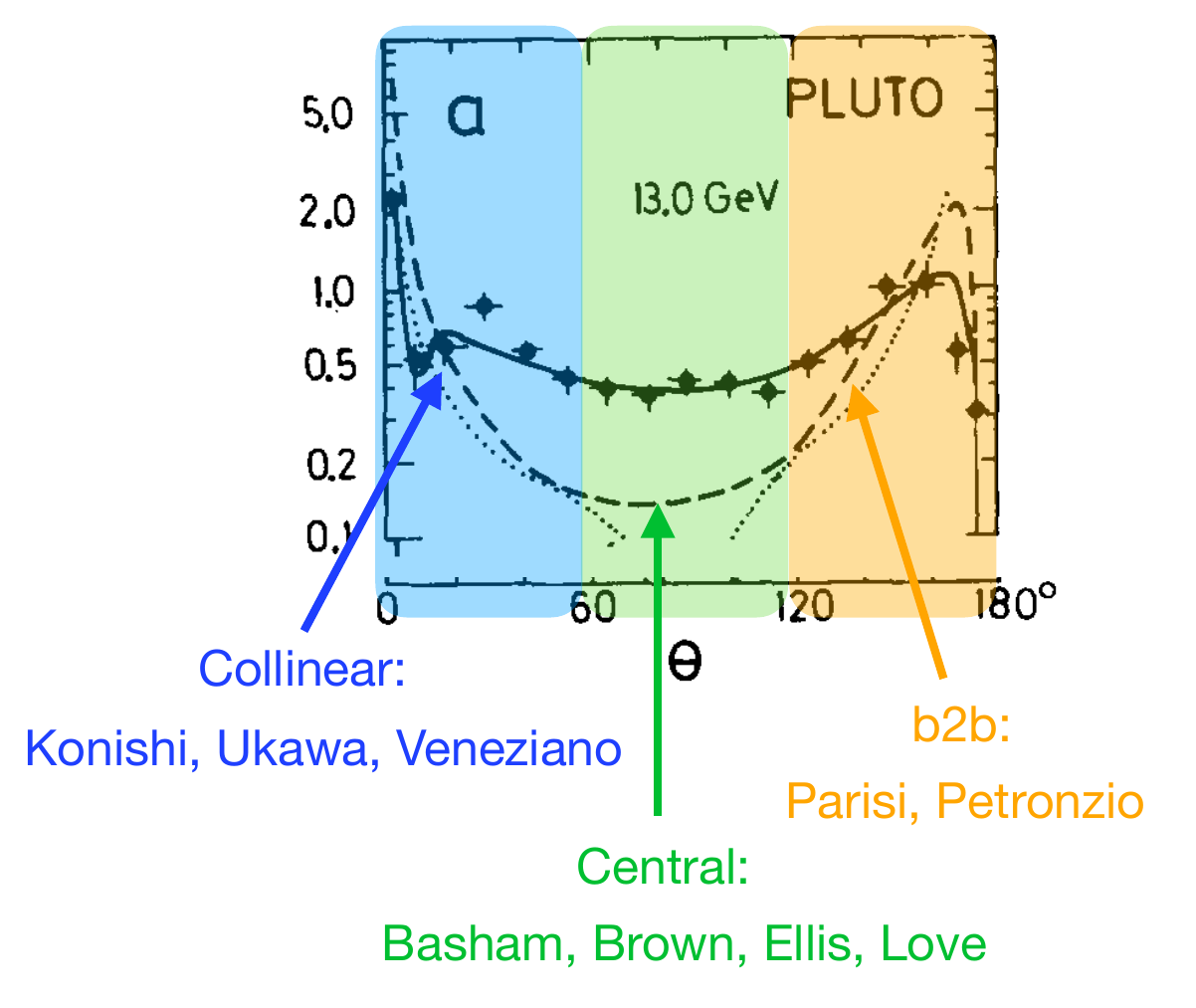}
 \includegraphics[width=0.8\linewidth]{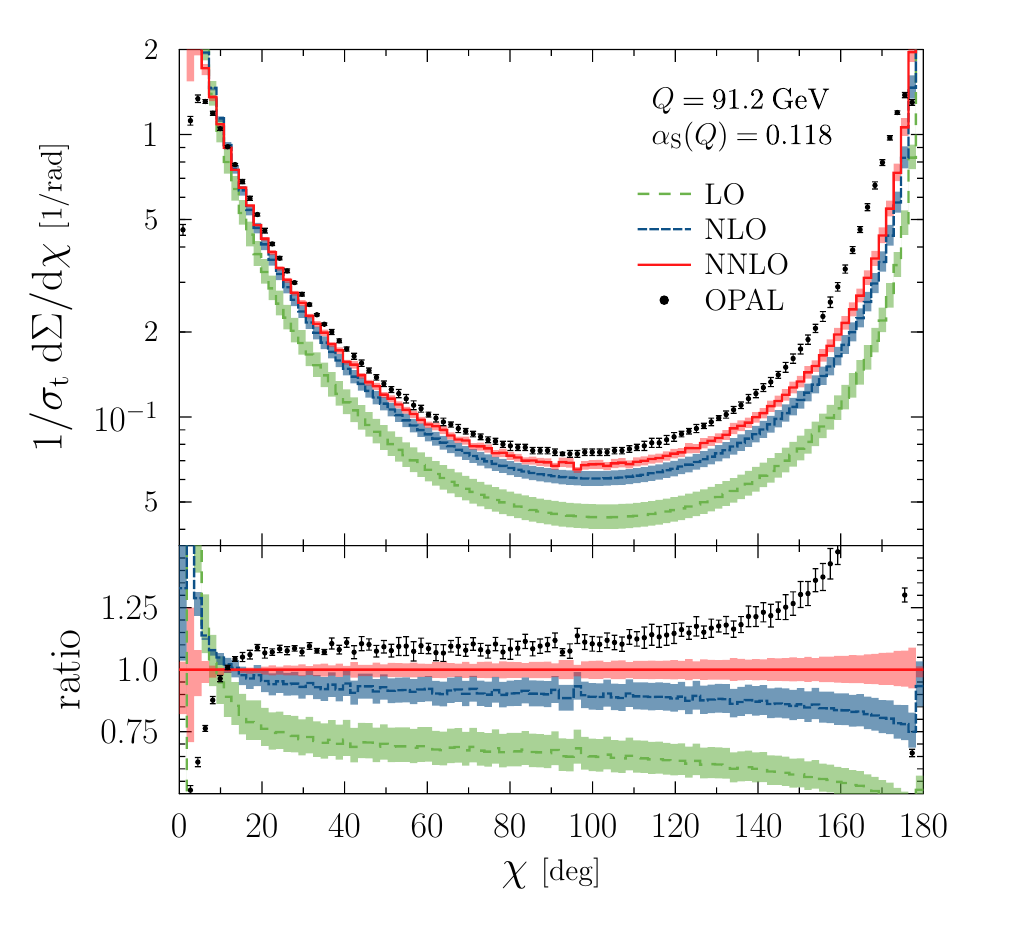}
\caption{Upper panel: The first measurement of the two-point energy correlator by the PLUTO experiment at 13 GeV. The agreement between perturbative calculations and data is poor due to the low energy of the collider.
Lower panel: A comparison of the OPAL data at 91.2 GeV with state of the art perturbative calculations at NNLO. Figures adapted from \cite{PLUTO:1981gcc} and \cite{Tulipant:2017ybb}.
}
\label{fig:pluto_single}
\end{figure}

Multi-point correlators were introduced in  \cite{Basham:1979gh,Basham:1978zq,Basham:1978bw,Basham:1977iq}
\begin{align}
\int \df ^4x\, e^{\img Q\cdot x}\, \langle0| J(x) \cE(\hat n_1) \cE(\hat n_2)  \cdots \cE(\hat n_k) J(0)|0 \rangle \nn \\ \equiv \langle \cE(\hat n_1) \cE(\hat n_2)  \cdots \cE(\hat n_k) \rangle \,,
\end{align}
and are referred to generically as ``energy correlators".
In terms of correlation functions of local operators, these are related to multi-point functions of the stress tensor, $\langle J(x) T(x_1) T(x_2) \cdots T(x_k) J(0)\rangle$, which encode highly non-trivial dynamical information about the theory. Correlation functions of energy flow operators  provide a clear strategy for characterizing the energy flux in collider physics experiments by measuring multi-point energy correlators, and relating them to properties of the underlying QFT.

In the case that the theory has a particle interpretation, such as in QCD, one can use the action of the energy flow operator in Eq. \ref{eq:ANEC_op} on asymptotic states to express the energy correlators as a weighted cross section. For example, for the two-point correlator, we have
\begin{align}
  \label{eq:EECdef}
  \text{EEC}(z)= \frac{1}{\sigma}\sum_{i,j}\int d\sigma\ \frac{E_i E_j}{Q^2} \delta\left(z - \frac{1 - \cos\chi_{ij}}{2}\right) \,,
\end{align}
where $\chi_{ij}$~(throughout this review, we will also interchangeably use $\theta_{ij}$; when it is unambiguous, we will also use $\chi$ or $\theta$ for short) is the polar angle between two particles $i$ and $j$. 
Energy correlator observables are infrared and collinear safe, allowing them to be systematically computed in perturbation theory.
Using this expression,  \cite{Basham:1979gh,Basham:1978zq,Basham:1978bw,Basham:1977iq} were able to analytically compute the leading order two-point correlator in perturbative QCD. While this definition is computationally, and experimentally useful in QCD, the operator definition applies even in the absence of asymptotic states, giving energy correlators a life beyond QCD.

The first measurement of the two-point energy correlator was performed by the PLUTO \cite{PLUTO:1981gcc} experiment in $e^+e^-$ collisions, and is shown in \Fig{fig:pluto_single}.  The calculation throughout the entire phase space required the development of collinear and Sudakov resummation techniques, for which the EEC was one of the first testing grounds. The resummation of leading logarithms in the collinear limit of the EEC was achieved first, due to its similarity to the (at that time) already understood DGLAP evolution of structure functions~\cite{Gribov:1972ri,Dokshitzer:1977sg,Altarelli:1977zs}, which was achieved using the jet calculus \cite{Konishi:1978ax,Konishi:1979cb,Konishi:1978yx}. Following the development of the resummation of Sudakov double logarithms~\cite{Dokshitzer:1978yd,Parisi:1979se,Parisi:1979xd,Ellis:1980my,Ellis:1981sj,Kodaira:1981nh}, the back-to-back limit of the energy correlators was extensively studied in the early 1980s, and is tied with the development of transverse momentum dependent (TMD) factorization theorems. Seminal works include~\cite{Chao:1982wb,Soper:1982wc,Kodaira:1982az,Collins:1981va,Collins:1985xx,Collins:1985kw,Collins:1981zc}. 

The data in  \Fig{fig:pluto_single} is compared with resummed perturbative calculations at leading logarithmic accuracy. Due to the low energies achieved at these colliders, perturbative QCD does not provide a quantitative description of the data.  However, these early measurements and theoretical calculations laid the foundations of our understanding of energy flux in QCD. The energy-energy correlator was subsequently measured in a variety of increasingly high energy $e^+e^-$ colliders, which to our knowledge, include the following list of experiments
\begin{itemize}
\item PLUTO \cite{PLUTO:1985yzc,PLUTO:1979vfu}
\item CELLO \cite{CELLO:1982rca}
\item JADE \cite{JADE:1984taa}
\item MAC \cite{Fernandez:1984db}
\item MARKII \cite{Wood:1987uf}
\item TASSO \cite{TASSO:1987mcs}
\item AMY \cite{AMY:1988yrv}
\item TOPAZ\cite{TOPAZ:1989yod}
\item ALEPH\cite{ALEPH:1990vew}
\item L3 \cite{L3:1991qlf,L3:1992btq}
\item DELPHI \cite{DELPHI:1990sof}
\item OPAL \cite{OPAL:1990reb,OPAL:1991uui}
\item SLD \cite{SLD:1994idb}
\end{itemize}
These experiments cover much of the remarkable history of $e^+e^-$ colliders.
For a nice overview of the available measurements, along with comparisons to theoretical calculations and an extraction of the strong coupling constant, see \cite{Kardos:2018kqj}.

In the subsequent decades, significant effort was undertaken to improve the theoretical description of these measurements.
Tremendous theoretical progress in our understanding of perturbative QFT enabled the calculation of $e^+e^-$ observables at next-to-next-to-leading order (NNLO) in perturbation theory \cite{Garland:2001tf,Garland:2002ak,Campbell:1997tv,Weinzierl:2008iv}\cite{Gehrmann-DeRidder:2007foh,Gehrmann-DeRidder:2007vsv,DelDuca:2016ily,DelDuca:2016csb}. In the lower panel of \Fig{fig:pluto_single} we show a comparison of the OPAL data at 91.2 GeV with NNLO calculations computed using the CoLoRFulNNLO subtraction scheme \cite{DelDuca:2016ily,DelDuca:2016csb,Tulipant:2017ybb}. The perturbative results at LO, NLO and NNLO show a convergence towards the data, illustrating the importance of higher order perturbative corrections. This calculation does not include the resummation of higher order corrections in the collinear, $\chi \to 0$, or back-to-back, $\chi \to \pi$, limit. The physical interpretation and technical calculation of these logarithmic corrections in kinematic limits will be a primary focus of this review. This calculation also does not include non-perturbative corrections from hadronization. Measurements of energy correlators also drove progress in the understanding of the structure of their non-perturbative corrections~\cite{Korchemsky:1999kt,Korchemsky:1997sy,Korchemsky:1994is,Dokshitzer:1999sh}, which played an important role in the understanding of non-perturbative corrections to QCD event shapes more generally \cite{Lee:2006fn,Hoang:2007vb,Abbate:2010xh,Hoang:2014wka,Benitez:2024nav,Benitez-Rathgeb:2024ylc}. Such non-perturbative corrections are crucial for obtaining precise agreement with data, and will be discussed in \Sec{sec:non_pert}.

In their first incarnation, energy correlators were viewed as an observable for characterizing QCD in $e^+e^-$ colliders. While they were also generalized to the transverse EEC~(TEEC) for hadron colliders~\cite{Ali:1984yp},  the measurement of this generalization had to await the modern LHC era \cite{ATLAS:2015yaa}. Although they were highly successful in the characterization of energy flow in $e^+e^-$ colliders, to our knowledge, they were only computed in QCD with the goal of better describing measurements, and had not received widespread interest as a general observable for characterizing and learning about the structure of generic QFTs.

\subsection{Energy Correlators Beyond QCD}\label{sec:modern}

The study of energy correlator observables was revived, systematized, and generalized in the work of Hofman and Maldacena \cite{Hofman:2008ar}, which highlighted their role as an interesting observable for studying generic QFTs, and the central role of detector operators in QFT.

Beyond the calculation of observables in specific theories, a primary goal of theoretical physics is to constrain the space of QFTs using consistency conditions. This is done by studying observables in the theory, such as scattering amplitudes, or correlation functions of local operators, as illustrated in \Fig{fig:fig2}, which have well understood properties. Famous examples of this include the conformal bootstrap program, which constrains the space of CFTs by studying properties of correlation functions of local operators \cite{Rattazzi:2008pe,El-Showk:2012cjh,El-Showk:2014dwa,Poland:2018epd}, or the S-matrix bootstrap program \cite{Paulos:2017fhb,Paulos:2016but,Paulos:2016fap}, which constraints the space of QFTs using properties of the S-matrix.

A recurring lesson from these investigations, is that certain constraints are most easily seen ``through a Lorentzian lens", namely using observables sensitive to light-cone dynamics. Intuitively the power of Lorentzian observables is clear: as compared to their Euclidean counterparts, short distances on the light-cone provides access to dynamics at macroscopic scales.  This perspective was popularized in \cite{Adams:2006sv}, and has since seen numerous successes, ranging from the EFT-hedron \cite{Arkani-Hamed:2020blm} to bounds on the graviton three-point coupling \cite{Camanho:2014apa}.

In nearly all applications, the Lorentzian observable of choice is the S-matrix. While the S-matrix exists in a gapped theory, it does not in general exist in a CFT. This can be made even more extreme by considering the case of a CFT coupled to gravity, where there are neither local observables, nor an S-matrix. One would therefore like Lorentzian observables, analogous to the S-matrix, that are well defined in generic field theories.  For this purpose, correlation functions of detector operators provide a natural observable.  A beautiful example of this philosophy was provided in \cite{Maldacena:2011jn}, who proved that in the presence of a higher spin symmetry, the $N$-point energy correlators are equivalent to those in a free theory, providing an analogue of the Coleman-Mandula theorem \cite{Coleman:1967ad,Haag:1974qh}, for detector correlators.

Despite the fact that the analytic structure of multi-point detector correlators is not well understood, many interesting constraints on CFT data have been derived, including constraints on theories with higher spin symmetry \cite{Maldacena:2011jn,Maldacena:2012sf} and on CFTs with extremal values of $a/c$ \cite{Zhiboedov:2013opa}, as well as constraints on anomaly coefficients $a$ and $c$ (to be discussed in more detail shortly) \cite{Hofman:2008ar}, and other CFT data \cite{Li:2015itl,Cordova:2017zej,Cordova:2017dhq,Chowdhury:2017vel}. We therefore see that ``collider thought experiments" can place interesting constraints on QFTs, just as real world colliders place constraints on our description of nature.

The fact that correlation functions of the energy flow operator provide constraints on CFT data is perhaps not surprising. 
Indeed, correlation functions involving the stress tensor, particularly when the source are conserved currents, play a special role in CFTs. As a simple example, we can consider  $\mathcal{N}=1$ super-conformal field theories in $d=4$. We can consider the one-point energy correlator in a state produced by the R-current, $\tilde J$. This is related to the three-point function $\langle \tilde J T \tilde J \rangle$. These  three-point functions can be expressed in terms of the anomaly coefficients, which in $d=4$ are referred to as $a$, $c$ \cite{Osborn:1998qu,Anselmi:1997ys,Anselmi:1997am}.

These coefficients are interesting from two-perspectives. First, they characterize the trace anomaly when a CFT is placed on a curved manifold. In the case of $d=4$, the trace anomaly on a curved manifold takes the form
\begin{align}
T^\mu_\mu=\frac{c}{16 \pi^2}W_{\mu \nu \delta \sigma}W^{\mu \nu \delta \sigma} -\frac{a}{16 \pi^2}E\,,
\end{align}
where $E$ is the Euler density and $W$ is the Weyl tensor (see  \cite{Deser:1993yx} for a characterization in general dimensions).  Second, the coefficient $a$ plays an important role in the characterization of renormalization group flows in $d=4$. In $d=2$, Zamolodchikov famously proved the existence of a c-function \cite{Zamolodchikov:1986gt}, which monotonically decreases along RG flows to the IR, proving irreversibility of the RG, and providing a counting of the number of degrees of freedom. In $d=4$, the corresponding ``a-theorem", namely that
\begin{align}
a_{\text{IR}} \leq a_{\text{UV}}\,,
\end{align}
was conjectured by Cardy  \cite{Cardy:1988cwa}, and proven by Komargodski and Schwimmer \cite{Komargodski:2011vj},

The conformal collider bounds of \cite{Hofman:2008ar} place interesting constraints on the values of $a$ and $c$ at the fixed point. Considering again for simplicity the case of an $\mathcal{N}=1$ super-conformal field theory in $d=4$, the one-point energy correlator in a state produced by the R-current takes the form
\begin{align}
\langle \mathcal{E}(\theta) \rangle =1+ 3 \frac{c-a}{c} \left(\cos^2 \theta -\frac{1}{3} \right)\,,
\end{align}
where $\theta$ represents the angle between the polarization vector of the current, the direction of the detector. Using the simple condition that the expectation value of the energy flux is positive,
\begin{align}
\langle \mathcal{E}(\theta) \rangle \geq 0\,,
\end{align}
which is also referred to as the  average null energy condition (ANEC). While this result may seem obvious from our daily experience with collider physics, it was in fact only rigorously proven quite recently.  It has been proven using two different techniques, one based on causality  \cite{Hartman:2016lgu}, and one based on monotonicity of relative entropy \cite{Faulkner:2016mzt}. We find it quite remarkable that properties of energy flux at colliders are controlled by the entanglement structure of the vacuum.

Combining the ANEC with the explicit expression for the three-point function allows one to derive the constraint
\begin{align}
\frac{3c}{2}\geq a \geq 0\,.
\end{align}
While this has since been proven directly using the conformal bootstrap \cite{Hofman:2016awc}, it emphasized that energy correlators thought experiments can provide interesting constraints on the space of QFTs. The positivity of energy flux has since been used  to study a variety of other quantities, ranging from OPE coefficients \cite{Cordova:2017zej} to bounds on RG flows \cite{Hartman:2024xkw,Hartman:2023ccw,Hartman:2023qdn}. These many applications of light-ray operators for constraining the space of QFTs has greatly broadened their interest beyond the specific theory of QCD.

In addition to using energy correlator observables to constrain properties of the local correlation functions of the underlying theory, the direct link between energy correlator observables and correlation functions of local operators provides a new way of understanding collider physics observables through the properties of the underlying correlation functions of local operators \cite{Hofman:2008ar,Belitsky:2013ofa,Belitsky:2013bja,Belitsky:2014zha,Korchemsky:2015ssa}. In particular, it provides a non-perturbative definition of energy correlator observables in generic field theories. This opens the door to the study of these observables at strong or finite coupling in highly symmetric theories, providing a remarkable glimpse into the non-perturbative structure of energy flux observables. The first example of this was the calculation of the $n$-point energy correlator in a strong coupling expansion in planar $\mathcal{N}=4$ sYM, using the AdS/CFT correspondence~\cite{Maldacena:1997re,Witten:1998qj,Gubser:1998bc}. At strong coupling, the energy correlator was found to be uniform \cite{Hofman:2008ar,Goncalves:2014ffa,Belitsky:2013ofa,Korchemsky:2015ssa} $\langle \mathcal{E}(\hat{n}_1) \cdots \mathcal{E}(\hat{n}_n)\rangle=(Q/(4\pi))^n +\mathcal{O}(1/\lambda)$, up to calculable corrections in the inverse 't Hooft coupling, $\lambda$, whose explicit form will be discussed in \Sec{sec:heavy}. Since then, there has been significant recent progress in using the conformal bootstrap to obtain four point functions in a variety of theories, in particular the 3d Ising model \cite{El-Showk:2012cjh,El-Showk:2014dwa,Kos:2016ysd,Kos:2014bka,Chang:2024whx}, 3d supersymmetric ABJM models \cite{Chester:2024bij}, and $\mathcal{N}=4$ sYM \cite{Caron-Huot:2024tzr}. These will enable the non-perturbative determination of energy-energy correlators in these theories, effectively providing a solution to the calculation of the simplest collider observables in these theories.

\subsection{The Beauty of Physical Observables}\label{sec:phys_pert}

 The last decades have seen spectacular progress in our understanding of perturbative QFT. These include the development of new combinatorial/ geometric structures describing scattering amplitudes \cite{Arkani-Hamed:2017vfh,Arkani-Hamed:2013kca,Arkani-Hamed:2013jha,Arkani-Hamed:2017mur}, correlators \cite{Eden:2017fow}, and cosmological wavefunctions/ correlators \cite{Benincasa:2024leu,Arkani-Hamed:2017fdk,Arkani-Hamed:2024jbp}, along with the discovery of new symmetries, such as the dual super-conformal \cite{Alday:2007hr,Drummond:2008vq,Brandhuber:2008pf} and Yangian symmetries \cite{Drummond:2009fd} of scattering amplitudes in planar $\mathcal{N}=4$ sYM. In the particular case of planar $\mathcal{N}=4$ sYM, this has enabled calculations of form factors and amplitudes at a remarkable eight loops \cite{Dixon:2022rse,Dixon:2023kop}.  These advances have provided countless insights into the perturbative structure of real world QCD \cite{Henn:2020omi}.

 The direct relation between energy correlator observables and correlation functions of local operators also provides new approaches for exploring the perturbative structure of detector correlators, as well as optimism that they may exhibit elegant structures and hidden symmetries in perturbation theory. The observable of relevance for the study of energy correlators, the four-point correlator, is known in planar $\mathcal{N}=4$ sYM to 12 loops at the integrand level \cite{Bourjaily:2025iad,He:2024cej,Bourjaily:2016evz,Bourjaily:2015bpz,Ambrosio:2013pba,Eden:2012tu,Eden:2011we} and three loops at the integrated level \cite{Drummond:2013nda,Eden:1998hh,Eden:1999kh,Eden:2000mv,Gonzalez-Rey:1998wyj} (See \cite{He:2025rza,He:2025vqt} for progress at four loops.). This spectacular knowledge can be leveraged to improve our understanding of detector correlators in perturbation theory.

\begin{figure}
\includegraphics[width=0.755\linewidth]{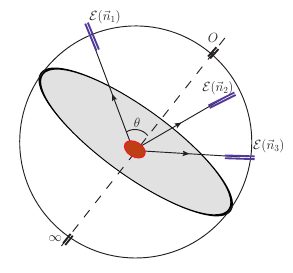}
\caption{Multi-point energy correlators, $\langle \mathcal{E}(n_1) \mathcal{E}(n_2) \cdots \mathcal{E}(n_k) \rangle$, define interesting functions of the positions of the detector operators. These correlation functions are beginning to be explored in perturbation theory, leading to elegant mathematical structures, and providing a playground for explorations of perturbative QFT. Figure from \cite{Yan:2022cye}.
}
\label{fig:multipoint_intro}
\end{figure}

\begin{figure*}
\includegraphics[width=0.655\linewidth]{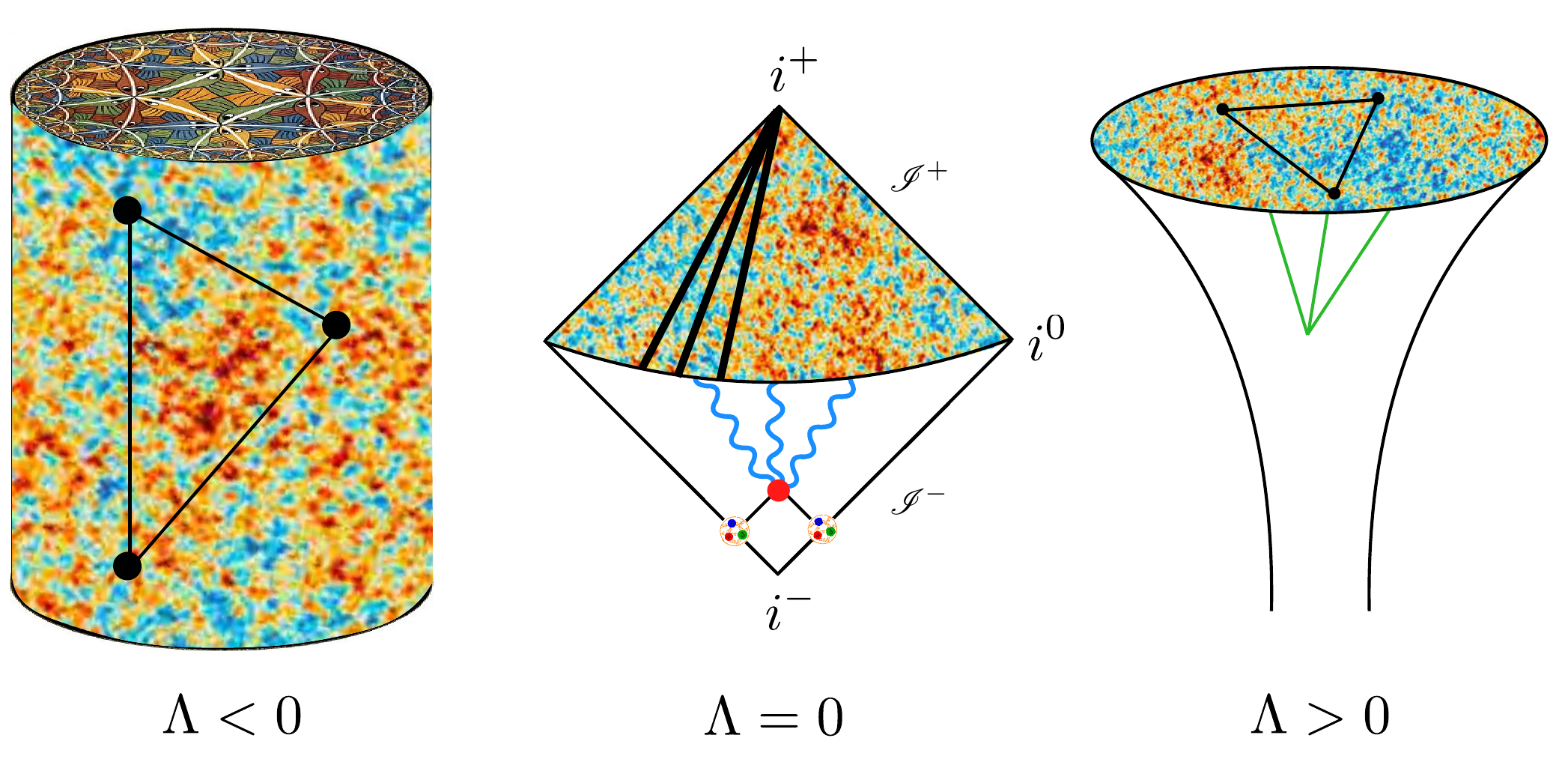}
\caption{Analogous to boundary local correlators for AdS space~($\Lambda <0$), and cosmological correlators in dS space~($\Lambda > 0$), energy correlators provide a ``canonical" flat space observable~($\Lambda = 0$), offering both a laboratory for theoretical exploration, as well as practical observable for collider experiments.
}
\label{fig:diff_correlators}
\end{figure*}

 The first calculation of energy correlators from local correlators was given  in \cite{Belitsky:2013ofa,Belitsky:2013bja,Belitsky:2014zha,Korchemsky:2015ssa}, enabling the first calculation of energy correlators at next-to-leading order \cite{Belitsky:2013ofa} in a state produced by two stress tensors in $\mathcal{N}=4$ super Yang-Mills (or by supersymmetry, any operators in the stress-tensor multiplet). This calculation is particularly noteworthy in that it performed the calculation directly from the correlator of local operators, bypassing issues with infrared divergences that appear in standard approaches. It also provided the first hint that energy correlator observables might have an interesting analytic structure. The result is expressed in terms of polylogarithms up to weight 2, and is given explicitly in Eq. \ref{eq:EEC_N4}. This approach was also used in QCD to compute the two-point correlator of the electromagnetic charge detector \cite{Chicherin:2020azt}.

Concurrently, there have been significant advances in techniques for the analytic calculation of amplitudes and cross sections in QCD, such as Integration-By-Parts (IBP) relations \cite{Chetyrkin:1981qh,Tkachov:1981wb}, differential equations \cite{Kotikov:1990kg,Gehrmann:2000zt,Henn:2013pwa}, and reverse unitarity \cite{Anastasiou:2002yz,Anastasiou:2003yy}. These facilitated the first analytic calculation of the EEC at NLO in QCD \cite{Dixon:2018qgp}, opening the door for a deeper understanding of QCD event-shape observables beyond LO, including their function space and asymptotic limits.

These calculations motivated the exploration of the perturbative structure of multi-point correlators.  Much like for cosmological correlators, correlation functions of detector operators are functions of the kinematics of the detector locations on the boundaries of spacetime, which in this case is the celestial sphere. This is shown in \Fig{fig:multipoint_intro} for the particular case of the three-point correlator. It is therefore interesting to explore the perturbative structure of these observables as a function of detector kinematics. Many of the questions we can ask for amplitudes or cosmological correlators can also be asked for detector operators. At the practical level, these include what is the space of functions needed to describe these correlators in perturbation theory, and what are their singularities.   At the more theoretical level, one would like to understand how properties of the bulk field theory, such as unitarity and causality, imprint themselves in the correlation functions of detector operators. We can also try and understand how symmetries of the underlying theory manifest on energy correlator observables. This has seen tremendous recent progress from the perspective of celestial holography (see e.g. \cite{Pasterski:2016qvg,Pasterski:2017ylz,Strominger:2017zoo}), and it is interesting to see how structures discovered there manifest in the study of energy correlators.  Additionally, one would like to understand if geometric or combinatorial structures discovered in amplitudes persist in detector correlators. Excitingly, answering these questions allows us to answer questions about physical observables used in collider experiments!

Surprisingly, nothing was known about the perturbative structure of multi-point correlators until quite recently. This is in stark contrast to the case of cosmological correlators which have been extensively studied \cite{Maldacena:2002vr,Arkani-Hamed:2015bza,Baumann:2021fxj,Baumann:2020dch,Arkani-Hamed:2018kmz,Baumann:2019oyu}. The first perturbative multi-point correlator was obtained in \cite{Chen:2019bpb}, which computed the three-point correlator in the collinear limit, finding that it was given by a simple expression in terms of polylogarithmic functions. These results have initiated a tremendous exploration of the perturbative structure of multi-point correlators, both theoretically and experimentally, resulting in calculations of the three and four-point correlators \cite{Yan:2022cye,Yang:2024gcn,Yang:2022tgm,Chicherin:2024ifn}. The observed simplicity suggests that much more can be learned, and that energy correlators are the simplest flat space observable. This provides an exciting opportunity to directly connect the study of physical observables with the remarkable combinatorial and geometric structures that have emerged in perturbative QFT. 

Combining all the remarkable features of energy correlator observables, along with their simultaneous relation to correlation functions of local operators and the conformal bootstrap on the one hand, and scattering amplitudes and on-shell techniques, on the other hand,  we believe that this makes detector observables arguably \emph{the} canonical flat space observables in generic QFTs, analogous boundary local correlators for AdS space and cosmological correlators for dS space (see \Fig{fig:diff_correlators}).

\subsection{Detector Operators Take on a Life of Their Own: The Story of Analyticity in Spin}\label{sec:analyticity}

Although detector/light-ray operators and their correlators are interesting observables in generic QFTs, one may think that they are only of interest in the specific application of collider observables and their abstractions. However, in recent years, they have been found to be central to organizing our understanding of QFTs, in particular CFTs, providing a unified description of many seemingly disparate phenomenon. Much like our historical review of energy flux in QCD, we attempt to provide some of the history of this fascinating evolution. We will see that this history has many parallels to the study of energy flux, with light-ray operators also originally being developing in QCD with specific phenomenological applications, before their implications for generic QFTs were understood.

The SLAC-MIT deep inelastic scattering (DIS) experiments \cite{Bloom:1969kc,Breidenbach:1969kd,Bodek:1979rx}, and the discovery of Bjorken scaling \cite{Bjorken:1968dy} played a crucial role in the development of QCD. For historical reviews, see \cite{Friedman:1972sy,Kendall:1991np,Taylor:1991ew}, and for an excellent review of the theoretical developments see \cite{Parisi:2025nob}. DIS involves a space-like momentum exchange, and can therefore be treated rigorously using an operator product expansion. As is common for processes dominated by light-cone dynamics \cite{Brandt:1970kg}, the leading operators appearing in this OPE are not the operators of lowest dimension, but rather the operators of lowest \emph{twist} \cite{Gross:1971wn}, defined as $t=d-s$, where $d$ is the dimension of the operator, and $s$ is the twist. In QCD, or other weakly coupled four-dimensional gauge theories, the lowest twist operators have twist 2 (up to quantum corrections). In QCD, the (unpolarized) twist-2 spin-J operators are given by
\begin{align}\label{eq:twist2_intro}
\mathcal{O}_q^{[J]}&=\frac{1}{2^J}\bar{\psi}\gamma^{+}(iD^+)^{J-1}\psi\,,\\
\mathcal{O}_{g}^{[J]}&=-\frac{1}{2^J} F_{c}^{i +}(iD^+)^{J-2}F_{c}^{i +}\,,
\end{align}
where $+$ denotes the light-cone component. We will discuss these operators in more detail in \Sec{sec:celestial_blocks}
These twist-2 operators play a central role in the study of light-cone dominated processes in four-dimensional gauge theories, and will also appear extensively in our study of the energy correlators.

Due to their phenomenological importance, the twist-2 operators in QCD have received much theoretical attention.\footnote{To our knowledge, the first calculation of twist-2 anomalous dimensions in any theory was performed in \cite{Christ:1972ms}. In this reference, the analogy between the Mellin transform of the twist-2 anomalous dimensions and the Sommerfeld-Watson transformation was noted.} Their anomalous dimensions were first computed to one-loop order in \cite{Gross:1974cs,Gross:1973ju,Gross:1973id,Georgi:1974wnj}. As local operators, these make sense only for integer spin. However, for studying the evolution of parton distribution functions, and fragmentation functions, these were extended to the so called DGLAP splitting functions \cite{Gribov:1972ri,Dokshitzer:1977sg,Altarelli:1977zs}. The Mellin moments of the splitting functions agree with the twist-2 anomalous dimensions for even integer $J$, but in perturbation theory, are analytic functions of the spin. For the particular case of pure Yang-Mills (which isolates the gluonic operator in Eq. \ref{eq:twist2_intro}), the one loop anomalous dimension reads
\begin{align}
  \label{eq:QCDAD}
\gamma_{gg}^{(0)}(J)&\ = -4 C_A \bigg[ \frac{1}{J (J-1)} + \frac{1}{(J+1) (J+2)}
\nn\\
&\ \quad  - (\Psi(J+1) + \gamma_E)  \bigg] 
- \beta_0 \,,
\end{align}
where $\beta_0$ is the one loop beta function, and the digamma function $\Psi(J)$ is a meromorphic function of spin $J$.

Following the seminal work of Polyakov \cite{Polyakov:1980ca} who advocated formulating the study of QCD (or more generally Yang-Mills theories) as the study of ``string operators", or what we would in a modern language call Wilson lines, Balitsky and Braun introduced what would now be called light-ray operators \cite{Balitsky:1987bk}.  They used them \cite{Balitsky:1990ck,Balitsky:1988fi} to understand fragmentation in $e^+e^-$ collisions. These operators are obtained as null integrals of a kernel $\phi(\alpha)$ over a string operator $\Phi(\alpha)$
\begin{align}
Q=\int d\alpha_i \phi(\alpha_i) \Phi(\alpha_i)\,.
\end{align}
Here $\Phi(\alpha)$ consists of fields strung along the light-cone, connected by Wilson lines. For example, we can consider a two-field string operator
\begin{align}
\Phi_\mu(\alpha_1, \alpha_2)=\bar \psi(\alpha_1) \gamma_\mu [\alpha_1, \alpha_2] \psi(\alpha_2)\,,
\end{align}
where the gauge link is given by
\begin{align}
[\alpha_1, \alpha_2]= P \exp\left[ i g \int \limits_{\alpha_2}^{\alpha_1} dt A_+(t)   \right]\,.
\end{align}
By computing the anomalous dimensions of these non-local operators, they were able to obtain the full analytic function of $J$. Note that for $J=2$, the anomalous dimension vanishes, and we have exactly the operator $\mathcal{E}$, however the link between these operators, and detector operators and their correlation functions, would have to wait.

These string operators were extensively studied, including at higher twist, for describing the evolution of parton distribution functions \cite{Braun:2001qx,Braun:2000yi,Derkachov:1999ze,Braun:2000av,Braun:1999te,Belitsky:1999bf,Belitsky:1999qh}. They were also studied in $\mathcal{N}=4$ sYM \cite{Derkachov:2013bda,Belitsky:2004sc,Belitsky:2005gr,Belitsky:2004yg,Belitsky:2003sh}.  For an excellent review \cite{Braun:2003rp}.  The particular case of the twist-2 operators are crucial in our understanding of QCD, since they control the evolution of the leading twist parton distribution functions. They were extensively studied, and their anomalous dimensions have been computed in QCD to three loops \cite{Moch:2004pa,Vogt:2004mw} and four-loop calculations are in rapid progress~\cite{Davies:2016jie,Moch:2017uml,Gehrmann:2023cqm,Falcioni:2023luc,Falcioni:2023vqq,Falcioni:2023tzp,Gehrmann:2023iah,Moch:2023tdj}.

  \begin{figure}[t!]
  \centering
  \includegraphics[width=0.45\textwidth]{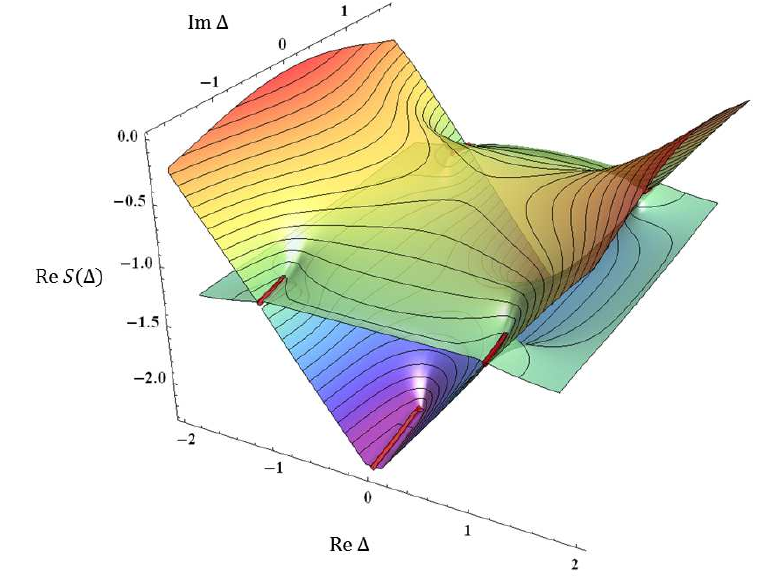}
  \caption{The Riemann surface for the leading Regge trajectory in $\mathcal{N}=4$ sYM computed using integrability \cite{Gromov:2015wca}.    }
  \label{fig:regge_N4}
  \end{figure}

Since the twist-2 anomalous dimensions describe the evolution of parton distribution functions, it was immediately understood that the pole at $J=1$ in Eq. \ref{eq:QCDAD} corresponds to so called ``small-x", or Regge physics associated with forward scattering. This limit had been extensively studied by BFKL \cite{Fadin:1975cb,Lipatov:1976zz,Kuraev:1976ge,Kuraev:1977fs,Balitsky:1978ic,Lipatov:1985uk}, who derived evolution equations in this limit.

In the context of $\mathcal{N}=4$ sYM, Lipatov and collaborators took seriously the analyticity in spin, arguing that one can in fact connect DGLAP and BFKL through analytic continuation \cite{Kotikov:2007cy,Kotikov:2004er,Kotikov:2002ab}, and that in the case of $\mathcal{N}=4$ sYM there should be a single Riemann surface defined by $\Delta(J)$. See also \cite{Jaroszewicz:1982gr} for early studies in this area. The analytic continuation has been well studied \cite{Kotikov:2005gr}, and further explored and elucidated in \cite{Brower:2006ea,Polchinski:2002jw} in the context of string theory. Ref. \cite{Brower:2006ea,Polchinski:2002jw} also explored a class of vertex operators, and their analytic continuation in spin.

With the discovery of the AdS/CFT correspondence, there was a tremendous focus on the spectrum of twist-2 operators, at weak, strong, and ultimately finite coupling. This ultimately led to the development of the quantum spectral curve \cite{Gromov:2014caa,Gromov:2013pga} enabling the calculation of operator dimensions at finite coupling in planar $\mathcal{N}=4$ super Yang-Mills. The results of \cite{Gromov:2015wca}, show that the anomalous dimensions are analytic in spin, and enable the calculation of the Riemann surface of the twist-2 operators shown in  \Fig{fig:regge_N4}. This enabled the NNLO calculation of the BFKL eigenvalue in planar $\mathcal{N}=4$ sYM \cite{Alfimov:2014bwa,Gromov:2015vua}. This has been extended and developed to higher twist \cite{Klabbers:2023zdz,Ekhammar:2024neh,Brizio:2024nso,Julius:2024ewf} more recently.

The link between these light ray operators and detector operators was provided in \cite{Hofman:2008ar}, through the study of the light-ray OPE. Ref. \cite{Hofman:2008ar} argued that detector operators should exhibit an operator product expansion in the small angle limit. Schematically, this takes the form for two energy operators~\cite{Hofman:2008ar,Kologlu:2019mfz,Chang:2020qpj}
\begin{align}
\mathcal{E}(\hat n_1)\mathcal{E}(\hat n_2) &\sim \sum_i \#\, (n_1\cdot n_2)^{\frac{\tau_i-4}{2}} \mathbb{O}_{i}^{[J=3]} (\hat n_2) \nn \\
&+ \text{transverse derivatives},
\label{eq:lightray_OPE}
\end{align}
where the transverse derivatives encode contributions from descendants and are captured by celestial blocks.
It is important that due to the light-ray integrals, this OPE does not follow from the standard Euclidean OPE of local operators. 

Interestingly, the leading operator appearing in the light-ray OPE is a twist-2 spin-3 operator. This does not exist as a local operator, and is exactly the string operator of  \cite{Balitsky:1987bk}, placed at asymptotic infinity. This provided the first connection between detector operators and string operators. It also greatly broadened the space of detector operators, since iteratively applying the OPE, one gets into a broad world of light-ray operators.  OPE breaths life into the space of detectors, giving them a life of their own. This OPE provides a completely new perspective on how to study correlation functions of detector operators in colliders.

From a phenomenological perspective, this equation is quite remarkable, as it enables one to project the measurement of a multi-point correlator onto operators with definite scaling properties, which can be measured in experiment. More conceptual, it provides a relation between the scaling of correlations in asymptotic energy flux, and anomalous dimensions of operators in the theory. This is exactly the goal of collider physics studies, namely understanding the relation between asymptotic fluxes and the underlying microscopic QFT description. As such, this OPE has been central in the renewed interest and application of energy correlators at collider experiments.

Although of clear practical interest, this OPE remained poorly understood, particularly beyond perturbation theory. Renewed study of this OPE and of light-ray operators came from an unexpected direction, namely the rejuvenation of the conformal bootstrap \cite{Rattazzi:2008pe,El-Showk:2012cjh,El-Showk:2014dwa}. In the study of the conformal bootstrap, one can study specific kinematic limits. As with the ``Lorentzian lens" discussed above, studying specific Lorentzian limits allows one to access information that is otherwise challenging from Euclidean correlators. One example is the light-cone limit, where the bootstrap crossing equations were studied in \cite{Komargodski:2012ek,Fitzpatrick:2012yx}, and is referred to as  ``light-cone bootstrap".  The light-cone limit is dominated by large spin operators in the CFT, enabling $1/J$ to be used as an expansion parameter in the so called ``large spin perturbation theory"   \cite{Alday:2015eya,Alday:2015ota,Alday:2016jfr,Alday:2016mxe,Kaviraj:2015xsa,Kaviraj:2015cxa,Alday:2015ewa,Alday:2016njk,Simmons-Duffin:2016wlq}.  The large spin limit also provides a direct connection to the EEC in the back-to-back limit, as will be discussed later. In specific cases such as the 3d Ising model, this expansion was shown to work to very low values of the spin \cite{Simmons-Duffin:2016wlq}. One way of understanding the success of this perturbative expansion is if CFT data is analytic in spin, raising numerous questions.

Analyticity in spin is familiar in the context of Regge theory and forward scattering. In that context, the Froissart-Gribov formula \cite{Regge:1959mz,Gribov:1962fw,Gribov:2003nw} expresses the analyticity in spin of partial amplitudes for the S-matrix of a relativistic QFT. This hinted at the possibility of a Froissart-Gribov formula for CFT data, in the context of generic CFTs. Important work in this direction includes the conformal Regge theory \cite{Costa:2012cb}. In  \cite{Caron-Huot:2017vep} introduced the ``Lorentzian Inversion" formula, providing a Froissart-Gribov formula for conformal correlators in which the OPE data is analytic in spin (For another derivation, see \cite{Simmons-Duffin:2017nub}). This left open the question as to the operator interpretation of this non-integer spin data in general CFT. This cycle was closed in \cite{Kravchuk:2018htv} where it was shown that they correspond to light-ray operators which are analytic in spin. Therefore, in the context of CFTs, we now have a complete picture linking all these areas!

We therefore see that detector/light-ray operators play a central role in giving structure to CFT data, organizing them into Regge trajectories. They show  that analyticity in spin is a general phenomenon of CFTs, beyond the particular case of $\mathcal{N}=4$, and explain the remarkable structures observed perturbatively in QCD.  In addition, these explorations have greatly developed our understanding of the space of light-ray operators that appear in the light-ray OPE, making it a practical tool for studying energy correlator observables, even beyond perturbation theory.

\subsection{Theory Meets Practice}\label{sec:obs}

\begin{figure}
\includegraphics[width=0.955\linewidth]{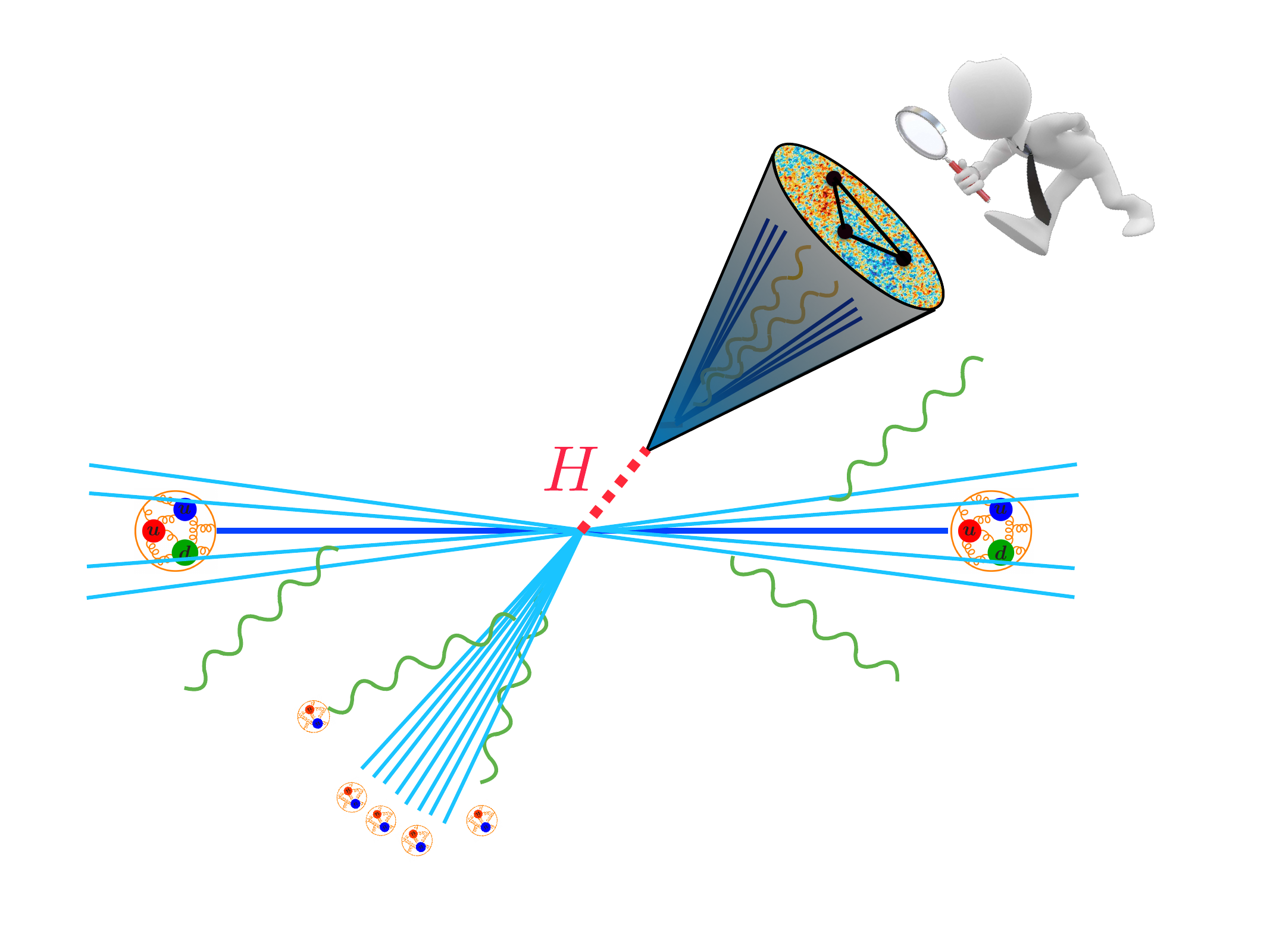}
\caption{Jet substructure as a new search channel at the LHC. Patterns in energy flux within a jet are used to identify the underlying microscopic physics, in this case a Higgs boson.
}
\label{fig:jss}
\end{figure}

Despite this tremendous progress in understanding detector observables in formal QFT, this progress remained quite disconnected from collider phenomenology.\footnote{Here we focus only on the interplay with collider physics experiments. Why observables of this form are/were not studied in condensed matter experiments would be interested to understand, as well as to understand how they can be measured in the future.} The reasons for this are simple: 1. QCD, the theory accesible at collider physics is not conformal, exhibiting confinement in the IR. Calculations of energy correlators require the development of factorization theorems which rigorously separate perturbative and non-perturbative contributions, allowing first principles calculations in QCD. 2. The highest energy colliders, which enable QCD to be probed in the asymptotically free regime, are hadron-hadron colliders. The reliable calculation of energy correlators in hadronic colliders requires high order perturbative calculations of multi-point scattering amplitudes and cross sections, combined with robust experimental techniques to identify high energy jets. 3. Unlike their theoretical counterparts, detectors in the real world do not have perfect resolution. In particular, hadronic calorimeters, which are necessary to measure the total energy flux, as defined by the $\mathcal{E}$ operator, have quite poor angular resolution, inhibiting the measurement of energy correlators in kinematic limits, and multi-point correlators.

With the starting of the Large Hadron Collider (LHC), a proton-proton machine, whose collisions are dominated by the physics of QCD, it was realized that the ability to identify short lived particles via their imprint in QCD radiation could greatly extend the physics program of the LHC, see \Fig{fig:jss}. This gave rise to the field of jet substructure  which uses the detailed patterns of energy flux inside individual jets to search for signals of UV physics. It was originally introduced in \cite{Butterworth:2008iy} as a means to search for the $H\to b\bar b$ decay. It was rapidly generalized, with important early works including \cite{Ellis:2009su,Ellis:2009me,Krohn:2009th,Thaler:2011gf,Thaler:2010tr,Thaler:2008ju,Kaplan:2008ie,Plehn:2009rk},  and has since seen widespread use in new physics searches. For modern reviews, see \cite{Larkoski:2017jix,Kogler:2018hem}.

Jet substructure was made possible by robust, theoretically calculable, and computationally efficient jet reconstruction algorithms, \cite{Cacciari:2008gp,Cacciari:2005hq,Salam:2010nqg,Cacciari:2011ma}, which built on early infrared and collinear safe jet definitions at hadron colliders \cite{Ellis:1993tq,Catani:1993hr} \footnote{We wish to highlight that the energy correlator itself \cite{Basham:1978bw}, successive combination jet algorithms at hadron colliders \cite{Ellis:1993tq}, and the inclusive one-jet cross section \cite{Ellis:1990ek}, three key ingredients which played a central role in enabling the study of energy correlators at hadron colliders, were all first introduced/computed by Stephen Ellis.}. These completely transformed what is possible with jets at hadron colliders, and made possible the detailed experimental study of jets and their substructure.

While the original motivation was to study new physics, the ability to study the substructure of jets provides an exciting new opportunity to study the physics of QCD at new energy scales, and at a level of detail that was never possible before. This has also had a significant impact on the study of nuclear physics at collider experiments, see e.g.~\cite{AbdulKhalek:2021gbh}.

These advances motivated a tremendous theoretical effort to understand QCD in general, and jets and their substructure in particular, at hadron colliders, at a level matching the new experimental capabilities. While we cannot do justice to the progress that has occurred, we wish to highlight several of the key advances that enable the challenges enumerated above to be overcome, enabling precise calculations of energy correlators in real world collider environments.

Calculations of energy correlators at hadron colliders are complicated due to their multi-scale nature, requiring a simultaneous description of the dynamics at the hard scale of the scattering, the scale of the jet radius, the scale of the energy correlator measurement, and the scale of $\Lambda_{\text{QCD}}$. The ability to theoretically describe complicated  multi-scale processes has been transformed by the development of effective field theory techniques, in particular the Soft-Collinear Effective Theory (SCET) \cite{Bauer:2000ew,Bauer:2000yr,Bauer:2001ct,Bauer:2001yt,Rothstein:2016bsq,Beneke:2002ph}, which enables renormalization group based approaches and operator based definitions of non-perturbative matrix elements. Combined with the factorization theorems of Collins-Soper-Sterman  \cite{Collins:1981ta,Bodwin:1984hc,Collins:1985ue,Collins:1988ig,Collins:1989gx,Nayak:2005rt,Collins:2011zzd}, SCET has enabled the precision calculation of jet substructure observables at the LHC, and in particular, has allowed perturbative calculations of energy correlators in specific kinematic limits to be translated into realistic calculations in the LHC environment.

Second, precision calculation at hadron colliders require at a minimum NLO or NNLO calculations of multi-parton amplitudes, as well as the ability to integrate them over phase space taking into account infrared divergences.  At NLO this was achieved through the advent of generalized unitarity \cite{Bern:1994cg,Bern:1994zx}, and its automation into computer codes \cite{Cascioli:2011va,Buccioni:2019sur}, combined with subtraction schemes \cite{Frixione:1995ms,Catani:1996vz}, and the ability to combine them with parton shower resummation \cite{Alwall:2014hca,Alwall:2011uj,Alioli:2010xd,Frixione:2007vw}. At NNLO, while no complete solution exists, tremendous efforts have enabled the calculation of two- \cite{Gehrmann-DeRidder:2013uxn,Currie:2017eqf} and three-jet production \cite{Czakon:2021mjy}. Recent public programs which enable NNLO calculations of processes involving jets at the LHC include \cite{Czakon:2023hls,NNLOJET:2025rno}. These calculations form the basis of all higher order jet substructure calculations at hadron colliders.

Finally, effective field theory techniques, combined with improved understanding of the renormalization group, have enabled the calculation of correlation functions of a variety of different detector operators, incorporating hadron information. Most importantly, this enables the calculation of observables that utilize charge information. For measurements on charged particles, experiments can use tracking detectors, with extremely good angular resolution, allowing the measurement of higher point correlators, and correlators in kinematic limits. The ability to systematically compute energy correlator observables on tracks was developed in \cite{Chang:2013iba,Chang:2013rca,Jaarsma:2023ell,Chen:2022muj,Chen:2022pdu,Jaarsma:2022kdd,Li:2021zcf}.

Although we will not discuss them in detail in this review, we wish to highlight two other areas which are crucial in our ability to provide a complete description of energy correlators in hadron collisions. The first are determinations of universal non-perturbative functions, namely  parton distribution functions (PDFs) \cite{Martin:2009iq,Lai:2010vv,Dulat:2015mca,Hou:2019efy,NNPDF:2014otw,NNPDF:2017mvq}, nuclear PDFs \cite{Eskola:2009uj,Eskola:2021nhw,AbdulKhalek:2022fyi}, and fragmentation functions \cite{deFlorian:2007aj,deFlorian:2007ekg,Bertone:2017tyb,deFlorian:2014xna,deFlorian:2017lwf,Hirai:2016loo,Sato:2016wqj,Gao:2024dbv}. The second are the range of parton shower simulations which allow the modeling of observables in a wide variety of hadronic collisions  \cite{Sjostrand:2007gs,Sjostrand:2014zea,Gleisberg:2008ta,Bellm:2019zci,Bahr:2008tf,Bahr:2008pv,Bierlich:2018xfw,Jung:1993gf,Charchula:1994kf,Ke:2023xeo,Zapp:2013vla,Zapp:2012ak,JETSCAPE:2020mzn,Gyulassy:1994ew,Wang:1991hta,vanBeekveld:2023ivn}. While we focus primarily on analytic approaches, simulations have been crucial for exploring the behavior of energy correlator observables in complicated hadronic environments, and the modern era of energy correlator studies would not have been possible without them.

With all these advances in place, the final step in moving from theory to practice was the proposal of experimentally realizable energy correlator based observables at hadron colliders. While this may seem like a simple generalization, the hadron collider environment makes it quite non-trivial. In particular, in most cases at hadron colliders, it is necessary to measure observables on the constituents of an identified highly energetic jet, as is illustrated in \Fig{fig:jss}. This is significantly more complicated than the case of $e^+e^-$ collisions where observables can be measured on the full energy distribution. In this more complicated situation, it was not clear which observables would behave well, enabling precise theoretical descriptions, and there was a significant exploratory period where the physics of jets and their substructure was understood. This goes beyond the scope of this review, and detailed discussions can be found elsewhere \cite{Salam:2010nqg,Larkoski:2017jix}.

\begin{figure}
\includegraphics[width=0.455\linewidth]{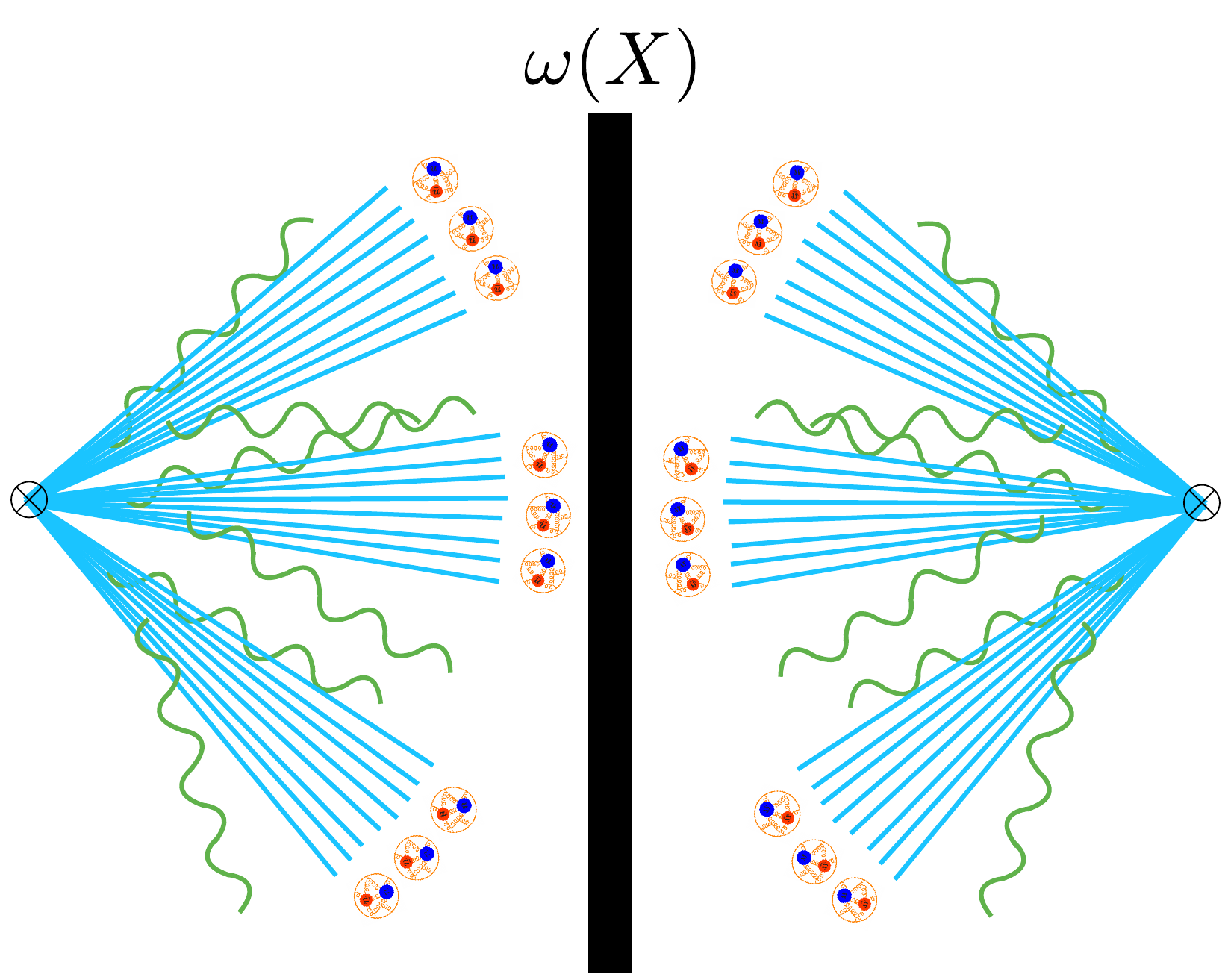}
\includegraphics[width=0.455\linewidth]{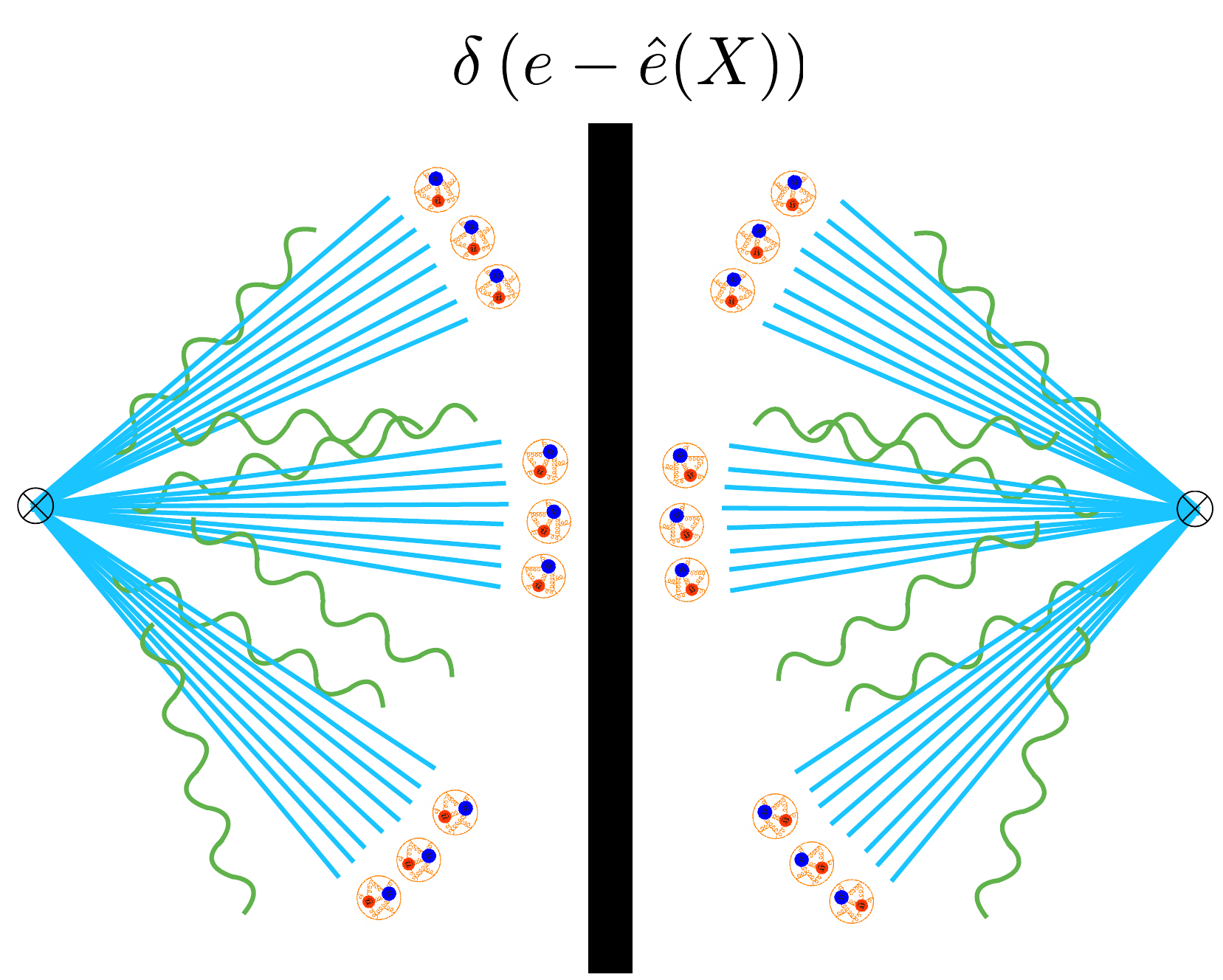}
\caption{A comparison of standard jet substructure ``shape" observables used at the LHC, and weighted cross sections defined by the energy correlators. Figure from \cite{Chen:2020vvp}.
}
\label{fig:weight_vs_delta}
\end{figure}

Due to the original motivation of identifying Higgs decays within jets, the classes of observables that were studied at the LHC were primarily so called ``jet shape" observables, modeled after the original Sterman-Weinberg jet definition \cite{Sterman:1977wj}, and the famous thrust observable \cite{Farhi:1977sg}, and its generalizations \cite{Berger:2003iw}. Since these observables can be used to constrain radiation in an event into particular jet configurations, they generalized naturally to studies of jets at hadron colliders. To our knowledge, the first such calculation was by Seymour \cite{Seymour:1997kj}. Other important early calculations of jet substructure observables include \cite{Almeida:2008yp,Dasgupta:2013ihk,Ellis:2010rwa}. These observables were extensively studied both phenomenologically, and theoretically, and generalized to identify particular energy flows. Examples include \cite{Thaler:2011gf,Thaler:2010tr,Larkoski:2014gra,Moult:2016cvt}. Such jet shape observables are defined by a particular operator expressed in terms of the energy flow operator integrated over a profile $f_e$
\begin{align}
\hat e |X\rangle =\frac{1}{Q}\int d\theta d\phi \, f_e(\theta, \phi) \mathcal{E}(\theta,\phi)|X\rangle\,.
\end{align}
The jet shape observable is then given by the expectation value
\begin{align}\label{eq:obs_def}
\frac{d\sigma}{de}= \int d^4x e^{iQ\cdot x}  \langle 0 | \cO (x)  \delta(e-\hat e) \cO^\dagger (0) |0 \rangle\,.
\end{align}
Note the crucial difference as compared to energy correlator observables, namely we are computing the expectation value of the $\delta$-function of an observable.
The operator valued $\delta$ function in this expression is formally defined by its moments
\begin{align}
\delta(e-\hat e)=\delta(e)+\hat e \delta^{(1)}(e)+\cdots + \frac{\hat e^n}{n!} \delta^{(n)}(e)+\cdots\,.
\end{align}
Jet shape observables are therefore sensitive to an infinite set of energy correlators. Physically, this arises due to the fact that they constrain the shape of all the radiation within a collision, as opposed to correlating fluxes at specific angles, while remaining inclusive over the remaining radiation. This distinction is shown in \Fig{fig:weight_vs_delta}. They are connected through that
moments of jet shape observables
\begin{align}
\int e^n \frac{d\sigma}{de}= \int d^4x e^{iq\cdot x}  \langle 0 | \cO (x)  \hat e^n \cO^\dagger(0) |0 \rangle\,,
\end{align}
are related to integrals of $n$-point energy correlators.

Due to the fact that these observables involve an infinite number of correlators, they are difficult to study using the operator based language of energy flux operators. Indeed, they require the knowledge of all correlation functions, which is well beyond what is achievable even in our best understood QFTs in $d=4$. This hints that they are inherently more complicated than the fundamental energy correlators, and will have a correspondingly more complicated perturbative and non-perturbative structure. Furthermore, the sum over an infinite number of correlators destroys the symmetry properties of the individual correlators, which will prove essential to their analysis.

\begin{figure}
\includegraphics[width=0.955\linewidth]{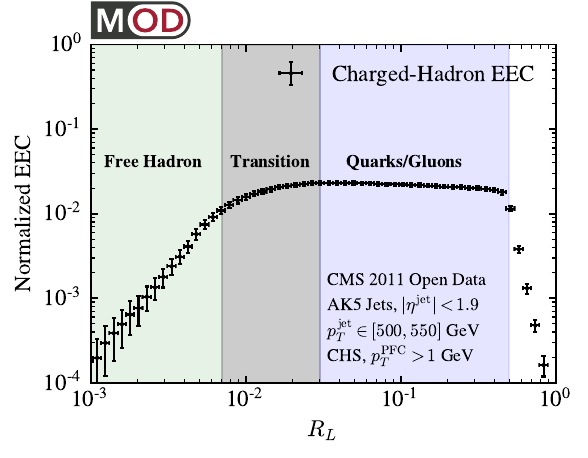}
\caption{The first study of the two-point energy correlator within high energy jets at the LHC, using CMS open data. The result is shown as a function of the opening angle, $R_L$ between the detector operators. Three regions are clearly seen: a perturbative region associated with asymptotically free quarks and gluons, a transition region, and a region associated with free hadrons in the deep IR. Figure from \cite{Komiske:2022enw}.
}
\label{fig:open_data}
\end{figure}

However, with the experimental advances of the jet substructure revolution, combined with theoretical developments, it now became possible to revive the study of energy correlators, but now computed \emph{inside} high energy jets at hadron colliders.  This requires combining the understanding of energy correlator observables with the developments of the jet substructure program to enable theoretical predictions for energy correlators inside high energy jets, and their experimental measurement. This program was proposed in \cite{Chen:2020vvp}, which proposed energy correlators inside high energy jets as a means of studying jet substructure, and the first multi-point correlator in the collinear limit, as relevant for jet substructure, was computed in \cite{Chen:2019bpb}.

The first study of energy correlators inside jets at a hadron collider was performed in \cite{Komiske:2022enw} using publicly available CERN Open Data \cite{CMS:JetPrimary2011A,CMS:JetPrimary2010B}, processed in MIT Open Data format \cite{Tripathee:2017ybi}. This is shown in \Fig{fig:open_data}, which measures the two-point energy correlator in the small angle region within $500$ GeV jets at the LHC. This uses two of the key ingredients discussed above, namely tracks (charged hadrons) to enable high precision angular resolution, and robustly identified anti-$k_T$ jets.

Using the high energies, combined with the exceptional resolution of tracking detectors, the two-point energy correlator is able to reveal in a remarkably clear way the different phases of QCD. While we will discuss in detail the behavior of the energy correlator in later sections, here we wish to briefly describe the physics illustrated in this plot: A remarkable feature of the energy correlators in QCD, is that the angular scale of the correlator is associated with an inverse time scale. Small angles correspond to long time scales, while larger angles correspond to short time scales. In \Fig{fig:open_data}, we see at early times a beautiful power law scaling with a non-integer power associated with the interactions of (nearly) asymptotically free quarks and gluons, then the abrupt confinement transition to an scaling behavior associated with non-interacting hadrons. Therefore we see that energy correlators allow us to ``image" QCD as a function of time scale in a particularly clean manner, with scales in the underlying physical problem imprinting themselves as modifications in the scaling behavior of the observable.

\begin{figure}
\includegraphics[width=0.855\linewidth]{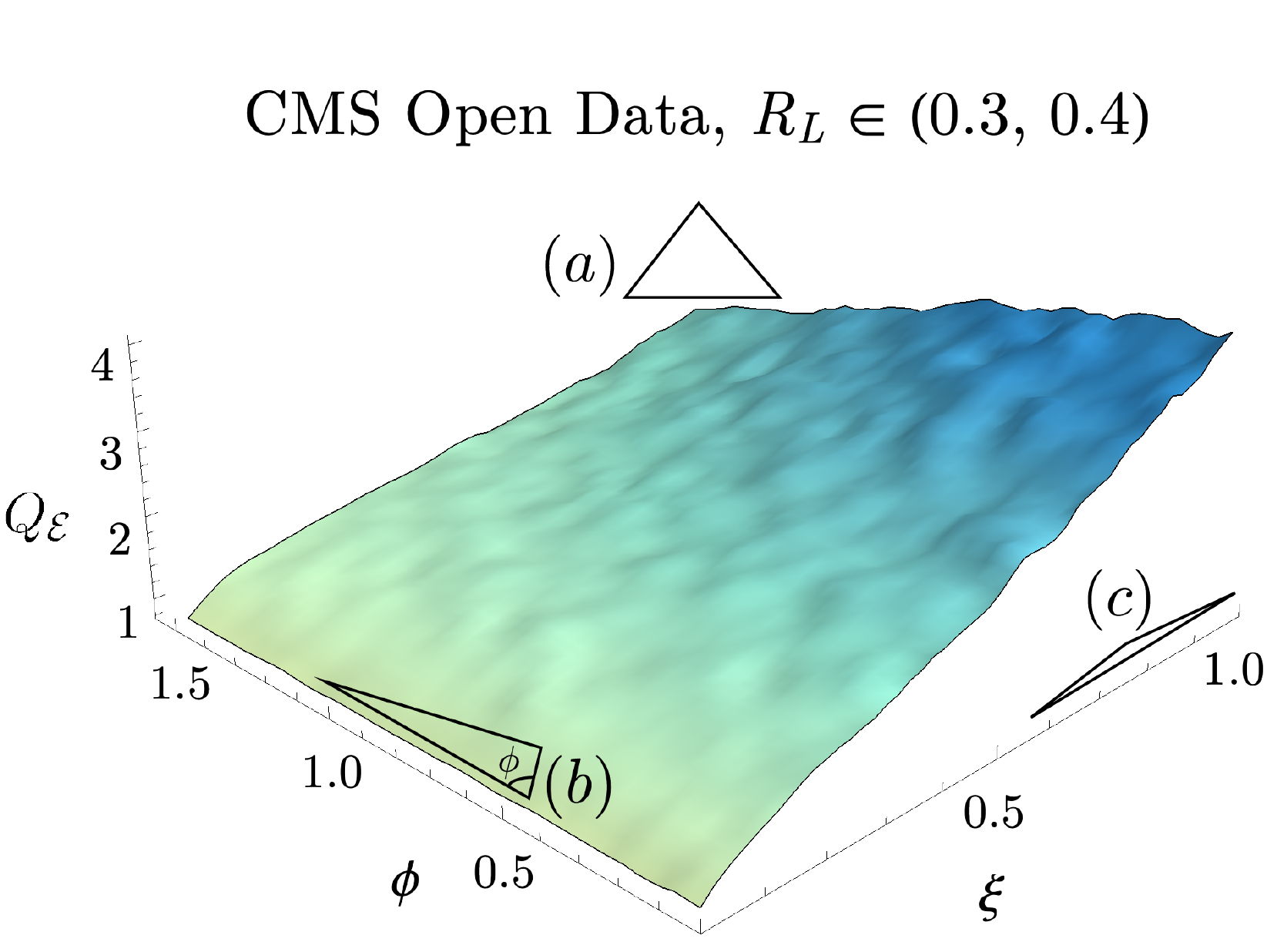}\\
\includegraphics[width=0.855\linewidth]{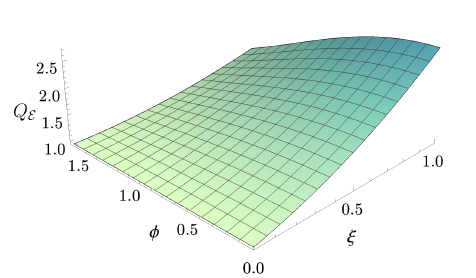}
\caption{A measurement of the non-gaussianity of the three point energy correlator, $Q_\mathcal{E}$ (defined in the text) inside high energy jets in CMS Open Data, along with an analytic calculation. This illustrated for the first time the ability to measure multi-point correlation functions of detector operators. Figures from \cite{Chen:2022swd}.
}
\label{fig:decorated_opendata}
\end{figure}

We will describe the rigorous factorization theorems for the collinear limit of the energy correlators in \Sec{sec:QCD}. This will allow us to make this statement sharp, as well as to emphasize the distinction between energy correlators and standard jet shape observables used in the study of jet substructure. A particularly beautiful aspect of the energy correlators in the collinear limit, is that the measurement introduces the dependence on a single scale, namely $\theta Q$. This is in contrast to many jet shape observables where one introduces multiple (soft and collinear scales). This allows for the simple mapping between the scaling of this single function, and the measurement, and makes sharp that the energy correlators probe the physics at a particular scale. This, combined with their theoretical understanding, makes the energy correlators extremely useful for imaging scales of the underlying QFT, and will find many applications in nuclear and particle physics.

The CMS Open Data was also used to perform the first study of a multi-point energy correlator, in this case the three-point energy correlator, again inside a high energy anti-$k_T$ jets, and using charged particles. This is shown in \Fig{fig:decorated_opendata}. To parameterize the correlator, we denote the long, medium, and small sides of the triangle spanned by the operators as $(R_L, R_M, R_S)$, and we define the coordinates:
\begin{align}
\xi=\frac{R_S}{R_M} \,, \qquad \phi&=\arcsin \sqrt{1 - \frac{(R_L-R_M)^2}{R_S^2}}
\,.
\label{eq:transf}
\end{align}
This parametrization blows up the OPE region into a line, with $\xi$ and $\phi$ the radial and angular coordinates about the OPE limit, respectively. In \Fig{fig:decorated_opendata} we show a non-gaussianity, introduced in \cite{Chen:2022swd}, which is defined as a particular ratio of the three and two-point energy correlators
\begin{align}\label{eq:NG_def}
Q_{\mathcal{E}}(\hat n_1, \hat n_2, \hat n_3)=\frac{\langle\mathcal{E}(\hat{n}_1)\mathcal{E}(\hat{n}_2)\mathcal{E}(\hat{n}_3)\rangle ~ \langle\mathcal{E}^2(\hat{n}_1)\rangle}
{\langle\mathcal{E}(\hat{n}_1)\mathcal{E}(\hat{n}_2)\rangle ~ \langle\mathcal{E}^2(\hat{n}_1)\mathcal{E}(\hat{n}_3)\rangle}\,. 
\end{align}
This particular ratio is chosen so that in the squeezed limit, where the three-point correlator factorizes into a two-point correlator, it reduces to unity. Due to the sufficiently high energies of the LHC, this measurement is made in a regime where the three-point correlator is described by perturbation theory, allowing it to be compared with perturbative calculations of multi-point correlators, and motivating the theoretical study of these objects. The corresponding perturbative calculation in QCD is shown in the lower panel. One of the beautiful aspects of the energy correlators, is that one directly calculates the observable, such as the three-point correlator, that is measured in the experiment. This provides a direct link between experimental measurements, and theoretical calculations.

\begin{figure*}
\includegraphics[width=0.485\linewidth]{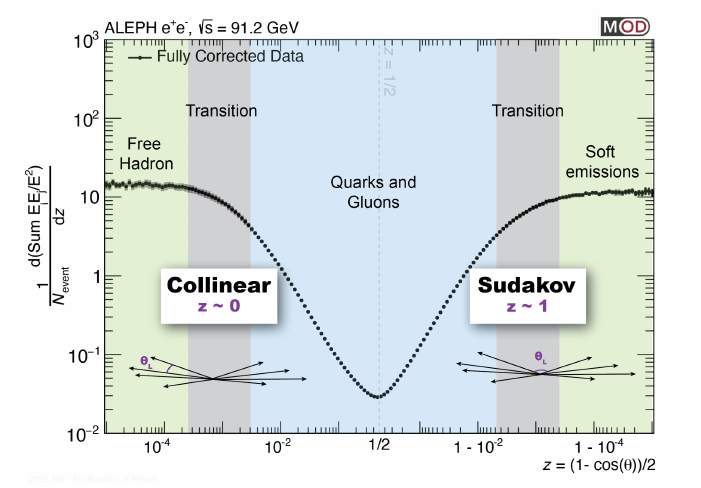}
\includegraphics[width=0.455\linewidth]{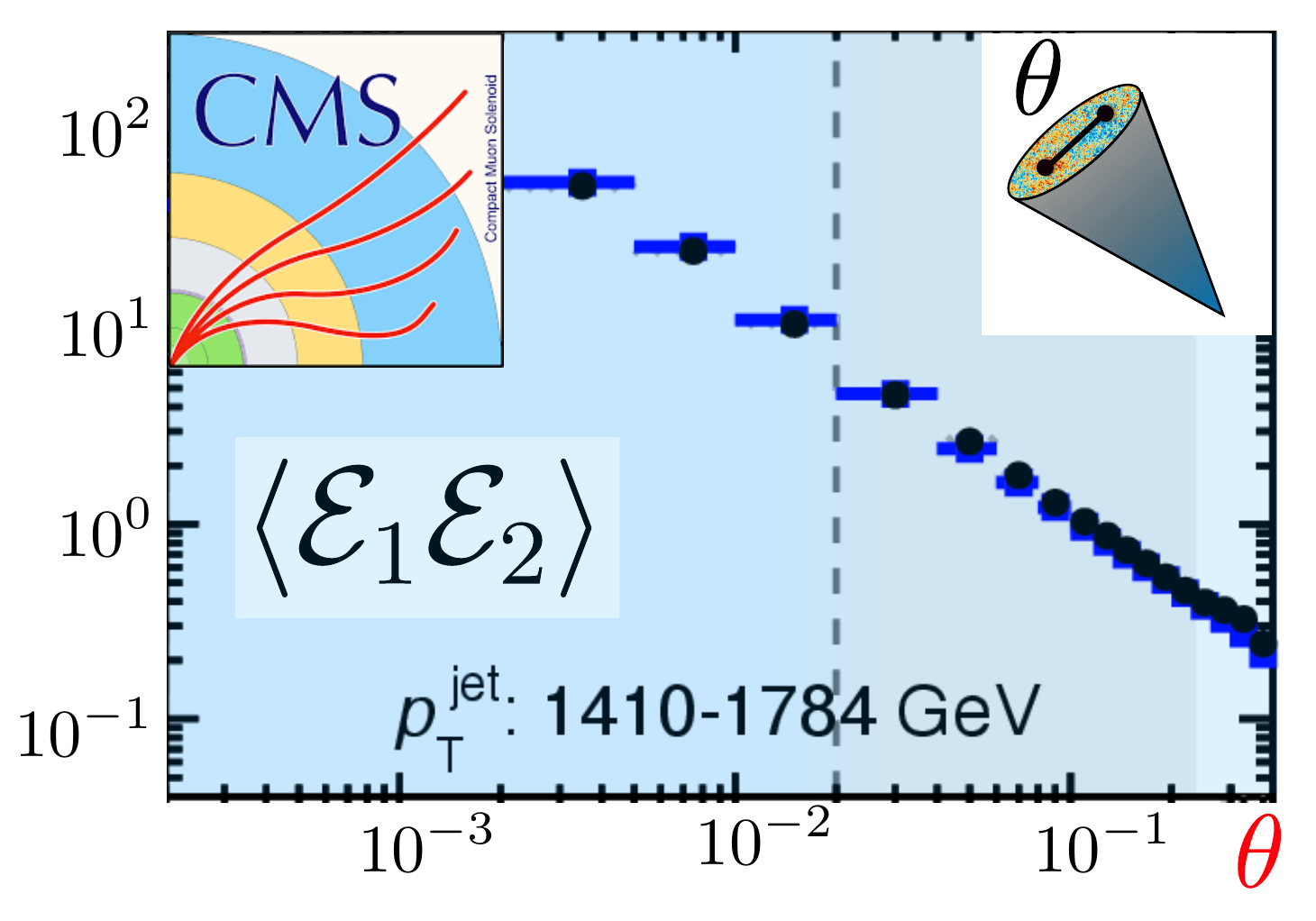}
\caption{Measurements of the energy correlators with extremely high angular resolution providing a clear view of the phases of QCD (left), and in the small angle limit at extreme energies providing precision measurements of scaling behavior (right), have transformed the interaction between theory and experiment.  Figures adapted from \cite{Bossi:2024qeu,Bossi:2025xsi} and \cite{CMS:2024mlf}.
}
\label{fig:CMS_scaling}
\end{figure*}

The combination of energy correlator observables that can be measured inside high energy jets, with robust jet algorithms, and effective field theory techniques to  perform calculations in complex hadronic collisions, laid the groundwork for a renewed and expanded interaction between theory and experiment in the study of energy correlators.

\subsection{Why Now?}\label{sec:now}

In the last year, energy correlators have been measured across a number of collider systems, in different kinematic limits, and in different states. For convenience we summarize them here
\begin{itemize}
\item $e^+e^-$ Colliders: Re-analysis using ALEPH archival data \cite{Bossi:2024qeu,Bossi:2025xsi}.
\item $e p$ Colliders: Measurement by HERA \cite{HERA}.
\item Proton-proton Colliders: Measurements by the CMS \cite{CMS:2024mlf}, ALICE \cite{ALICE:2024dfl,ALICE:2025igw,talk_Hwang} and STAR \cite{Tamis:2023guc,STAR:2025jut,talk_Shen} collaborations.
\item Proton-Pb Colliders: Measurement by the ALICE collaboration  \cite{talk_Anjali,talk_Anjali2}.
\item Pb-Pb Colliders: Measurements by the ALICE \cite{talk_Ananya} and CMS \cite{CMS-PAS-HIN-23-004,CMS:2025jam,CMS:2025ydi} collaborations.
\end{itemize}
These cover a tremendous range of physics across both particle and nuclear experiment.\footnote{It is also amusing to note that the energy correlators have been measured  on cosmic rays with $E> 5$ EeV (1 EeV =$10^{18}$ eV \cite{Schiffer:2011mva,Schiffer:2011zz,PierreAuger:2014tos}!} These measurements have achieved a number of successes ranging from a precision extraction of the strong coupling constant, to new ways to probe hot and dense nuclear matter. We will discuss in detail these different measurements, and their physics goals in \Sec{sec:exp_opp}.

Many of these measurements are in extreme kinematic limits, or extreme energies, never previously accessed in energy correlator measurements.  In \Fig{fig:CMS_scaling} we highlight two examples of this. In the left panel, we show a high angular resolution measurement of the two-point energy correlator at an $e^+e^-$ collider (ALEPH) \cite{Bossi:2024qeu,Bossi:2025xsi}, using tracks (charged particles) to access the collinear and back-to-back kinematic limits. We will describe the physics of this correlator in detail in this article, but this is a direct measurement of the $\langle J \mathcal{E} \mathcal{E} J \rangle$ correlator.  In the right panel of \Fig{fig:CMS_scaling}, we show a measurement by the CMS collaboration of the collinear limit of the two-point correlator at the TeV energy scale \cite{CMS:2024mlf}, revealing for the first time the asymptotically free scaling behavior of the energy correlator at small angles. We find it quite remarkable to compare this with the first measurement of the EEC in \Fig{fig:pluto_single}. In particular, we wish to emphasize the difference in the angular scale, with the entire CMS measurement being contained within a single bin of the PLUTO measurement.

The remarkable feature of \Fig{fig:CMS_scaling} is that we observe the scaling behavior of perturbative quarks and gluons over several orders of magnitude. The fact that this scaling behavior was measured at the LHC, and our ability to compute it precisely was motivated by the corresponding developments in formal theory, and represents a genuine interplay between these two areas. We are reminded by the quote by Polyakov ``\emph{I wanted to learn about elementary particles by studying boiling water,}” and we believe that this is an instance where understanding of ``collider physics" in CFTs has had a genuine impact on real world collider phenomenology.

These measurements completely transform the possibility for interaction between theory and experiment. They are of interest not just for phenomenology, but reveal universal aspects of QFT of interest to a general quantum field theorist.  The simultaneous, but completely disconnected, developments of formal theoretical tools for the study of asymptotic fluxes in QFT, and phenomenological/ experimental tools to study jet substructure at colliders, makes it an exciting time to bridge these different fields, from formal theory, to particle physics, to nuclear physics, to develop new ways of learning about the real world at collider physics experiments.

\subsection{Overview and Guide to the Reader}\label{sec:overview}

One of the most exciting aspects of energy correlators is their ability to bridge many fields, ranging from formal theory, to measurements in particle physics and nuclear experiment. It is therefore impossible in a single review to describe in detail (or even in a cursory fashion), all the interesting aspects of detector operators and their correlators. Rather, the goal of this review is to highlight the connections between the many different areas, so that the interested reader can further explore specific areas of interest. 

In particular, we wish to highlight to a formal audience how energy correlator observables are being used in real world collider experiments, and what properties are of interest to phenomenologists. Similarly, we wish to highlight to experimentalists the many deep connections of these observables to fundamental aspects of QFT. We hope that this will lead to further exchanges between these communities.

An outline of this review is as follows. In \Sec{sec:E_QFT} we provide an overview of light-ray/ detector operators and their correlators in generic QFTs. In \Sec{sec:QCD} we highlight aspects of energy correlators specific to the theory of QCD, including modifications due to the presence of confinement or intrinsic mass scales. In \Sec{sec:exp_opp} we provide an overview of how energy correlators can be studied at different collider experiments. This section is intended to be for a general audience, with no background knowledge of experimental collider physics, to illustrate at a big picture level what is possible with modern colliders, and to highlight some of the main measurements that have invigorated the study of energy correlators. We provide a survey of $e^+e^-$, $e^-p$, $pp$ and nucleus-nucleus colliders, highlighting the physics goals of each, and how energy correlators can be applied in new ways.  In \Sec{sec:particle} and \Sec{sec:nuclear}, we then provide more detailed discussions of physics applications of energy correlators in both particle and nuclear experiment. In \Sec{sec:open} we discuss a number of future directions closely related to the study of energy correlators, where we hope to see progress in the near future. We conclude in \Sec{sec:conc}.

\section{Energy Operators and Their Correlators in Quantum Field Theory}\label{sec:E_QFT}

In this section we provide an overview of detector operators and their correlation functions in generic QFTs. Topics related to the specific case of QCD will be discussed in \Sec{sec:QCD}.

\subsection{Light-Ray Operators and Their Regge Trajectories} \label{sec:lightray}

Light-ray operators appear in a wide ranging contexts in QFT, ranging from the study of renormalization group flows \cite{Hartman:2024xkw,Hartman:2023ccw,Hartman:2023qdn}, to the emergence of bulk locality in the AdS/CFT correspondence \cite{Caron-Huot:2022lff}, to the study of entanglement entropy and modular hamiltonians \cite{Balakrishnan:2019gxl,Casini:2017roe}, to detector operators in colliders. In this section we provide a review of basic properties of light-ray operators with an eye towards their application in collider physics. In particular, we will cover the topics of the quantum numbers of light-ray operators, Regge trajectories, reciprocity, and Regge intercepts.

\begin{figure}
  \centering
  \includegraphics[width=0.3\textwidth]{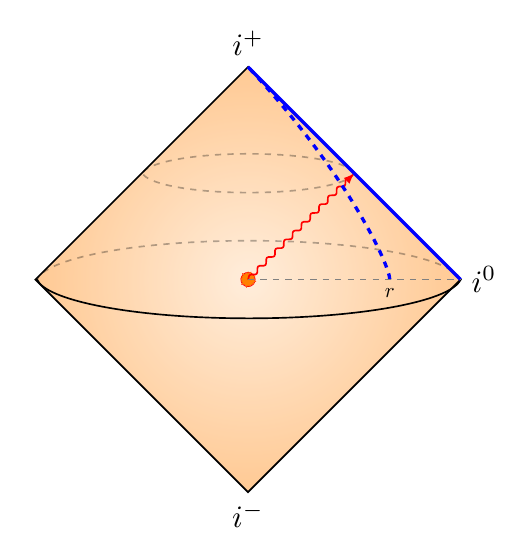}
  \caption{An illustration of an energy operator in a Penrose diagram for Minkowski spacetime. In a generic QFT, the detector operator is defined using a limiting procedure, illustrated by the dashed line. In a CFT, the operator can be directly placed at null infinity, illustrated by the solid blue line.}
  \label{fig:ANE}
  \end{figure}

The energy operator as defined in \eqref{eq:ANEC_op} is expressed in terms of an $r \to \infty$ limit, making it difficult to study. Since the most important property of the energy operator discussed in this review is its scaling behavior under boosts, it is convenient to map the Minkowski spacetime into a finite region using a Penrose diagram. This is represented in Fig.~\ref{fig:ANE}, where the full spacetime is mapped into a finite region by a conformal transformation that preserves the causal structure. For the sake of illustration, we have only shown two spatial dimensions. The detector sitting at $r$ and  and integrated over the time component is represented by the dashed blue line. Light-like radiation from the origin in the direction of the detector will be recorded by the detector. 

From a theoretical perspective, it is convenient to express the energy operator in a manifestly covariant form~\cite{Belitsky:2013bja}. For a light-like particle propagating in the $\hat{n}$ direction, we introduce two null reference vectors $n^\mu = (1, \hat{n})$ and $\bar{n} = (1, - \hat{n})$. Using these reference vectors, we can define light-cone coordinates $x_+ = x^- = \bar{n} \cdot x$, $x_- = x^+ = n \cdot x$, in terms of which any four vector can be decomposed as
\begin{equation}
x^\mu = \frac{x_+}{n \cdot \bar n} n^\mu + \frac{x_-}{n \cdot \bar n} \bar{n}^\mu + x_\perp^\mu \,.
\end{equation}
The large $r$ limit should be understood as taking $x_+ \to \infty$ while keeping $x_-$ fixed. We can then write the energy operator as
\begin{equation}
  {\cal E}(n) = \lim_{x_+ \to \infty} \frac{(x_+)^2}{(n \cdot \bar n)^4} \int\limits_{-\infty}^\infty dx^+ T_{++}(x) \,.
  \label{eq:ANE_covariant}
\end{equation}
This is represented as the solid blue line at future null infinity in Fig.~\ref{fig:ANE}.
The definition in \eqref{eq:ANE_covariant} make manifest the symmetry of the energy operator under boosts. Under a scaling $n^\mu \to \rho n^\mu$, ${\cal E}(\rho n) = \rho^{-3} {\cal E}(n)$~\cite{Hofman:2008ar,Belitsky:2013bja}. Since energy operators are a physical observable, the manifestation of boost symmetry is significant because it allows us to understand certain properties of measurements using symmetry, notably the scaling behavior when two detectors become close~\cite{Hofman:2008ar,Kologlu:2019mfz,Chang:2020qpj,Chen:2020adz}, as in Eq.~\ref{eq:lightray_OPE}.

The transformation that maps a local operator to a null-integrated operator is referred to as the light transform~\cite{Kravchuk:2018htv}. The resulting operator is commonly called a light-ray operator.  A number of other integral transforms were also introduced in ~\cite{Kravchuk:2018htv}. To characterize a light-ray operator, it is useful to consider its Lorentzian collinear spin $J_L$, which corresponds to the quantum number under boosts, and its Lorentzian scaling dimension $\Delta_L$. For the energy operator, these quantities are given by
\begin{equation}
J_L = 1 - \Delta = -3 \,, \quad \Delta_L = 1 - J = -1 \,,
\end{equation}
where $\Delta = 4$ and $J = 2$ represent the scaling dimension and spin, respectively, of the local energy-momentum tensor operator.

Among light-ray operators, the energy operator stands out as one of the most significant example.  Perhaps its most important property is that expectation values of the energy operator are non-negative in any state in a QFT,
\begin{equation}
\langle {\cal E}(n) \rangle_\psi \geq 0 \,,
\label{eq:ANEC}
\end{equation}
also known as the average null energy condition (ANEC). In general relativity, the violation of ANEC can lead to the construction of time machines from wormhole solutions~\cite{Morris:1988tu}. As a result, significant effort has has been devoted to rigorously proving the ANEC in ~\eqref{eq:ANEC}.

In a free field theory, the ANEC is straightforward to understand because it corresponds to the measurement of the energy of freely propagating particles in a specific direction $\hat{n}$, as shown in Eq.~\eqref{eq:ANEC_particle}, where $k^0$ is manifestly positive.
For an interacting, Lorentz-invariant QFT, the ANEC has been proven using two distinct approaches. The first approach relies on the microcausality of the theory~\cite{Hartman:2016lgu}, while the second employs methods from quantum information theory, particularly the monotonicity of relative entropy~\cite{Faulkner:2016mzt}. In the context of the AdS/CFT correspondence the ANEC also plays a crucial role: causality in the bulk implies the ANEC on the boundary \cite{Kelly:2014mra,Hofman:2008ar}, while the ANEC in the bulk implies focusing \cite{Hartman:2022njz}.

Positivity of the ANEC has been used to prove a number of basic results in QFT. These include the conformal collider bounds, and their generalizations \cite{Hofman:2008ar,Cordova:2017zej,Cordova:2017dhq,Chowdhury:2017vel}, constraints on RG flows \cite{Hartman:2024xkw,Hartman:2023ccw,Hartman:2023qdn}, and constraints on modifications to the three-point graviton coupling \cite{Hartman:2022njz}.

\begin{figure}[t!]
  \centering
  \includegraphics[width=0.5\textwidth]{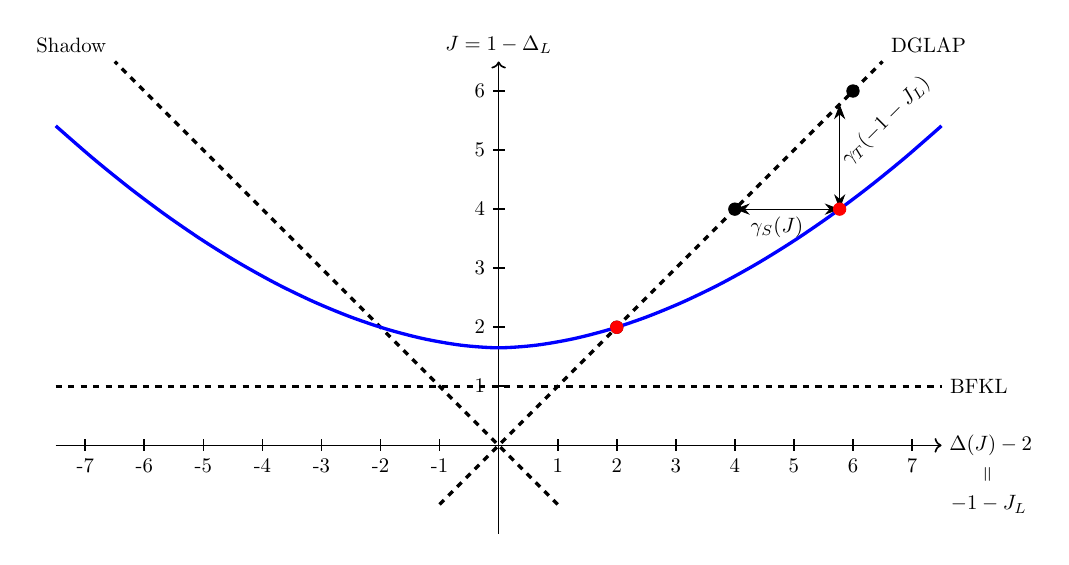}
  \caption{A schematic Regge trajectory in a weakly-coupled conformal gauge theory in 4D, highlighting the free DGLAP, shadow DGLAP and BFKL trajectories (dashed), and the leading Regge trajectory (solid blue). Both the space-like($\gamma_S)$ and time-like ($\gamma_T$) anomalous dimensions, as well as their relationship, are also shown.}
  \label{fig:regge}
  \end{figure}

While the energy operator plays a central role in this review, it is just one example of a light-ray operator. Understanding the space of light-ray operators proves to be important in CFTs and QCD, since light-ray operators provide the analytic continuation in spin of local operator \cite{Kravchuk:2018htv}. The light-ray operators in the theory organize operators into analytic trajectories, called Regge trajectories, as illustrated in \Fig{fig:regge}. This understanding leads to numerous insights, including space-like  time-like reciprocity, and an understanding of high energy scattering, with many phenomenological implications.

One way of motivating this generalization is to start with the energy operator, and ask how to construct light-ray operators in interacting theories that measures the energy raised to a power. In perturbative gauge theories such as QCD, this was considered in \cite{Chen:2021gdk} due to the appearance of these operators in the light-ray OPE. Instead of applying the light-transform to the stress-tensor, we can consider applying it to a twist-2 operator, $O_{+\cdots +}(x)$. The explicit form of the twist-2 operators in QCD is reviewed in \Sec{sec:celestial_blocks}, but will not be important here. We are therefore led to define the operator
\begin{equation}\label{eq:spin2_traj}
{\cal D}_{J_L} (n) = \lim_{x_+ \to \infty} \frac{(x_+)^{\Delta(J) - J}}{(n \cdot \bar n)^{\Delta(J)}} \int \limits_{-\infty}^\infty dx^+ O_{\underbrace{+\cdots +}_{J}}(x) \,.
\end{equation}
By expanding this operator in terms of creation and annihilation operators, one finds that its action on a physical state is
\begin{equation}
  {\cal D}_{J_L}(n) | X \rangle = \sum_{k \in X} (k^0)^{-2 - J_L} \delta^{(2)}(\Omega_{\hat n} - \Omega_k) | X \rangle \,.
  \label{eq:E_n}
\end{equation}
In a generic weakly-coupled conformal theory, the construction and renormalization properties of the detector operator ${\cal D}_{J_L}(n)$ has been systematically discussed in \cite{Caron-Huot:2022eqs}.

This operator in Eq. \ref{eq:spin2_traj} is quite interesting. First, its quantum numbers are
\begin{equation}
J_L = 1 - \Delta(J) \,, \quad \Delta_L = 1 - J \,,
\end{equation}
where $\Delta(J)$ is a scaling dimension for a local twist operator with spin $J$. In particular, this operator has non-integer spin. Note that since it is non-local, this agrees with Mack's classification \cite{Mack:1975je}.  This simple exercise raises numerous questions from both the formal and phenomenlogical perspective: How do we understand continuous spin operators in general? An experimentalist can measure arbitrary generic powers of the energy: $J_L$ serves as a tunable parameter whose value typically does not correspond exactly to the Lorentz spin of the light-ray operator of an integer-spin local operator. One might also argue that measuring the energy of final-state hadrons effectively requires extending operators of even integer spin $J$ to continuous spin values~\cite{Balitsky:1990ck,Hofman:2008ar,Kravchuk:2018htv}.

By performing the light-transform of local operators, we have obtained a set of operators that measure energy raised to a very specific set of powers, namely where $J_L = 1 - \Delta(J)$. How do we construct these more general operators in an interacting CFT?

To answer these questions, and motivate light-ray operators a little more generally, we must first understand what it means for an operator to have continuous spin~\cite{Kravchuk:2018htv}. To do this, it is convenient to introduce an index free notation,
where we contract an operator
\begin{align}
\mathcal{O}(x,z) \equiv \mathcal{O}^{\mu_1 \cdots \mu_J}(x) z_{\mu_1}\cdots z_{\mu_J}\,,
\end{align}
with a null, $z^2=0$, polarization vector. The spin is encoded in the homogeneity under rescalings of the polarization vector. This can easily be extended to non-integer $J$
\begin{align}
\mathbb{O}(x,\lambda z) =\lambda^J \mathbb{O}(x,z)\,.
\end{align}
However, by Mack's classification \cite{Mack:1975je}, any such non-integer spin operators are necessarily non-local, and annihilate the vacuum
\begin{align}
\mathbb{O}(x,z) |\Omega \rangle =0\,, \qquad J \notin \mathbb{Z}\,.
\end{align}
The operators in Eq. \ref{eq:spin2_traj} are an example of such non-local operators. 

We can generalize the construction used for the ANEC, and the twist-2 operators in Eq. \ref{eq:spin2_traj}, by defining the light-transform of a general operator with quantum numbers $(\Delta, J)$, as
\begin{align}
\bold{L}[\mathcal{O}](x,z) \equiv \int \limits_{-\infty}^\infty d\alpha (-\alpha)^{-\Delta -J} \mathcal{O}\left( x-\frac{z}{\alpha},z \right)\,.
 \end{align}
 We typically take the reference point to be at infinity $\bold{L}[\mathcal{O}(-\infty z, z)]$. This light-transform maps the quantum numbers
 \begin{align}
 \bold{L} : (\Delta, J) \to (1-J, 1-\Delta)\,.
 \end{align}
 
 As such, these specific light-ray operators are only the light-transforms of local operators. However, we are now in a position to perform an analytic continuation in spin. We can illustrate this with the simple case of a free scalar field. 
 We can consider the operators
 \begin{align}
 [\phi \phi]_J (x) \equiv : \phi (x) \partial_+^J \phi (x) +\cdots
 \end{align}
 where the $\cdots$ denote derivatives to make the operator a spin-$J$ primary. These operators have quantum numbers $(\Delta=2\Delta_\phi +J, J)$. We can now define the operator
 \begin{widetext}
 \begin{align}
 \mathbb{O}_J(0,-\infty) = \frac{i \Gamma(J+1)}{2^J} \int\limits_{-\infty}^\infty dx_+   \int\limits_{-\infty}^\infty  \frac{ds}{2\pi}\left(  \frac{1}{(s+i\epsilon)^{J+1}} + \frac{1}{(-s+i \epsilon)^{J+1}} \right) : \phi (0,x_+ + s, x_\perp) \phi(0, x_+-s,x_\perp):\,.
 \end{align}
 \end{widetext}
 When $J$ is an even integer, this operator collapses to the light transform of a local operator 
 \begin{align}
 \mathbb{O}_J (0,-\infty) = \int\limits_{-\infty}^\infty dx_+ : \phi \partial_{+}^J \phi: (0,v) = \bold{L}[[\phi \phi]_J] (0, -\infty)\,.
 \end{align}
 For non-even integer-$J$, this defines a genuinely non-local operator. We have therefore defined a family of operators defined by a continuous parameter $J$, which reduce to local operators (more precisely their light-transforms at integer values of $J$. These light-ray operators (and their generalizations) therefore provide a general understanding of the analytic continuation in spin of local operators. It is also easy to check that these detector operators can be used to measure generic powers of the energy  \cite{Caron-Huot:2022eqs}.

The relationship between local operators and detector operators is best understood in terms of the so called Chew-Frautschi plot of the theory, as shown in Fig.~\ref{fig:regge}. The Chew-Frautschi plot of the theory is simply a plot of the dimensions and spins of the operators in the theory. In this plot, local operators in the free theory are represented by black dots along the $45^\circ$ dashed line, also known as the DGLAP trajectory. If one considers only the local operators of the theory, they are simply disconnected points in the Chew-Frautschi plot. However, the analytic continuation in spin provided by light-ray operators connects these into analytic families, called Regge trajectories.\footnote{The language of Regge trajectories and the Chew-Frautschi plot are borrowed from the early studies of hadrons, which in turn borrowed the language from the properties of poles in non-relativistic potential scattering \cite{Regge:1959mz}. In the context of hadrons, Regge trajectories are plots of spin, $J$, versus mass, $M^2$, for hadrons of specific quantum numbers. Chew and Frautschi conjectured that all hadrons lie on Regge trajectories \cite{Chew:1961ev,Chew:1962eu}. Experimentally, these trajectories are observed to be nearly linear, and non-intersecting. For a summary of current data, and a comparison to calculations from an effective string theory description of the QCD flux tube, see  \cite{Cuomo:2024gek}.} For the DGLAP branch in the free theory, this is illustrated by the dashed line.  When interactions are turned on, the DGLAP trajectory evolves into the blue solid line, which passes through the point $(2, 2)$, where the energy-momentum tensor resides.

This organization into Regge trajectories already allows us to understand interesting features of the theory. For integer-spin local operators, their scaling dimensions receive quantum corrections,
\begin{equation}
\Delta(J) = \Delta_0(J) + \gamma_S(J),
\end{equation}
where the subscript $S$ indicates that the anomalous dimension corresponds to the operator's properties in the deep Euclidean (space-like) region. This difference is illustrated in the Chew-Frautschi plot as the horizontal distance between the interacting DGLAP trajectory and the free one. When $J$ is not an integer, $\gamma_S$ no longer represents the anomalous dimension of a local operator but rather of ``anomalous spin" of a light-ray operator. Alternatively, one can measure the vertical departure of the interacting DGLAP trajectory from the free one. This is labeled as $\gamma_T(-1 - J_L)$ in Fig.~\ref{fig:regge}, where the subscript $T$ denotes \emph{time-like} anomalous scaling. The fact that this shift is indeed what is commonly referred to as the time-like anomalous dimension in QCD will be discussed in \Sec{sec:QCD_CF}.

From simple triangle geometry, one immediately derives the relation~\cite{Caron-Huot:2022eqs}:
\begin{equation}
\gamma_S(J) = \gamma_T(-1 - J_L) = \gamma_T(J + \gamma_S(J)). \label{eq:GLBK}
\end{equation}
This is the Gribov-Lipatov-Basso-Korchemsky (GLBK) reciprocity relation~\cite{Basso:2006nk}.  Here it emerges almost trivially once we have understood the structure of the Regge trajectory. We will revisit this relation when discussing its interpretation in QCD. Note that this relation is naturally defined for continuous values of the spin $J$.

\begin{figure}
\includegraphics[width=0.755\linewidth]{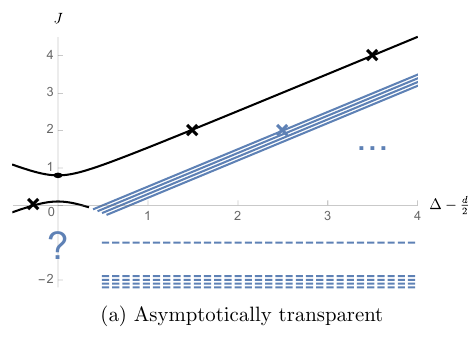}
\includegraphics[width=0.755\linewidth]{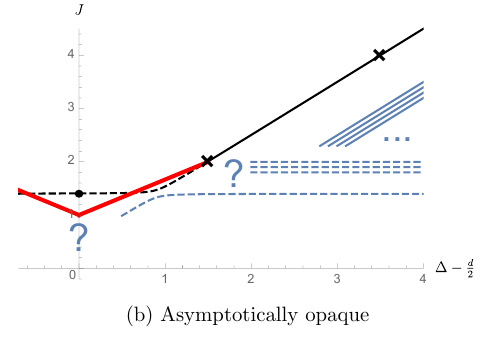}
\caption{An illustration of the Chew-Frautschi plots in an asymptotically transparent theory, as exmplified by the Wilson-Fisher theory, and an asymptotically opaque theory, as exemplified by a perturbative gauge theory. Figure from \cite{Caron-Huot:2020ouj}.
}
\label{fig:CF_schematic}
\end{figure}

A final important property of Regge trajectories, which provides significant insight into the physics of the theory, is the Regge intercept. The leading Regge trajectory is symmetric about the y-axis due to ``Shadow symmetry" \cite{Kravchuk:2018htv}, and is also convex \cite{Costa:2017twz,Kundu:2020gkz,Komargodski:2012ek}. This makes the intercept of the leading trajectory, particularly interesting. This intercept is referred to as the Regge intercept, and is denoted either $j_*$, or for historical reasons, $\alpha(0)$.

The Regge intercept controls the Regge limit of correlators or scattering process in the theory. A correlator will behave as $e^{(\alpha(0)-1)\eta}$, where $\eta$ is the boost, while a scattering process will have a cross section behaving as $s^{\alpha(0)-1}$, where $s$ denotes the Mandelstam $s$. This leads to different behavior of the theories in the high energy limit: Theories with $\alpha(0)<1$ are referred to as ``transparent" meaning that the correlator or cross-section decays in the high energy limit, while theories with $\alpha(0)>1$ are referred to as ``opaque", with correlators/ cross-sections that increase at high energies. Non-perturbatively, unitary CFTs have $\alpha(0)<1$, which is referred to as Regge boundedness \cite{Caron-Huot:2017vep}.

\begin{figure}
\includegraphics[width=0.655\linewidth]{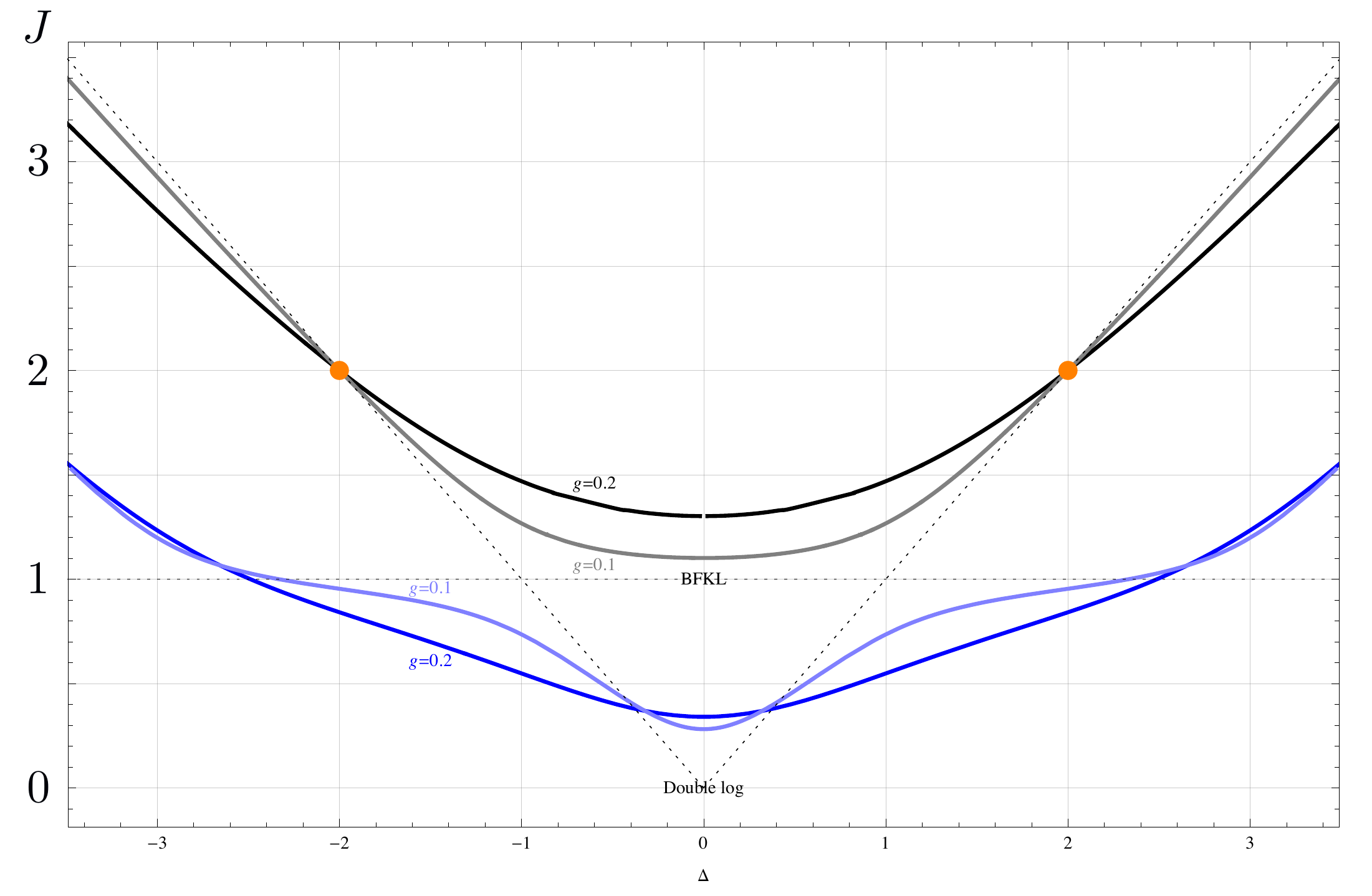}\\
\hspace{0.8cm}\includegraphics[width=0.755\linewidth]{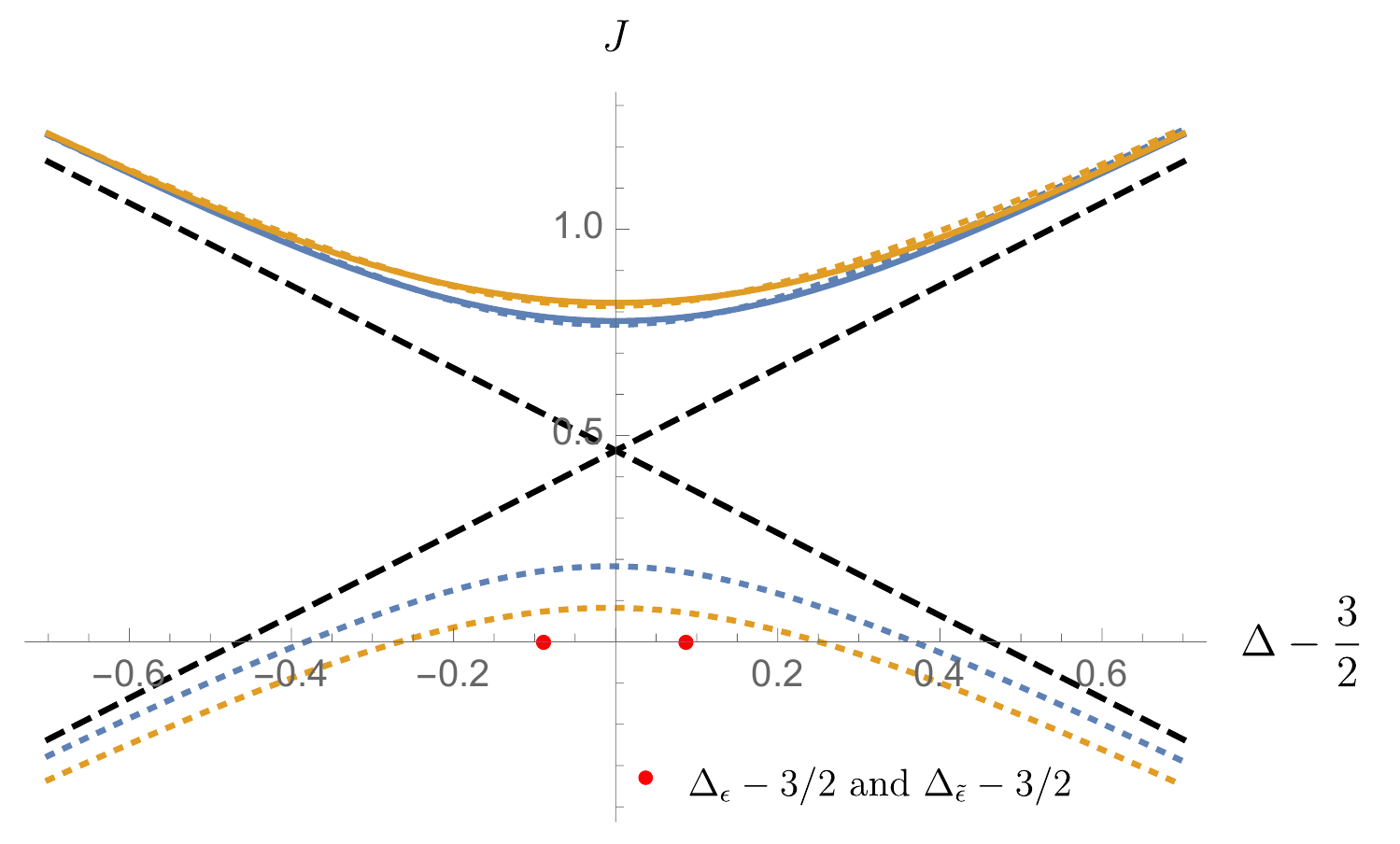}
\includegraphics[width=0.755\linewidth]{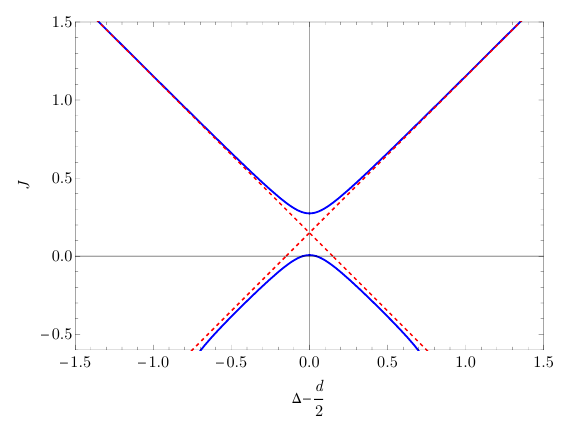}
\caption{Computations of the leading Regge trajectories in $\mathcal{N}=4$ sYM, the 3d Ising Model, and the 4d Wilson-Fisher theory. These provide explicit realizations of the features in \Fig{fig:CF_schematic}. Figures from \cite{Gromov:2015wca}, \cite{Caron-Huot:2020ouj}, and \cite{Caron-Huot:2022eqs}.
}
\label{fig:CF_assorted}
\end{figure}

In perturbative theories, the Regge intercept has a clear physical interpretation. In addition to the DGLAP trajectory, there is also a horizontal trajectory, known as the BFKL trajectory \cite{Lipatov:1985uk,Balitsky:1978ic,Kuraev:1977fs,Kuraev:1976ge,Lipatov:1976zz,Fadin:1975cb}. In the free theory, it arises at the spin of the highest spin particle of the theory: 0 for a scalar, 1 for a gauge theory, 2 for gravity. This horizontal trajectory is shown for the specific case of a gauge theory in Fig.~\ref{fig:regge}, and for illustrative opaque and transparent theories in \Fig{fig:CF_schematic}. Here the DGLAP trajectories are shown as solid blue lines, and the BFKL trajectories as dashed blue lines.  In the interacting theory, the DGLAP and BFKL trajectories mix, leading to the full trajectory and its Regge intercept, as shown in blue in \Fig{fig:regge}, and black in the examples of \Fig{fig:CF_schematic}. This perturbative mixing has been studied in detail in the 4d Wilson-Fisher theory \cite{Caron-Huot:2022eqs}, and very recently in critical O($N$) model~\cite{Li:2025knf} and in perturbative QCD~\cite{Chang:2025zib}. In fact, much of the understanding of Regge trajectories in perturbative QCD and $\mathcal{N}=4$ sYM has been driven by interest in elucidating this mixing~\cite{Jaroszewicz:1982gr,Kotikov:2002ab,Brower:2006ea}.

It is interesting to illustrate these general features in specific theories. In the top panel of \Fig{fig:CF_schematic}, we show the leading Regge trajectory in planar $\mathcal{N}=4$ sYM computed using integrability \cite{Gromov:2015wca}. At weak coupling it exhibits a Regge intercept $\alpha(0) \sim 1 +\mathcal{O}(g^2)$, as expected for a gauge theory. In the lower panel of \Fig{fig:CF_schematic} we show the leading Regge trajectory in the 4d Wilson-Fisher theory ($\phi^4$ theory) \cite{Caron-Huot:2022eqs}, which exhibits a Regge intercept $\alpha(0)\sim 0 +\mathcal{O}(\epsilon)$ (here $\epsilon=0.3$ has been chosen for illustrative purposes). In the middle panel of \Fig{fig:CF_schematic} we show an extraction of the leading trajectory for the 3d Ising Model. Since the 3d Ising Model does not have a perturbative description, it is much less clear a-priori where its Regge intercept should lie. Remarkably, using data from the conformal bootstrap \cite{El-Showk:2012cjh,El-Showk:2014dwa}, it is possible to extract the leading trajectory \cite{Caron-Huot:2020ouj}, and show that the theory is transparent, namely it lies between a weakly coupled scalar theory, and a weakly coupled gauge theory. The yellow and blue lines illustrate two different extractions, and their difference should be viewed as an uncertainty.  The red dotes denote the dimension of the $\epsilon$ operator extracted from the conformal bootstrap. The leading Regge trajectories have also been studies in the  3d $O(2)$ model \cite{Liu:2020tpf}, 6d (2,0) theories \cite{Lemos:2021azv}, and conformal fishnet theory \cite{Caron-Huot:2020nem}.

An important unresolved issue about the structure of Regge trajectories in gauge theories relates to the fact that Regge intercept of the leading trajectory is positive, namely the BFKL theory \cite{Lipatov:1985uk,Balitsky:1978ic,Kuraev:1977fs,Kuraev:1976ge,Lipatov:1976zz,Fadin:1975cb}  predicts $\alpha(0)^{\text{transient}}=1+\mathcal{O}(\alpha_s) >1$, which can also be seen in \Fig{fig:CF_schematic}. Here we have added the superscript ``transient", since it is understood that at intermediate values of boost/Mandelstam s, one can have transient behavior. The bound on chaos \cite{Maldacena:2015waa} implies that $\alpha(0)^{\text{transient}}\leq 2$. While perturbative gauge theories are far from saturating this bound, it is saturated by holographic CFTs \cite{Brower:2006ea,Shenker:2013pqa}, or the SYK model \cite{Maldacena:2016hyu,Kitaev:2017awl}. However, in weakly coupled gauge theories, the non-perturbative resolution of this growth remains a mystery. In a gauge theory one has horizontal BFKL trajectories of increasing spin, as illustrated in \Fig{fig:CF_schematic}. By Regge boundedness, and convexity of the leading trajectory, these must recombine in some highly non-trivial manner, illustrated schematically in red. The mechanism for this is unresolved. However, it is important, because it controls the behavior of the forward scattering cross section in QCD, which is an open problem, as discussed in \Sec{sec:open}. There has been significant progress in understanding the structure of BFKL trajectories in planar $\mathcal{N}=4$ sYM \cite{Ekhammar:2024neh,Klabbers:2023zdz,Brizio:2024nso,Gromov:2015wca,Gromov:2015vua,Alfimov:2014bwa}, making us optimistic for progress in this area.

Beyond the topics, covered in this short overview of light-ray operators, there has recently been significant progress in understanding the space of light-ray operators from several different perspectives. On the one hand, there has been progress exploring higher twist light-ray operators and their Regge trajectories, both perturbatively, and using integrability \cite{Henriksson:2023cnh,Homrich:2024nwc,Ekhammar:2024neh,Homrich:2022cfq,Klabbers:2023zdz,Brizio:2024nso}.  In a completely different direction, different classes of light-ray operators incorporating time \cite{Korchemsky:2021htm,Korchemsky:2021okt}, or other weightings \cite{Belin:2020lsr}. These exhibit interesting algebras \cite{Korchemsky:2021htm,Belin:2020lsr}, generalizing those found for the standard light-ray operators in \cite{Cordova:2018ygx,Belin:2020lsr,Gonzo:2020xza}. These topics go beyond what we can cover in this review, but we look forward to the developments in these areas in the near future.

\subsection{Correlation Functions of Detector Operators}\label{sec:det_correlator}

Although detector operators annihilate the vacuum, they have interesting correlation functions, $\langle \mathcal{E}(n_1) \mathcal{E}(n_2)\cdots \mathcal{E}(n_k) \rangle_\psi$, which characterize the state $\psi$. These correlation functions are precisely what can be measured in colliders experiment. Beyond the one-point detector correlator, these contain  genuine dynamical information about the theory, and are therefore difficult to compute in generic theories. In this section, we provide an overview of what is known about correlation functions of detector operators. In \Sec{sec:two_point}, we first describe the one and two point functions, which take on a special role as the simplest and most studied correlation functions. We then discuss explorations of higher point correlation functions in \Sec{sec:multipoint}. Here we provide a detailed overview both at strong and weak coupling, as well as in specific kinematic limits where we have improved control.

\subsubsection{The One and Two-Point Energy Correlators}\label{sec:two_point}

As highlighted in the introduction, the one and two-point detector correlators are arguably the simplest observables for characterizing energy flux. They have therefore been studied extensively in both QCD, and related gauge theories. In this section we provide an overview of calculations of these observables, highlighting both historical aspects of their calculation, as well as recent theoretical progress.

The simplest detector correlator is the one-point point function. For the specific case of the energy detector, it takes the form originally introduced by Sterman~\cite{Sterman:1975xv}
\begin{align}
\text{EC}(\hat n) &=
\frac{1}{\sigma_{\rm tot} Q} \int \df^4x\, e^{\img Q\cdot x} \langle0| J(x) \cE(\hat n) J(0)|0 \rangle 
\,.
\end{align}
Using the definition of the energy operator on on-shell particle states,  it can be written as the following weighted cross-section
\begin{align}
  \label{eq:ECdef}
  \text{EC}(\hat n) = \sum_{i}\int d\sigma\ \frac{E_i}{\sigma_{\rm tot} Q} \delta\left(\hat n - \hat p_i \right) \,
\end{align}
where $i$ sum over all particles in the final state. This allows it to be computed using standard perturbative techniques in weakly coupled theories. 

While the one-point correlator has been extensively studied in CFTs, where it is fixed in terms of anomaly coefficients, it is surprisingly under explored in non-conformal theories. For the particular case of $e^+e^-$ colliders, one couples to the conserved electromagnetic current, so the generic form of the one-point function can be written as
\begin{align}
\langle \mathcal{E}(\vec n) \rangle&= \frac{1}{4\pi}\left[  1+a_2(Q^2) \left(\frac{| \vec \epsilon \cdot \vec n|^2}{|\vec \epsilon |^2}-\frac{1}{3}   \right) \right] \nonumber \\
&=\frac{1}{4\pi} \left[1+a_2(Q^2) \left(\cos(\theta)^2-\frac{1}{3} \right) \right]\,.
\end{align}
Here $\theta$ denotes the angle between the vector $\hat n$ and the polarization vector for the electromagnetic current. Integrating this result over all angles gives unity, as fixed by energy conservation. In perturbative QCD, we have \cite{Basham:1977iq}
\begin{align}
a_2=-\frac{3}{2}+18\frac{\alpha_s}{4\pi}+\cdots\,,
\end{align}
which agrees with the conformal collider bounds \cite{Hofman:2008ar}. This is arguably one of the cleanest theoretical observables beyond the total cross section.
To our knowledge, it has not been precisely measured, nor has there been much theoretical attention to calculation it to higher perturbative orders. We believe it should be studied in significantly more detail, and we hope that the recent interest in energy correlators will motivate such measurements.

With the definition in Sec.~\ref{sec:lightray}, it is natural to consider the correlation function of multiple energy operators sitting at different angles. The simplest such correlator is of two energy operators in a given state $\psi$:
\begin{equation}
\langle {\cal E}(n_1) {\cal E}(n_2) \rangle_\psi \,.
\label{eq:one-point}
\end{equation}
This is referred to as the ``energy-energy" correlator.
One interesting consequence of the boost symmetry of the ANE operator, discussed in Sec.~\ref{sec:lightray}, is that the functional form of \eqref{eq:one-point} is heavily constrained. For a scalar source $\Phi$ in a momentum eigenstate with momentum $q$, it takes the form:
\begin{equation}
  \langle {\cal E}(n_1) {\cal E}(n_2) \rangle_\Phi =  \frac{c_\Phi}{(n_1 \cdot n_2)^3} {\cal F}(z) \,,
\end{equation}
where $z$ is invariant under scaling in $n_1$ or $n_2$,
\begin{equation}
z = \frac{n_1 \!\cdot\! n_2 \, q^2}{2 n_1 \! \cdot\! q \, n_2 \!\cdot\! q} = \frac{1 - \cos\theta}{2} \,,
\end{equation}
where the second equality holds in the rest frame of $q$ with $\theta$ the angle between the two detectors.

The two-point correlator can be computed in a variety of different quantum field theories. Beyond QCD, it is particularly interesting to study its behavior in related supersymmetric gauge theories. In particular, one can ask whether the huge amount of perturbative data for four-point function in  ${\cal N}=4$ sYM can be utilized to obtain predictions for the EEC. Since the EEC is a physical observable, this provides an extremely interesting connection between formal studies in ${\cal N}=4$ sYM and CFT to real world collider experiments. Such efforts have been initiated in \cite{Belitsky:2013ofa,Belitsky:2013bja,Belitsky:2013xxa}, in which a systematic investigation of the relationship between correlation functions and event shape observable was performed. 

\begin{figure}
  \includegraphics[width=0.9\linewidth]{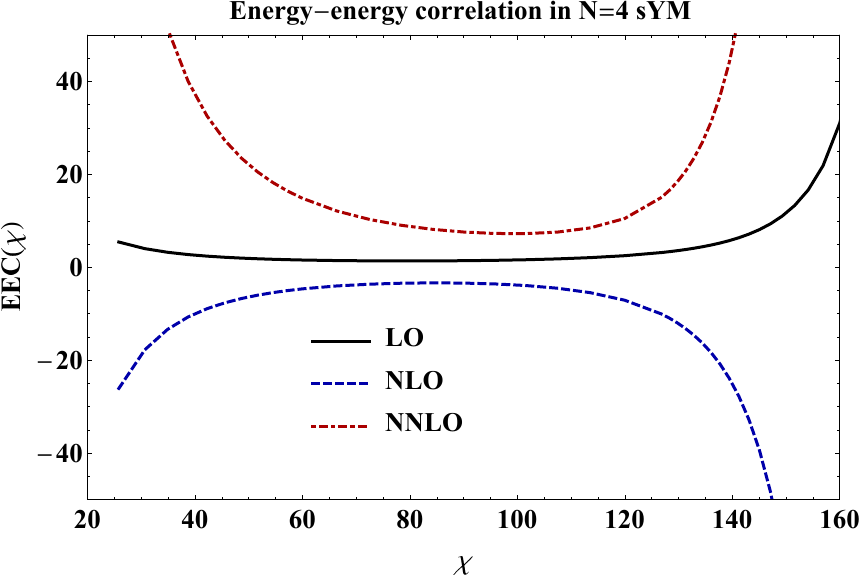}
  \caption{The LO, NLO and NNLO contributions to the energy-energy correlator in $\mathcal{N}=4$ sYM. Figure from \cite{Henn:2019gkr}.
  }
  \label{fig:EECFOv2}
\end{figure}

In \cite{Belitsky:2013ofa,Belitsky:2013bja,Belitsky:2013xxa}, the two-point energy correlator was studied in a momentum-$q$ state created by scalar half-BPS operators, which mimics the production of two colored scalars from the vacuum. By definition, the EEC corresponds to the light transform and Fourier transform of a four-point local Wightman function $\langle O^\dagger T_{\mu\nu} T_{\rho\sigma} O \rangle$ in Minkowski space. Perturbative data for the scalar four-point function in $\mathcal{N}=4$ sYM are available up to three loops in Euclidean signature \cite{Drummond:2013nda}. However, analytically continuing these results to Minkowski space remains challenging. Two successful approaches have been proposed in the literature to address this issue.

In the first method~\cite{Belitsky:2013ofa}, the Mellin amplitude for the four-point function is utilized to streamline the analytic continuation. The Mellin amplitude is related to the four-point function by
\begin{equation}
  \Phi(u, v; a) = \int_{-\delta - i \infty}^{-\delta + i \infty} \frac{d j_1 d j_2}{(2 \pi i)^2} M\left(j_1, j_2; a\right) u^{j_1} v^{j_2} \,,
\end{equation}
where $\Phi(u, v; a)$ represents the scalar four-point function up to an overall factor, and $u$ and $v$ are the standard conformal cross-ratios in CFT. The key observation is that all dependence on the signature and prescription in $u$ and $v$ is explicitly factored out. With this simplification, all dynamical information is encoded in the Mellin amplitude, allowing the light transform and Fourier transform to be performed explicitly.

In the second approach~\cite{Henn:2019gkr,Chicherin:2020azt}, a dispersion relation is found, which relates EEC to the integrals of triple discontinuity of a four-point Euclidean correlators. The key observation is that the energy operator annihilates the vacuum, ${\cal E}|\Omega \rangle = 0$~\cite{Epstein:1965zza,Hartman:2016lgu,Kravchuk:2018htv}. Therefore, instead of considering the light transform of the four-point function, one can instead consider the light transform of its double commutator, $\langle [O^\dagger, T_{\mu\nu}] [T_{\rho\sigma}, O] \rangle$, which in turn can be cast into a double discontinuity of the Euclidean correlator.

Given the scarcity of analytic results for event shape observables, we briefly discuss the results for $\text{EEC}_{{\cal N}=4}$ below. Following the normalization convention in \cite{Belitsky:2013ofa},  the EEC in ${\cal N}=4$ sYM can be expressed as
\begin{equation}
\text{EEC}_{{\cal N}=4} = \frac{1}{4 z^2 (1-z)} F(z; a) \,,
\end{equation}
where $a$ is the 't Hooft coupling, defined as $a = g_{\text{YM}}^2 N / (4 \pi)^2$. Expanding the coefficient function $F(z; a)$ in powers of $a$,
\begin{equation}
  F(z ; a)=a F_1(z)+a^2\left[(1-z) F_2(z)+F_3(z)\right] \,,
\end{equation}
the LO term is given by:
\begin{equation}
F_1(z) = - \ln  (1 - z ) \,.
\end{equation}
The NLO terms are given by:
\begin{widetext}
\begin{align}
  F_2(z)= & 4 \sqrt{z}\left[\text{Li}_2(-\sqrt{z})-\text{Li}_2(\sqrt{z})+\frac{\ln z}{2} \ln \left(\frac{1+\sqrt{z}}{1-\sqrt{z}}\right)\right]+(1+z)\left[2 \text{Li}_2(z)+\ln ^2(1-z)\right]+2 \ln (1-z) \ln \left(\frac{z}{1-z}\right)+z \frac{\pi^2}{3} \,, \\
  F_3(z)= & \frac{1}{4}\left\{(1-z)(1+2 z)\left[\ln ^2\left(\frac{1+\sqrt{z}}{1-\sqrt{z}}\right) \ln \left(\frac{1-z}{z}\right)-8 \text{Li}_3\left(\frac{\sqrt{z}}{\sqrt{z}-1}\right)-8 \text{Li}_3\left(\frac{\sqrt{z}}{\sqrt{z}+1}\right)\right]-4(z-4) \text{Li}_3(z)\right. \nonumber \\
  & +6\left(3+3 z-4 z^2\right) \text{Li}_3\left(\frac{z}{z-1}\right)-2 z(1+4 z) \zeta_3+2\left[2\left(2 z^2-z-2\right) \ln (1-z)+(3-4 z) z \ln z\right] \text{Li}_2(z) \nonumber \\
  & \left.+\frac{1}{3} \ln ^2(1-z)\left[4\left(3 z^2-2 z-1\right) \ln (1-z)+3(3-4 z) z \ln z\right]+\frac{\pi^2}{3}\left[2 z^2 \ln z-\left(2 z^2+z-2\right) \ln (1-z)\right]\right\} \,.
  \label{eq:EEC_N4}
\end{align}
\end{widetext}
Remarkably, the results are simple enough to be expressed in terms of polylogarithmic functions up to weight $3$, and can be compactly written in just a few lines. It is worth noting that this represents the first analytic result for an event shape observable beyond LO, albeit in ${\cal N}=4$ sYM. Additionally, while the results explicitly contain $\sqrt{z}$, all odd powers of $\sqrt{z}$ cancel out in an expansion around $z = 0$.

Using the triple discontinuity method, the NNLO prediction has also been obtained analytically, where only harmonic polylogarithms and elliptic functions are involved~\cite{Henn:2019gkr}. Qualitative features of the perturbative predictions up to three loops can be seen in Fig.~\ref{fig:EECFOv2}, where the coefficient functions $F(z;a)$ expanded up to three loops are shown. One can clearly see the collinear singularity in $\theta = \chi \to 0$, which is a result of emergence of jet at small angle~\cite{Konishi:1978ax,Konishi:1978yx,Konishi:1979cb,Richards:1983sr,Hofman:2008ar,Dixon:2019uzg,Kologlu:2019mfz,Korchemsky:2019nzm}, and the back-to-back singularity in $\theta = \chi \to \pi$, which is a consequence of both soft and collinear singularities~\cite{Chao:1982wb,Soper:1982wc,deFlorian:2004mp,Moult:2018jzp,Kardos:2018kqj,Duhr:2022yyp}. Both of these singularities are of substantial theoretical and phenomenological interests in QCD, and will be discussed in detail below.

In QCD, perturbative calculations of the EEC have followed a very different route. Instead of starting from the local four-point function, which is not available in QCD, the calculation is performed by inserting a complete set of Fock space states:
\begin{equation}
\text{EEC}_{\text{QCD}} \propto \sum_X \langle \Omega | O^\dagger(-q) {\cal E}(n_1) {\cal E}(n_2) | X \rangle \langle X|  O(q) | \Omega \rangle \,,
\label{eq:EEC_def_QCD}
\end{equation}
where $\sum_X$ includes summation over different Fock space states with the same quantum number, as well as an integral over continuous phase space. In fact, \eqref{eq:EEC_def_QCD} is how the EEC is defined in the first place, where $O(q)$ is chosen as the electromagnetic current operator $J^\mu(q)$. At lowest order, $X$ consists of a $q\bar{q}$ state. Then $\langle q\bar q| J^\mu(q) | \Omega\rangle$ is nothing but the well-known Sudakov form factor in QCD. Using the defining property of the ANE operator in \eqref{eq:ANEC_particle}, Eq.~\eqref{eq:EEC_def_QCD} is simply the two-particle angular correlation weighted by the product of the two-particle energies:
\begin{equation}
\text{EEC}_{\rm QCD} =  \sum_{ab, X'} \int d\sigma_{\gamma^* \to abX'} \frac{E_a E_b}{\sigma_{\rm tot} Q^2} \delta( z - \frac{1 - \cos\chi_{ab}}{2}) \,.
\label{eq:amp_approach}
\end{equation}
The advantage of this representation is that it allows the use of many cutting-edge techniques developed for computing on-shell quantities in gauge theory. In particular, on-shell form factors are much easier to compute than off-shell correlation functions.

An important property of energy correlators are a set of integral constraints that they satisfy, known as energy and momentum conservation sum rules. Starting from the definition of the energy correlators in terms of correlation functions of local operators, these arise as a consequence of the Ward identities for correlation functions of the stress tensor. Starting from the definition of the energy correlator in terms of particle states, they arise as a consequence of energy and momentum conservation. For a detailed discussion of the sum rules from these different perspectives, see the appendix of \cite{Chen:2024iuv}. Here, for simplicity, we phrase these sum rules in terms of the particle definition of the energy correlators.

Using the normalization convention in Eq.~\eqref{eq:amp_approach}, the energy conservation sum rule can be written as
\begin{equation}
  \int_0^1 dz\, \text{EEC}(z) = 1 \,.
\end{equation}
This sum rule can be easily derived by noting that total energy conservation implies $\sum_{a,b} E_a E_b = Q^2$, where the summation takes into account contact terms, $a=b$~\cite{Basham:1978bw}. The momentum conservation sum rule only applies in the massless limit~\cite{Kologlu:2019mfz,Korchemsky:2019nzm}, and can be written as
\begin{equation}
  \int_0^1 dz\, \text{EEC}(z) 2 z = 1 \,.
\end{equation}
This can be derived by using that $\sum_{ab} 2 E_a E_b z = \sum_{a,b} (p_a + p_b)^2  = Q^2$.

Note that the energy and momentum conservation sum rules apply in both conformal field theory~\cite{Kologlu:2019mfz,Korchemsky:2019nzm} and in perturbative QCD (or more generally, any theory with massless particles)~\cite{Dixon:2019uzg}, since they only depend on the presence of spacetime symmetry. At Born level, $\text{EEC}(z) = (\delta(z) + \delta(1-z))/2$, both sum rules are trivially satisfied. Starting from ${\cal O}(\alpha_s)$, they become non-trivial and can be used to cross check or derive new perturbative results. In particular, they can be used to relate contact terms in the collinear limit $\delta(z)$, with those in the back-to-back limit $\delta(1-z)$. The sum rule has also inspired the IRC-safe definition of higher points projected energy correlators~\cite{Chen:2020vvp}. The energy conservation sum rule continues to hold non-perturbatively in QCD, which should impose interesting constraints on non-perturbative corrections, although its implications remain to be explored. 

\begin{figure}
  \includegraphics[width=0.9\linewidth]{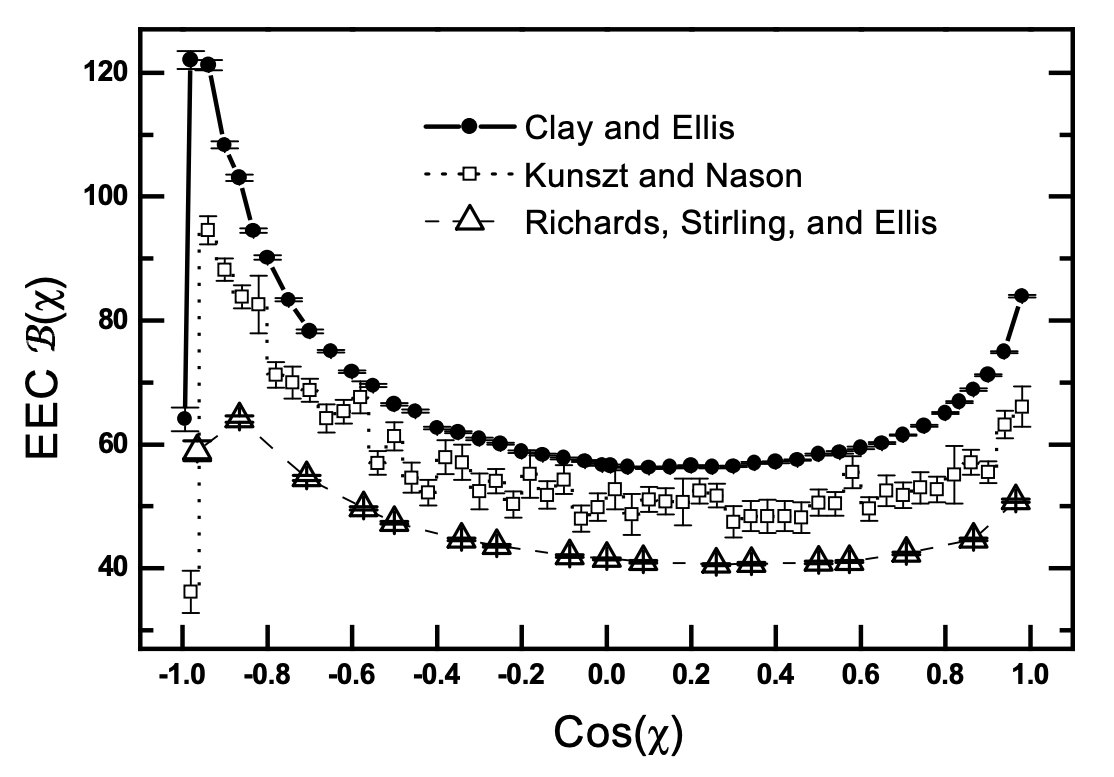}
    \includegraphics[width=0.8\linewidth]{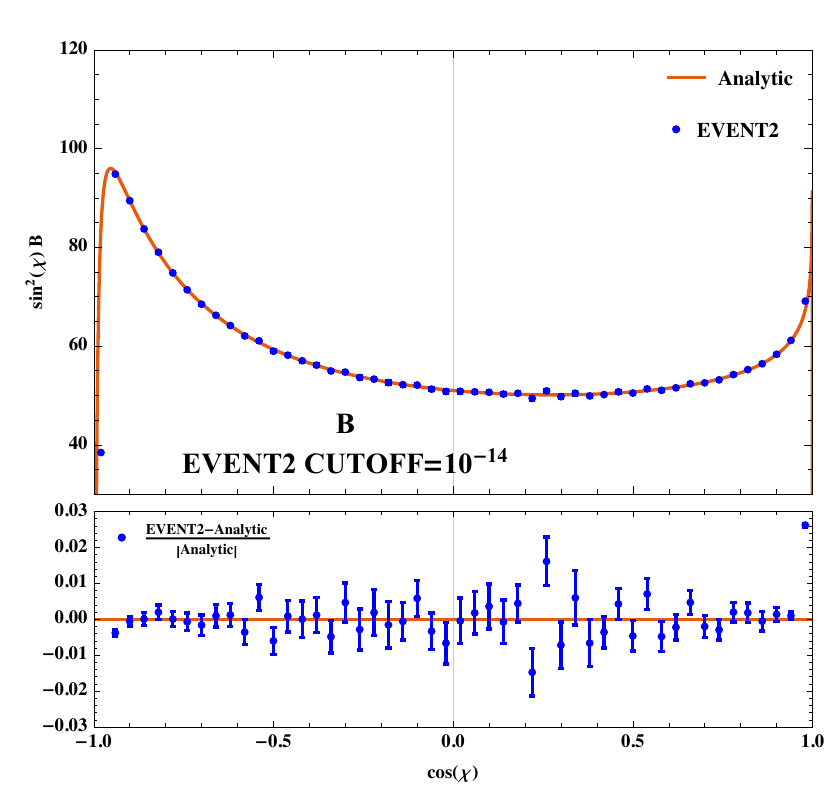}
  \caption{Upper Panel: Discrepancies between numerical results for the EEC at NLO  in early calculations. Lower Panel: A comparison of the analytic NLO calculation of the EEC in QCD with numerical results from \texttt{Event2}. Figures from \cite{Clay:1995sd} and \cite{Dixon:2018qgp}.
  }
  \label{fig:NLO_diff}
\end{figure}

The EEC is an infrared and collinear~(IRC) safe observable. However, the decomposition in the fixed-multiplicity Fock space state is not IRC safe in massless gauge theory. As a result, one must carefully cancel the infrared singularities between virtual and real corrections to achieve finite results.
At LO, the EEC requires at least one radiation to span the phase space region $0< z <1$, and it is manifestly safe from IRC singularities in this region. The LO result in QCD can be written as
\begin{align}
  & \text{EEC}_{\rm QCD}=\frac{\alpha_s(\mu)}{4 \pi} C_F \frac{3-2 z}{4(1-z) z^5}  \\
  & \times\left[3 z(2-3 z)+2\left(2 z^2-6 z+3\right) \ln (1-z)\right]+\mathcal{O}\left(\alpha_s^2\right)\nn \,,
  \end{align}
where $\alpha_s(\mu)$ is the running strong coupling constant.

Starting from NLO, it is essential to develop methods for extracting infrared and collinear  poles in real radiative corrections. Interestingly, while the first NLO QCD corrections to event shape observables were obtained numerically in the 1980s in the seminal work of Ref.~\cite{Ellis:1980wv}, numerical calculations for the EEC at NLO took significantly longer to achieve (see, e.g., \cite{Schneider:1983iu,Falck:1988gb,Glover:1994vz,Kramer:1996qr,Ali:1982ub,Ali:1984gzn,Richards:1982te,Richards:1983sr,Kunszt:1989km,Clay:1995sd}). For a period of time, discrepancies were observed among results obtained using different methods, hindering the effectiveness of using the EEC for precision SM measurements. This issue is illustrated in Fig.~\ref{fig:NLO_diff}. This discrepancy was fully resolved only after the introduction of the Catani-Seymour subtraction method~\cite{Catani:1996jh,Catani:1996vz}.

A similar situation arose at NNLO. While NNLO calculations for classic event shape observables, such as thrust, were completed in 2007~\cite{Gehrmann-DeRidder:2007nzq,Weinzierl:2008iv}, the corresponding NNLO corrections for EEC were only achieved in 2016~\cite{DelDuca:2016csb}. This delay suggests that the current numerical subtraction schemes may not be optimal for EEC-type statistical observables. In Fig.~\ref{fig:pluto_single}, we present the numerical predictions for EEC from LO to NNLO, along with a comparison to OPAL experimental data.
It is intriguing to observe that the qualitative behavior of higher-order QCD corrections for EEC resembles that of ${\cal N} = 4$ sYM theory. In both cases, the corrections exhibit peaks in the collinear and back-to-back limits. Furthermore, the numerical corrections are substantial across a significant portion of the phase space. Even at NNLO, it remains unclear whether the perturbative predictions have stabilized, highlighting the need for further investigation.

In addition to numerical calculations, a remarkable feature of the EEC is that analytic results can be obtained, even in QCD. This is achieved by calculating the different multiplicity contribution to \eqref{eq:amp_approach} analytically in dimensional regularization. In contrast to numerical calculations, analytic calculations for the EEC turn out to be simpler than for other classical event shape, as the resulting phase space integrals are amendable to modern techniques for Feynman diagram calculation, such as IBPs~\cite{Tkachov:1981wb,Chetyrkin:1981qh} and differential equation~\cite{Kotikov:1990kg,Gehrmann:2000zt,Henn:2013pwa}. The full analytic results at NLO can be found in \cite{Dixon:2018qgp} for $e^+e^- \to$ jets, and for Higgs decay to jets in \cite{Luo:2019nig,Gao:2020vyx}. The analytic results are in excellent agreement with results from Catani-Seymour subtraction, as shown in the lower panel of Fig.~\ref{fig:NLO_diff}. It's interesting to point out that the results can be expressed by the same set of transcendental functions that also appear in the ${\cal N}=4$ sYM results. This provides helpful guide for future NNLO analytic calculation in QCD.

It is useful to briefly summarize the status of perturbative calculations for EEC so far. The EEC stands out as a unique observable, for which several distinct approaches enable its computation to higher orders. On one hand, the correlation function representation of the EEC allows it to be computed from an IRC-finite local correlation function in position space, eliminating the need to expand the perturbative observable into contributions from different multiplicities. In fact, this feature was one of the motivations behind \cite{Belitsky:2013xxa,Belitsky:2013bja}. On the other hand, the EEC is an asymptotic observable, which permits its computation using the on-shell S-matrix, typically characterized by simpler perturbative structures. This duality highlights the dual nature of EEC from both on-shell and off-shell perspectives, making it an ideal candidate for further theoretical and experimental investigation using methods from both sides.

As one can see from the discussion above, the EEC has been almost exclusively studied in the specific case of four dimensional QCD, as well as closely related supersymmetric theories. However, it is also interesting to study it in other theories. One interesting target is the 3D Ising model for which the exceptional data from the conformal bootstrap
\cite{El-Showk:2012cjh,El-Showk:2014dwa,Kos:2016ysd,Kos:2014bka,Chang:2024whx} should enable a relatively precise extraction. It would be interesting to explore the two-point energy correlator in other $2+1$D CFTs, and see if it is possible to realize such measurements in the laboratory.

\begin{figure}[t!]
  \includegraphics[width=0.9\linewidth]{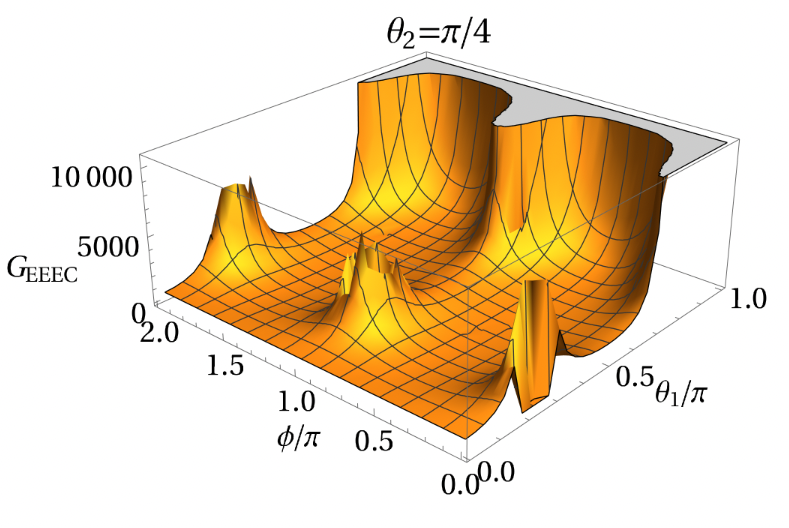}
  \caption{A plot of the three-point correlator as a function of the detector kinematics, showing an intricate pattern of energy flux, with interesting azimuthal and angular correlations. The angles used to parameterize the detector configurations are defined in the text.
  }
  \label{fig:3point_plot}
\end{figure}

\subsubsection{Exploring the Perturbative Structure of Higher Point Correlators}\label{sec:multipoint}

In this section we survey the exploration of multi-point correlators in perturbative QFT. As compared to scattering amplitudes, correlation functions of local operators, or even cosmological correlators, relatively little is understood about the structure of energy correlators in perturbative QFT. This motivates their systematic exploration, which has so far been pursued both in QCD, and $\mathcal{N}=4$ sYM. One appealing feature of energy correlators is that they are infrared finite, enabling the opportunity for calculations directly in $d=4$. 

Much like for the two-point correlator, there are two ways of computing multi-point correlators in perturbation theory. Either we can perform the light-transform of a known multi-point correlator of local operators, or we can integrate a form factor over phase space. Unfortunately, the knowledge of higher-point correlators of local operators is limited. Therefore, beyond the two-point energy correlator, all examples have been obtained by phase space integration of multi-point form factors.

We first consider the three-point correlator, which is defined perturbatively
\begin{widetext}
\begin{align}
\frac{1}{\sigma} \frac{d^3 \sigma}{dx_1 dx_2 dx_3}=\sum_{ijjk} \int d\sigma \frac{E_i E_j E_k}{Q^3} \delta\left(x_1 -\frac{1-\cos \theta_{jk}}{2} \right) \delta\left(x_2 -\frac{1-\cos \theta_{ik}}{2} \right)   \delta\left(x_3 -\frac{1-\cos \theta_{ij}}{2} \right) \,,
\end{align}
\end{widetext}
and is illustrated  in \Fig{fig:multipoint_intro}. Momentum conservation constrains the allowed values of the variables $x_i$. This correlator contains a number of different kinematic limits, including back-to-back and collinear limits, whose physics will be discussed extensively throughout this review. The explicit perturbative results for this correlator provides valuable data for understanding factorization properties in these limits. 

The three-point correlator was computed analytically, first in $\mathcal{N}=4$ sYM \cite{Yan:2022cye}. It was found to exhibit an elegant analytic structure expressed in terms of weight-2 polylogarithms. Using this understanding of the function space, it was calculated in QCD in \cite{Yang:2022tgm,Yang:2024gcn}. This is the first analytic calculation of a three-point event shape in QCD, and illustrates the power of understanding the perturbative structure of the observable.  In \Fig{fig:3point_plot} we show a plot of the shape of a particular slice of the three-point correlator. To make this plot, we have parameterized the directions of the detectors as
\begin{align}
\hat n_1 &= (\sin \theta_1, 0, \cos \theta_1) \,, \nn \\
\hat n_2 &= (\sin \theta_2 \cos \phi, \sin \theta_2 \sin \phi, \cos \theta_2) \,, \nn \\
\hat n_3 &= (-\sin \theta_2 \cos \phi, -\sin \theta_2 \sin \phi, \cos \theta_2) \,. 
\end{align}
We can then fix the opening angle between two of the detectors, and rotate the pair of detectors in $\phi$. This is shown in \Fig{fig:3point_plot} for a fixed opening angle, $\theta_2 =\pi/4$. We see an intricate structure illustrating azimuthal correlations, as well as divergences in both back-to-back and collinear limits. These divergences would be cured by resummation. It will be particularly interesting to measure such higher point energy correlators in $e^+e^-$ colliders.

\begin{figure}
\includegraphics[width=0.955\linewidth]{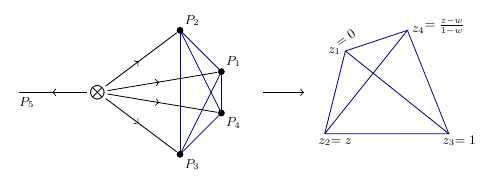}
\caption{Multi-point correlators in the collinear limit provide an interesting class of infrared finite integrals. Here we show the specific case of the four-point energy correlator in the collinear limit, along with its parameterization. Figure from \cite{Chicherin:2024ifn}.
}
\label{fig:four_point_collinear}
\end{figure}

For generic detector kinematics, as illustrated in \Fig{fig:multipoint_intro}, the energy correlator is a function of all the $\zeta_{ij} = (1 - \cos\theta_{ij})/2$, and is dependent on the details of the source. From the perspective of a perturbative amplitudes calculation, the phase space integration is made difficult by the full momentum conserving delta function. A simplifying kinematic limit, which retains a rich structure while greatly simplifying the structure of the results, is the multi-collinear limit. This is defined by scaling $\zeta_{ij}\to 0$, but for generic ratios $\zeta_{ij}/\zeta_{kl}$, and is illustrated in \Fig{fig:four_point_collinear}.  In addition to its simplicity in perturbation theory, this kinematic limit is also motivated by measurements of multi-point correlators inside high energy jets at the LHC, which are performed in the collinear limit.

If we focus on the tree level contribution to the $N$-point correlation function in the collinear limit, we can write it as 
\begin{align}
 {\rm E^N C}\overset{\rm coll.}{=} 
 \hskip-1.0mm  \int_0^1 \hskip-.5mm d x_1 \cdots d x_N \, \delta(1- \sum_i x_i) \,  (x_1 \cdots x_N)^2\, \mathcal{P}^{(0)}_{1\rightarrow N} \,.
\end{align}
As compared to the expression for the energy correlator with generic kinematics, we see two main simplifications arising from the expansion in the collinear limit. First, the integration is now over the canonical simplex, and second, the integrand is the universal splitting function, $\mathcal{P}^{(0)}_{1\rightarrow N}$, which is independent of the source.  The dependence on the positions of the detectors appears in $\mathcal{P}^{(0)}_{1\rightarrow N}$, which is a rational function of Mandelstam invariants. The Mandelstam invariants can be expressed in terms of the energy fractions, and the detector positions $z_i$, as $s_{ij}=x_i x_j |z_i-z_j|^2$. This shows that the integrand for the energy correlator will be expressed in term of products of quadrics. In the case of a single quadric, the structure of such integrals has been extensively studied \cite{Arkani-Hamed:2017ahv}, and there is progress towards an understanding for multiple quadrics \cite{Gong:2022erh}. Furthermore, due to the definition of the energy correlator, this integral is infrared finite. It therefore defines an interesting class of finite integrals in perturbative quantum field theory, which are an interesting target for exploring the perturbative structure of physical observables.

There is significant perturbative data for the universal splitting functions, $\mathcal{P}^{(0)}_{1\rightarrow N}$. In QCD, the $1\to 3$ splitting functions have long been known \cite{Catani:1998nv,Campbell:1997hg}, and the $1\to 4$ splitting functions were recently computed  \cite{DelDuca:2020vst,DelDuca:2019ggv}. In $\mathcal{N}=4$ sYM, the splitting functions were computed at three-points in \cite{Chen:2019bpb}, four points in \cite{Chicherin:2024ifn}, and up to 11 points \cite{He:2024hbb}. Using \cite{Bourjaily:2025iad}, this can be extended to twelve! Beyond the exploration of the integrated results, this motivates the exploration of the combinatorial structure of these splitting functions, which are one of the simplest squared amplitude observables.

\begin{widetext}

To illustrate the explicit structure of a three-point correlator, we consider for simplicity the case of  $\cN=4$ sYM. Writing the correlator as
\beq\label{crosssec}
J_{\rm EEEC}^{{\cal N}=4, (1)} (\zeta_{12}, \zeta_{23}, \zeta_{31}) = \frac{g^4}{32 \pi^5 \zeta_{12} \zeta_{23} \zeta_{31}} \frac{|z|^2 |1-z|^2}{|z - \bar z|} G_{{\cal N}=4}(z) \,,
\eeq
the function $G_{\cN=4}(z)$ takes the simple form \cite{Chen:2019bpb}
\begin{align}\label{eq:three_point_N4}
&G_{\cN=4}(z)=\frac{1+u+v}{2uv}(1+\zeta_2)-\frac{1+v}{2uv}\log(u)-\frac{1+u}{2uv}\log(v)\nn \\
&-(1+u+v)(\partial_u +\partial_v)\Phi(z)+\frac{(1+u^2+v^2)}{2uv}\Phi(z) +\frac{(z-\bar z)^2(u+v+u^2+v^2+u^2v+uv^2)}{4u^2 v^2}\Phi(z)\nn \\
&+\frac{(u-1)(u+1)}{2u v^2}D_2^+(z) +\frac{(v-1)(v+1)}{2u^2 v}D_2^+(1-z)+\frac{(u-v)(u+v)}{2uv}D_2^+\left( \frac{z}{z-1} \right)\,,
\end{align}
where we have used the well-known one-loop box function
\begin{align}
  \Phi(z) =\frac{2}{z-\bar z} \left(  {\rm Li}_2(z) - {\rm Li}_2(\bar z) + \frac{1}{2} \left(\log(1-z) - \log(1 - \bar z) \right) \log (z \bar z) \right)\,,
\end{align}
and a transcendental weight-2 function
\begin{align}
  \label{eq:2}
  D_2^+(z) = {\rm Li}_2(1 - |z|^2) + \frac{1}{2} \log(|1-z|^2) \log(|z|^2)\,,
\end{align}
which is even under $z \leftrightarrow \bar z$. 
\end{widetext}

This result is remarkably simple for a physical observable, and has a number of interesting features. First, its symbol alphabet is
\begin{align} 
\mathcal{A}_{123} =   \left\{ z,  \bar z , 1-z, 1-\bar z , 1- |z|^2 \right\} \,. \nn
\end{align} 
It satisfies a first entry condition  \cite{Gaiotto:2011dt}, namely that the first entry of the symbol is a Mandelstam invariant, $|z_{ij}|^2$, and its leading singularity is $1/(z-\bar z)$.  Importantly, there is no physical singularity at $z\to \bar z$. The only physical singularities are the OPE singularities at $z\to 0$ and $z\to 1$. Although we have expressed the result in terms of polylogarithmic functions, we can use the Feynman-parameter-like representation to map it to Feynman integrals, in which case one can express it in terms of the three-mass triangle integral and the three-massive, one massless box integral. The result in QCD can be expressed in terms of the same transcendental functions but with more complicated rational pre-factors. The remarkable aspect is that this distribution can be directly measured by experiment, as was shown in \Sec{sec:intro}.

The simplicity of the three-point correlator motivates the further exploration of physical infrared safe observables in perturbation theory. To do so, we must better understand both the integrand, and the integration. Much like for scattering amplitudes, we would like to understand how symmetries of the underlying theory simplify, or manifest, in squared observables measured in experiments. One of the most remarkable symmetries of $\mathcal{N}=4$ sYM is its dual conformal symmetry of both scattering amplitudes
\cite{Drummond:2006rz,Drummond:2007aua,Drummond:2008vq}, and form factors \cite{Alday:2007he,Maldacena:2010kp,Bork:2014eqa, Bianchi:2018rrj,Ben-Israel:2018ckc}. The implications of this for squared observables was first investigated in \cite{Chicherin:2024ifn}. We can compute the universal splitting functions from the collinear limit of squared form factors. Explicitly, 
\begin{align}\label{splitting}
{\cal P}^{(0)}_{1 \to N} &=  {N_c^{N-1} \over (x_1 \ldots x_N)^2 |z_{12} \ldots z_{N-1 N}|^2}\,  \mathcal{G}_N \nn \\
&+ \text{perm}(1,2,\ldots,N) \,,
\end{align}
where
\begin{align}
\quad  \mathcal{G}_{N} := \lim_{1||2\dots ||N} { \mathbb{F}\overline{\mathbb{F}}_{N+1} \over \mathbb{F}\overline{\mathbb{F}}^{\rm MHV}_{N+1}}\ .
\end{align}
In $\mathcal{N}=4$ sYM, form factors are dual to periodic Wilson loops. 
 To make manifest the dual conformal symmetry of the theory, we can introduce dual coordinates $y_i$ defined by the rule $p_i=y_i-y_{i-1}$, and express the results in terms of the dual conformal cross-ratios $(a,b,c,d)={y^2_{ab} y^2_{cd}}/({y^2_{ac}y^2_{bd}})$. For example, for the three-point splitting functions, we can write, 
\begin{align*}
\hspace{-0.25cm}
\hskip-0.5mm
 \frac{ \mathbb{F}\overline{\mathbb{F}}^{\rm NMHV}_{4}}{\mathbb{F}\overline{\mathbb{F}}^{\rm MHV}_{4}} \overset{\rm coll.} {=}    ({ -1},1,2, { 4}) + ({ -1},3,2,0) + (3,1,0, { 4}) \,. 
\end{align*}
Similarly, a compact dual conformal invariant form of the four-point splitting function was found
\begin{align}
&\frac{\mathbb{F} \overline{\mathbb{F}}^{\rm NMHV}_{5}}{\mathbb{F}\overline{\mathbb{F}}^{\rm MHV}_{5}}  \overset{\rm coll.}{=}   -1+  ({ -1},1,  2, { 5})  +   ( { -1}, 2, 3,  { 5} )   \nn \\
&+
({ -1}, 4,  3,0) +  (4, 1, 0 ,{5}) 
  + ({ -1}, 3,  2,0) +  (4, 2, 1 ,{ 5})   \nn \\
  &+(0,4, 3, 1) 
  + (0,4, 3, 1)  ({ -1},1, 3,{ 5}) 
+   ({ -1}, 4, 3,1) ( 3,1, 0, { 5})  \nn \\
& + ({ -1}, 4, 2, 0)( 0,2,3 , { 5}) 
 +   ({ -1}, 1, 2, 4)( 4,2,0 , { 5}) \nn \\
 &+   ({ -1}, 3, 2, 0)( 4,2,0 , { 5})  +    ({ -1}, 4, 2, 0)( 4,2, 1 , { 5})  \nn \\
 &+ ({ -1}, 4, 3,0) ({ -1}, 1,2,4) 
  + (4,1,0, { 5})(  0,2,3,{5})  \nn \\
&+   ({-1}, 4, 3,1) ({ -1}, 4,2,0) +  (3,1,0, { 5})(  4,2,0,{ 5})  \nt 
& + ({ -1}, 4, 3, 1)( { -1}, 1,2 , { 5})  +   ( 3,1,0, {{5}})( { -1}, 2,3 , { 5}) \nn \\
&+  ({ -1}, 1,2,4)( { -1}, 1,3 , {5}) +  (0,2,3, {{5}})( { -1}, 1,3 , { 5}) \,. 
\end{align}

\begin{figure}
\includegraphics[width=0.655\linewidth]{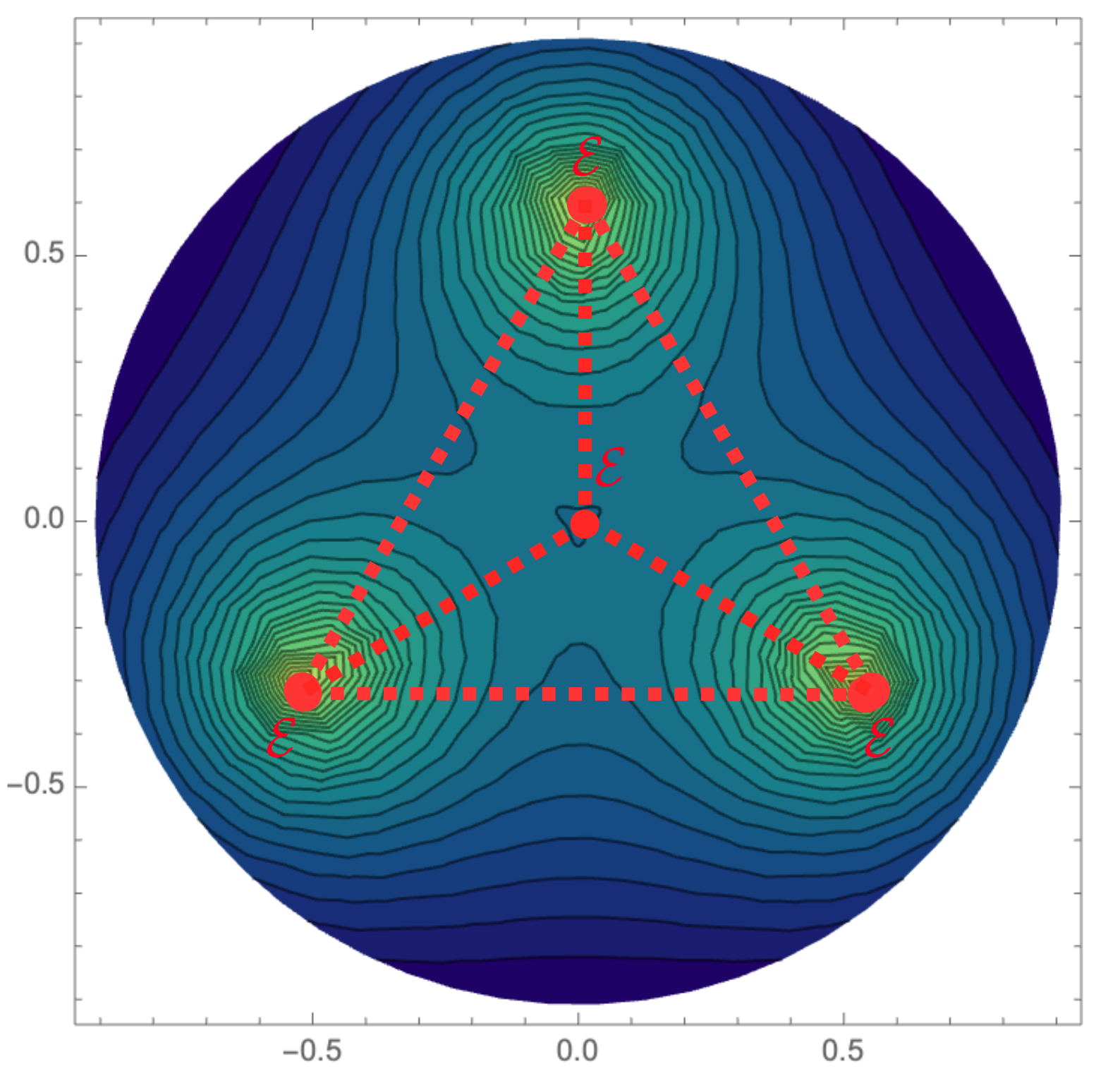}
\caption{A plot of the four-point energy correlator in $\mathcal{N}=4$ sYM in the collinear limit. Three points in the equilateral triangle are kept fixed, while the fourth point is moved. The OPE limits are clearly visible in the heat map plot.
}
\label{fig:four_point_figure}
\end{figure}

Using this compact result for the splitting function, one can integrate over the energy fractions to obtain a compact result for the four-point correlator. This required the development of advanced integration-by-parts techniques for finite integrals in Feynman paramter space \cite{Chicherin:2024ifn}. To express the result, we use the parameterization  (see \Fig{fig:four_point_collinear}) 
\begin{align}\label{eq:zwparam}
(z_1, z_ 2, z_3, z_4)  = (0  , \,   z  , \,  1   , \,  \frac{z-w }{1-w} )\,.
\end{align} 
It was found that the result can be expressed in terms of weight three polylogarithms multiplied by rational functions
\begin{align} 
\hskip-2.5mm
{\rm{E^4C}}^{\rm LO, coll}_{\mathcal{N}=4} (z_{ij})& = \frac{ \sum\limits_{I=1}^{51} R_I(z,\omega)  \, F_I(z,\omega)}{|z_{12}|^2 |z_{23}|^2 |z_{34}|^2}   +  \text{perm}(1,2,3,4)\,. \nn  
\end{align}
A plot of this result is shown in \Fig{fig:four_point_figure}.
This result also satisfies a first-entry condition \cite{Gaiotto:2011dt}, namely that the first entry of the symbol is $|z_{ij}|^2$, and is a single-valued function whose branch cuts cancel on the Euclidean sheet. However, its symbol alphabet is much more complicated than in the three-point case.

In addition to the letters for the three-point correlator
\begin{align} 
\mathcal{A}_{123} =   \left\{ z,  \bar z , 1-z, 1-\bar z , 1- |z|^2 \right\} \,, \nn
\end{align} 
it contains letters drawn from the following alphabet
\begin{align}
 \overline{\mathcal{A}}_{123} \cup  \overline{\mathcal{A}}_{124} \cup  \overline{\mathcal{A}}_{134} \cup  \overline{\mathcal{A}}_{234} 
  \cup \mathcal{A}_{\text{Quad}}\,,  \nn
\end{align}
where $\overline{\mathcal{A}}_{123} \equiv \mathcal{A}_{123} \cup \{z-\bar z \}  $ and 
\begin{align}
 \mathcal{A}_{\text{Quad}} & \equiv  \{   w \bar z - \bar w z , \,  w \bar z - \bar w , \,  z \bar w -  w ,   \nn   \\
& 1- w - \bar w +  z \bar w ,  \,  1- w- \bar w + w \bar z  \} \,.   \nn 
\end{align}
 Here the permutations $\overline{\mathcal{A}}_{124}, \overline{\mathcal{A}}_{134},\overline{\mathcal{A}}_{234}$ can be generated from $\overline{\mathcal{A}}_{123}$ by an $S_4$ symmetry transformation. In addition, the symbol alphabet contains  roots of a cubic polynomial, $(a,b,c)$, defined by 
\begin{align}\label{eq:abc_def}
& a\, b\, c = -|z|^2 |w|^2 , \quad a+ b+ c = 1- w - \bar w - z -\bar z\,, \notag \\ 
& a ^{-1}+b^{-1} +c^{-1}  = 1-z^{-1} - \bar{z}^{-1}- w^{-1} - \bar{w}^{-1} \,. 
\end{align}
The additional letters in the symbol alphabet are given by
\begin{align}
&  \mathcal{A}_{\text{Tri}^2} \cup \{ 1 \leftrightarrow 4 ,    2 \leftrightarrow 3    \}  \nn   , 
\end{align}
where
\begin{align}\label{ATri2}
\mathcal{A}_{\text{Tri}^2}  =& \left\{|z|^2- |w|^2 ,  z- |w|^2 , \bar z - |w|^2\,,    \right.  \\
& \;  \left. 
   \frac{a}{b},  \frac{a+ |w|^2}{ b+ |w|^2} ,   \frac{a+ w}{ b+ w}   \frac{b+ \bar w}{ a + \bar w}   \right\}\cup \{a\rightarrow b, b\rightarrow c \}\,, \nn 
\end{align}
 and the \emph{reflection} maps $(z, w) $ to  $ (1/w, 1/z)$ and $(a,b,c) $ to $ (1/a,1/b,1/c)$.  All these singularities can be found using an analysis of the Landau equations \cite{Correia:2025yao}. 
 
 The presence of cubic root letters is non-standard as compared to many examples in scattering amplitudes, and arises in this case due to the fact that we are integrating an \emph{amplitude squared}. The origin of the cubic root letters is identified as a maximal-cut condition ${s_{123}} = {  s_{234} } = x_{1234}=0$. When projected onto the $(x_2, x_3)-$plane,  this define a singular cubic curve $ (a\,  x_2 - x_3) (b \,x_2 - x_3)  (c\, x_2 - x_3) =0 $.  Despite the cubic-root dependence,  the master integrals are single-valued functions whose branch cuts cancel on the Euclidean sheet \cite{Brown:2004ugm,Dixon:2012yy,Bourjaily:2022vti}. The appearance of more complicated algebraic varieties is a general feature of higher point correlators, and is important to systematically understand.

\begin{figure}
\includegraphics[width=0.955\linewidth]{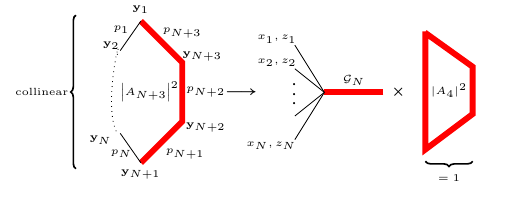}
\caption{Multi-point splitting functions, which describe the universal collinear dynamics in gauge theories, are interesting ``squared amplitude" level observables which can be extracted to high points from correlators or amplitudes. Figure from \cite{He:2024hbb}.
}
\label{fig:song_splitting}
\end{figure}

Using high loop data for the four-point correlator in $\mathcal{N}=4$ sYM, the 11 point splitting amplitude was obtained in a dual conformal invariant form in \cite{He:2024hbb}, by taking an appropriate light-like limit, see \Fig{fig:song_splitting}. Additionally, some of the integrals appearing in the calculation of higher point correlators were investigated providing strong evidence that they become elliptic, or even Calabi-Yau. Beyond the collinear limit, progress has been made on the full-angle four-point correlator in $\mathcal{N}=4$ sYM, where the relevant master integrals have been analyzed using the method of differential equations~\cite{Ma:2025qtx}. Additionally, a bootstrap program has been developed for form factors in $\mathcal{N}=4$ sYM, which is relevant for higher-point energy correlators~\cite{He:2025zbz}.

There are many directions that would be interesting to pursue at both the level of the integrand, and the integrated results. At the level of the integrand, while there has been tremendous progress in novel geometric approaches to determining the integrands of scattering amplitudes, in particular the amplituhedron \cite{Arkani-Hamed:2013jha,Arkani-Hamed:2013kca,Arkani-Hamed:2017vfh}, it is interesting to study geometries associated with squared amplitudes. Such objects are natural, since the pairing with the conjugate amplitude removes the little group weight. Squared amplituhedron geometries have been studied in \cite{Eden:2017fow,Dian:2021idl,He:2024hbb}.  There has also been recent progress in studied similar objects in the context of cosmology \cite{Arkani-Hamed:2024jbp}. It would be interesting to pursue these directions further, or to develop direct combinatorial approaches for computing the splitting functions.

At the level of the integrand, it would be interesting to more systematically study the integrals appearing in the energy correlators using recent developments in the study of the Landau equations \cite{Mizera:2021icv,Fevola:2023fzn,Fevola:2023kaw,Correia:2025yao}, intersection theory \cite{Mizera:2017rqa,Mizera:2019vvs,De:2023xue}, 
 as well as the structure of integrals directly in Feynman parameter space \cite{Arkani-Hamed:2017ahv,Gong:2022erh,Arkani-Hamed:2022cqe,Bourjaily:2020wvq,Britto:2023rig,Artico:2023bzt,Artico:2023jrc,Chen:2019mqc,Lee:2014tja}. We anticipate that these techniques can provide a more complete understanding of the structure of energy correlator observables in perturbation theory.

\subsubsection{Celestial Blocks and the Light-Ray OPE}\label{sec:celestial_blocks}

\begin{figure}
\includegraphics[width=0.75\linewidth]{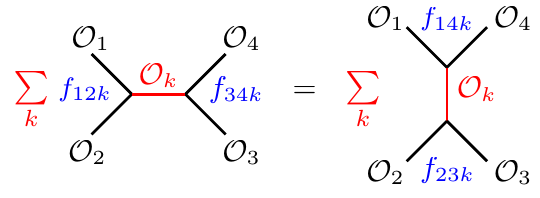}
\caption{Correlation functions of local operators can be decomposed using the OPE and knowledge of the structure constants $f_{ijk}$. The structure of the theory is strongly constrained by the crossing equations. The light-ray OPE allows these techniques to be extended to correlation functions of light-ray operators, providing new ways to compute and understand them in colliders. Figure from \cite{Poland:2018epd}.
}
\label{fig:CFT_4point}
\end{figure}

One of the exciting aspects of interplay between CFTs and energy correlators has been new techniques for simplifying the calculation of detector operator observables. In the case of correlation functions of local operators, a central role is played by the operator product expansion
\begin{align}
\mathcal{O}_i(x_1) \mathcal{O}_j(x_2)= \sum_k f_{ijk}(x_1 - x_2) \mathcal{O}_k(x_2)\,.
\end{align}
Iteratively applying this OPE allows one to compute any multi-point correlation function knowing the spectrum of operators (quantum numbers) and their three-point functions (referred to as structure constants). This is illustrated in \Fig{fig:CFT_4point}. This reduces the solution of the theory to the calculation of these quantities. Much in analogy with how scattering amplitudes can be decomposed into partial waves in terms of the Legendre polynomials (functions with definite quantum numbers under SO(3) symmetry of massless $2\to 2$ scattering in the center of mass frame in $D=4$)
\beq
A(s,t) = \sum_{l = 0}^{\infty} (2 l + 1) a_l(s) P_l(\cos\theta) \,,
\eeq
multi-point correlators of local operators can be expanded in terms of the CFT data, and kinematical factors, called conformal blocks \cite{Ferrara:1972kab,Ferrara:1972uq,Ferrara:1974nf,Ferrara:1974ny}, which are fully fixed by symmetries. For example, for the four-point correlator, we can write
\begin{equation}\label{eq:OPE_local}
    G(z,\bar{z})=\sum_{\delta,j} c_{\delta,j} G_{\delta,j}(z,\bar{z})\,.
\end{equation}
Here the $c_{\delta,j}$ encode the dynamics of the theory, and the $G_{\delta,j}(z,\bar{z})$ encode the kinematics.

\begin{figure}
\includegraphics[width=0.95\linewidth]{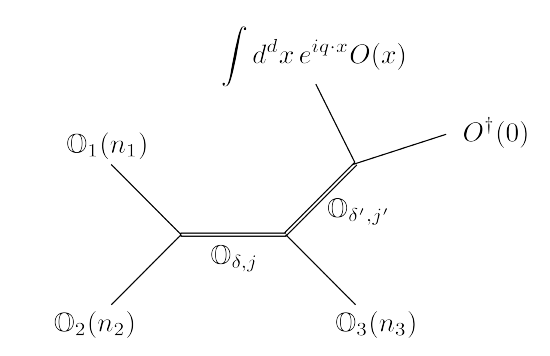}
\caption{The light-ray OPE allows multi-point detector correlators to be systematically expanded in terms structure constants and celestial blocks. Here we show a schematic of the iterated OPE for the three-point correlator of detector operators, $\mathbb{O}_i$, whose decomposition is discussed in detail in the text. Figure from \cite{Chen:2022jhb}.
}
\label{fig:block_OPE}
\end{figure}

The light-ray OPE \cite{Hofman:2008ar} enables a similar approach to the calculation of detector operators. Multi-point correlators can be expressed in terms of the spectrum of light-ray operators of the theory, as well as their structure constants, and certain ``celestial blocks"~\cite{Kologlu:2019mfz}, which are specific partial waves on the celestial sphere encoding constraints of symmetry.\footnote{Amusingly,  Parisi seems to have been involved both in the introduction of conformal partial waves \cite{Ferrara:1972kab,Ferrara:1972uq}, and in the introduction of perturbative splitting functions \cite{Altarelli:1977zs}. In this section we will see how they can be combined to understand jet substructure. } This is illustrated in \Fig{fig:block_OPE} for the three-point correlator, whose celestial blocks will be derived in this section. While the complete knowledge of the spectrum of light-ray operators of the theory is not currently known, we will see that this approach still provides a conceptually appealing way for organizing our understanding of detector correlators. 

\begin{figure}
\includegraphics[width=0.85\linewidth]{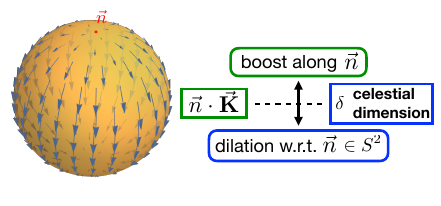}
\includegraphics[width=0.85\linewidth]{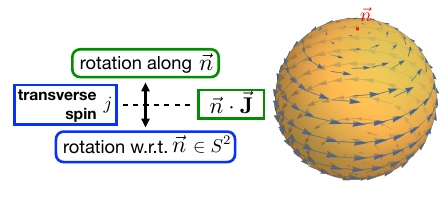}
\caption{The action of Lorentz symmetry on the celestial sphere. Operators on the celestial sphere can be characterized by their celestial dimension, and transverse spin. Multi-point correlators can be decomposed into ``celestial blocks", which can be thought of as partial waves incorporating these symmetries. Figure from \cite{Chen:2022jhb}.
}
\label{fig:block_symmetry}
\end{figure}

Before proceeding to the technical details, it is interesting to rephrase this in the usual language of jet substructure observables.  The difficulty of jet substructure observables, as compared to standard fragmentation observables studied since the early days of QCD, is in the description of the transverse structure of the jet. However, this transverse structure introduces the physical scale into the energy correlators, and provides them with their phenomenological power. The light-ray OPE allows us to reduce this transverse structure, ultimately reducing the calculation to a three-point function of a general light-ray operator in a source, $\langle \mathcal{O}^\dagger \mathbb{O}^{[J]}_{\delta,j} \mathcal{O} \rangle$,
which is just generalized fragmentation. The light-ray OPE therefore provides an ideal approach to studying jet substructure by effectively expanding the transverse structure of the jet, reducing it to the study of fragmentation type operators with well defined scaling laws. Higher point correlators also require the knowledge of structure constants of the light-ray operators themselves. Much less is known about these in general. See for example \cite{Balitsky:2018irv,Balitsky:2015tca,Balitsky:2015oux}. Additionally, the use of conformal blocks provides an approach to robustly characterizing imprints of physical effects into the correlator. For example, if we measure the energy correlator in the quark gluon plasma, we can hope to characterize the modifications by how the QGP modifies the coefficients of celestial blocks of different quantum numbers.

In this section we provide a high level overview emphasizing the utility of the light-ray OPE and the celestial block decomposition to a broader audience. Our presentation is not designed to be rigorous, but rather to introduce these techniques to a broader audience. For a rigorous discussion in the context of CFTs, we refer the reader to \cite{Kologlu:2019mfz,Chang:2020qpj,Kologlu:2019bco}.

Local operators in a CFT are classified by their dimension and spin.  Light-ray operators are classified by the action of the Lorentz group on the celestial sphere. This is illustrated in \Fig{fig:block_symmetry}. The most important symmetry generators are the boost along the light-like direction, and rotations about the light-like direction. The quantum numbers under these transformations are the celestial dimension, which we will refer to as $\delta$, and the transverse spin $j$.

The general light-ray OPE takes the form
\begin{align} \label{eq: schematic_ope_ansatz}
\mathbb{O}^{[J_1]}_{\delta_1}(n_1) \mathbb{O}^{[J_2]}_{\delta_2}(n_2) &\sim \sum \# \, (n_1\cdot n_2)^\kappa \mathbb{O}^{[J]}_{\delta,j}(n_2,\varepsilon)\nn \\
& + \text{transverse derivatives}\,,
\end{align}
Much like for the OPE of local operators we can fix the values of $\kappa$ and $J$ by matching the quantum numbers of both sides. First, we recall that the light-ray operator ${\cal E}_{J_L} = \mathbb{O}^{[J]}$ has dimension $J-1$. This implies the constraint
\beq \label{eq: dimension_constraint}
(J_1-1)+(J_2-1)=J-1.
\eeq
The value of $\kappa$ is then fixed by demanding that both sides transform in the same manner under boosts along the collinear direction. Boosts along the collinear direction correspond to dilatations on the celestial sphere, and therefore this correspond to performing dimensional analysis on the celestial sphere. This fixes $\kappa$ as
\beq \label{eq: boost_constraint}
\delta_1+\delta_2=-2 \kappa +\delta \quad \Rightarrow \quad \kappa=\frac{\delta-\delta_1-\delta_2}{2}.
\eeq
For the particular case of two ANEC operators, this gives the formula
\begin{align}\label{eq:EE_OPE}
\mathcal{E}(n_1)\mathcal{E}(n_2) &\sim \sum_i \#\, (n_1\cdot n_2)^{\frac{\tau_i-4}{2}} \mathbb{O}_{i}^{[J=3]} (n_2) \nn \\
& + \text{transverse derivatives}.
\end{align}
This OPE is structurally different than the OPE of two local stress tensor operators. In particular, the selection rule to $J=3$ imposes that the identity operator does not contribute. Furthermore, the operator appears on the right-hand-side is intrinsically non-local.

The OPE expansion decomposes the two-point function into operators with well defined scaling under dilations, i.e. that exhibit specific power laws for the physical distribution measured in experiment. 

In the specific case of weakly coupled four dimensional gauge theories, such as QCD at high energies, this formula is particularly useful since the operators at low twist are relatively sparse, and therefore by measuring this distribution, we can easily project out the contribution from the lowest twist operators, namely the twist-2 operators.

While Eq. \ref{eq:EE_OPE} is quite abstract, it is interesting to make it  concrete for the specific case of perturbative QCD, which is our end goal for applications. Recall that for a dimension-$\Delta$, spin-$J$, transverse spin-$j$ local primary operator $\mathcal{O}^{\mu_1\dots\mu_J;\nu_1\dots\nu_j}$, we can define a corresponding light-ray operator as \cite{Kravchuk:2018htv,Chen:2021gdk}
\begin{align}
 \mathbb{O}^{[J]}(\hat{n},\varepsilon) =  &\
 \lim_{r\to \infty} r^{\Delta-J}\int_{0}^{\infty}\! dt\; \mathcal{O}^{\mu_1\dots\mu_J;\nu_1\dots\nu_j}(t,r\hat{n})
\nonumber
\\
&\ \times \bar{n}_{\mu_1}\dots\bar{n}_{\mu_J}\varepsilon_{\nu_1}\dots\varepsilon_{\nu_j}\,,
\end{align}
where $n^\mu=(1,\vec{n}), \bar{n}^\mu=(1,-\vec{n})$, and $\varepsilon$ is a polarization vector that satisfies $\varepsilon^2=\varepsilon\cdot n = \varepsilon\cdot \bar{n}=0$. Perturbatively this detector operator can be viewed as measuring some particle state, the specific nature of which depends on the nature of the operator $\mathcal{O}$. At weak coupling, the operators that will dominate are the twist-2 operators, which can be viewed as detecting single quark or gluon states. The twist-2 operators in QCD are
\bea
\mathcal{O}_q^{[J]}&=&\frac{1}{2^J}\bar{\psi}\gamma^{+}(iD^+)^{J-1}\psi\,,\\
\mathcal{O}_{g}^{[J]}&=&-\frac{1}{2^J} F_{c}^{i +}(iD^+)^{J-2}F_{c}^{i +}, \\
\mathcal{O}_{\tilde{g},\lambda}^{[J]}&=&-\frac{1}{2^J} F_{c}^{i +}(iD^+)^{J-2}F_{c}^{j +} \varepsilon_{\lambda,i} \varepsilon_{\lambda,j} \,.
\eea
Performing the light-transform, the corresponding operators in the free theory are given by~\cite{Chen:2021gdk}
\bea
\mathbb{O}_q^{[J]}(\hat n)&=&\sum_{s}\int\!\! \frac{E^2 dE}{(2\pi)^3 2E}  E^{J-1} 
\nonumber
\\
&& \times 
\left(b_{p, s}^{\dagger}b_{p,s}+(-1)^J d_{p,s}^{\dagger} d_{p,s} \right)\,, \label{eq: unpolarized_quark} \\
\mathbb{O}_{g}^{[J]}(\hat n)&=&\sum_{\lambda, c} \int\!\! \frac{E^2 dE}{(2\pi)^3 2E} E^{J-1} a^{\dagger}_{p,\lambda,c} a_{p,\lambda,c}\,, \label{eq: unpolarized_gluon}\\
\mathbb{O}_{\tilde{g},\lambda}^{[J]}(\hat n)&=&-\sum_{\lambda} \int\!\! \frac{E^2 dE}{(2\pi)^3 2E}  E^{J-1} a^{\dagger}_{p,\lambda,c} a_{p,-\lambda,c}\,. \label{eq: polarized_gluon}
\eea
Here $p^\mu = E n^\mu$ and $\lambda$ is the helicity of the polarization vector $\varepsilon_\lambda$, and we have used the following conventions for the mode expansions
\bea
\psi(x)&=&\sum_s \int\frac{dp^+ d^2p_{\perp}}{(2\pi)^3 2p^+} 
\\
&& \times
\left(u_s(p) b_{p,s} e^{-i p\cdot x} + v_s(p) d^{\dagger}_{p,s} e^{i p\cdot x}\right)\,,\nn \\
A^{\mu}_{c}(x)&=&\sum_{\lambda} \int\frac{dp^+ d^2p_{\perp}}{(2\pi)^3 2p^+}
\\
&& \times
\left(\varepsilon^{\mu}_{\lambda}(p) a_{p,\lambda,c} e^{-ip\cdot x} +{\varepsilon^{*}_{\lambda}}^{\mu}(p) a^{\dagger}_{p,\lambda,c} e^{ip\cdot x} \right)\,. \nn
\eea
These operators describe the leading power scaling behavior of $n$-point energy correlator observables in QCD \cite{Hofman:2008ar,Chen:2021gdk}. This is intuitively clear, since one is just decomposing a general detector measurement into quark and gluon detectors.  

\begin{widetext}
The calculation of the OPE coefficients at one-loop was performed in \cite{Chen:2021gdk}, providing the explicit form of the OPE at leading twist
\beq \label{eq: EE_OPE}
\begin{split}
&\mathcal{E}(\hat n_1)\mathcal{E}(\hat n_2)=\\
&-\frac{1}{2\pi} \frac{1}{2(n_1\cdot n_2)} \left\{ 
\left[ (\gamma_{qq}(2)-\gamma_{qq}(3))+(\gamma_{gq}(2)-\gamma_{gq}(3))\right]\mathbb{O}_q^{[3]}(\hat n_2) +\left[ (\gamma_{gg}(2)-\gamma_{gg}(3))+2 n_f(\gamma_{qg}(2)-\gamma_{qg}(3))\right]\mathbb{O}_g^{[3]}(\hat n_2) \right.\\
&\qquad\qquad \left. +\frac{1}{2} \left[(\gamma_{g\tilde{g}}(2)-\gamma_{g\tilde{g}}(3))+2 n_f (\gamma_{q\tilde{g}}(2)-\gamma_{q\tilde{g}}(3)) \right]
\left( e^{2i\phi_S} \mathbb{O}_{\tilde{g},-}^{[3]}(
  \hat n_2) + e^{-2i\phi_S} \mathbb{O}_{\tilde{g},+}^{[3]}(\hat n_2)\right)
\right\}+\mathcal{O}((n_1\cdot n_2)^0)\,.
\end{split}
\eeq
\end{widetext}
We do not provide here the full set of OPE coefficients, which can be found in \cite{Chen:2021gdk}. However to highlight there structure, we provide a single example.
Writing $\gamma_{ab}(J)$ perturbatively as $\gamma_{ab}(J)= \alpha_s/(4\pi) \gamma_{ab}^{(0)}(J)+\mathcal{O}\left(\alpha_s^2\right)$ we have
\begin{align}
\gamma_{qq}^{(0)}(J)&=C_F\left( 4\left(\psi^{(0)}(J+1)+\gamma_E\right)-\frac{2}{J(J+1)}-3\right)\,,
\end{align}
where $\psi^{(0)}(z) = \Gamma'(z)/\Gamma(z)$ is the digamma function, and $\beta_0 = 11/3 C_A - 4/3 n_f T_F$ is the one-loop beta function in QCD. These OPE coefficients exhibit an interesting analytic structure in $J$, which would be interesting to explore further.

\begin{figure}
\includegraphics[width=0.85\linewidth]{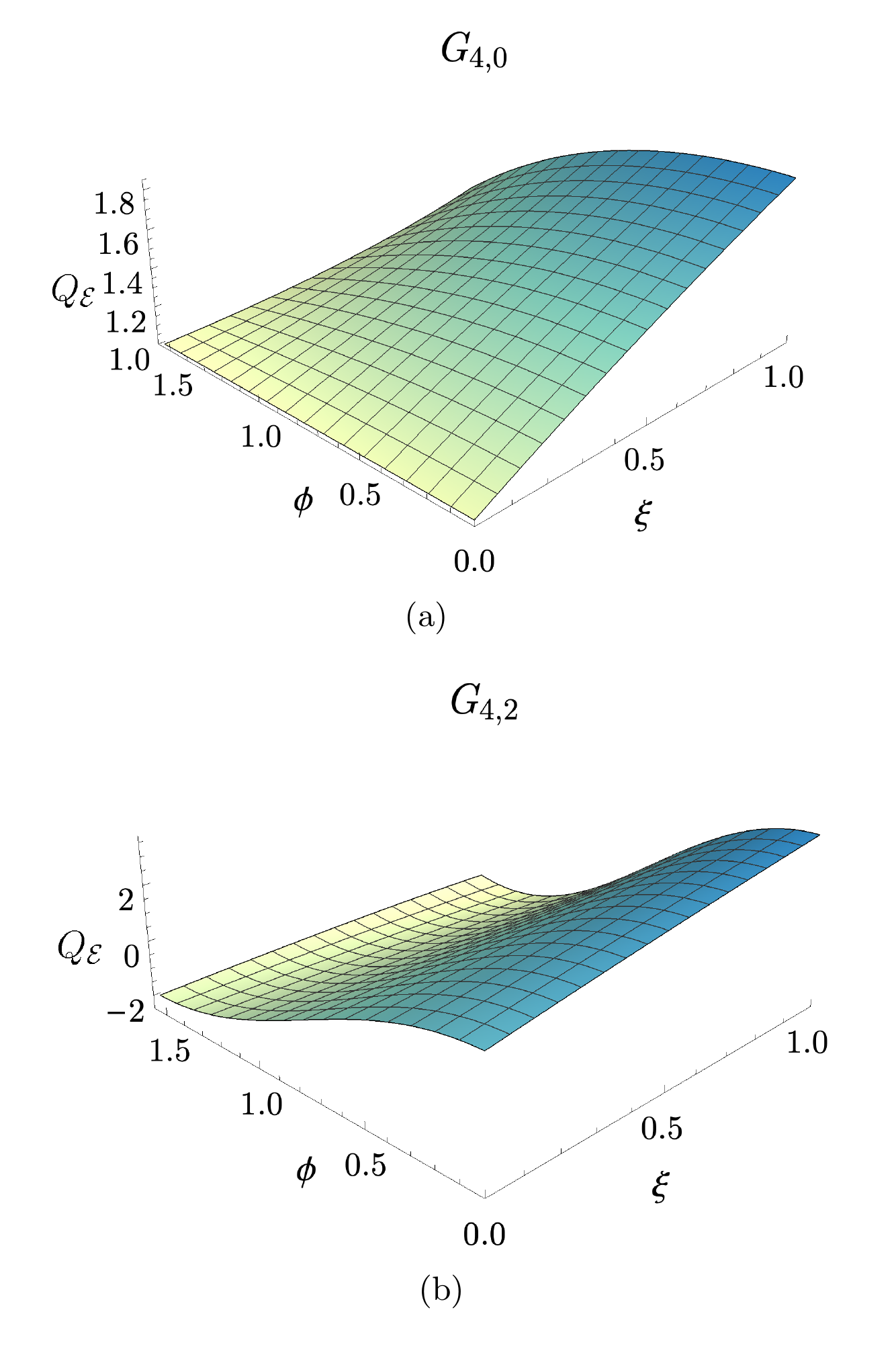}
\caption{Multi-point correlators of detector operators can be decomposed into ``celestial blocks". Here we show the twist-4, transverse spin-0 and 2 blocks. The coordinates $\phi$ and $\xi$ are the same as used in the measurement of the three-point correlator in \Fig{fig:decorated_opendata}.  Figure from \cite{Chen:2022jhb}.
}
\label{fig:block_plots}
\end{figure}

Beyond identifying specific power laws, the light-ray OPE can be applied iteratively to allow the analysis of higher point correlation functions, exactly as in the study of correlation functions of local operators. Due to the fact that this approach is quite unfamiliar to the QCD community, we wish to introduce some of the techniques, and relevant references in this section. More generally, we hope that this further motivates the use of symmetry based approaches in the study of energy correlator observables.

We will illustrate these techniques on the three-point energy correlator in the collinear limit.  For concreteness, we will choose the collinear anti-quark field $\chi$ as the source field
\begin{equation}
\int dt\; e^{i t \bar{n}\cdot P}\mae{\Omega}{\bar{\chi}(t\bar{n}) \frac{\slashed{\bar{n}}}{2}  \mathcal{E}(n_1)\mathcal{E}(n_2)\mathcal{E}(n_3)\chi(0)}{\Omega}\equiv \vev{\mathcal{E}_1\mathcal{E}_2\mathcal{E}_3}_\chi\,.
\end{equation}
Lorentz covariance implies that this function can be written as
\begin{equation}
\vev{\mathcal{E}_1\mathcal{E}_2\mathcal{E}_3}_\chi = \frac{(\bar{n}\cdot P)^5}{(n_1\cdot n_2)^3 (n_3\cdot\bar{n})^4} \left(\frac{n_1\cdot n_3}{n_1\cdot \bar{n}}\right) G(u,v)\,, \label{eq: collnear_EEEC_ansatz}
\end{equation}
where the cross ratios are defined by
\begin{align}
&u=z\bar{z}=\frac{(n_1\cdot n_2)(n_3\cdot \bar{n})}{(n_1\cdot n_3)(n_2\cdot \bar{n})}\,, \\
&v=(1-z)(1-\bar{z})=\frac{(n_1\cdot\bar{n})(n_2\cdot n_3)}{(n_1\cdot n_3)(n_2\cdot \bar{n})},
\end{align}
and describe the shape of the energy detectors on the celestial sphere.

The dimensionless component of Eq. \ref{eq: collnear_EEEC_ansatz} has an identical structure to a four-point  correlator of local scalar operators in a CFT
\begin{align}
&\vev{\mathcal{O}_1(x_1) \mathcal{O}_2(x_2) \mathcal{O}_3(x_3) \mathcal{O}_4(x_4)} =\\
&\frac{1}{(x_{12}^2)^\frac{\Delta_1+\Delta_2}{2}} \frac{1}{(x_{34}^2)^\frac{\Delta_3+\Delta_4}{2}}\left(\frac{x_{14}^2}{x_{24}^2}\right)^{\frac{\Delta_2-\Delta_1}{2}}  \hspace{-0.2cm}\left(\frac{x_{14}^2}{x_{13}^2}\right)^{\frac{\Delta_3-\Delta_4}{2}}\hspace{-0.2cm} \mathcal{G}(u,v),
\end{align}
after we make the identification
\begin{equation}
x_{ij}^2 \to n_i\cdot n_j,\quad x_{i4}^2\to n_i\cdot \bar{n},\quad \Delta_1,\Delta_2,\Delta_3\to 3,\quad \Delta_4\to 5. 
\end{equation}
This relationship has tremendous physical consequences, as it implies that the wealth of techniques for understanding correlation functions of local operators can be extended to understanding energy correlators. 

One of the most important implications is that we can expand the three-point correlator into functions with well defined quantum numbers under the action of the Lorentz group on the celestial sphere, much like Eq. \ref{eq:OPE_local} in the case of local correlators. These functions are referred to as celestial blocks \cite{Kologlu:2019mfz}. One of the remarkable features of the energy correlators is that since we can directly measure the correlators in experiment, we can see these ``harmonics of the Lorentz group" by eye in the energy distribution imprinted in the detector!

A detailed discussion of the decomposition of the three-point function into celestial blocks is given in\cite{Chen:2022jhb,Chang:2022ryc}. Here we only state the result.  The non-trivial function, $G(z,\bar z)$, describing the shape dependence of the EEEC  satisfies an expansion in celestial blocks,
\begin{equation}
    G(z,\bar{z})=\sum_{\delta,j} c_{\delta,j} G_{\delta,j}(z,\bar{z})\,,
\end{equation}
which cleanly separates kinematics $g_{\delta,j}$ from dynamics $c_{\delta,j}$. Note that having made the relation between the structure of the three-point energy correlator, and a point function of local scalar operators, this equation is \emph{identical} to Eq. \ref{eq:OPE_local}. Using the approach of Casimir differential equations \cite{Dolan:2003hv}, one can show that the celestial blocks are given by
\begin{equation}\label{2dBlock}
G_{\delta,j}(u,v)
=\frac{1}{1+\delta_{j,0}}\Big(k_{\frac{\delta-j}{2}}(z)k_{\frac{\delta+j}{2}}(\bar{z})+k_{\frac{\delta+j}{2}}(z)k_{\frac{\delta-j}{2}}(\bar{z})\Big)\,,
\end{equation}
where 
\begin{equation}
k_{h}(x)\equiv x^{h} \left._2 F_1 \right.\left(h+a,h+b,2h,x\right).
\end{equation}
These are simply the standard global conformal blocks of at 2d CFT. This is not surprising, since in the collinear limit we have expanded the celestial sphere into a plane, and boosts along the null direction act as dilations on the sphere. For the specific example of the EEEC with a collinear quark source, we set $a=0, \, b=-1$.

An interesting aspect of the energy correlators are that these celestial blocks correspond to physical patterns on the detector. Several examples of celestial blocks are shown in \Fig{fig:block_plots}. We believe that in the future a powerful approach to computing multi-point correlators may be through their approximation by the leading celestial blocks, much as is done for the case of correlation functions of local operators in CFTs.

\begin{figure}
\includegraphics[width=0.65\linewidth]{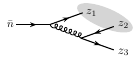}
\includegraphics[width=0.65\linewidth]{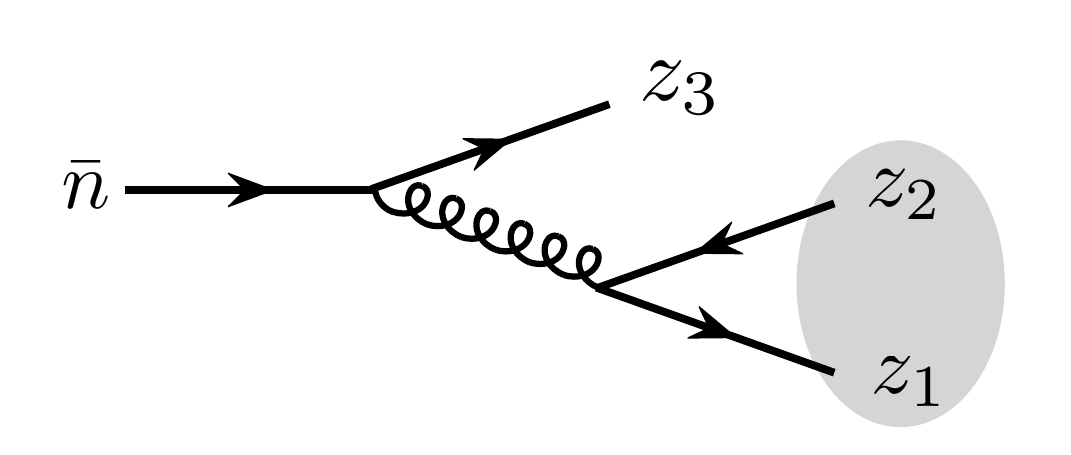}
\caption{The two OPE channels for the three-point correlator of detector operators. The equivalence of these decompositions leads to crossing equations. Figure from \cite{Chen:2022jhb}.
}
\label{fig:block_crossing}
\end{figure}

Equipped with this technology, we can begin to study correlation functions of light-ray operators through a much more sophisticated lens, applying many of the techniques used in the study of correlation functions of local operators. One of the interesting aspects of the multipoint energy correlators that we can study is the structure of the OPE coefficients for higher tranverse spin (higher twist) operators. Typically in QCD it is extremely challenging to study observables beyond leading twist. However, we will see that this can be achieved by studying crossing symmetry of the correlators, and ultimately the Lorentzian inversion formula  \cite{Caron-Huot:2017vep}, which will show that remarkably, the OPE data for higher transverse spin contributions is analytic in transverse spin! This generalizes the story of analyticity in Sec.~\ref{sec:analyticity} to transverse spin on the celestial sphere.

\begin{figure}
  \includegraphics[width=0.95\linewidth]{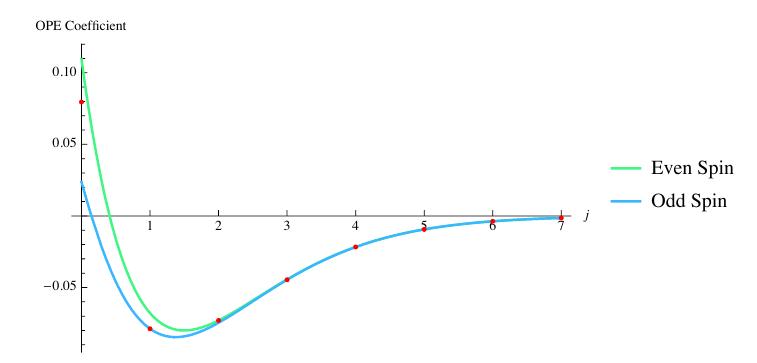}
  \caption{The even and odd-spin data for the three-point energy correlator in $\mathcal{N}=4$ sYM, as extracted using the Lorentzian inversion formula, compared with that from an explicit perturbative calculation. Here we observe a Regge intercept in transverse spin of $j=1$. Figure from \cite{Chen:2022jhb}.
  }
  \label{fig:block_analytic}
  \end{figure}

A particularly interesting feature is that multi-point correlators of energy flow operators obey a crossing equation, namely that expansions in different OPE channels must give the same result. This is shown in \Fig{fig:block_crossing} for the case of the three-point energy correlator. The solution of the crossing equations is at the core of the conformal bootstrap program. It will be extremely interesting to better understand it for correlation functions of detector operators. 

So far crossing has been studied \cite{Chen:2022jhb,Chang:2022ryc} in the analog of the light-cone bootstrap limit \cite{Komargodski:2012ek,Fitzpatrick:2012yx}. Here the existence of the OPE singularity in one channel indicates the presence of an infinite number of transverse spin operators propagating in the other channel. This can be solved in the large spin expansion. 

In perturbation theory, the energy correlator is bounded in the Regge limit, enabling the application of the  Lorentzian inversion \cite{Caron-Huot:2017vep, Simmons-Duffin:2017nub} to understand the spectrum of higher transverse spin (higher twist) light-ray operators. For a detailed review of this technique, and its application to the energy correlators, see \cite{Chen:2022jhb}. The Lorentzian inversion formula shows that the OPE data for the energy correlators is in fact an analytic function of the transverse spin! This is quite remarkable, as in QCD, it is typically extremely challenging to understand properties of higher twist contributions to event shape observables. 

The even and odd spin OPE data for the correlator are shown in \Fig{fig:block_analytic}. We find the fact that the OPE data is analytic in spin quite remarkable, hinting at much more unexplored structure. In particular it would be interesting to combine these constrains with an understanding of the function space to bootstrap the energy correlators. More generally, we believe that these tools are only just beginning to be exploited, and provide optimism for new avenues in real world calculations.

\subsubsection{The Back-to-Back Limit}\label{sec:b2b_general}

\begin{figure}
\includegraphics[width=0.6\linewidth]{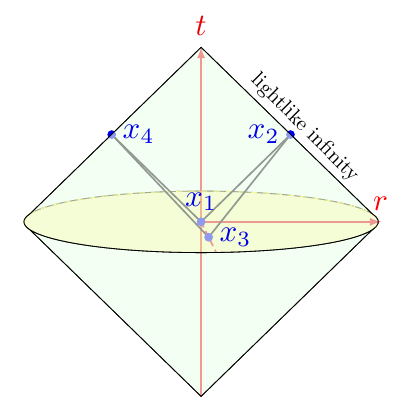}
\caption{In a gauge theory, the leading contribution to the back-to-back limit of the energy correlator, is described by a classical gauge flux propagating between the sources and the detectors. Figure from \cite{Chen:2023wah}.
}
\label{fig:wilson_line_OPE}
\end{figure}

In addition to the collinear limit, which we found was controlled by the low spin operators of the theory, the other kinematic limit we can consider is the so called ``back-to-back", or $z\to 1$ limit. This has generalizations to higher-point correlators, which we will discuss in the context of QCD.

While the light-ray OPE gives a universal understanding of the collinear limit in any conformal (or nearly conformal) theory, the dynamics in the back-to-back limit depend strongly on the particular theory. They are most interesting in the case of a conformal (or nearly conformal) \emph{gauge} theory, where they probe the high spin physics of Wilson lines. In more general CFTs much less is understood about the back-to-back limit. However, since our ultimate goal is to understand QCD, we will focus on the gauge theory case.

There are a number of different ways of understanding the physics of the back-to-back limit, but at its core, it is extremely elegant, and illustrated in \Fig{fig:wilson_line_OPE}. In the back-to-back limit, the integrated correlator defined the EEC localizes and form a polygon with all operators light like separated. This is a particular manifestation of relation between Wilson loops and correlators  \cite{Alday:2010zy}. In the case of the energy correlators, this is best understood using the Mellin amplitude technique~\cite{Belitsky:2013xxa,Korchemsky:2019nzm,Chen:2023wah}. In any theory with a conserved flux, the leading dynamics of this correlator in this limit is described by a classical flux propagating between these sources, i.e. a Wilson line. The Wilson loop picture is intimately connected to the large spin sector of local operators.

This relation allows one to write down an all orders formula for the leading singular behavior in the back-to-back limit in a conformal gauge theory
\begin{widetext}
\begin{align}
\text{EEC}(z)=\frac{H(a)}{8 (1-z)} \int\limits_0^\infty \df b\, b\, J_0(b) \exp \biggl[ -\frac{1}{2}\Gamma_{\text{cusp}}(a)\ln^2 \Bigl( \frac{e^{2\gamma_E}\, b^2}{4(1-z)} \Bigr)+2B_\delta(a) \ln \Bigl( \frac{e^{2\gamma_E}\, b^2}{4(1-z)} \Bigr)     \biggr]\,.
\end{align}
\end{widetext}
The double logarithms in this result are expressed in terms of the cusp anomalous dimension  \cite{Polyakov:1980ca,Korchemsky:1987wg}, $\Gamma_{\text{cusp}}(a)$, and $B_\delta(a)$, which govern the large spin behavior of the twist-2 anomalous dimensions \cite{Korchemsky:1988si,Korchemsky:1992xv,Belitsky:2006en}
\begin{align}\label{eq:large_spin_def}
\Delta-S= \Gamma_{\text{cusp}}(a) (\log S+\gamma_E) + B_\delta(a) +\mathcal{O}(1/S)
\end{align}
In planar $\mathcal{N}=4$ sYM, both the cusp anomalous dimension \cite{Eden:2006rx,Beisert:2006ez}, and  $B_\delta$ \cite{Freyhult:2007pz,Freyhult:2009my,Fioravanti:2009xt} can be computed exactly using integrability. The analogous expression in a non-conformal gauge theory, such as QCD, will be given in \Sec{sec:QCD}.

This formula can be arrived at from a number of different perspectives. It can be argued from general grounds in any conformal theory with a conserved flux \cite{Alday:2007mf}, it can be derived \cite{Korchemsky:2019nzm} using the duality between correlators and Wilson loops \cite{Alday:2010zy}, or it can be derived using effective field theory techniques \cite{Moult:2018jzp}.

Note that as compared to the case of the collinear limit, this result is not a pure power law. This can be interpreted as arising from the  resummation of contributions of twist-two operators with large spin propagating in different OPE channels.  This simple observation has important phenomenological consequences. One of the appealing features of the energy correlators in the collinear limit is that they isolate the scaling behavior of a single operator, leading to a simple power law. The presence of additional scales in QCD then imprint themselves in an extremely clear manner as a break in the power law. On the other hand, this is not the case for the energy correlator in the back-to-back limit. The double logarithmic behavior leads to a famous Sudakov peak, however, this peak is not associated with a physical scale in the theory, and is instead generated dynamically, through the resummation of contributions from multiple operators.

It is particularly interesting to explore higher powers in the expansion about the back-to-back limit, and to develop a systematic OPE onto Wilson lines dressed with local operators. Sub-leading power corrections were studied from the perspective of large spin perturbation theory in \cite{Chen:2023wah}, and from the perspective of effective field theory in \cite{Moult:2019vou}.

Note that as compared to the collinear limit, the back-to-back limit is sensitive to the source. This enters through the fact that we have used the particular configuration of Wilson lines in \Fig{fig:wilson_line_OPE}. This means that while the collinear limit will behave universally in different collider setups, the back-to-back limit will not. This fact can be used in collider physics experiments, particularly those involving hadronic particles in the collisions, where the back-to-back limit will enable sensitivity to properties of the colliding particles.

\subsubsection{Strong Coupling, Large Charge and Heavy States}\label{sec:heavy}

In addition to kinematic limits, we can also gain control over the energy correlators by considering specific states. So far, we have primarily focused on results of the energy correlators in perturbative states consisting of only a couple of excitations above the vacuum. However, it is interesting to develop an intuition for these observables in non-trivial states far from the vacuum and perturbative regime. Motivations for this comes from the study of heavy ion collisions and nuclear tomography. More generally, calculations of detector operators in different states provide intuition for their state dependence, which is useful for interpreting experimental measurements.

One interesting limit is the strong coupling limit in $\mathcal{N}=4$ sYM, which can be computed using the AdS/CFT correspondence~\cite{Maldacena:1997re,Witten:1998qj,Gubser:1998bc}. The $n$-point energy correlator was computed in $\cN=4$  by Hofman and Maldacena in \cite{Hofman:2008ar} in a strong coupling expansion, 
\begin{align}
    &\langle \mathcal{E}(\hat{n}_1) \cdots \mathcal{E}(\hat{n}_n)\rangle=\left(\frac{Q}{4\pi}\right)^n
    \big[
        1+\sum_{i<j}\frac{6\pi^2}{\lambda}\left[ (\hat{n}_i\cdot\hat{n}_j)^2-\frac{1}{3}\right] 
   \nn \\
    &+\frac{\beta}{\lambda^{3/2}} [\sum_{i<j<k} (\vec n_i \cdot \vec n_j)(\vec n_j \cdot \vec n_k )(\vec n_i \cdot \vec n_k)+\cdots ] \big] + \cO \left(\frac{1}{\lambda^{2}}\right)\,,
\end{align}
where $\beta$ is a non-vanishing numerical coefficient. As expected, that the result is uniform in angle, up to small corrections. However, surprisingly, the fluctuations, $\delta=(\mathcal{E}-\langle \mathcal{E} \rangle)/\langle \mathcal{E} \rangle$, scale like $1/\sqrt{\lambda}$, so that the three-point function of the fluctuations is not suppressed relative to the two-point function. The fluctuations are thus highly non-gaussian! This is in complete contrast to the weak coupling result. This strong coupling result has been computed in two distinct ways. First, in \cite{Hofman:2008ar}, it was computed by directly identifying the dual description of a light-ray operator as a shockwave \cite{tHooft:1987vrq,Dray:1984ha}. This perspective was studied in more detail in \cite{Kologlu:2019bco}, and used to derive ``super-convergence sum rules". Energy correlators have also been directly computed from the strong coupling expansion of the local four-point correlator \cite{Goncalves:2014ffa,Belitsky:2013ofa,Korchemsky:2015ssa}. The leading quantum gravity corrections in this limit has been explored in \cite{Chen:2024iuv}.

\begin{figure}
\includegraphics[width=0.955\linewidth]{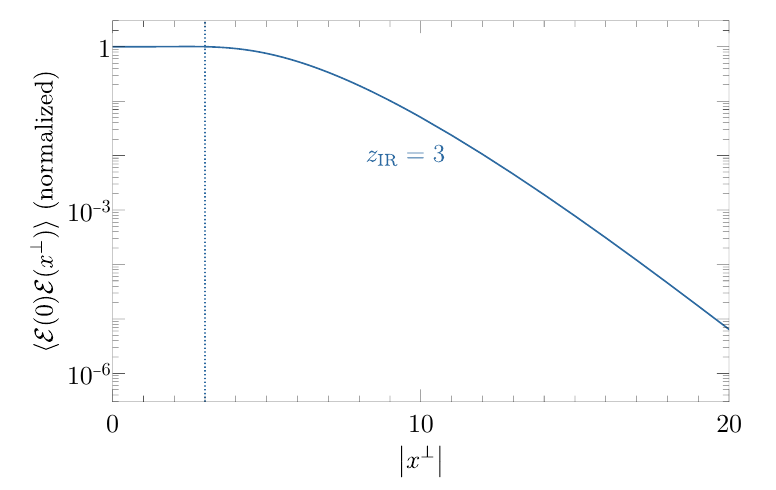}
\caption{The two-point correlator of light-ray operators as a function of separation $x_\perp$ in the transverse plane, as computed in a confining (hard wall) background. At short distances the result agrees with the uniform result of pure AdS, while at long distances the result is modified. Figure from \cite{Csaki:2024joe}.
}
\label{fig:hard_wall}
\end{figure}

Ref.  \cite{Hofman:2008ar} also provided a description of the small angle limit in terms of a world-sheet OPE. They predicted that the small angle limit is given by 
\begin{align}
 \mathcal{E}(\theta_1)  \mathcal{E}(\theta_2)  \sim \theta_{12}^{\Delta-6} \langle \mathcal{U}_{3-1} (\theta_2) \cdots \rangle
\end{align}
with $\Delta \sim \sqrt{2} \lambda^{1/4}+\cdots$,
and is controlled by certain analytically continued vertex operators
\begin{align}
( \partial_\alpha y^+ \partial_\alpha y^+)^{j/2} \delta(y^+) e^{ik\cdot y}\,.
\end{align}
Similar operators appear in \cite{Brower:2006ea} in the description of the Regge limit. It would be interesting to explore the world sheet OPE description further, and develop to the same level that the light-ray OPE is understood.

Recently, this calculation has been extended to other AdS like geometries, which incorporate a modification in the infrared to model the effects of confinement \cite{Csaki:2024zig,Csaki:2024joe}. In \Fig{fig:hard_wall} we show a result for the calculation of a matrix element of two light-ray operators separated by a distance $x_\perp$ in the transverse plane in a Randall-Sundrum geometry, with a hard wall cutoff in the IR \cite{Randall:1999ee}. We clearly see the modification arising from confinement at large distances. This result is interesting as a model for confinement in QCD, and it would be nice to explore similar calculations in more general backgrounds, for example Klebanov-Strassler \cite{Klebanov:2000hb}. The anomalous dimensions of twist-2 operators have been computed in Klebanov-Strassler theories \cite{Kruczenski:2003wz}. One difficulty, is that to model real world QCD, one needs to incorporate the effects of string breaking \cite{Csaki:2006ji}.

\begin{figure}
  \includegraphics[width=0.655\linewidth]{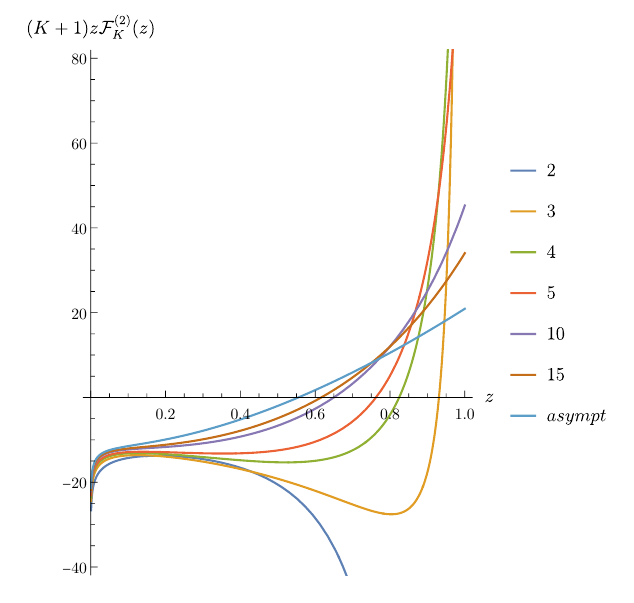}
  \caption{The two loop perturbative corrections to the energy correlator in $\mathcal{N}=4$ sYM, in the state produced by the operator $ \text{tr}[\phi^K(x)$, shown as a function of $K$. Figure from \cite{Chicherin:2023gxt}.
  }
  \label{fig:heavy}
  \end{figure}

Another approach to studying energy correlators in non-trivial states is to use heavy operators to source the state. In \cite{Chicherin:2023gxt}, the EEC was computed at weak coupling in heavy states created by the $1/2$ BPS operator $O_{\text{H}}(x)=\text{tr}[\phi^K(x)]$ with dimension $\Delta_{\text{H}}=K$. Remarkably, they were able to perform the calculation to two-loop in perturbation theory for generic $K$. The result is shown in \Fig{fig:heavy} as a function of $K$. A remarkable feature of this calculation, is that since it is performed at weak coupling, it provides a test of the light-ray OPE in wide variety of states.  In the OPE limit, \cite{Chicherin:2023gxt} found
\begin{align}
\langle \mathcal{E}(n_1) \mathcal{E}(n_2) \rangle \sim \frac{ \langle \mathbb{O}_i^{[3]} \rangle_{\text{H}}   }{\theta^{2-\gamma(3)}} + 1 \,,
\end{align}
with $\langle \mathbb{O}_i^{[3]} \rangle_{\text{H}} \lesssim 1/\Delta_{\text{H}}$ for large $K$. This reveals the critical angle $\theta_*^{2-\gamma(3)}\sim 1/\Delta_{\text{H}}$,  determining the radius of convergence of the light-ray OPE. For $\theta< \theta_*$ we observe the characteristic twist-2 scaling, while for $\theta> \theta_*$, the uniform scaling characteristic of a twist-4 operator dominates. This is an interesting lesson that we will use when studying energy correlators in heavy ion collisions: while the scaling behavior is always universal at small angles, the details of the state are imprinted in the relative OPE coefficients of the twist-2 and twist-4 operators (and in general the higher twist operators).

\begin{figure}
\includegraphics[width=0.35\linewidth]{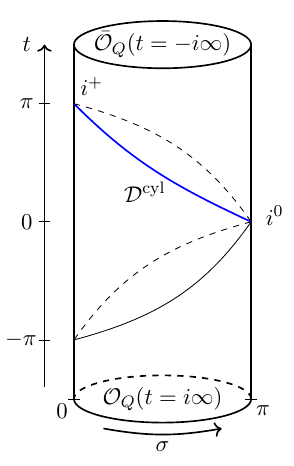}
  \includegraphics[width=0.755\linewidth]{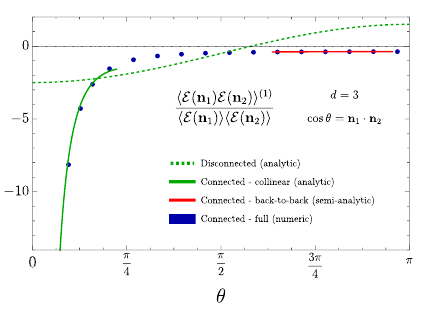}
\caption{Using an effective field theory description of large charge states, correlation functions of detector operators can be computed using a superfluid description. The two-point energy correlator computed such states shows an enhancement in the collinear limit corresponding to a ``sound jet". Figures from \cite{Cuomo:2025pjp}.
}
\label{fig:large_charge_cylinder}
\end{figure}

Finally, there has been significant progress in understanding states produced by large charge states using effective field theory techniques \cite{Hellerman:2015nra,Monin:2016jmo,Alvarez-Gaume:2016vff}. These are far from vacuum states, but can be described using superfluid EFT \cite{Son:2002zn}. This is illustrated in \Fig{fig:large_charge_cylinder}, where the large charge operator can be viewed as creating a specific superfluid state, in which the detector operators are measured.  Flux correlators in large charge states were first studied in  \cite{Firat:2023lbp}, who computed the leading term in the large $Q$ expansion. It was shown that multi-point correlators factorize to leading order in the $1/Q$ expansion, namely
\begin{align}
\langle \mathcal{E}(\theta_1) \mathcal{E}(\theta_2) \cdots \mathcal{E}(\theta_k) \rangle &=\langle \mathcal{E}(\theta_1) \rangle  \langle \mathcal{E}(\theta_2) \rangle  \cdots \langle \mathcal{E}(\theta_k) \rangle \nn \\
& +\mathcal{O}(1/Q).
\end{align}
The leading order correction in the large charge expansion, which exhibits non-trivial behavior, was computed in \cite{Cuomo:2025pjp}. The result is shown in \Fig{fig:large_charge_cylinder}. This result exhibits a number of remarkable features, including the presence of ``sound jets".

It will be interesting to further explore energy correlators in other non-trivial states. Of particular interest for phenomenology would be those that resemble heavy ion collisions, perhaps the ``plasma ball" states of \cite{Aharony:2005bm}. Interesting implications of the ANEC for thermal states were explored in \cite{Delacretaz:2018cfk}.

\subsection{Detector Operators and Their Correlators in Quantum Gravity}\label{sec:QG}

\begin{figure}
\includegraphics[width=0.655\linewidth]{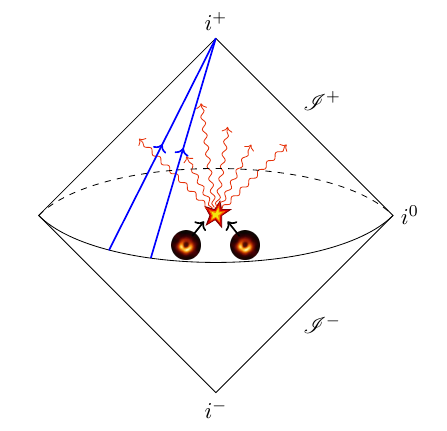}
\includegraphics[width=0.955\linewidth]{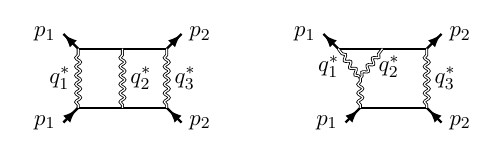}
\caption{Correlation functions of detector operators provide an interesting class of asymptotic observables in quantum gravity. Top Panel: Two detectors measure correlations in gravitons produced through the annihilation of two black holes. Bottom Panel: In perturbation theory, we can study an analogous process, namely the annihilation of two scalars into graviton radiation, whose amplitudes can be obtained as a t-channel cut of $2\to 2$ massive scattering. Figures from \cite{Herrmann:2024yai}.
}
\label{fig:penrose_BH}
\end{figure}

Although the focus of this review is on detector operators and their correlators in QFT, in this section we briefly mention detector operators and their correlators in perturbative quantum gravity.  In quantum gravity all observables are asymptotic observables. This suggests that detector correlators are the natural language for discussing observables in quantum gravity. Despite this, they remain essentially unexplored. 

There are a number of reasons to believe that detector correlators in gravity might exhibit remarkable features.  First, asymptotically flat spacetimes famously have an enhanced symmetry group, namely the the BMS group \cite{Bondi:1962px,Sachs:1962zza,Sachs:1962wk}. These have been found to have striking implications for gravitational scattering amplitudes \cite{Strominger:2013jfa,Strominger:2014pwa,Strominger:2017zoo}. What is their implication for detector correlators? Second, $\mathcal{N}=8$ supergravity is famously claimed to be the ``simplest quantum field theory" \cite{Arkani-Hamed:2008owk}. Does this lead to a simple perturbative structure of multi-point correlators? Third, color kinematics duality \cite{Bern:2008qj,Bern:2010ue,Bern:2019prr} famously relates gravitational and gauge theory amplitudes. What are its implications for detector correlators? While there has been progress in the understanding for scattering amplitudes and form factors \cite{Boels:2017ftb,Yang:2016ear,Boels:2012ew,Boels:2017skl}, it is of interest to extend it to more off-shell observables. Light-ray operators provide a link between off-shell observables (correlation functions of local operators) and on-shell observables, providing an interesting opportunity for furthering our understanding of the duality. All these remarkable features of asymptotic observables in gravity hint that detector correlator in gravity will have many interesting surprises.

From the other direction, the tools developed for the study of light-ray operators may prove useful for resolving open questions in gravity.  For example, in classical gravity, many observables used to characterize binary inspirals, such as the radiated energy \cite{Kosower:2018adc,Herrmann:2021lqe,Herrmann:2021tct}, waveforms \cite{Cristofoli:2021vyo,Herderschee:2023fxh,Bini:2024rsy}, and memory effects \cite{Elkhidir:2024izo} can be expressed as detector operators. As a concrete example, the gravitational waveform can be expressed as a generalized detector \cite{Korchemsky:2021okt}
\be
{\bold{L}}_\omega[h](\infty,z)=\int_{-\infty}^\infty d\alpha\, e^{i\alpha \omega}\, h(\alpha,z)\,,
\ee 
in terms of the asymptotic graviton field, $h$. This suggests that many of the tools developed to study light-ray operators in QFT could be used to improve our understanding of observables in classical gravity. Another area where we believe the language of detector operators could prove useful relates to Regge scattering in gravity. This is an old subject, but there remain many open questions related to Regge scattering in gravity \cite{Giddings:2011xs}. For recent progress in this direction, see \cite{Haring:2024wyz,Raj:2024xsi,Raj:2023irr,Rothstein:2024nlq}. In the case of CFTs, the light-ray perspective has unified the description of Regge scattering and detector operators, and we can hope that it can provide similar insights in gravity. All of these interesting directions motivate the exploration of detector correlators in perturbative quantum gravity. 

In QFT, the existence of concrete perturbative results for the energy correlators has been invaluable. Therefore, we can begin to explore detector correlators in perturbative quantum gravity by generating perturbative data. Multi-point detector correlators in perturbative gravity were considered in \cite{Gonzo:2023cnv,Gonzo:2020xza}. They showed that in the classical limit the detector correlators factorize $\langle \mathcal{E}(n_1) \mathcal{E}(n_2) \cdots \mathcal{E}(n_k) \rangle= \langle \mathcal{E}(n_1) \rangle \langle \mathcal{E}(n_2) \rangle \cdots \langle \mathcal{E}(n_k) \rangle$.

%

\begin{figure}
\includegraphics[width=0.955\linewidth]{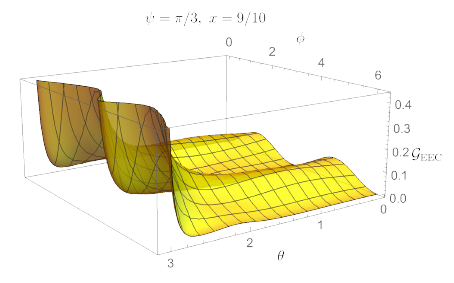}
\includegraphics[width=0.755\linewidth]{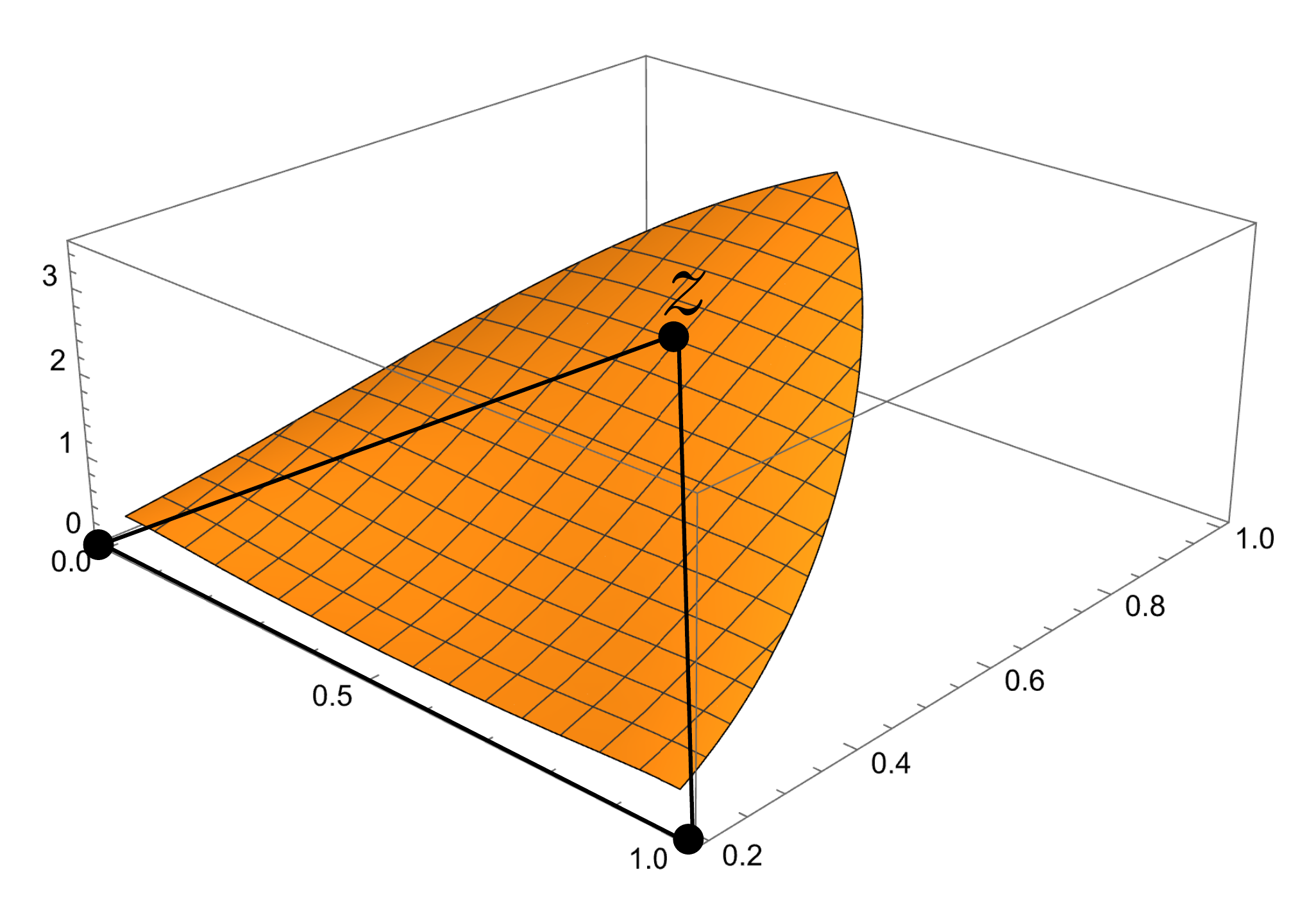}
\caption{Upper Panel: The full angle two-point correlator in pure Einstein gravity. Unlike in a gauge theory, it is regular in the collinear limit, and exhibits a single soft divergence in the back-to-back limit. The azimuthal dependence arises from the spin-2 nature of the graviton. Lower Panel: The three-point energy correlator in the collinear limit in $\mathcal{N}=8$ SUGRA, which is also regular in the collinear (OPE) limits. Figure from \cite{Herrmann:2024yai}.
}
\label{fig:GR_EEC_2point}
\end{figure}

To extend the study of detector correlators to perturbative quantum gravity, we consider linearized gravity $\mathfrak{g}_{\mu\nu} = \eta_{\mu\nu} + \kappa\, h_{\mu\nu}$, with $\kappa = \sqrt{32 \pi G_N}$. We define the asymptotic graviton field as
\begin{align}
\label{eq:covariant definition of the asymptotic shear}
h_{\mu\nu}(\alpha,z) =  \lim_{L\to \infty} L^{\Delta_h} \, h_{\mu\nu}(x+Lz)\, 
\end{align}
where $\Delta_h=\frac{d-2}{2}$, $\alpha = 2x\cdot z$ is the retarded detector time, and $z^\mu$ is a future pointing null-vector. The graviton energy detector can then be written in terms of the asymptotic metric field as
\begin{align}
\label{eq:GR_energy_detector_definition_covariant}
\mathcal{E}_h(z) = 2 \int_{-\infty}^\infty d\alpha : (\partial_\alpha h_{\mu\nu}(\alpha,z)) (\partial_\alpha h^{\mu\nu}(\alpha,z)) :\, .
\end{align}
This definition agrees \cite{Gonzo:2020xza} with $\alpha\alpha$ component of the effective stress tensor for gravitational waves \cite{Maggiore:2007ulw}
\begin{align}
T^{\rm eff, GW}_{\mu\nu} = \langle \partial_\mu h^{\alpha \beta} \partial_\nu h_{\alpha \beta} \rangle \,,
\end{align}
where $\langle\cdot\rangle$ denotes the Isaacson averaging prescription \cite{Isaacson:1968hbi,Isaacson:1968zza}. In this review we will focus on the case of the energy detector, but the leading trajectory of detector operators was constructed in \cite{Herrmann:2024yai}.

We can now consider correlation functions of these detector operators in different states. In the case of QFT, we were able to compute such detector correlators starting from either correlation functions of local operators, or using perturbative form factors. Neither of these exist in gravity. Instead, we must compute detector correlators in scattering states. An illustration of this is shown in \Fig{fig:penrose_BH}, where we imagine the annihilation of two black holes into a complicated multi-graviton state, whose correlations we measure with detector operators. 

In perturbation theory we can consider a simpler process, namely the annihilation of two massive scalars into N-gravitons (or two gravitons into N-gravitons). The calculation of the energy correlators requires the calculation of the spin-summed squared amplitude for the $2\to N$ process. This can be efficiently computed by considering cuts of the $2\to 2$ amplitude. A subset of the relevant diagrams required for the calculation of $2\to 3$ process are shown in \Fig{fig:penrose_BH}.

We can consider the specific case of a scalar minimally coupled to Einstein gravity,
\begin{align}
\label{eq:grav_action}
S_{{\rm EH} + \Phi}= \int d^dx \sqrt{-\mathfrak{g}} \left( 
      \frac{1}{16 \pi G_N} R 
    + \frac{1}{2} \Phi\left( \Box - m^2 \right)  \Phi
\right)\,.
\end{align}
The relevant integrands can be computed using the color-kinematics duality \cite{Bern:2008qj,Bern:2010ue} and the associated gauge theory amplitudes \cite{Edison:2020ehu}. In \Fig{fig:GR_EEC_2point} we show the two-point correlator in Einstein gravity \cite{Herrmann:2024yai}. It is finite in the collinear limit, due to the absence of collinear singularities in gravity \cite{Akhoury:2011kq}, and has a soft divergence in the back-to-back limit. The azimuthal dependence arises from the spin of the graviton. 

\begin{figure*}[t!]
  \includegraphics[width=0.6\linewidth]{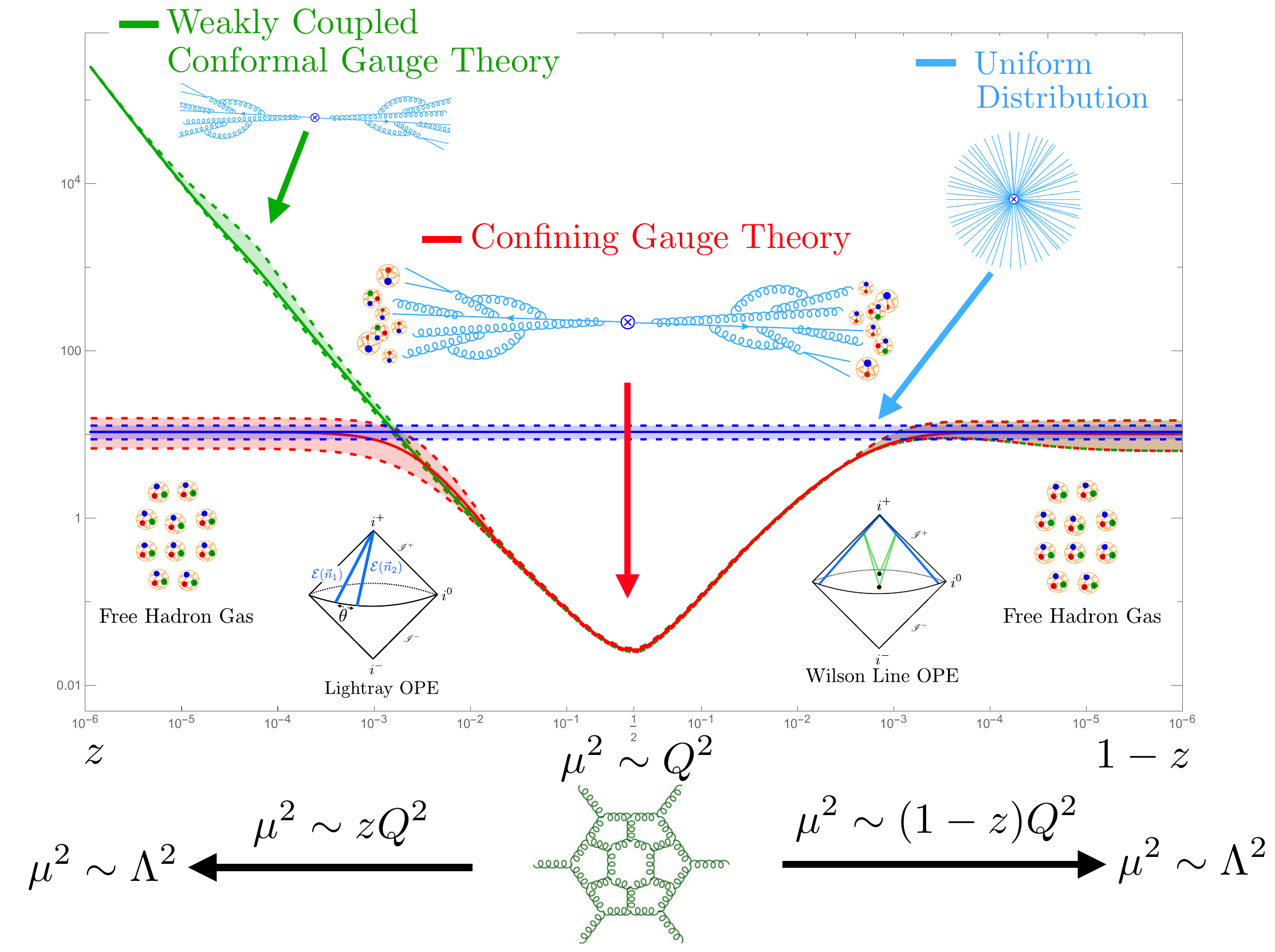}
  \caption{The rich structure of QCD, as exemplified by the two-point energy correlator. Starting from the UV, probed at a scale $\mu^2\sim Q^2$, the two-point correlator probes two distinct RG flows to the IR: one described by the light-ray OPE and low spin operators (collinear physics), and one described by the Wilson loop OPE and high spin operators. Both end with a free hadron gas, which exhibits a uniform scaling behavior. Results for the two-point energy correlator in a uniform energy distribution, and a conformal gauge theory, are shown for comparison. Note that we have chosen to normalize the value of the value of the energy correlator in the back-to-back region to be the same for illustration. In reality, the distribution is constrained by the sum rule, and will have different heights in these different theories. The height of the plateaus in the collinear and back-to-back limit are also generically distinct. However, for the specific case of QCD at the Z-pole, they are approximately equal, so we have used this case for illustrative purposes.
  }
  \label{fig:QCD_schematic}
\end{figure*}

This approach to the calculation of the energy correlators in perturbative gravity is efficient due to the availability of high loop perturbative data, particularly in the case of $\mathcal{N}=8$ super gravity.  The $2\to 2$ scattering amplitudes in  $\mathcal{N}=8$ are known to five loops \cite{Bern:2008pv,Bern:2009kd,Bern:2010tq,Bern:2012uc}, allowing for the efficient calculation of the integrand for multi-point energy correlators. As an illustration, in the lower panel of \Fig{fig:GR_EEC_2point} we show the collinear limit of the three-point energy correlator in $\mathcal{N}=8$. The result is expressed in terms of the same functions as in $\mathcal{N}=4$ super Yang-Mills, however, as compared to the gauge theory case, it is completely regular in all OPE limits, as can be seen in  \Fig{fig:GR_EEC_2point}.

Beyond the simple perturbative states considered in this section, a tantalizing target is the calculation of detector operators in non-trivial gravitational states. Beyond the case of colliders, the other most famous example of an energy distribution is Hawking radiation \cite{Hawking:1975vcx}. Would it be possible to calculate energy correlators in Hawking radiation? For some recent progress in understanding Hawking radiation from an amplitudes perspective, see \cite{Aoude:2024sve}.

These results of this section provide only a glimpse into the relatively unexplored subject of detector correlators in perturbative quantum gravity. We hope that these topics are explored more in the future, so that a unified language in both field theory and gravity can be achieved.

\section{The Theory of Energy Correlators in QCD}\label{sec:QCD}

In this section we discuss in more detail aspects of the energy correlator observables in the specific theory of QCD. Our goals in this section are two-fold. On the one hand, we wish to emphasize at a technical level how improvement in our understanding of detector operators and their correlators are improving our understanding of QCD. On the other hand, we wish to emphasize how particular aspects of QCD manifest in the correlators. This intuition is essential for understanding experimental measurements of the energy correlators, and for understanding in which kinematic regimes they can be understood using perturbative techniques.

The primary differences between QCD and the conformal theories discussed previously, are the $\beta$ function in the perturbative regime, the confinement transition, as well as the presence of heavy quark masses. We will illustrate all of these within the context of the energy correlators, and emphasize how these enable the study of different detector operators.

Our control over QCD is best in the perturbative regime, and this is where much of the progress in understanding detector operators has occurred. However, there has also been progress understanding aspects of the energy correlators non-perturbatively. Quite excitingly, in specific kinematic limits field theoretic objects appearing in the description of the energy correlators can be computed on the lattice, giving us an insight into non-perturbative aspects of QCD through the lens of the energy correlators. The lessons learned here will generalize to understanding more general detector correlators.

An illustrative representation of the two-point energy correlator in different theories is shown in \Fig{fig:QCD_schematic}. Here we show in blue, the result of the energy correlator in a free scalar theory, in a state produced by a $\phi^k$ operator with $k\to \infty$, producing a uniform energy distribution. This is compared with the distribution for a conformal gauge theory in green, and finally, a confining gauge theory, such as QCD, in red. These highlight the differences, and similarities between the correlators in the different theories, which we would like to understand.  To understand the distribution in QCD, we will begin with a discussion of the Regge trajectory in perturbative QCD, highlighting the operators present in the theory. We will then describe the collinear and back-to-back limits of the two-point correlator, showing that they probe low spin and high spin operators, respectively.  We will emphasize that in both cases, the energy correlator is probing the infrared of the theory, where ultimately we have confinement, and sensitivity to the IR dynamics of the theory. This is particularly clear in the collinear limit, where we see a large difference between the distributions in \Fig{fig:QCD_schematic}, arising from the presence of the confinement scale.  We will then discuss how these effects can systematically be treated, beginning with quark masses, which provide a perturbatively calculable modification of the infrared, and then transitioning to a description of genuine non-perturbative effects.

Our ability to accurately describe the energy correlators at high orders in perturbation theory relies on the knowledge of different anomalous dimensions to high loop orders. We provide a summary of the knowledge of anomalous dimensions in QCD in \Sec{sec:QCD_CF} along with our discussion of the Chew-Frautschi plot in perturbative QCD.

\subsection{The Chew-Frautschi Plot in Perturbative QCD} \label{sec:QCD_CF}

The Chew-Frautschi plot, whose theoretical underpinnings were reviewed in \Sec{sec:lightray}, provides one of the most important connections between CFT and QCD. In this section, we strengthen this connection, discussing the Chew-Frautschi plot in the specific case of QCD, as well as relating some of the language used to discuss the Chew-Frautschi plot in \Sec{sec:lightray} to the language more commonly used in QCD. For simplicity, we shall consider a pure Yang-Mills theory in this subsection. 

In collider physics, two of the most important quantities are the space-like and time-like DGLAP kernels, $P_S(z)$ and $P_T(z)$, which govern the evolution of parton distribution functions (PDFs) in the initial state and fragmentation functions (FFs) in the final state, respectively. Since we consider a pure Yang-Mills theory for simplicity, we have ignored the flavor label for the splitting functions.

There is a long history of attempts to derive relationships between the space-like and time-like splitting functions, motivated by a crossing symmetry between initial and final state processes. At tree level, the splitting function satisfy the Gribov-Lipatov relation \cite{Gribov:1972ri}
\begin{align}\label{eq:gribov}
P_T(z) = P_S(z)\,,
\end{align}
and the
Drell-Levy-Yan relation \cite{Drell:1969wb,Drell:1969jm}
\begin{align}
P_T(z) = -z P_S(1/z)\,.
\end{align}
These have been extensively studied in the QCD literature, and found to be violated at loop level \cite{Curci:1980uw,Furmanski:1980cm,Floratos:1981hs,Floratos:1980hk,Floratos:1980hm,Stratmann:2001pt,Stratmann:1996hn,Blumlein:2000wh}. By properly understanding the space-like and time-like anomalous dimensions as properties of the Chew-Frautschi plot of QCD, we will be able to understand the correct quantum generalizations of these statements.

In \Fig{fig:regge} we discussed the structure of the Chew-Frautschi plot in a general CFT, highlighting the space-like and time-like anomalous dimensions as horizontal and vertical shifts of operators relative to their positions on the trajectory of the free theory. Here we show how these are related to the space-like and time-like splitting functions in QCD.

In the case of the space-like splitting functions, which govern the evolution of PDFs, it has been well known since their introduction, that moments of PDF string operators collapse them to matrix elements of local twist-$2$ operators. Indeed this enables moments of the PDFs to be computed on the lattice, since they are in fact Euclidean \cite{Martinelli:1987zd,Kronfeld:1984zv,Detmold:2005gg}. This also implies that the moments of the space-like splitting functions
\begin{equation}
  \gamma_S(J) = - \int_0^1 \! dz \, z^{J-1} P_S(z) \,,
\end{equation}
are precisely the anomalous dimensions of the twist operators. We have made implicit the dependence on running coupling $\alpha_s(\mu)$ on both side of the equation.  In terms of the Chew-Frautschi plot these space-like anomalous dimensions measure the horizontal deviation (shift in dimension) of the leading Regge trajectory.

Similarly, one can define a time-like anomalous dimension from the the time-like splitting function,
\begin{equation}
  \gamma_T(J) = - \int_0^1 \! dz \, z^{J-1} P_T(z) \,.
  \label{eq:timelike_ad}
\end{equation} 
To connect this time-like anomalous dimension with the vertical shift in the Chew-Frautschi plot illustrated in \Fig{fig:regge} , we can consider the one-point function of detector operators $\langle \Omega | O^{\dagger}(-q) {\cal E}_{J_L} O(q) | \Omega \rangle$ in a protected scalar source. Using the on-shell representation for the correlation function of detector operators, we can write the matrix element as
\begin{equation}
\langle \Omega | O^{\dagger}(-q) {\cal E}_{J_L} O(q) | \Omega \rangle = \frac{1}{4 \pi}\sum_{a X'} \int d\sigma_{\Phi \to aX'} E_a^{-2 - J_L} \,,
\label{eq:onepoint_partonic}
\end{equation}
where $a$ and $X'$ are partonic states. In perturbative QCD, the quantity on the RHS is also known as the single-inclusive hard annihilation coefficient. For generic $J_L$, the partonic cross section in \eqref{eq:onepoint_partonic} is not IRC safe and requires a renormalization factor $Z_{J_L}$ to render it finite. Since $O(q)$ is a protected operator, the same renormalization factor is responsible for the renormalization of the detector operator ${\cal D}_{J_L}$, namely it is related to the quantum corrections to $\Delta_L$ for the detector operator. The time-like anomalous dimension in \eqref{eq:timelike_ad} is related to the renormalization constant as
\begin{equation}
\frac{d \ln Z_{ J_L}}{d\ln \mu} \propto \gamma_T(-1 - J_L) \,.
\end{equation}
We have therefore related the standard language of QCD, namely the space-like and time-like splitting functions, to the general language of used in \Sec{sec:lightray} and \Fig{fig:regge}. One aspect that this relation makes immediately clear, is why time-like anomalous dimensions are not the anomalous dimension of a local operator in QCD. By reciprocity, the time-like anomalous dimension is related to the space-like anomalous dimension of a non-integer spin operator, which is an intrinsically non-local light-ray operator. This makes it much harder to understanding fragmentation on the lattice, as compared to parton distribution functions.

Having understood the QCD space-like and time-like anomalous dimensions in the more general context of a Chew-Frautschi plot, we can now exploit its general structure to learn about QCD. One of the important features highlighted in  \Fig{fig:regge} was the reciprocity relation between time-like and space-like anomalous dimensions, which provides the proper quantum generalization of Eq. \ref{eq:gribov}. This relationship is particularly fascinating, since  the time-like anomalous dimension governs the scale evolution of the single-inclusive hard annihilation coefficient, and therefore the partonic fragmentation function through factorization \cite{Collins:1981uw}, while the space-like anomalous dimension is related to the evolution of the parton densities. The geometric relation in the Chew-Frautschi plot therefore relates two seemingly distinct physical processes in an interesting way. As mentioned, while such relations had been extensively explored in the long history of study of PDFs and FFs in QCD, and the GLBK reciprocity relation in \eqref{eq:GLBK} was discovered for the non-singlet parton evolution~\cite{Dokshitzer:2005bf,Basso:2006nk,Dokshitzer:2006nm}, this general understanding puts it on firm theoretical footing as a general consequence of the analyticity in spin of light-ray operators, and should hopefully allow for its generalization. We can also understand that the appropriate quantum generalization of the Drell-Levy-Yan relation is the shadow symmetry \cite{Kravchuk:2018htv} of the Regge trajectory, illustrated by its symmetry about the y-axis in \Fig{fig:regge}.

Such relations are of concrete computational use in perturbative QCD, since it is easier to compute the space-like anomalous dimensions. For example, they have been used to determine the large $x$ non-singlet time-like splitting functions in higher order perturbation theory~\cite{Moch:2017uml}. In the singlet part of QCD, the splitting functions contains mixing between quark and gluon flavor, and is actually a matrix. In this case, the GLBK-like relation has also been discovered for the eigenvalues of the splitting matrix \cite{Chen:2020uvt}. This enabled the unambiguous calculation of the three-loop time-like anomalous dimensions in QCD  \cite{Chen:2020uvt}. We hope to see further investigations of relations between time-like and space-like processes in QCD.

While we have focused on the conceptual underpinnings of the Chew-Frautschi plot in QCD, the quantitative calculation of components of the twist-2 Regge trajectory in QCD has a long history. These quantities play a critical role in our ability to precisely describe the scaling laws discussed in this review. 

The Dokshitzer-Gribov-Lipatov-Altarelli-Parisi (DGLAP) anomalous dimensions were computed at one loop in \cite{Gribov:1972ri,Dokshitzer:1977sg,Altarelli:1977zs}, at two-loops in \cite{Curci:1980uw,Furmanski:1980cm}, at three-loops in \cite{Moch:2004pa,Mitov:2006ic}. Low moments and phenomenological approximations for the four loop results are known \cite{Ruijl:2016pkm,Moch:2017uml,Herzog:2018kwj,Moch:2018wjh,Falcioni:2023luc,Falcioni:2023vqq,Falcioni:2023tzp,Moch:2023tdj,Falcioni:2024xyt,Gehrmann:2023cqm,Gehrmann:2023iah,Gehrmann:2023ksf}, and the calculation will be completed in the near future. The three-loop time-like anomalous dimensions were computed in \cite{Mitov:2006ic,Mitov:2006wy,Chen:2020uvt}.

Since QCD is a gauge theory, we are also interested in the large spin limit of the twist-2 trajectory \cite{Korchemsky:1988si,Korchemsky:1992xv,Belitsky:2006en}
\begin{align}
\Delta-S= \Gamma_{\text{cusp}}(a) (\log S+\gamma_E) + B_\delta(a) +\mathcal{O}(1/S)\,.
\end{align}
The cusp anomalous dimension in QCD is known analytically to four loops \cite{Moch:2018wjh,Moch:2017uml,Davies:2016jie,Henn:2019swt}, and approximately at 5 loops \cite{Herzog:2018kwj}. For a detailed discussion of the status and relation between different anomalous dimensions see \cite{Moult:2022xzt,Falcioni:2019nxk,Dixon:2017nat}.

Although we will not focus on it in this review, we should also emphasize that there has been a tremendous effort to understand the Pomeron intercept in QCD using the BFKL theory \cite{Fadin:1975cb,Lipatov:1976zz,Kuraev:1976ge,Kuraev:1977fs,Balitsky:1978ic,Lipatov:1985uk}. After a tour-de-force calculation, the NLO Pomeron intercept in QCD was computed \cite{Fadin:1998py,Camici:1997ij,Ciafaloni:1998kx,Ciafaloni:1998gs,DelDuca:1998cx,DelDuca:1998kx,Fadin:1997hr,Fadin:1996nw,Fadin:1996yv,Fadin:1994fj,Fadin:1996tb,Fadin:1989kf,Fadin:1993wh,Fadin:1997zv,Catani:1990xk,Catani:1990eg,DelDuca:1995ki,DelDuca:1996nom}.

Finally, 50 years of calculations have provided the QCD $\beta$ function to five loops
\cite{Gross:1973id,Politzer:1973fx,Jones:1974mm,Caswell:1974gg,Egorian:1978zx,Tarasov:1980au,Larin:1993tp,vanRitbergen:1997va,Czakon:2004bu,Baikov:2016tgj}.

Combined, these results for these anomalous dimensions provide the backbone for our ability to quantitatively describe collider physics in QCD, and will appear frequently throughout this review.

\subsection{The Physics of the Collinear Limit} \label{sec:EC_coll}

In this section we discuss the collinear limit of the energy correlator in QCD. This limit is interesting both phenomenologically, since it can be studied at collider experiments, and as a theoretical laboratory. The small angle asymptotics of the energy correlator can be studied using several different theoretical frameworks, including traditional QCD factorization, and the light-ray OPE. The equality of the results in the two cases have provided significant insight into QCD factorization, with broad implications. Furthermore, having multiple formulations enables us to relate the results to extract ingredients previously computed in perturbative QCD.

\begin{figure}
  \includegraphics[width=0.755\linewidth]{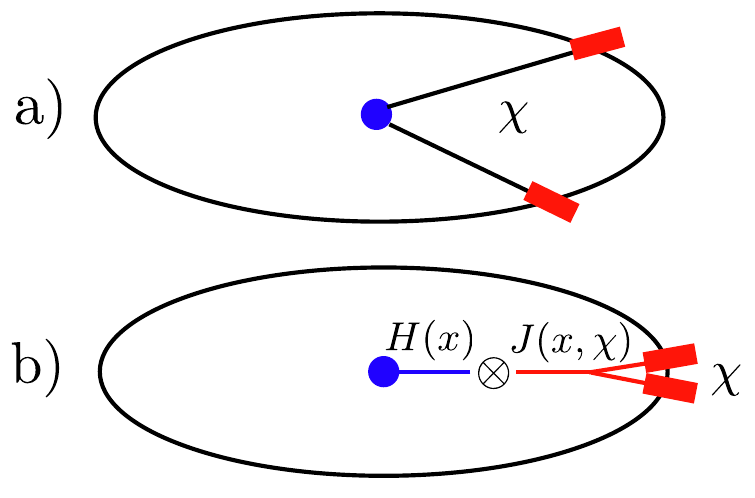}
  \caption{The factorization of the energy correlator in the collinear (small angle) limit. As $\chi\to 0$ we can factorize at leading twist onto a single partonic state, enabling the cross section to be expressed in terms of an inclusive single particle hard function and a jet function. Figure from \cite{Dixon:2019uzg}.
  }
  \label{fig:collinear_fact}
\end{figure}

We begin by illustrating the treatment of the collinear limit of the energy correlator within a factorization approach using effective field theory and renormalization group techniques.  A factorization theorem for the leading power asymptotics of the energy correlator observable in QCD was presented in \cite{Dixon:2019uzg}. It is most naturally expressed in terms of the cumulant of the energy correlator, defined as
\begin{align}
  \label{eq:cumulant}
\Sigma\Bigl(z, \ln \frac{Q^2}{\mu^2} , \mu\Bigr)\
\equiv\ \frac{1}{\sigma_0}
\int^z_0 dz' \, 
\frac{d\sigma}{dz} \Bigl(z', \ln\frac{Q^2}{\mu^2}, \mu\Bigr) \,,
\end{align}
where $z = (1 - \cos\chi)/2$.
We can factorize onto intermediate time-like states. In the small angle limit, and at weak coupling, the single particle states will dominate, allowing us to express the cumulant of the energy correlator in terms of a hard function, $H$, which describes the production of single particle states, and a jet function, $J$, which describes the measurement of the energy correlator in the jet seeded by the single-particle state. This is illustrated in \Fig{fig:collinear_fact}, and can be expressed mathematically as
\begin{align}
\label{eq:fact}
 \Sigma(z, \ln \frac{Q^2}{\mu^2} , \mu)
= \int_0^1 dx\, x^2 \vec{J} (\ln\frac{z x^2 Q^2}{\mu^2},\mu)
   \cdot  \vec{H} (x,\frac{Q^2}{\mu^2},\mu) \,.
\end{align}
The hard function depends on the source, or state, in which the energy correlator is studied, while the jet function is universal. This factorization is particularly convenient, since these hard functions are well studied objects in perturbative QCD. Indeed, for the case of $e^+e^-$, the hard function is now known to three-loops \cite{He:2025hin}, and for many processes at the LHC, it is known to two-loops \cite{Czakon:2021ohs,Czakon:2024tjr}.

A particularly beautiful aspect of this factorization theorem, and of the energy correlators in the collinear limit, is that the measurement of the energy correlators introduces a \emph{single} physical scale into the problem, namely the scale $Q\sqrt z$. This is made clear by the above factorization theorem. This factorization theorem therefore makes precise that we are probing the theory at the scale $Q\sqrt z$, and it also allows us to cleanly measure the evolution of the jet function as a function of the scale in a  clean manner. It is this elegant scaling that is observed in the experimental plots of the collinear limit.

The hard function obeys a renormalization group evolution equation, which is referred to as the ``time-like DGLAP equation"
\begin{align}
  \label{eq:hardRG}
  \frac{d \vec{H} (x, \frac{Q^2}{\mu^2},\mu)}{d \ln \mu^2} = - \int_x^1 \frac{dy}{y} \widehat P_T(y,\mu) \cdot \vec{H}\left(\frac{x}{y}, \frac{Q^2}{\mu^2},\mu\right) \,,
\end{align}
where $\widehat P_T$ is the singlet time-like splitting kernel matrix
\begin{align}
  \label{eq:splitK}
  \widehat P_T = 
  \begin{pmatrix}
    P_{qq} &  P_{qg}
\\
    P_{gq} & P_{gg}
  \end{pmatrix} \,.
\end{align}
The time-like anomalous dimensions are expressed in terms of these splitting kernels by a Mellin transform
\begin{equation}
\label{eq:gammaPTrelation}
\gamma_T(N)\ \equiv\ - \int_0^1 dy \, y^{N-1} \, \widehat{P}_T(y).
\end{equation}
Renormalization group invariance then implies that the jet function obeys the RG equation
\begin{align}
  \label{eq:jetRG}
\frac{d \vec{J}(\ln\frac{z Q^2}{\mu^2}, \mu) }{d \ln \mu^2} = \int_0^1 dy\, y^2 \vec{J} (\ln\frac{z y^2 Q^2}{\mu^2}, \mu) \cdot \widehat P_T(y,\mu) \,.
\end{align}

It is a simple exercise to see how this factorization theorem reproduces the famous leading logarithmic results of the jet calculus \cite{Konishi:1979cb}. Indeed, solving the renormalization group evolution equations for the jet function at LL accuracy, one finds
\begin{align}
  \label{eq:jetLLresum}
  \vec{J}_{\rm LL}(\ln\frac{z Q^2}{\mu^2}, \mu) = (1,1) \cdot V \left[\left(\frac{\alpha_s(\sqrt{z} Q)}{\alpha_s(\mu)} \right)^{-\frac{\vec{\gamma}_T^{(0)}}{\beta_0}}\right]_D \! V^{-1} \,,
\end{align}
where $\beta_0 = (11C_A-2n_f)/3$, $V$ is the matrix that diagonalizes $\gamma_T^{(0)}$, and $\vec{\gamma}_T^{(0)}$ is the diagonal vector of the diagonalized matrix.  Therefore, we see, as expected, that in addition to the time-like anomalous dimension, we also need the $\beta$ function of the theory. This result can  systematically be extended to higher perturbative orders.

This formulation generalizes straightforwardly to other collider systems by changing the hard function, and has been key to achieving a precision understanding of the collinear limit of the energy correlators at hadronic collider experiments. This formula also emphasizes an important aspect about the collinear limit of energy correlators, which will be reiterated numerous times. The strong coupling constant in Eq. \ref{eq:jetLLresum} is evaluated at the scale $\sqrt{z}Q$, emphasizing that the OPE is an infrared OPE, and that in the limit $z\to 0$ we are probing the infrared of the theory. This is a crucial difference between QCD and a conformal field theory. This also illustrates an important aspect of the energy correlators in QCD, namely that perturbative predictions, even incorporating resummation, are only reliable when $\sqrt{z}Q > \Lambda_{\text{QCD}}$. The factorization formula remains valid in the non-perturbative regime, however, we are no longer able to compute the jet function.

In data, we see that in the deep collinear limit of QCD we eventually transition to a uniform scaling behavior associated with free hadrons. This is not something that we can capture in QCD using perturbation theory. A calculable model where this turn over occurs is when there are massive quarks in the theory, which we will study in \Sec{sec:heavy_quarks}.

This formulation may seem quite distinct from that in terms of the operator product expansion. However, we can see that they are in fact the same in a conformal theory. In the case of a CFT, the only scale is $z Q^2$. We can then make a power law ansatz
\begin{align}
  \label{eq:Jneqfour}
  J(z Q^2,\mu) = C_J(\alpha_s) \left(\frac{z Q^2}{\mu^2} \right)^{\gamma_J^{{\cal N}=4}(\alpha_s)} \,,
\end{align}
where the anomalous dimension $\gamma^{{\cal N}=4}(\alpha_s)$ can be determined by substituting into the jet function evolution equation~\eqref{eq:jetRG}. Explicitly, using the definition~\eqref{eq:gammaPTrelation}, we find
\begin{align}
  \label{eq:gammaNeqfour}
2 \gamma_J^{{\cal N}=4}(\alpha_s) = &\, - 2 \int_0^1 dy\, y^{2 + 2 \gamma_J^{{\cal N}=4}(\alpha_s)} P_{T, \text{uni.}}(y,\alpha_s)
\nn\\
=&\, 2 \gamma_T^{{\cal N}=4}(1 + 2 \gamma_J^{{\cal N}=4}, \alpha_s) \,,
\end{align}
where $\gamma_T^{{\cal N}=4}(N,\alpha_s)$ is the Mellin $N+2$ moment of the universal splitting kernel $P_{T, \text{uni.}}(x, \alpha_s)$. Note that in the $\cN=4$ case we use a shifted argument, since performing the sum $\sum_j \gamma_{j\phi}(N) = \sum_j \gamma_{j\lambda}(N) = \sum_j \gamma_{jg}(N) = \gamma_{T, \rm uni.}(N-2)$ shifts the argument by two units in Mellin space. Therefore, for the scalar $\cN=4$ universal anomalous dimension, although it is evaluated at $N=1$, we will still refer to it as the twist two spin three anomalous dimension.

\begin{figure}
\includegraphics[width=0.955\linewidth]{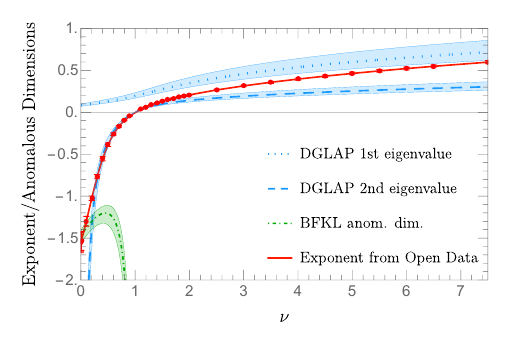}
\caption{Measurements of the scaling exponent of the $\nu$-point correlator using CMS open data, compared with theoretical predictions. Figure from \cite{Budhraja:2024tev}.
}
\label{fig:nu_correlator}
\end{figure}

We can combine Eq.~\eqref{eq:gammaNeqfour} with the reciprocity relation between time-like and space-like anomalous dimensions \cite{Mueller:1983js,Dokshitzer:2005bf,Marchesini:2006ax,Basso:2006nk,Dokshitzer:2006nm} discussed in \Sec{sec:lightray}. For the particular case of $\mathcal{N}=4$ this relationship reads,
\begin{align}\label{eq:reciprocity}
2 \gamma_S^{{\cal N}=4}(N , \alpha_s)=  2 \gamma_T^{{\cal N}=4}(N + 2 \gamma_S^{{\cal N}=4}, \alpha_s), 
\end{align}
which then implies
\begin{align}
 \gamma_J^{{\cal N}=4}(\alpha_s) = \gamma_S^{{\cal N}=4}(1,\alpha_s).
 \end{align}
 This provides a beautiful illustration of the equivalence of the different approaches. 
 
 In the case of a non-conformal theory, the $J=3$ selection rule in the light-ray OPE is broken by $\beta$ function corrections. This appears as a dependence on derivatives of the anomalous dimensions about $J=3$. These are naturally incorporated in the renormalization group approach, and can also be incorporated in the OPE approach \cite{Chen:2023zzh}. A detailed discussion of the relation between the two approaches can be found in \cite{Chen:2023zzh}.

This factorization formula can be generalized to $N$-point projected energy correlators, as
\begin{align}
\label{eq:fact_N}
 \Sigma(z, \ln \frac{Q^2}{\mu^2} , \mu)
= \int_0^1 dx\, x^N \vec{J} (\ln\frac{z x^2 Q^2}{\mu^2},\mu)
   \cdot  \vec{H} (x,\frac{Q^2}{\mu^2},\mu) \,.
\end{align}
Although we will not describe it in detail here, non-integer point correlators can be given an algorithmic definition \cite{Chen:2020vvp}, which can be measured experimentally. A study of the $\nu$ point correlators using CMS open data, is shown in \Fig{fig:nu_correlator} along with theoretical calculations.

\begin{figure}
\includegraphics[width=0.95\linewidth]{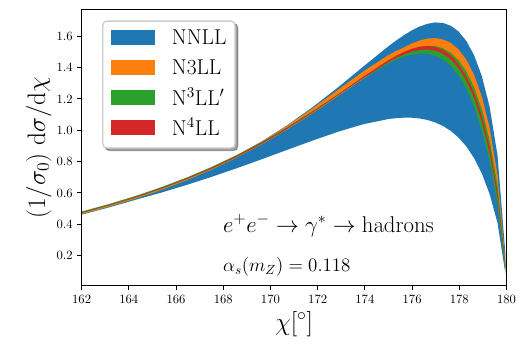}
\caption{A calculation of the EEC in the back-to-back limit at increasing perturbative orders. The back-to-back limit of the energy correlator is the most precisely computed event shape, achieving a remarkable N$^4$LL accuracy.  Figure from \cite{Duhr:2022yyp}.
}
\label{fig:N4LL}
\end{figure}

An interesting feature of viewing the energy correlators as a function of $\nu$, is that not only the anomalous dimensions, but also the jet function constants, are analytic functions of $\nu$, at least in perturbation theory. The one-loop $\nu$-point correlator was computed in $\mathcal{N}=4$ super Yang-Mills theory in \cite{Chen:2020vvp}, and takes the simple form
\begin{align}
  2^\nu J_1^{\cN = 4, [\nu]} = - 8 N_c (\Psi(\nu) + \gamma_E) \left( \frac{1}{\epsilon} - \ln \frac{x_L Q^2}{\mu^2} \right)
\nn\\
-4 N_c [ \pi^2 + 2 (\Psi(\nu) + \gamma_E)^2 - 6 \Psi'(\nu) ] + \mathcal{O}(\epsilon) \,.
  \label{eq:jet_Neq4}
\end{align}
Much like the twist-2 anomalous dimensions in $\mathcal{N}=4$ super Yang-Mills, the constant also exhibits universal transcendental weight. This has also been observed for the DIS structure constants in $\mathcal{N}=4$  \cite{Bianchi:2013sta}. It would be interesting to explore this in more detail to see if this could be used to be bootstrap the result to higher perturbative orders. To achieve this, it will also be important to understand the behavior in the asymptotic limits. An interesting example is the study of the $\nu \to \infty$ limit in \cite{Dai:2024wff}. It would also be intriguing to see if there is any connection with integrability, or to understand how integrability manifests in the structure of the $\nu$ point correlators.

\subsection{The Physics of the Back-to-Back Limit} \label{sec:EC_b2b}

Since QCD in the perturbative regime is a nearly conformal gauge theory, we expect that it exhibits double logarithmic asymptotics in the back-to-back limit, much like in the case of a weakly coupled conformal gauge theory discussed above \cite{Alday:2007mf}. The back-to-back limit therefore provides the opportunity to study the high spin limit, and the dynamics of Wilson lines experimentally~\cite{Alday:2007mf}. Due to the similarities with the conformal case presented above, we will here focus on the key differences, which arise from the perturbative running of the coupling, as well as from confinement.

An all orders factorization theorem describing the leading power asymptotics of the energy correlator in the back-to-back limit in a weakly coupled (not necessarily conformal) gauge theory was presented in  \cite{Moult:2018jzp}. It is formulated within SCET \cite{Bauer:2000ew,Bauer:2000yr,Bauer:2001ct,Bauer:2001yt,Rothstein:2016bsq,Beneke:2002ph}, using the rapidity renormalization group \cite{Chiu:2012ir,Chiu:2011qc}, as implemented with the regulator of \cite{Li:2016axz}. This factorization builds on the seminal work of \cite{Collins:1981uk,Collins:1981zc,Collins:1981va}. Resummation in the back-to-back limit has also been studied in other formalisms \cite{deFlorian:2004mp,Aglietti:2024xwv,Kardos:2018kqj,Tulipant:2017ybb}.

For the particular case of a form factor composed of two quark fields, it can be written as
\begin{widetext}
\begin{align}
  \label{eq:resformula}
\frac{d\sigma}{dz} = &\; \frac{1}{4} \int\limits_0^\infty db\, b
  J_0(bQ\sqrt{1-z})H(Q,\mu_h) j^q_\text{EEC}(b,b_0/b,Q) j^{\bar
  q}_\text{EEC}(b,b_0/b,Q) S_\text{EEC}( b,\mu_s, \nu_s) 
   \left(\frac{Q^2}{\nu_s^2}\right)^{\gamma^r_\text{EEC}(\alpha_s(b_0/b))}
\\
&\; \cdot
  \exp \left[ \int\limits_{\mu_s^2}^{\mu_h^2}
  \frac{d\bar{\mu}^2}{\bar{\mu}^2} \gcusp(\alpha_s(\bar \mu)) \ln
  \frac{b^2\bar{\mu}^2}{b_0^2}
+\int\limits_{\mu_h^2}^{b_0^2/b^2}\frac{d\bar{\mu}^2}{\bar{\mu}^2}
  \left(\gcusp(\alpha_s(\bar \mu)) \ln\frac{b^2 Q^2}{b_0^2} +
   \gamma^H (\alpha_s(\bar \mu)) \right) -
  \int\limits_{\mu_s^2}^{b_0^2/b^2}\frac{d\bar{\mu}^2}{\bar{\mu}^2}
   \gamma^s_\text{EEC} (\alpha_s(\bar \mu))  \right]\,. \nn
\end{align}
\end{widetext}
It is easy to check that assuming the coupling does not run, this result reduces to the case for a conformal theory above. This provides interesting relations on the anomalous dimensions appearing in the factorization theorem, and is discussed in detail in \cite{Moult:2022xzt}.

\begin{figure}
\hspace{0.6cm}  \includegraphics[width=0.75\linewidth]{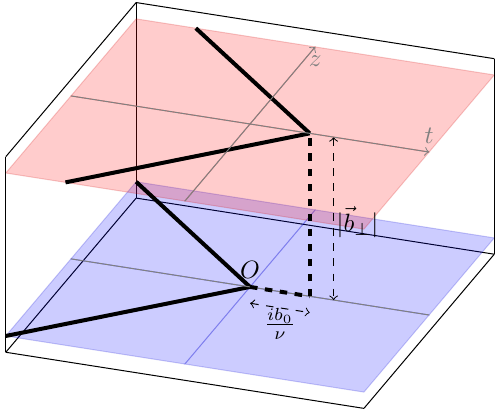}
  \includegraphics[width=0.89\linewidth]{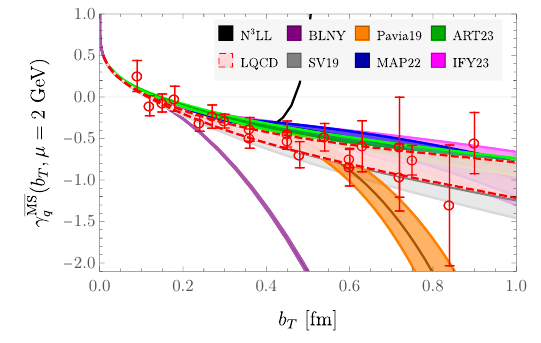}
  \caption{Top Panel: Cusped Wilson lines appearing in the description of the energy-energy correlator in the back-to-back limit. Lower Panel: The scale evolution of this Wilson line configuration is controlled by the Collins-Soper kernel, which can be computed using lattice QCD. Comparisons to phenomenological fits and perturbative calculations are also shown.  Figures from \cite{Moult:2018jzp} and \cite{Avkhadiev:2024mgd}. 
  }
  \label{fig:b2b_wilson}
  \end{figure}

It is interesting to contrast this factorization theorem with the factorization theorem describing the collinear limit of the energy correlator. A crucial difference is the appearance of two distinct functions sensitive to the scale of the measurement, namely the jet and the soft functions, and two scales, namely $Q$ and $Q\sqrt{1-z}$. The final result is expressed as a convolution of the dynamics at these two different scales. Alternatively, in the CFT interpretation this  resummation of contributions of twist-two operators with large spin propagating in different OPE channels. In either way, it is not determined by a single operator. It is this feature which leads to its non-power law behavior. While the study of the back-to-back limit of the energy correlators is extremely interesting, and enables access to universal features of gauge theories, the fact that it probes multiple physical scales at the same time makes it more difficult to use to study the presence of new physical scales. In the collinear limit, new physical scales imprint themselves as the breaking of a power law. In the back-to-back limit, the observable is not a power law, and so the manner in which new scales e.g. from nuclear physics, imprint themselves, is more non-trivial.

This factorization theorem has enabled extremely high precision calculations of the energy correlator in perturbative QCD. In particular, it has enabled the highest perturbative accuracy achieved for a QCD event shape, An illustration of the perturbative convergence is shown in \Fig{fig:N4LL}, showing the remarkable improvement in the perturbative uncertainty as higher order ingredients are included in the calculation. This calculation uses a number of state of the art perturbative ingredients including the form factor at three-loops \cite{Baikov:2009bg,Lee:2010cga,Gehrmann:2010ue}, the rapidity anomalous dimension at four loops \cite{Moult:2022xzt,Duhr:2022yyp}, and the three loop jet  \cite{Echevarria:2016scs,Luo:2019hmp,Luo:2019bmw} and soft functions \cite{Li:2016ctv}.

A much more drastic difference in the behavior of the energy correlators in QCD arises due to confinement. The double logarithmic asymptotics of the energy correlator is governed by a matrix element of Wilson lines, shown in \Fig{fig:b2b_wilson}. The same Wilson loop also appears in Drell-Yan process at small transverse momentum~\cite{Li:2016ctv}. Here, the separation between the two cusped Wilson lines, $b_\perp$, is conjugate to $1-z$, so that the energy correlators in the extreme $z\to 1$ limit probe the behavior of this Wilson line matrix element as $b_\perp$ increases. In the perturbative regime, there is a conserved gauge flux, and this matrix element of Wilson lines gives rise to double logarithmic asymptotics. String breaking via fermions is a non-perturbative effect, and is therefore suppressed in the UV. However, as $b_\perp$ is increased, confinement occurs, and we expect a drastically different behavior of this matrix element. Remarkably, this physics is accessible using the energy correlators!

It is conventional to split the anomalous dimension governing the energy correlators into a perturbative and a non-perturbative component
\begin{align}
    \gamma_\nu^q(b_\perp,\mu)
    &=
    \gamma_\nu^{q,\text{NP}}(b_\perp)
    +2\gamma^q_r[a_s(\mu_0)]\nn \\
    &
    \qquad \qquad -2\int_{\mu_0}^{\mu}\df\ln\mu^{\prime2}\,
    \gammacusp\bigl[a_s(\mu^\prime)\bigr]\,.
\end{align}
The perturbative rapidity anomalous dimension is known to 4 loops \cite{Davies:1984hs,Davies:1984sp,deFlorian:2000pr,Li:2016axz,Vladimirov:2016dll,Moult:2022xzt,Duhr:2022yyp}. Recently, it has become possible to compute this on the lattice \cite{Ji:2013dva,Lee:2013mka,Izubuchi:2018srq,Ji:2020ect}, which is being accurately pursued by numerous groups. Here we will show the results of  \cite{Avkhadiev:2024mgd,Avkhadiev:2023poz,Shanahan:2021tst,Shanahan:2020zxr,Shanahan:2019zcq}, but we refer the reader to the review of \cite{Ji:2020ect}  for a description of the techniques to compute it on the lattice, as well as a more complete set of references.

In the lower panel of \Fig{fig:b2b_wilson} we show a lattice extraction of this evolution kernel, compared with several phenomenological extractions. The lattice extraction is also compared with a perturbative calculation. We observe good agreement between the lattice and the perturbative result to $\sim 0.4$ fm, and the lattice result allows the knowledge of this evolution kernel to be extended to $\sim 1$ fm. We expect significant improvement in the ability to compute this quantity on the lattice in the coming years. We find it quite remarkable that using the energy correlators, we are able to make contact with lattice results in the description of final state energy flux.

\begin{figure}
\includegraphics[width=0.755\linewidth]{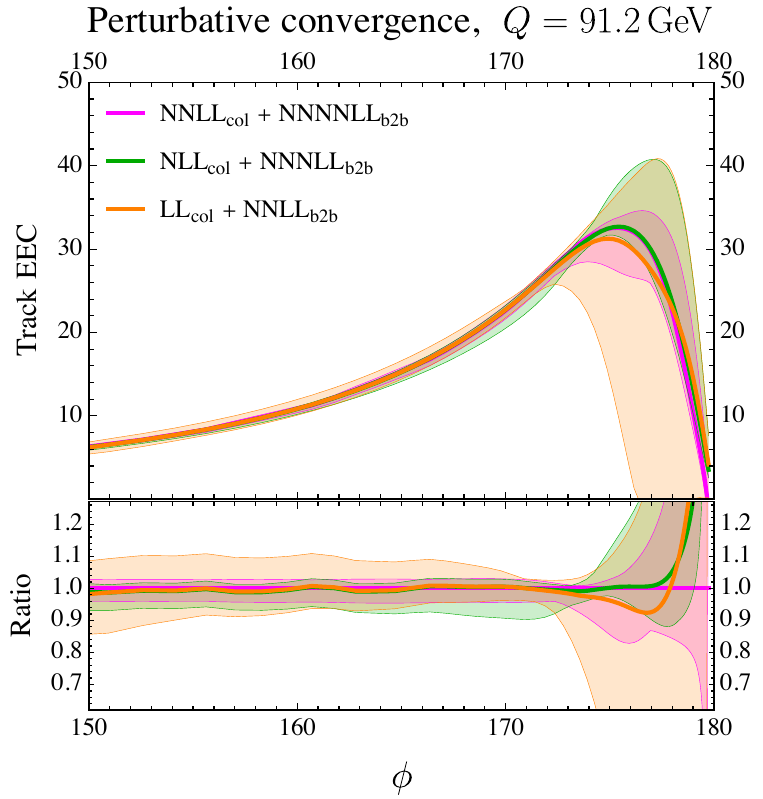}
\caption{The back-to-back limit of the energy correlator computed on charged particles (tracks), and incorporating non-perturbative corrections, as well as uncertainty estimates from the knowledge of these non-perturbative corrections. The effect of non-perturbative corrections grows as $1-z \to (\Lambda_{\text{QCD}}/Q)^2$. Figure from \cite{talk_Max}.
}
\label{fig:b2b_us}
\end{figure}

Using this knowledge of the non-perturbative corrections to the energy correlator, in \Fig{fig:b2b_us} we show a calculation of the back-to-back limit of the energy correlator, taking into account all state of the art perturbative ingredients, but also performing a complete error analysis of non-perturbative corrections. Due to the good understanding of these non-perturbative corrections, the uncertainty remains under control, however, it is significantly larger than that in \Fig{fig:N4LL}, showing that more work is required to improve our understanding of the non-perturbative physics in the back-to-back limit of the energy correlator. 

It would be interesting to explore non-perturbative calculations of the Collins-Soper kernel in different theories, to understand more precisely how different features of confinement manifest in its long distance structure. For example, if one compute the Collins-Soper kernel in pure Yang-Mills, as compared to Yang-Mills with fundamental quarks, can one see the effects of string breaking? This would enable experimental studies of these features, and an improved intuitive understanding of the energy correlators in this deep non-perturbative region. We leave the investigation of these questions to future studies.

\begin{figure}
\hspace{0.95cm}\includegraphics[width=0.655\linewidth]{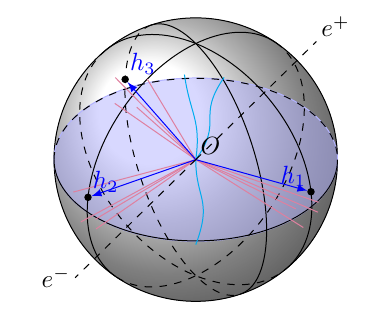}
\includegraphics[width=0.755\linewidth]{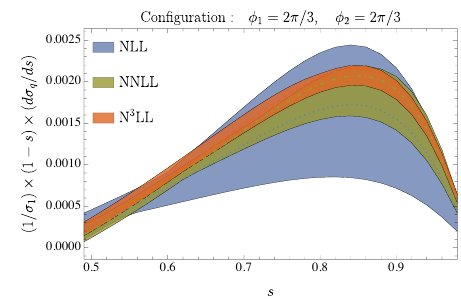}
\caption{The co-planar limit of the three-point correlator function exhibits a factorization into matrix elements of Wilson lines. This gives rise to a Sudakov that depends on the particular kinematics of the detector operators, encoded in the particular configuration of Wilson lines. Figures from \cite{Gao:2024wcg}.
}
\label{fig:coplanar}
\end{figure}

It is important to emphasize that the particular structure of cusped Wilson lines in the back-to-back limit depends on the underlying form factor. One way of generalizing it is to consider the ``co-planar limit" of the three-point energy correlator. This is shown in \Fig{fig:coplanar}. In this case one has a generalized factorization involving two sets of three-Wilson lines \cite{Gao:2024wcg}, nevertheless it can be treated in a similar way. The back-to-back limit of this correlator is also shown in \Fig{fig:coplanar} for a particular configuration of the detector operators. These more general configurations of Wilson lines will play an important role in hadron colliders.

In this section we have focused on the description of the back-to-back limit of the energy correlator using effective field theory approaches. It will be of interest to develop techniques used to study this limit in CFTs to the case of QCD. In particular, this limit can be analyzed \cite{Korchemsky:2019nzm} using the duality between correlators and Wilson loops \cite{Alday:2010zy}. It would be interesting to systematically explore the expansion onto Wilson loop operators beyond the leading power. This limit can also be explored using large spin perturbation theory, as was considered in the case of $\mathcal{N}=4$ sYM theory in \cite{Chen:2023wah}. It would be interesting to develop this approach in QCD. Finally, the dynamics of Wilson loops are extremely well understood in $\mathcal{N}=4$ sYM due to the pentagon OPE \cite{Basso:2014jfa,Basso:2013aha,Basso:2014koa,Basso:2013vsa}. It would be interesting to understand if their are kinematic limits of energy correlators that could be understood using this OPE.

\subsection{Massive Quarks}\label{sec:heavy_quarks}

Another interesting feature of QCD is the presence of massive quarks. In particular, the $c$ and $b$ quarks have masses $m_{c,b} \gtrsim \Lambda_{\text{QCD}}$.  Energy correlators on charmed and b-hadrons can be directly measured in experiment, as we will see in \Sec{sec:exp_opp}, providing access to intrinsic mass effects. Additionally, the description of heavy quark jets is phenomenologically important, for example for studying the substructure of jets in $H\to b \bar b$.

Apart from the phenomenological motivation, one of the reasons that masses are interesting from a purely theoretical perspective is due to the infrared nature of the light-ray OPE. A Euclidean OPE is a short-distance, or UV OPE, and therefore the presence of masses does not modify the leading behavior of the OPE, or the correlation function at short distances. On the other hand, the light-ray OPE is an IR OPE. Masses are relevant in the infrared, and will therefore imprint themselves as a scale in the light-ray OPE, much like confinement. However, they provide a fully calculable example of this phenomenon. We will see that they act to cut off the scaling behavior in the infrared.

\begin{figure}
\includegraphics[width=0.90\linewidth]{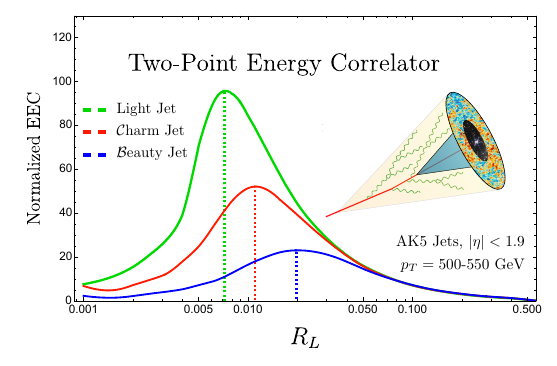}
\includegraphics[width=0.85\linewidth]{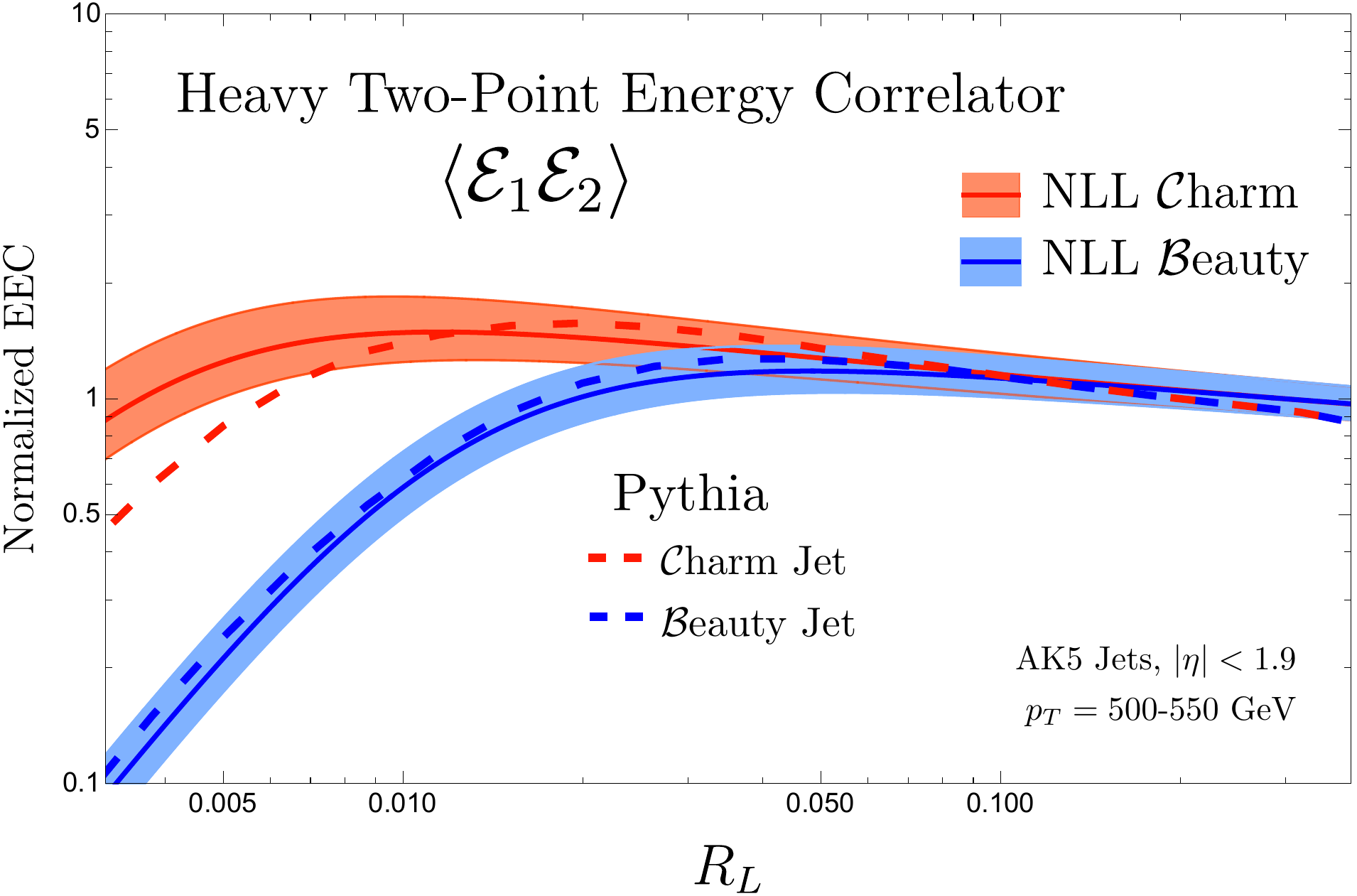}
\caption{The collinear limit of the two-point energy correlator for light quark jets, compared with charm quark and b-quark jets. All exhibit massless scaling at angles above their intrinsic mass, and a transition at a scale set by their mass. In the lower panel we show a zoom in of the transition region compared with analytic predictions. Particularly for the $b$-quark, the energy correlator exhibits a transition in a regime described by perturbation theory.  Figures from \cite{Craft:2022kdo}.
}
\label{fig:bquark_vacuum}
\end{figure}

Energy correlators on massive quarks were studied in \cite{Craft:2022kdo}, which extended the factorization formula in the collinear limit to incorporate mass effects. In the upper panel of \Fig{fig:bquark_vacuum} we show a simulation of the energy correlators in the collinear limit on massless jets, and $c$ and $b$ jets. The mass scales imprints in the correlator, which exhibits a characteristic turnover. In the lower panel of \Fig{fig:bquark_vacuum} we show a comparison of the analytic calculations with parton shower simulations. Particularly for the $b$-quark, the transition due to the finite b-quark mass is in a regime which is under perturbative control.

To investigate the structure of the jet function, we can compute it analytically, as an exact function of  $\delta = \frac{iM}{p_TR_L}$. At one loop, it is found to be
\begin{align}\label{eq:heavy_quark}
J_Q^{[2]}|_{R_L\neq 0}=&\frac{\alpha_s C_F}{4\pi}\left\{\left[\delta^4-4\delta^3+2\delta^2-3\right]\ln\left(\frac{\delta}{1+\delta}\right) \right.\nn \\
&\left. -\frac{1}{2}\left(9\delta^2+\frac{31}{6}\right)\right\}+c.c\,.
\end{align}
This result is quite interesting. For $\theta \gg M/p_T$, we expect to be able to perform a conformal perturbation theory in the mass, similar to the treatment of non-perturbative corrections. Expanding the result in $M/p_T$, we indeed observe that the leading term is the expected massless scaling.  On the other hand, we can also expand about $\theta=0$. Here the leading behavior is a uniform scaling, as also observed for non-interacting hadrons. Therefore, we see that the presence of an explicit mass scale, or more generally Higgsing, provides a calculable example of a transition, which shares a number of the features of the hadronization transition. 

One interesting feature of having the exact result in Eq. \ref{eq:heavy_quark} is that we can study its analytic structure, and the radius of convergence of these two different expansions, namely the expansions in $\theta$ and $M/p_T$. Quite interestingly, we see that there is a pole in the complex plane, which prevents the expansions from converging through the transition point at $\theta \gg M/p_T$. This illustrates some genuine non-analytic behavior in the transition.  It would be interesting to understand if this is a more general feature of correlation functions of light-ray operators in the presence of an additional scale.

The effect of heavy quarks on the resummation in the back-to-back limit has also been studied in \cite{vonKuk:2024uxe,Aglietti:2024zhg}.

\subsection{Non-Perturbative Power Corrections}\label{sec:non_pert}

\begin{figure}
  \includegraphics[width=0.95\linewidth]{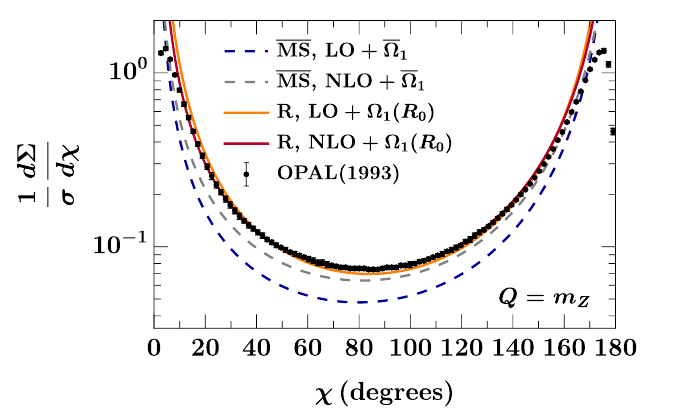}
  \caption{The two-point correlator in the bulk of distribution (away from the endpoints), incorporating both perturbative and leading non-perturbative corrections. The separation between perturbative and non-perturbative contributions can lead to renormalons. The convergence of the series is shown in both the MSbar scheme, and the renormalon free R-scheme, which illustrates improved convergence. Figure from \cite{Schindler:2023cww}.
  }
  \label{fig:renormalon_2point}
  \end{figure}

Measurements of energy correlators, or any other detector operators in QCD are ultimately made on confined hadrons. The ability to calculate any such observable relies on factorization theorems to rigorously separate perturbative short distance contributions, which can be computed in perturbation theory, from long-distance non-perturbative matrix elements. In certain cases these non-perturbative matrix elements are universal, appearing in multiple observables, allowing them to be extracted from experiment and used to make predictions.

Before discussing more detailed aspects of non-perturbative corrections, it is first important to get an understanding of their size, and in which kinematic limits of the energy correlators they contribute. For infrared and collinear safe observables, such as the energy correlators, non-perturbative corrections are suppressed by the ratio of the scales probed by the observable to the intrinsic non-perturbative scale in QCD, which we denote $\Lambda_{\text{QCD}}$. In the case of the energy correlators, we have seen that the relevant virtuality scales are $Q \sqrt{z}$ in the collinear limit, and $Q\sqrt{1-z}$ in the back-to-back limit. We therefore expect the leading non-perturbative power corrections to scale as $\Lambda_{\text{QCD}}/(Q \sqrt{z})$ and $\Lambda_{\text{QCD}}/(Q\sqrt{1-z})$. In kinematic regimes where these ratios are small, we can perform an expansion, and incorporate the non-perturbative corrections order by order. However, as $Q \sqrt{z} \sim \Lambda_{\text{QCD}}$, or $Q \sqrt{1-z} \sim \Lambda_{\text{QCD}}$, this separation of scales breaks down and we must resum the full set of non-perturbative power corrections. In such kinematic regions the energy correlator observables become completely non-perturbative.

\begin{figure}
\includegraphics[width=0.95\linewidth]{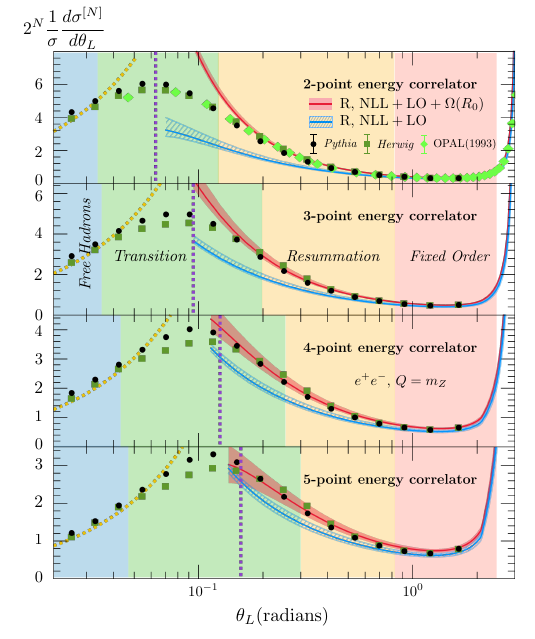}
\caption{Impact of the leading non-pertubative power corrections in the collinear limit of the projected energy correlators. Perturbative results are shown in blue, and results incorporating the leading non-perturbative corrections are shown in red. The leading non-perturbative corrections lead to an enhanced scaling at small angles, and are crucial to describe data/simulation. Figure from \cite{Lee:2024esz}.
}
\label{fig:renormalon_Npoint}
\end{figure}

We begin by considering the simplest case in the bulk of the energy correlator distribution, away from any kinematic limits. In this case, $z\sim 1/2$, so that there are only two scales, $\Lambda_{\text{QCD}}$ and $Q$. The leading non-perturbative power corection is linear, scaling like $\Lambda_{\text{QCD}}/Q$.  A remarkable feature of the energy correlator is that the  functional form in $z$ of the leading power correction is fixed, as originally shown in the seminal work of refs.~\cite{Korchemsky:1999kt,Korchemsky:1997sy,Korchemsky:1995zm,Korchemsky:1994is,Belitsky:2001ij}. The result is that we can write the two-point energy correlator, including the leading non-perturbative correction, as
\begin{align} \label{eq:nonp}
\frac{1}{\sigma_0} \text{EEC}(z) =\frac{1}{\sigma_0} \frac{\df \hat \sigma}{\df z}+
\frac{1}{2}\frac{\bar \Omega_{1q}}{Q (z(1-z))^{3/2}}\,.
\end{align}
Here $\hat \sigma$ is the perturbative contribution to the energy correlator, and the magnitude of the leading non-perturbative correction is quantified by the parameter $\bar \Omega_{1q}$ (here the bar denotes that this is the MS-bar scheme, and the q emphasizes that this is for a source $J^\mu = \bar \psi \gamma^\mu \psi$.)

Similar to non-perturbative condensates appearing in the OPE, the parameter $\bar \Omega_{1q}$ can be given a field theory definition \cite{Korchemsky:1999kt,Korchemsky:1997sy,Korchemsky:1995zm,Korchemsky:1994is,Belitsky:2001ij,Lee:2006fn}. Instead of the expectation value of a local operator, it is now the expectation value of the energy flux operator in a state produced by Wilson lines 
\begin{align}
\Omega_{1q}=\frac{1}{N_c} \langle 0 | \tr \bar Y_{\bar n}^\dagger Y_n^\dagger \cE_T(0) Y_n \bar Y_{\bar n} |  0 \rangle\,.
\end{align}
Here $Y_{n,\bar n}$  are Wilson lines in the fundamental representation. In principle this non-perturbative parameter could be directly computed, for example on the lattice. However, this is currently not possible. Instead, we rely on the fact that it appears in the calculation of other $e^+e^-$ event shape observables from which it has been extracted \cite{Abbate:2010xh}. More precisely, there is a small non-universality from the treatment of hadron mass effects in different observables \cite{Salam:2001bd,Mateu:2012nk}, but this is understood, and can be compensated for.

One physically interesting feature of the leading non-perturbative power correction is that it is even more singular as $z\to0,1$ than the perturbative contribution. Higher non-perturbative power corrections become further singular. Therefore, to get an accurate description of the energy correlator in the resummation region we must also understand how the non-perturbative power correction behaves in the collinear and back-to-back region, which will be discussed shortly.

While Eq. \ref{eq:nonp}, which establishes a separation of the energy correlator into a perturbative component and a non-perturbative component is conceptually satisfactory, its technical implementation is more challenging. In particular, depending on the renormalization scheme that is used, attempts to divide the overall cross-section, which is a non-perturbatively well defined quantity, into a perturbative and a non-perturbative component, can introduce renormalon ambiguities (For a detailed review of renormalons, we refer the reader to \cite{Beneke:1998ui}) into the perturbative and non-perturbative components of the cross section. While the renormalon cancels between the perturbative and non-perturbative components, neither is non-perturbatively well defined. At a practical level, this can manifest as a poor convergence of the perturbative series, making it difficult to achieve precision predictions. 

\begin{figure}
  \includegraphics[width=0.8\linewidth]{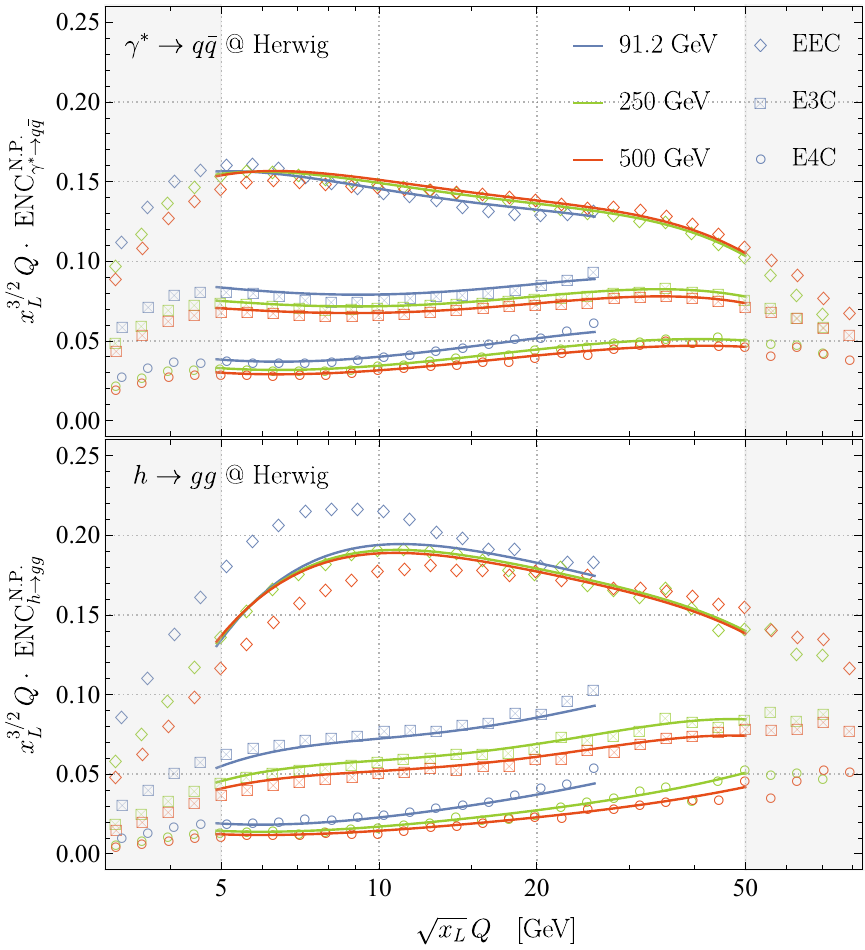}
  \caption{A comparison of the scale evolution of the non-perturbative matrix element, $\vec{D}_N$, as predicted by light-ray OPE and simulations. Good agreement is found, except for the gluon EEC shown in the lower panel. This is a hint of small-$x$ physics in gluonic power corrections. Figure from \cite{Chen:2024nyc}.
  }
  \label{fig:evolution-quarkgluon}
  \end{figure}

To resolve this problem, renormalon free schemes, for example the R-scheme \cite{Hoang:2007vb,Hoang:2008fs,Hoang:2009yr,Hoang:2017suc,Bachu:2020nqn}, have been developed for the treatment of a wide variety of problems.  They have been applied to the study of the energy correlators in refs.~\cite{Schindler:2023cww,Lee:2024esz}. While renormalons are not the primary focus of this review, we want to mention this aspect, since it is conceptually interesting, and further, the most precise extractions of the non-perturbative parameter $\Omega_{1q}$ are performed in the R-scheme~\cite{Abbate:2010xh}

The structure of the renormalon is universal for any observable with the same leading non-perturbative parameter. Using the known structure of the renormalon, we can convert to a renormalon free scheme by performing a renormalon subtraction
\begin{align}
\Omega_{1\kappa}(R)=\bar \Omega_{1\kappa}-R \sum\limits_{n=1}^\infty d_{\kappa n}(\mu/R)\, a_s^n(\mu)\,.
\end{align}
The coefficients $d_{\kappa n}$ are known to two loops \cite{Schindler:2023cww}. Since the overall cross section is renormalon free, this acts to shift contributions between the perturbative and non-perturbative components of the result. In particular, it shifts the perturbative contribution by
\begin{align}
\frac{1}{\sigma}\frac{d\hat \sigma (R)}{dz}=\sum \limits_{n=1}^\infty \left\{  c_n\left(z,\frac{\mu}{Q} \right)+\frac{R}{2Q} \frac{d_{n}(\mu/R)}{[z(1-z)]^{3/2}}     \right \} \left[ \frac{\alpha_s(\mu)}{4\pi}  \right]^n\,.
\end{align}
The non-perturbative parameter is set by physics at the confinement scale, but can be renormalization group evolve to the scale $Q$. Based on extractions from the thrust observable  \cite{Abbate:2010xh}, ref.~\cite{Schindler:2023cww} provided an extraction of the non-perturbative parameter for the EEC
\begin{align}
\Omega_1(R_0)=0.7895 \pm 0.054 \text{GeV}\,,
\end{align}
where $R_0=2$ GeV.
Results for the perturbative expansion of the energy correlator with and without the renormalon subtraction are shown in \Fig{fig:renormalon_2point}. Here we see that the series converges rapidly in the R-scheme, achieving good agreement with the data already with the NLO perturbative result.

In the collinear and back-to-back limits, the non-perturbative corrections must be combined with resummation. An advantage of the factorization formulas for the collinear and back-to-back limits presented in the previous sections is that the leading non-perturbative corrections are incorporated in a single function, enabling the resummation framework to remain unchanged. Here we briefly discuss how this occurs for the two limits of the energy correlator. For simplicity of presentation, we do not include the renormalon subtractions, however, they can be performed in the same manner.

\begin{figure}
  \includegraphics[width=0.8\linewidth]{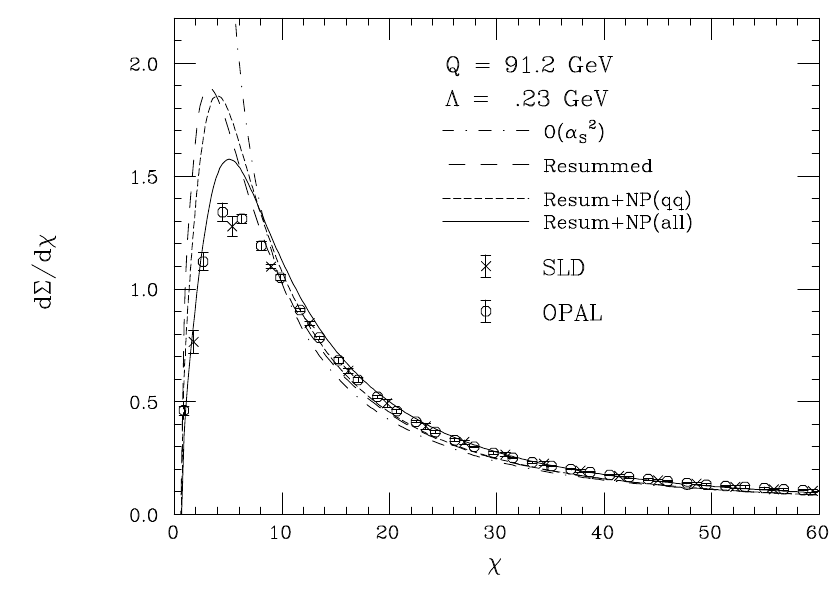}
  \caption{Illustration of the non-perturbative corrections to the energy correlator in the back-to-back region. Such corrections are essential for achieving agreement with data. Figure from \cite{Dokshitzer:1999sh}.
  }
  \label{fig:marchesini_NP}
  \end{figure}

In the factorization formula for the collinear limit of the energy correlator, the only dependence on the observable arises in the jet function. The leading non-perturbative correction therefore also arises in the jet function. For the particular case of the two-point energy correlator, one finds that it takes the form  \cite{Lee:2024esz,Chen:2024nyc}
\begin{align}
J\Bigl(\ln \frac{z x^2 Q^2}{\mu^2},\mu   \Bigr) = \hat J \Bigl(\ln \frac{z x^2 Q^2}{\mu^2},\mu   \Bigr) - \frac{\bar \Omega_{1,q}}{\sqrt{z} x Q}
\,.\end{align}
Note that the leading non-perturbative correction is again set by the same universal non-perturbative parameter, $\Omega$. In \Fig{fig:renormalon_Npoint} we show a comparison of calculations of the small angle limit of the $N$-point projected energy correlators, with simulation \cite{Lee:2024esz}. The leading non-perturbative corrections are crucial to describe the simulation, and lead to an enhanced scaling at small angles

Remarkably, the non-perturbative corrections in the small angle limit can also be studied using light-ray OPE techniques~\cite{Chen:2024nyc}. 
The leading non-perturbative corrections for $N$-point correlator $\text{ENC}$ can be expressed as a product of a perturbative Wilson coefficient $\langle\overrightarrow{\mathbb{O}}_{\tau=2}^{[J=N]}\rangle$ and a non-perturbative matrix-element $\vec{D}_N$
\begin{align}
    & \text{ENC}^{\mathrm{N.P.}}\left( \theta Q, Q\right)=\Lambda_{\mathrm{QCD}} \nn \\
    & \times \vec{D}_N\left(\frac{\theta^2 Q^2}{\mu^2}, \frac{\Lambda_{\mathrm{QCD}}^2}{\mu^2}\right) \cdot \frac{\langle\vec{\mathbb{O}}_{\tau=2}^{[J=N]}(\hat n, \mu)\rangle}{(4 \pi)^{-1} \sigma Q^{N-1}} \,.
\end{align}
Crucially, the difference with leading power light-ray OPE is that the spin of the light-ray operator has lowered by one, due to the appearance of another scale $\Lambda_{\text{QCD}}$. The scale evolution of the non-perturbative matrix element $\vec{D}_N$ is governed by anomalous dimension of local operator with spin $N$. The comparison between theory and Monte-Carlo simulation is shown in \Fig{fig:evolution-quarkgluon}. It would be interesting study the non-perturbative matrix element using real data.

In the back-to-back limit, the linear power correction arises in the jet function. Following the seminal work of \cite{Dokshitzer:1999sh}, one can show that the leading non-perturbative correction is a linear shift to the jet function in $b$ space, namely
\begin{align}
J(b_\perp)&\to J(b_\perp)+ J_{\rm NP}(b_\perp) = J(b_\perp) -b_\perp \Omega_{1,q}\,.
\end{align}
This can then be dressed with resummation in the standard fashion. This non-perturbative correction acts primarily to shift the location of the Sudakov peak. An illustration of this is shown in \Fig{fig:marchesini_NP}. We see that the non-perturbative corrections are essential for achieving agreement with data. As compared to the non-perturbative power corrections in the collinear limit, which have been studied extensively using recent collider data, there has been less work on studying the non-perturbative power corrections in the back-to-back limit of the energy correlator. We hope that recent measurements of the energy correlators using archival ALEPH data \cite{Bossi:2025xsi,Bossi:2024qeu} will lead to a significant improved understanding of the non-perturbative power corrections in the back-to-back limit.

\begin{figure}
  \includegraphics[width=0.6\linewidth]{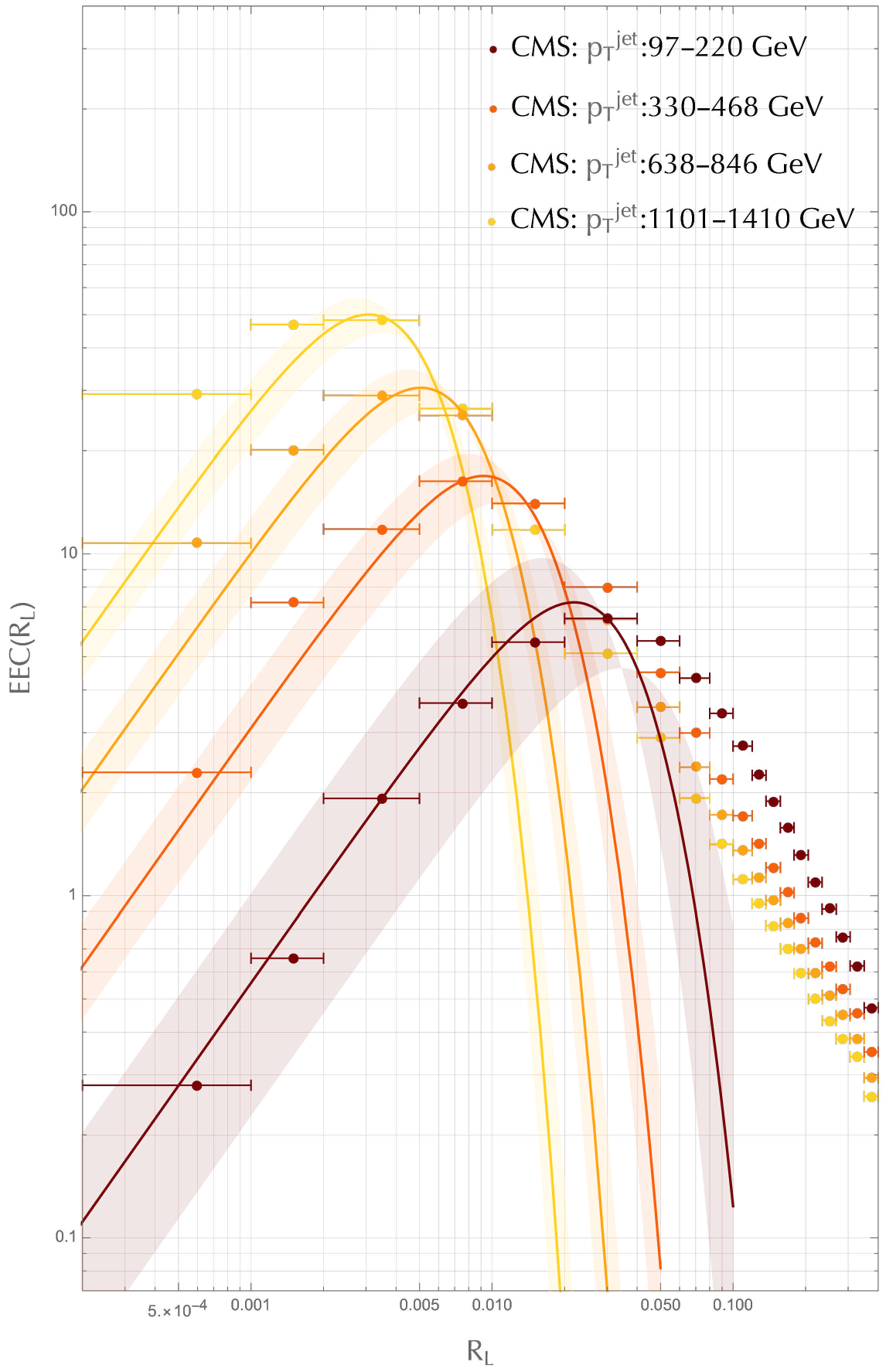}
  \caption{Comparision of the TMD-based model for the description of back-ward transition from free hadron to quarks and gluons and CMS data. Figure adpated from \cite{Liu:2024lxy}.
  }
  \label{fig:near_side}
  \end{figure}

While we have illustrated how the leading non-perturbative corrections can be incorporated, it will be important to go beyond this. In particular, describing the transition in the collinear region will require an improved understanding of the dynamics of confinement. A non-perturbative TMD fragmentation model has been proposed to describe the transition from the regime of free hadrons to that of quarks and gluons \cite{Liu:2024lxy}. Its comparison to CMS data is shown in Fig.~\ref{fig:near_side}, where good agreement is found. Nevertheless, a unified treatment connecting the full dynamics from quarks to hadrons remains elusive. It would be interesting to develop a proper framework for understanding this transition. It highlights the shift from perturbative quarks and gluons to hadrons, somewhat analogous to a phase transition. However, it is not a phase transition in the standard sense, as no external parameter is being varied. Perhaps it can be described using the language of dynamical phase transitions \cite{Heyl:2018jzi}. We leave this to future work.

\subsection{Detector Functions}\label{sec:det_func}

\begin{figure}
\includegraphics[width=0.75\linewidth]{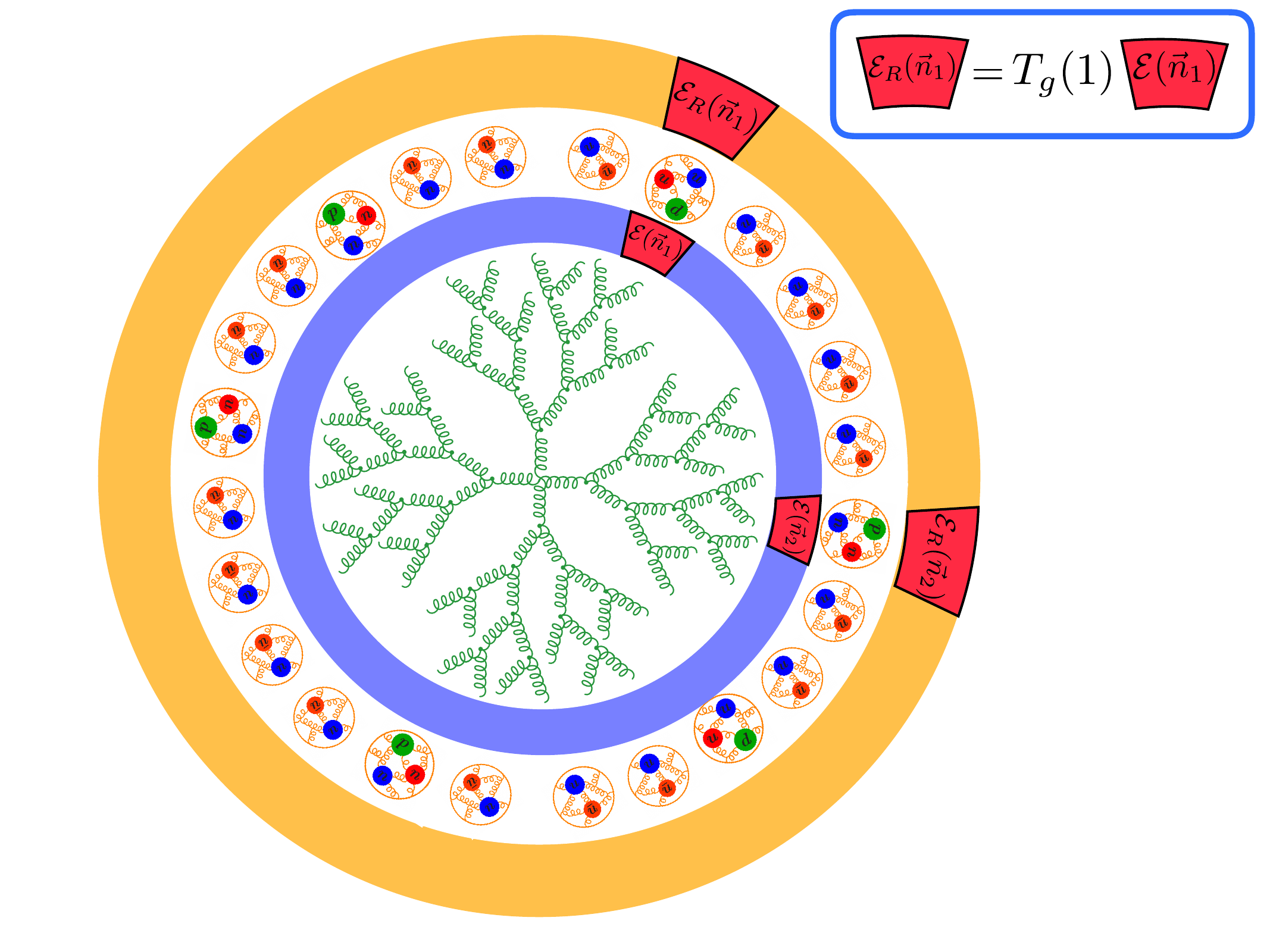}
 \includegraphics[width=0.98\linewidth]{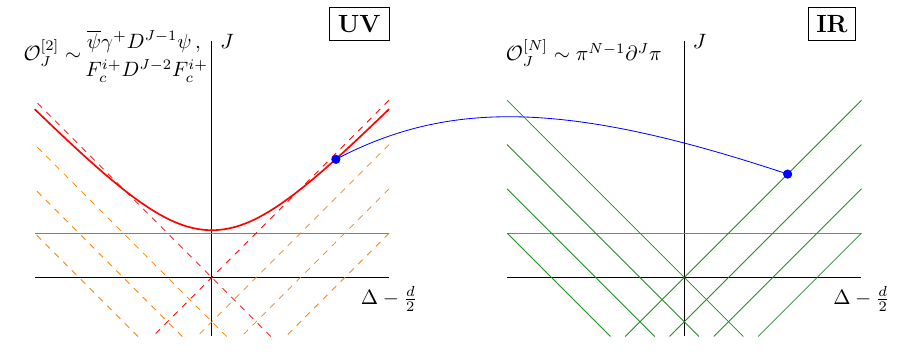}
\caption{Detector functions act as a matching between UV and IR detectors, enabling an understanding of more general detectors which incorporate IR properties of hadrons, such as charge. Figure from \cite{Jaarsma:2022kdd}.
}
\label{fig:detector_function_schematic}
\end{figure}

One of the most interesting features of QCD, which distinguishes it from the case of a CFT, is that the degrees of freedom change between the UV, and the IR. In the previous sections, we have phrased the detector operators of QCD in terms of quarks and gluons. The Regge trajectories of these operators place constraints on the detectors that are well defined in the theory. In reality, detectors are in the IR of the theory. In the case of QCD, this means that they are expressed in terms of free hadrons. The space of detector operators of these free hadrons is distinct from the detector operators in the UV of the theory. In particular, one can study detector operators which measure properties of the hadrons which are not in general well defined quantities in the UV. 

The study of these more general detectors is interesting for studying properties of non-perturbative QCD. For example, non-perturbatively QCD exhibits chiral symmetry breaking. One would like to be able to study chiral symmetry breaking using asymptotic energy flux. Since it is inherently non-perturbative, this must be formulated in terms of detectors involving the IR hadrons. Similarly, many questions such as how charge or other quantum numbers are distributed in the hadronization process require the study of more general IR detectors, beyond the energy flow operator.

Understanding more general detectors that incorporate properties of hadrons requires a mapping between UV detectors and IR detectors. This is illustrated schematically in \Fig{fig:detector_function_schematic}. This mapping can be referred to generically as a ``detector function", and is inherently non-perturbative. The complete mapping is of course not well understood, but it is well understood on the leading twist-2 trajectory, for $J$ above the Regge intercept of the theory.

In general, we can perform a matching of our IR detectors onto UV detectors expressed in terms of quark and gluon fields, and non-perturbative detector functions
\begin{align}\label{eq:match_general}
\mathcal{D}_{\text{IR}}=  \sum_i \mathcal{D}^{(i)}_{\text{UV}} (\mu) \cdot F^{(i)}_{\mathcal{D}}(\mu)\,.
\end{align}
In doing so, we introduce a renormalization group scale, which leads to the evolution of both the detector and the non-perturbative functions. 

\begin{figure}
  \includegraphics[width=0.85\linewidth]{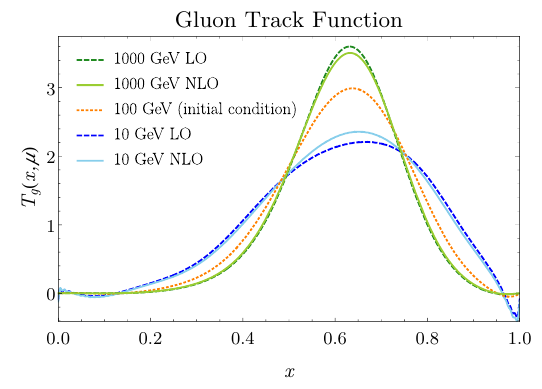}
  \caption{The gluon track function, as extracted from Pythia, and its perturbative evolution. The peak at $x\sim 2/3$ is understood as the fraction of charged pions. The distribution narrows in the UV, ultimately collapsing to $\delta$ function. Figure from \cite{Chen:2022muj}.
  }
  \label{fig:track_1d}
  \end{figure}

\begin{figure}
\includegraphics[width=0.65\linewidth]{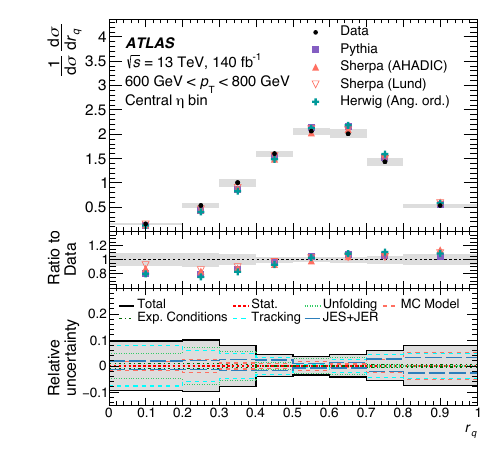}
\includegraphics[width=0.65\linewidth]{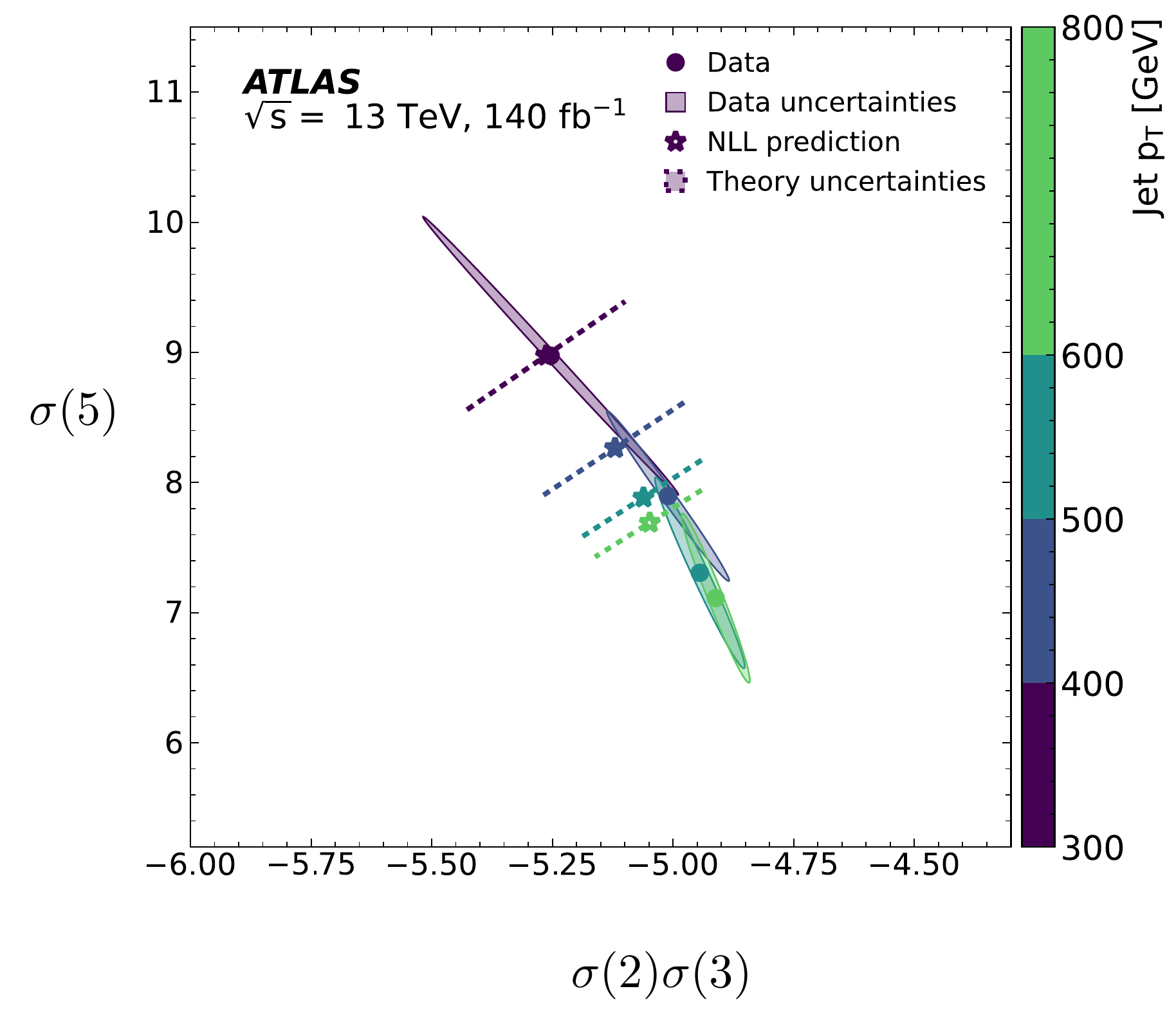}
\caption{A measurement of the track function by the ATLAS collaboration, along with the renormalization group evolution of its cumulants, illustrating mixing. Data is compared with results from parton shower programs. Figures from \cite{ATLAS:2025qtv}.
}
\label{fig:ATLAS_track}
\end{figure}

A simple and well known example is a fragmentation function, which maps a single particle detector in the UV to a single hadron detector in the IR. More generally, multi-hadron fragmentation functions map between single particle detectors in the UV, and multi-hadron detectors in the IR. More recently a number of more sophisticated detector functions have been understood. One example, which is particularly useful are detector operators which measure the energy flux on charged particles. We denote such detectors by $\cE_\text{tr}$. Such detectors are of significant practical importance, since charged particles leave ``tracks" in the detectors and can be measured much more precisely. For this reason the detector functions for the case of charged particle energy flux are called ``track functions". Track functions were originally defined in \cite{Chang:2013rca,Chang:2013iba}. For the case of a fragmenting quark, they are given by
\begin{align} \label{T_def}
T_q(x)&=\!\int\! \df y^+ \df ^{d-2} y_\perp e^{ik^- y^+/2} \sum_X \delta \biggl( x\!-\!\frac{P_R^-}{k^-}\biggr)  \frac{1}{2N_c}\\
&\text{tr} \biggl[  \frac{\gamma^-}{2} \langle 0| \psi(y^+,0, y_\perp)|X \rangle \langle X|\bar \psi(0) | 0 \rangle \biggr]\,,
\end{align} 
and a similar definition exists for a gluon track function. A plot of the gluon track function as extracted from Pythia is shown in \Fig{fig:track_1d}. A measurement of the track function, using a formalism developed in \cite{Lee:2023xzv}, was performed by the ATLAS collaboration, and is shown in \Fig{fig:ATLAS_track}. We will often work in terms of the moments of the track functions
\begin{align} \label{eq:T_mom}
T_a(n,\mu)=\int \limits_0^1 \df x~ x^n~ T_a (x,\mu)\,,
\end{align}
which will appear in calculations of the energy correlators.

Using track functions, we can now make the formula in Eq. \ref{eq:match_general} explicit. For the case of a one-point energy correlator which measures only the energy on tracks, we have
\begin{align}
\cE_\text{tr}(\vec n_1)=T_{\bar q}(1) \cE_{\bar q} (\vec n_1)+  T_q(1) \cE_q (\vec n_1)+  T_g(1) \cE_g (\vec n_1) \,.
\end{align}
Here we see that the first moment of the track functions appear as the matching coefficients. This is quite intuitive, since for the one-point function, we only need to know the average energy that goes into charged particles in the hadronization process. This is illustrated in \Fig{fig:detector_function_schematic}. In higher point correlation functions we will also get higher moments of the track functions, which will appear in the contact terms. For example, for the two-point correlator, we have the matching relation
\begin{align}
\langle \cE_{\text{tr}}(n_1) \cE_{\text{tr}}(n_2) \rangle &=\sum_{a_1,a_2}   T_{a_1}(1) T_{a_2}(1)  \langle \cE_{a_1} (\vec n_1) \cE_{a_2} (\vec n_2) \rangle\nn \\
& +\sum\limits_a T_a(2) \langle \cE_a^{(1,1)}(\vec n_1)\rangle \delta(\vec n_1-\vec n_2)\,,
\end{align}
and similarly for higher point correlators. This enables a systematic calculation of energy correlator observables on tracks.

Although track functions are non-perturbative objects, their renormalization group evolution can be computed in perturbation theory. As compared to fragmentation functions, the renormalization group structure of the track functions is more involved. It has been systematically explored in \cite{Chang:2013iba,Chang:2013rca,Chen:2022muj,Chen:2022pdu,Jaarsma:2022kdd,Li:2021zcf,Jaarsma:2023ell}. To study the renormalization group evolution of the track functions, it is convenient to use central moments
\begin{align}
\label{eq:centralmoment}
\sigma(n,\mu)=\int \limits_0^1 \df x~ (x - \langle x \rangle)^n~ T(x,\mu)\,,
\end{align}
The renormalization group evolution of the low central moments has a relatively simple form
\begin{align}
    \frac{\df}{\df \ln \mu^2} \sigma(3) &= \ga_3\, \sigma(3) \,,\nn\\
    \frac{\df}{\df \ln \mu^2} \sigma(4) &= \ga_4\, \sigma(4) + \ga_{22} \sigma(2)^2 \,,\nn \\
    \frac{\df}{\df \ln \mu^2} \sigma(5) &= \ga_5\, \sigma(5) + \ga_{32} \sigma(3) \sigma(2) \,,
\end{align}
(here we have dropped flavor indices for simplicity). Interesting, as compared to the case of DGLAP, these evolution equations are non-linear, involving mixing into products of lower central moments. Much like for the case of the DGLAP evolution equations, we can also derive a full evolution equation in $x$-space, that describes the evolution of all the moments. The result is relatively complicated, but can be found in \cite{Chen:2022muj,Chen:2022pdu}.

Using this evolution equation, we can evolve the track function distribution. This is shown in \Fig{fig:track_1d}, where we compare the evolved distribution starting from an initial value, with the extracted value from Pythia, finding good agreement. As the track functions are evolved to the UV, they collapse to a delta function. Quite remarkably, it is also possible to measure the renormalization group evolution of the moments of the track function to verify the non-linear evolution. This has been performed by the ATLAS collaboration, and is shown in \Fig{fig:track_1d} for the particular case of the mixing between $\sigma(5)$ and $\sigma(2) \sigma(3)$, along with the theoretical prediction. The ellipsis and lines show experimental and theoretical errors, respectively. Although there are large error bars, good agreement is seen.

\begin{figure}
\includegraphics[width=0.95\linewidth]{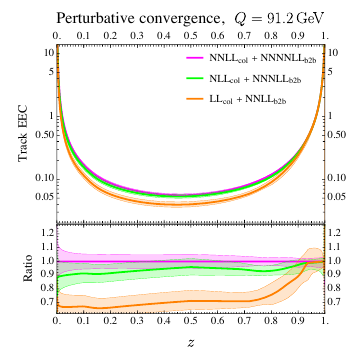}
\caption{A calculation of the EEC on tracks in $e^+e^-$ collisions at N$^4$LL in the back-to-back region and N$^2$LL in the collinear region. Figure from \cite{talk_Max}.
}
\label{fig:ee_us_prediction}
\end{figure}

Using the formalism of detector/ track functions, we are able to perform high precision calculations of energy correlator observables on tracks.  As an example, in \Fig{fig:ee_us_prediction}, we show a calculation of the energy correlator on tracks computed with N$^4$LL in the back-to-back region and N$^2$LL in the collinear region. This is the highest perturbative accuracy ever achieved for an event shape observable in QCD, and it is achieved on tracks!  This is transformative for comparison to experiment. For example, in the re-analysis of ALEPH data \cite{Bossi:2024qeu,Bossi:2025xsi} shown in \Fig{fig:CMS_scaling}, the  high resolution of the tracker was essential for enabling a view of all the different regions of the correlator. Similarly, many measurements of energy correlators in hadron colliders, are nuclear colliders rely on the use of tracks.

\begin{figure}
\includegraphics[width=0.855\linewidth]{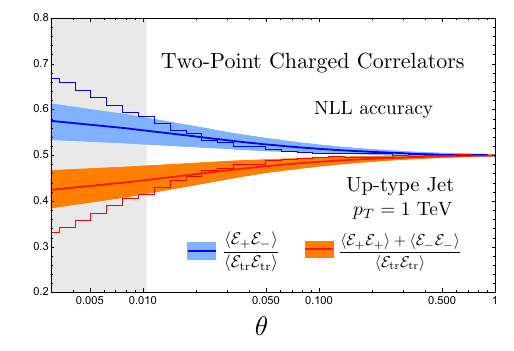}
\caption{A calculation of the two-point energy correlators incorporating charge, namely $\langle \mathcal{E}_+ \mathcal{E}_- \rangle$ and $\langle \mathcal{E}_+ \mathcal{E}_+ \rangle$. Enhanced correlations in oppositely charged energy flux are observed at small angles. Figure from \cite{Lee:2023npz}.
}
\label{fig:figEpEm}
\end{figure}

Another advantage of this formalism is that it allows for the calculation of a broader class of observables, which begin to probe interesting aspects of the hadronization transition. As an example, one can study correlators of energy flux on positively and negatively charged hadrons, $\langle \mathcal{E}_+ \mathcal{E}_- \rangle$ and $\langle \mathcal{E}_+ \mathcal{E}_+ \rangle$ \cite{Lee:2023npz,Lee:2023tkr}.  In a string like model for hadronization, string breaking will produce pairs of oppositely charged hadrons, and so one might anticipate enhanced correlations in energy flux of oppositely charged hadrons at small angles. In \Fig{fig:figEpEm} we show a calculation of these different correlators showing that this is indeed the case. The fact that these detectors are C-odd also has interesting implications in nuclear physics, as will be discussed in \Sec{sec:nuclear}.

\subsection{Broader Impact on QCD Phenomenology}\label{sec:impact}

Although the focus of this article is on energy correlators, many of the techniques developed in their study have a broader applicability. Indeed, one can view energy correlator observables, arguably the simplest collider observables in QFT, as a laboratory for the study and development of new techniques. For this purpose, the explicit perturbative calculations of the energy correlators, which are unavailable for other observables prove invaluable. Above we have emphasized how insights from energy correlators have improved our understanding of splitting functions, and track functions.  Here we highlight several examples where studies of the energy correlators have had a broader impact on phenomenology.

The study of the energy correlators in the collinear limit has greatly improved our understanding of collinear factorization, which applies in more general situations. One example is inclusive jet production at hadron colliders. This is one of the most fundamental processes studied at hadron colliders, and its accurate description plays an important role both in precision studies of QCD, and searches for new physics. In the limit that the jet radius is small, which is often of phenomenological interest, one can factorize the dynamics into a universal hard function, and a jet function, much in analogy with the factorization for the energy correlators presented in \Sec{sec:EC_coll}. However, in this case the jet function involves the clustering associated with the experimental analysis, making it challenging to study analytically. As a result, a factorization formula for this process had been conjectured \cite{Kang:2016ehg,Kang:2016mcy}, but it had never been explicitly tested beyond one-loop. 

\begin{figure}
\includegraphics[width=0.95\linewidth]{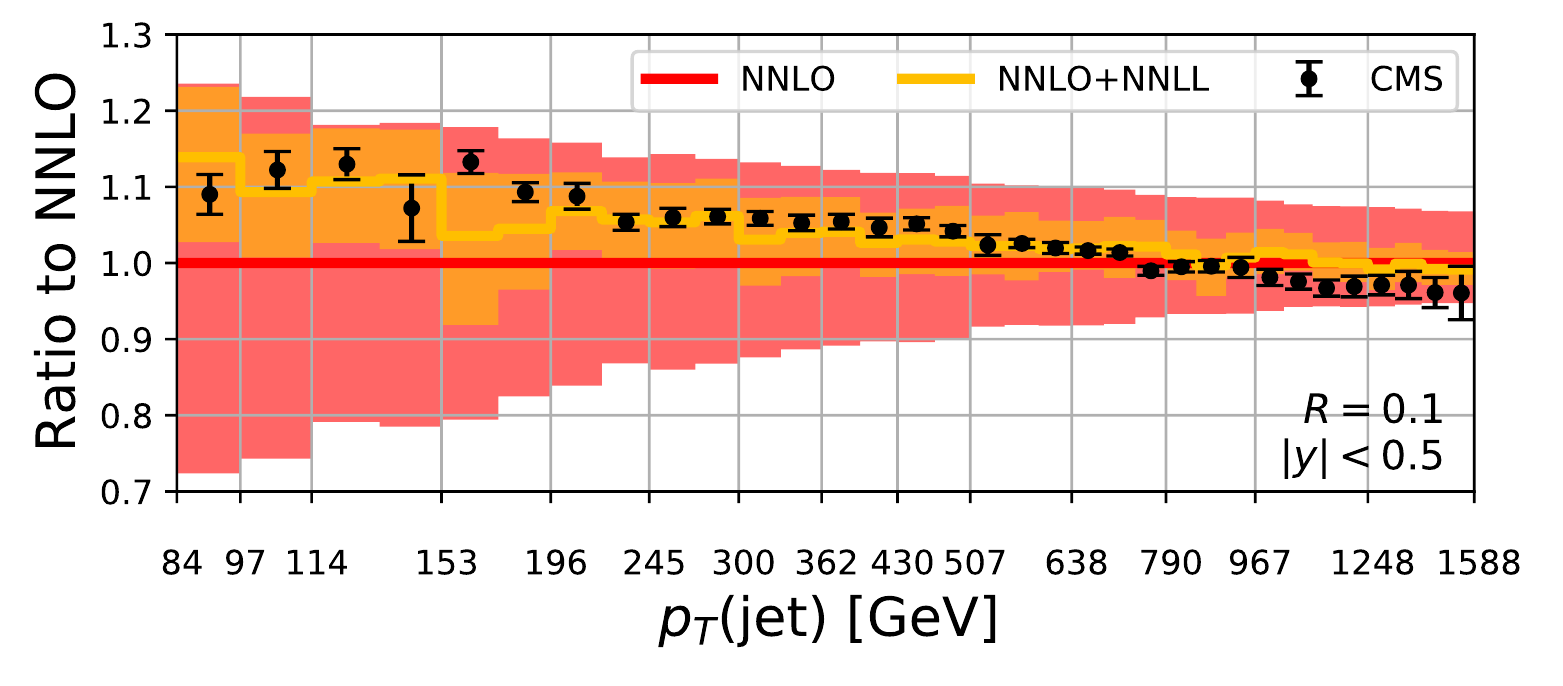}
\caption{Predictions for inclusive small-$R$ jet production at the LHC. Techniques developed in the study of the energy correlators enabled NNLL resummation of small-R logarithms, providing the most accurate predictions for this process. Figure from \cite{Generet:2025vth}.
}
\label{fig:smallR}
\end{figure}

In  \cite{Lee:2024tzc,Lee:2024icn}, a new factorization formula was developed based on the factorization for the collinear limit of the energy correlator, and it was shown that previously proposed formulas were in fact incorrect. This observation was also made in \cite{vanBeekveld:2024jnx,vanBeekveld:2024qxs}. The energy correlators, and the improved understanding of reciprocity, provide a conceptual reason for why previous results were incorrect, and the high order perturbative results for the energy correlator allowed theses ideas to be explicitly tested.

\begin{figure}
\includegraphics[width=0.55\linewidth]{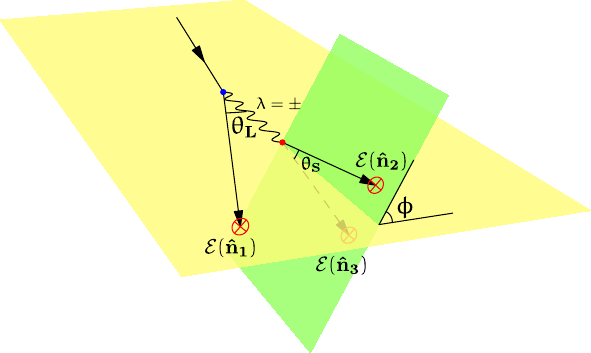}\\
\includegraphics[width=0.85\linewidth]{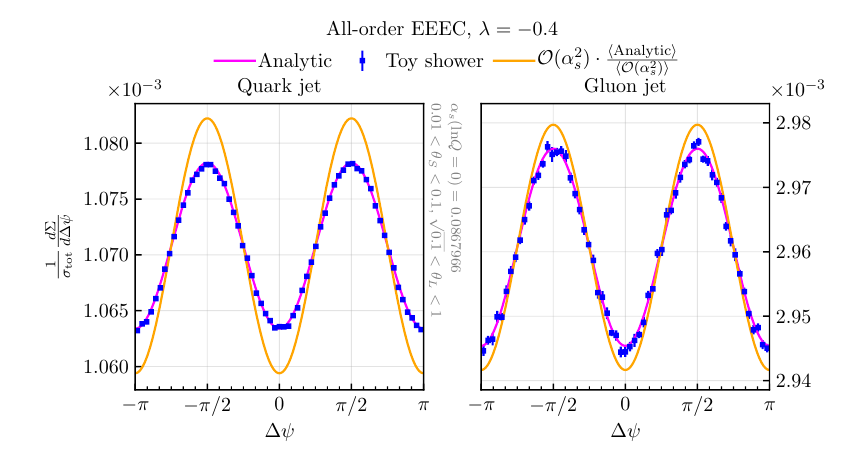}
\caption{The light-ray OPE in QCD enabled the resummation of transverse spin sensitive observables. As an example, we can consider the three-point correlator in the OPE limit, where there exists an azimuthal dependence as the pair of detectors are rotated.  Analytic results for this observable are compared with a parton shower implementation, providing a validation of this implementation, which can then be used for much more general hadron collider phenomenology. Figures from \cite{Chen:2020adz} and \cite{Karlberg:2021kwr}.
}
\label{fig:spin_shower}
\end{figure}

This new factorization approach enabled the first calculations of small-R resummation at NNLL \cite{Generet:2025vth}, using NNLO calculations from \cite{Czakon:2010td, Czakon:2014oma, Czakon:2019tmo,Czakon:2021ohs,Czakon:2024tjr}, as shown in \Fig{fig:smallR}. Here we show results for inclusive jet production with radius $R=0.1$, comparing perturbative calculations at various orders with CMS data. We see the importance of the resummation at NNLL for describing the CMS data. This new approach has enabled the most precise calculations of inclusive jet production at the LHC, and an understanding of the RG for this process will be important for improving jet substructure pheonomenology more generally. It will also be important in the study of jets in DIS, where there has been significant recent perturbative progress in the calculation of the necessary ingredients  \cite{Bonino:2024qbh,Bonino:2024wgg,Goyal:2024tmo,Goyal:2023zdi}. This illustrates how an improved understanding of reciprocity, and factorization, obtained in the well controlled theoretical environment of the energy correlators, can be applied more generally.

Another example involves the improved understanding of transverse spin from the light-ray OPE, and spin correlations in parton showers. Gluons in the parton shower induce spin correlations, which manifest in azimuthal modulations in energy correlator observables. One interesting feature of energy correlator observables is that they are sensitive to spin correlations as pairs (or groups) of detectors are rotated. While an approach to resumming spin correlations in parton showers had been introduced long ago \cite{Knowles:1988hu,Knowles:1987cu,Knowles:1988vs}, and implemented in the Herwig shower \cite{Bahr:2008pv,Bellm:2015jjp,Bellm:2019zci,Richardson:2018pvo} these effects had not been studied either analytically, or benchmarked in parton showers for logarithmic accuracy. In \cite{Chen:2021gdk,Chen:2020adz}, the light-ray OPE approach was used in QCD to analytically resum these transverse spin effects. Transverse spin effects were implemented in the parton shower \cite{Karlberg:2021kwr,Hamilton:2021dyz} with a guaranteed logarithmic accuracy. This implementation was validated by comparing the predictions of the parton shower with analytic calculations from the light-ray OPE. This is shown in \Fig{fig:spin_shower}. The parton shower can then be used to predict a wide variety of other observables useful for experimental analysis, again emphasizing how the energy correlators provide a well understood laboratory where techniques can be developed. There is currently tremendous progress in improving the perturbative accuracy of parton showers from a variety of different directions, see e.g. \cite{Dasgupta:2020fwr,Hamilton:2020rcu,Hoche:2017hno,Hoche:2017iem,Hoche:2024dee}.
We believe that energy correlator observables, where the resummation is understood precisely, will serve as important benchmarks for validating general purpose parton showers.

\begin{figure}
\includegraphics[width=0.5\linewidth]{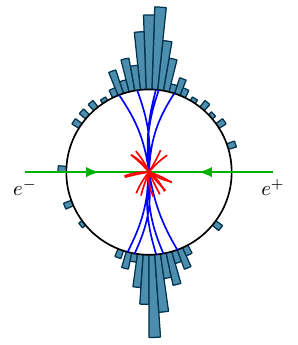}\\
\includegraphics[width=0.5\linewidth]{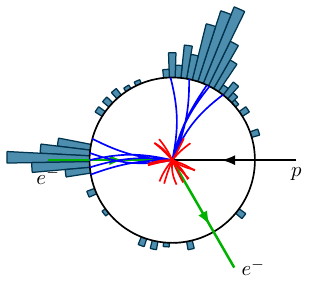}\\
\includegraphics[width=0.6\linewidth]{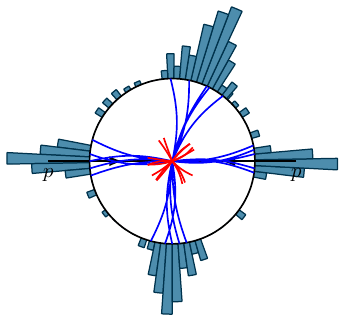}
\caption{A comparison of the different collider geometries, $e^+e^-$, $e p$ and hadron-hadron. The presence of the colliding hadronic states modifies the geometry of the collision, but also provides the opportunity to study the structure of the colliding nuclei.  Each of these collider geometries provides interesting opportunities for the study of energy correlators.
}
\label{fig:diff_colliders}
\end{figure}

\begin{figure*}
\includegraphics[width=0.65\linewidth]{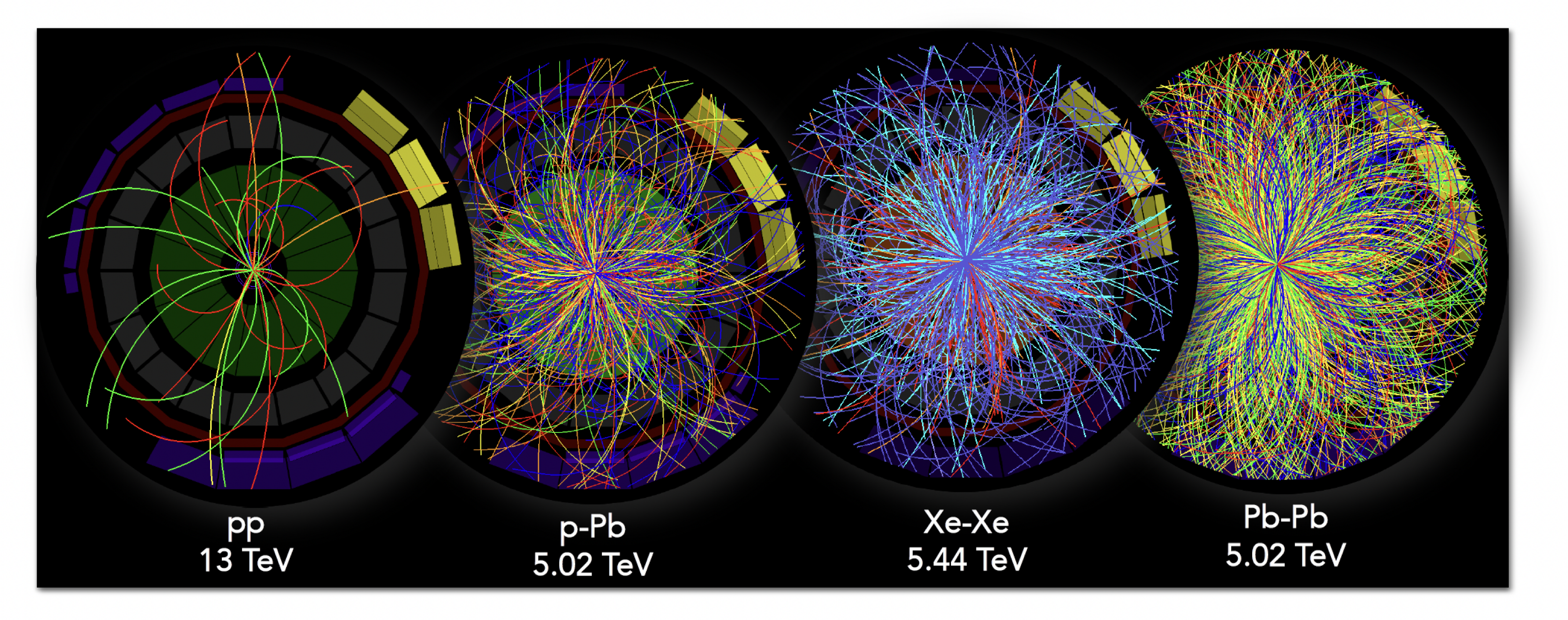}
\caption{Hadron-hadron colliders enable collisions of multiple hadronic species. Energy correlator observables have provided new techniques to robustly map the patterns in asymptotic energy flux in these complicated collisions to the underlying microscopic physics.
}
\label{fig:HIC_event_display}
\end{figure*}

As a final example, we emphasize the utility of having different approaches to understanding the asymptotics of observables, arising from having both a correlator perspective and an amplitudes perspective. As discussed above, in the study of the energy correlator in the back-to-back limit, there are two approaches that can be used to derive the all orders structure. On the one hand, one can use effective field theory techniques to derive a factorization theorem for its all orders asymptotics \cite{Moult:2018jzp}. Resummation of the functions in the factorization theorem then allows the result to be expressed in terms of the cusp anomalous dimension, and the rapidity anomalous dimension. On the other hand, in a conformal gauge theory, the end point asymptotics of the energy correlator were derived from the correlator perspective \cite{Korchemsky:2019nzm}, using the duality between Wilson loops and correlators \cite{Alday:2010zy}. In this case, the result is expressed in terms of the cusp anomalous dimension and $B_\delta$, defined in Eq. \ref{eq:large_spin_def}. The equality of these two approaches provides a proof of the equivalence, in a conformal gauge theory, of the rapidity and collinear anomalous dimensions \cite{Moult:2022xzt}. This allowed the anomalous dimension to be computed to four-loops \cite{Moult:2022xzt,Duhr:2022yyp}. This anomalous dimension has since been used in precision studies of the Z $p_T$ spectrum \cite{ATLAS:2023lhg}, enabling a precision extraction of the strong coupling constant, and again emphasizing how improved understandings of asymptotics of correlation functions can have broader impact. We hope for more exciting such applications in the future.
We also hope that different approaches to studying the asymptotics can also provide insights into other limits, for example the Regge limit, whose discussion is limited in this review.

\section{Experimental Opportunities Across Collider Systems}\label{sec:exp_opp}

In this section we discuss how energy correlators can be studied in real world colliders, highlighting how different collider setups provide unique avenues for studying the energy correlators. Our goal is to provide a high level overview emphasizing the connection between what is measured experimentally, and fundamental QFT  probed by the energy correlators. For example, as highlighted above, the energy correlator exhibits interesting kinematic singularities, for example in its collinear, or back-to-back limits, as well as interesting behavior in non-trivial states. We will discuss how all of these can be accessed using current collider systems, and detail their limitations. In doing so, our goal is on the one hand to highlight to formal theorists why different collider systems are useful, and why measurements are done in the specific manner that they are, and on the other hand, to illustrate to experimentalists the connection between the measurements that are being done, and fundamental physical principles. Our new understanding of the physics of energy correlators will allow us to reformulate many physics questions across collider systems, as sharp questions about energy correlators, leading to new precision measurements  in the Standard Model, and new ways to probe nuclear physics. These physics cases will be described in more detail in \Sec{sec:particle} and \Sec{sec:nuclear}, respectively.  For these reasons, we will purposely not discuss experimental details of the measurements. We will provide references to the experimental measurements for the interested reader.

An overview of the different types of colliders is illustrated in \Fig{fig:diff_colliders}. The physics goals, and possible measurements of each of these colliders is quite distinct. The theoretically simplest colliders are those where the initial state is not charged under QCD. The existing colliders of this form are $e^+e^-$ colliders, although there is interest in a future $\mu^+\mu^-$ collider~\cite{InternationalMuonCollider:2025sys}. In this case the incoming particles couple to a local operator charged under the strong force, enabling the collision to be described by the action of a simple local operator (e.g. the electromagnetic current) on the vacuum. These colliders enable clean studies of final states.

Despite the amazing simplicity and elegance of energy correlators in $e^+e^-$ colliders, they also have limitations, some theoretical and some practical. First, lepton colliders have a fixed state in which the energy correlators are measured. QCD exhibits a rich variety of interesting states, from nucleons, to dense nuclear matter, to the quark-gluon plasma, and one would like to be able to study energy correlators in these states. To do this requires different colliders. Second, existing lepton colliders have only run at relatively low energies. This means that hadronization corrections are large, making it difficult to identify precision scaling over large angular ranges, or to probe multi-point correlators. Third, achieving a high center-of-mass energy in $e^+e^-$ collisions is challenging compared to what can be attained in hadron colliders. This limitation restricts the range of scaling behavior that can be probed at small angular scales. This motivates us to explore opportunities to directly connect measurements with energy correlators in other collider systems.

There are two types of colliders involving nuclear states, namely lepton-hadron colliders, and hadron-hadron colliders. In both cases the presence of a hadron in the initial states opens up the opportunity to study the properties of the initial state using the collision. Such collisions are therefore of great interest in nuclear physics. Hadron-hadron colliders are themselves quite rich, enabling the collision of a variety of different nuclei. This is illustrated schematically in \Fig{fig:HIC_event_display}, which shows collisions ranging from the relatively simple proton-proton collisions, to more complicated Pb-Pb collisions where the quark-gluon plasma is formed.

In this section we will focus primarily on aspects of the energy correlators that have already been measured, and are representative of the measurements in different colliders. In \Sec{sec:particle} and \Sec{sec:nuclear}, we will highlight more sophisticated measurements that can be performed with more specific physics goals in mind.

\subsection{$e^+e^-$ Colliders}

\begin{figure}
\includegraphics[width=0.855\linewidth]{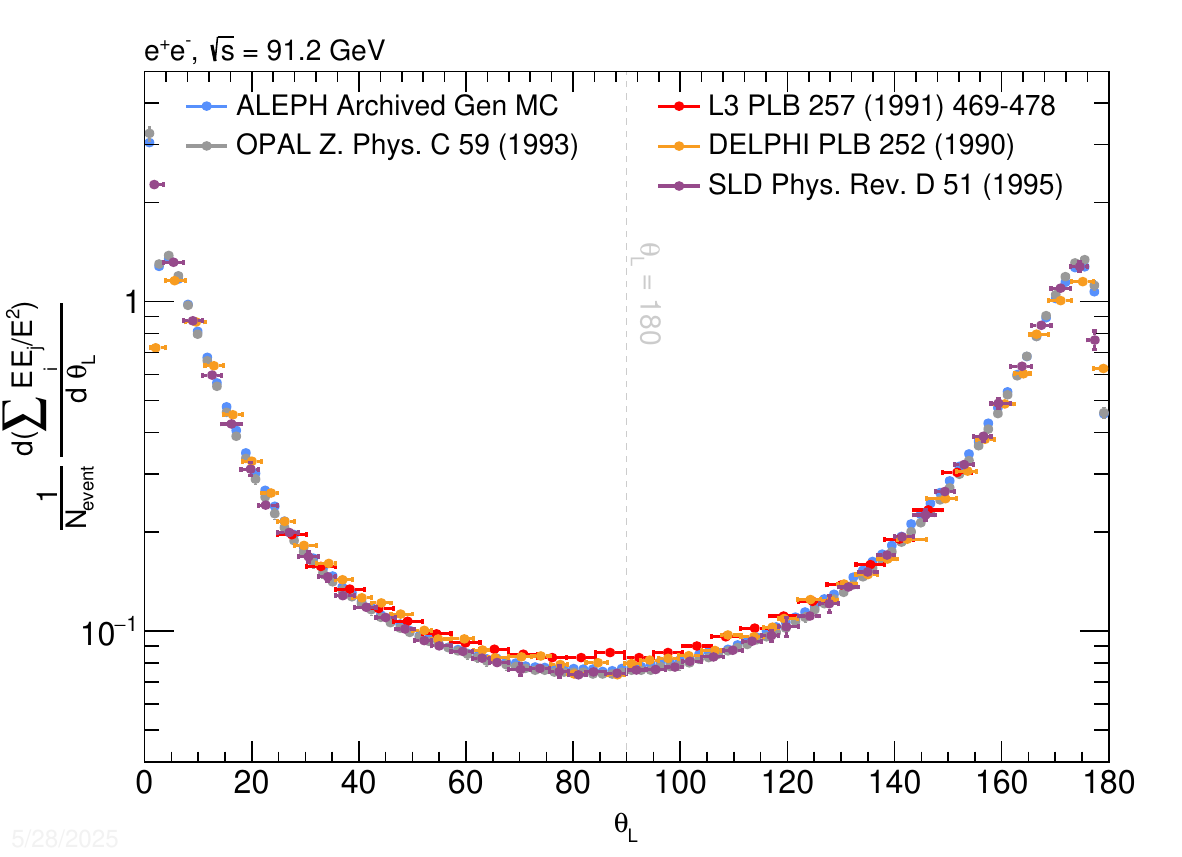}
\includegraphics[width=0.855\linewidth]{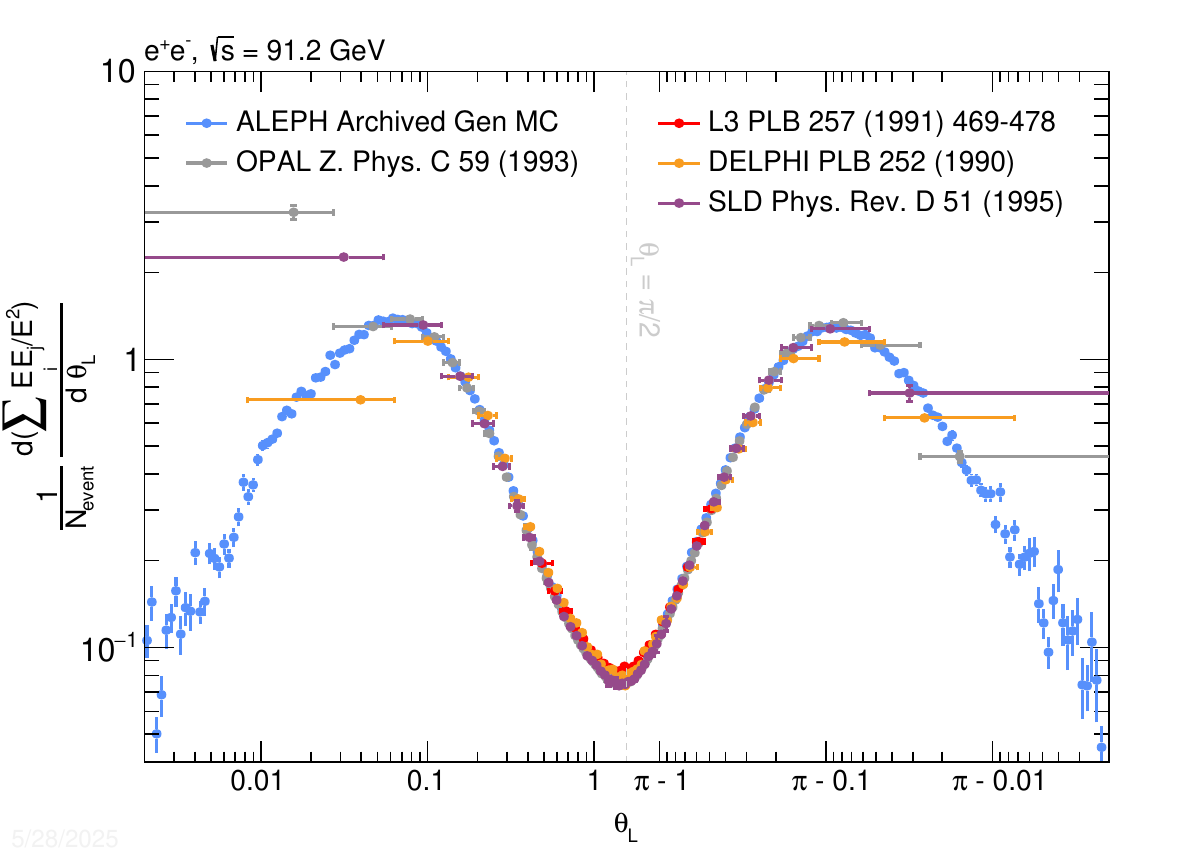}
\caption{Top Panel: Measurements of the energy correlators from the experiments that ran at the Z pole. Data from \cite{ALEPH:1990vew,L3:1991qlf,L3:1992btq,DELPHI:1990sof,OPAL:1990reb,OPAL:1991uui,SLD:1994idb}
combined in \cite{Bossi:2025xsi}. Lower Panel: A comparison of the angular resolution of the ALEPH re-analysis \cite{Bossi:2025xsi}, illustrated with the ALEPH archived MC, with other measurements at the Z-pole. The ALEPH re-analysis provides a transformative view of the asymptotics of the energy correlator.
}
\label{fig:Zpole_measurements}
\end{figure}

Lepton colliders provide the most direct connection between real world collider measurements and the underlying QFT definition of the energy correlators. As such, our discussion of them will be brief. We will focus in particular on advances in the use of track based measurements, which enable energy correlators to be measured with exceptional angular resolution.

For concreteness, we focus on the case of electron-positron colliders $e^+e^-$ where there is abundant data. The complete list of experiments that measured the energy correlator was provided in \Sec{sec:intro}. The majority of this data is from the LEP and SLD experiments which ran at the Z-pole, the highest energy $e^+e^-$ events in which the energy correlators have been measured. In addition, there are a number of currently running $e^+e^-$ colliders, including BELLE-II and BES-III. There are also future $e^+e^-$ colliders proposals under intense investigation, including the Circular Electron Positron Collider~(CEPC)~\cite{CEPCStudyGroup:2018ghi},  the Compact Linear Collider~(CLIC)~\cite{CLICPhysicsWorkingGroup:2004qvu}, the Future Circular electron-positron Collider~(FCC-ee)~\cite{FCC:2018evy}, the International Linear Collider~(ILC)~\cite{ILC:2007bjz}, and the Super Tau-Charm Factory~(STCF)~\cite{Achasov:2023gey}. Each of these has their own physics programs, but we hope that they will all be able to measure energy correlator observables, to give a complete picture over a wide range of energy scales. For an overview of studies of QCD at BELLE-II \cite{Accardi:2022oog}. Similar conclusions would also hold for possible future muon or photon colliders.

\begin{figure}
\includegraphics[width=0.555\linewidth]{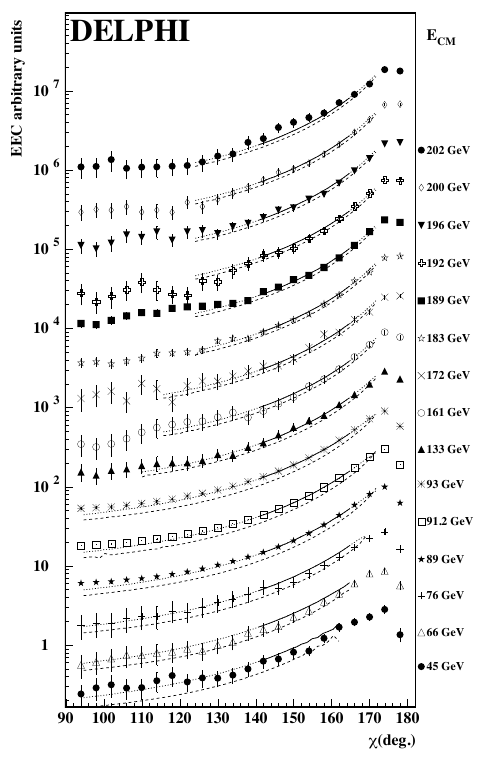}
\caption{The highest energy measurements of the energy corelator in $e^+e^-$ colliders, focusing on the back-to-back limit. Figure from \cite{DELPHI:2003yqh}
}
\label{fig:EEC_energy_scan}
\end{figure}

As discussed above, $e^+e^-$ colliders allow us to directly measure the multi-point functions 
\begin{align}
\frac{1}{\sigma Q^2} \int \df^4x\, e^{\img Q\cdot x} \langle0| J(x) \cE(\hat n_1) \cE(\hat n_2)\cdots  \cE(\hat n_k) J(0)|0 \rangle 
\,,
\end{align}
with $J^\mu=\bar \psi \gamma^\mu \psi$ the electromagnetic current.  Measurements so far have focused on the two-point correlator
\begin{align}
\text{EEC}(\hat n_1, \hat n_2) &=
\frac{1}{\sigma Q^2} \int \df^4x\, e^{\img Q\cdot x} \langle0| J(x) \cE(\hat n_1) \cE(\hat n_2) J(0)|0 \rangle 
\,.
\end{align}
We have already extensively discussed the rich physics of this correlator.

The beauty of $e^+e^-$ colliders is that there is almost no gap between theory, and experimental reality. This will provides the most clean probe into perturbative and non-perturbative regime of QCD. A collection of measurements of the energy correlator from the Z-pole measurements is shown in \Fig{fig:Zpole_measurements}. Finally, in \Fig{fig:EEC_energy_scan} we show measurements of the energy correlator up to 202 GeV. This is the highest energy at which the energy correlators have been measured in $e^+e^-$ collisions, and as such this data is extremely valuable for our understanding of QCD. The only experimental issue is the angular resolution with which the observable can be measured. In recent years, the ability to compute the energy correlators on tracks \cite{Jaarsma:2023ell,Chen:2022muj,Chen:2022pdu,Jaarsma:2022kdd,Li:2021zcf} enables for the first time the opportunity to compare high resolution track based measurements with first principles calculations. 

Remarkably, it is possible to re-analyze this data using modern techniques. See e.g. \cite{Badea:2019vey,Chen:2021iyj,Chen:2021uws}. This has enabled measurements of the energy correlator with extreme angular resolution \cite{Bossi:2024qeu,Bossi:2025xsi}.  An example is shown in \Fig{fig:CMS_scaling} from a re-analysis of ALEPH data at the Z-pole.  Such a precise measurement enables both precision extractions of the strong coupling constant, as well as studies of non-perturbative parameters of QCD, both of which will be discussed in \Sec{sec:particle} and \Sec{sec:nuclear}. It will be extremely interesting to extend this study to the maximal energy of 202 GeV.

\subsection{$ep$ and $eA$ Colliders}\label{sec:ep}

\begin{figure}[t]
  \centering
    \hspace{0.8cm}\includegraphics[width=0.35\textwidth]{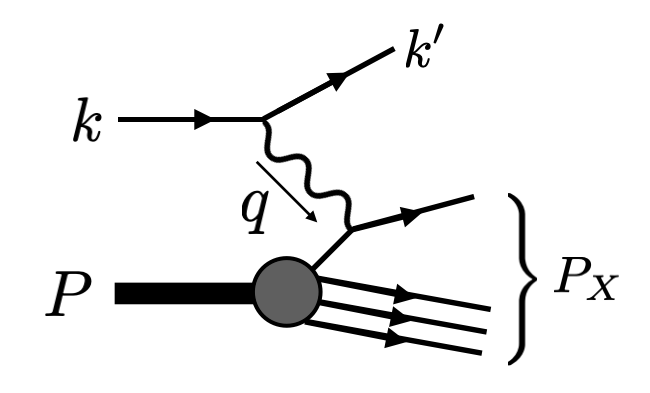}
  \includegraphics[width=0.4\textwidth]{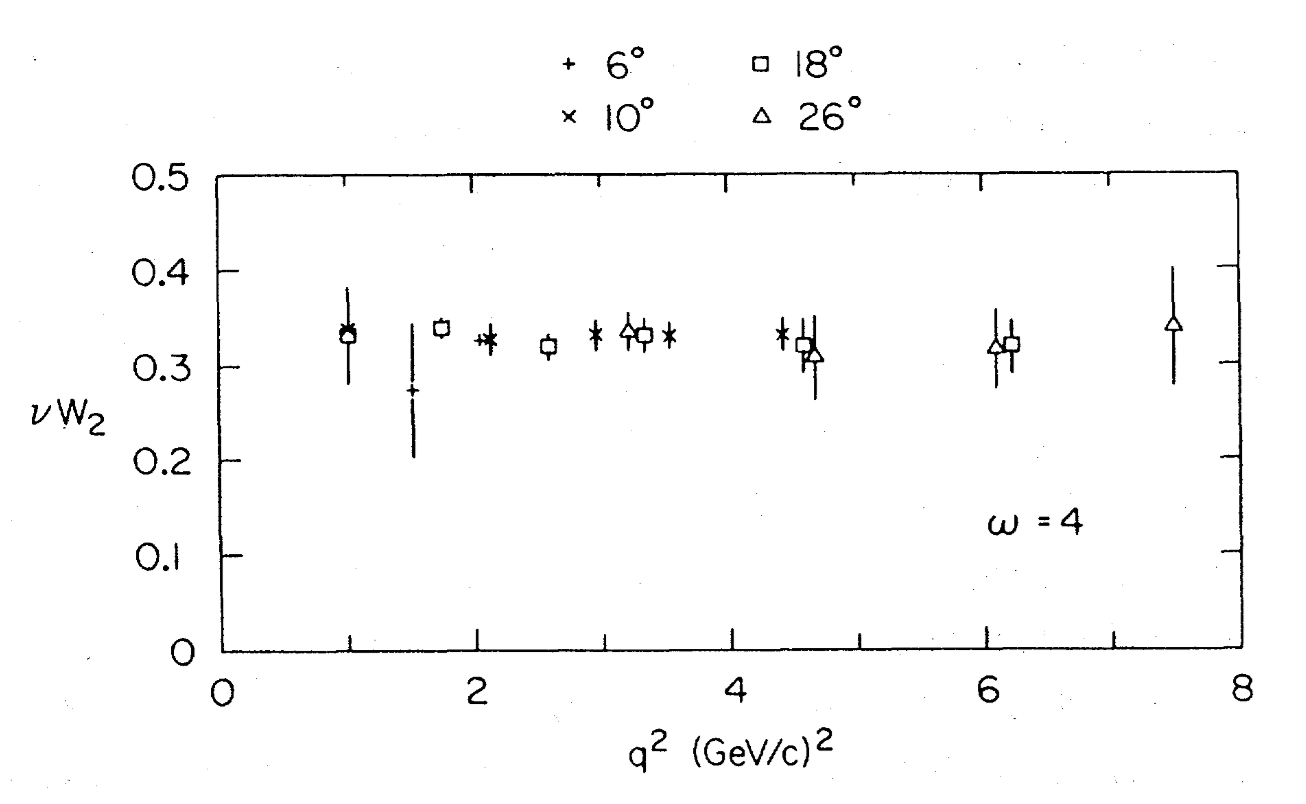}
  \caption{Top: An illustration of the kinematics in DIS. Bottom: Evidence of $q^2$ scaling (Bjorken scaling) from early SLAC-MIT experimental data~\cite{Bloom:1969kc,Breidenbach:1969kd,Bodek:1979rx}. Figure from \cite{Kendall:1991np}.}
  \label{fig:DIS_scaling}
\end{figure}

Electron-proton ($ep$) and electron-ion ($eA$) colliders occupy a unique and historically significant position in the landscape of high-energy physics, providing crucial insights into the structure of hadrons and nuclei. Their operation is analogous to the legendary Rutherford experiment, but instead of using alpha particles, these experiments employ highly off-shell virtual photons to probe the inner structure of nucleons, see \Fig{fig:DIS_scaling}. Deep Inelastic Scattering (DIS) experiments at $ep$ colliders were foundational in establishing QCD as the theory of strong interactions. The landmark discovery of Bjorken scaling \cite{Bjorken:1968dy} at the SLAC-MIT experiments~\cite{Bloom:1969kc,Breidenbach:1969kd,Bodek:1979rx} revealed an approximate scale invariance in the strong force at high energies, see \Fig{fig:DIS_scaling}, pointing to the point-like nature of partons within the proton. The concept of approximate scale invariance observed in DIS finds a conceptual parallel in the characteristic power-law scaling behavior exhibited by energy correlators in their collinear limit. This connection highlights how energy correlators can serve as complementary observables for probing the nearly conformal dynamics of QCD at short distances by studying the structure of final states produced in high energy collisions.

\begin{figure}[t!]
  \centering
  \includegraphics[width=0.4\textwidth]{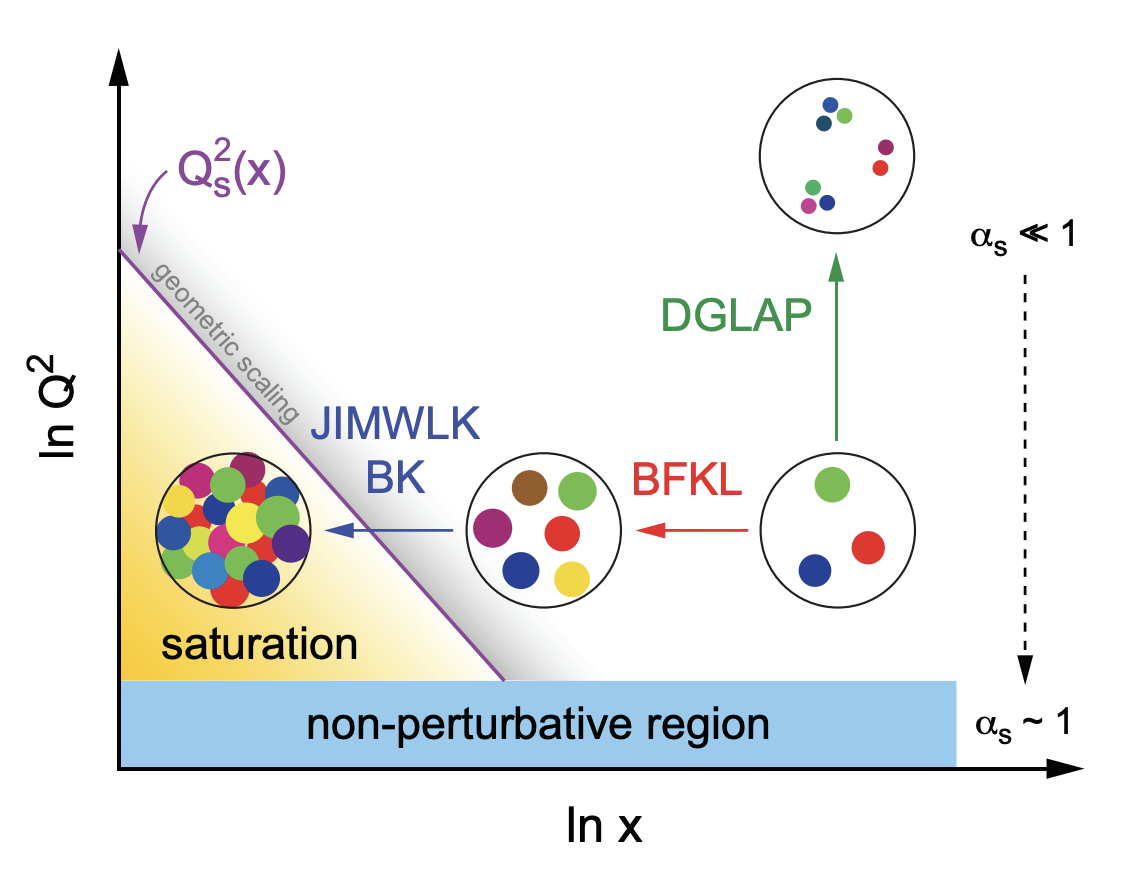}
  \caption{A representation of the landscape of QCD as viewed from the perspective of hadron structure, parameterized by the parton momentum fraction $x$ and the probe scale $Q$. Figure adapted from \cite{Accardi:2012qut}.}
  \label{fig:QCD_landscape}
  \end{figure}

A simplified view of the landscape of QCD studies concerning nucleon structure is often depicted in terms of the parton momentum fraction $x$ and the probe's resolution scale $Q^2 = -q^2$, as illustrated schematically in Fig.~\ref{fig:QCD_landscape}. The "high-energy frontier," characterized by evolution in $\ln Q^2$ governed by the DGLAP equations~\cite{Dokshitzer:1977sg, Gribov:1972ri, Altarelli:1977zs}, describes how parton distributions change with resolving power. Understanding this evolution is central to the precision physics programs at colliders like the LHC, underpinning extensive efforts in perturbative calculations, lattice QCD simulations, and global analyses of PDFs. Complementary to this is the "density frontier" at small $x$. Here, the evolution is expected to be governed by BFKL~\cite{Kuraev:1977fs, Balitsky:1978ic} and non-linear BK-JIMWLK~\cite{Balitsky:1995ub, Kovchegov:1999yj, Jalilian-Marian:1997jhx,Weigert:2000gi,Ferreiro:2001qy} dynamics along the $\ln(1/x)$ axis. A critical scientific goal in this domain is to understand and identify the saturation scale, the point where the rapid growth of gluon density within nucleons and nuclei at small $x$ is tamed by gluon recombination and non-linear QCD effects.

\begin{figure}[t]
    \centering
    \includegraphics[width=0.4\textwidth]{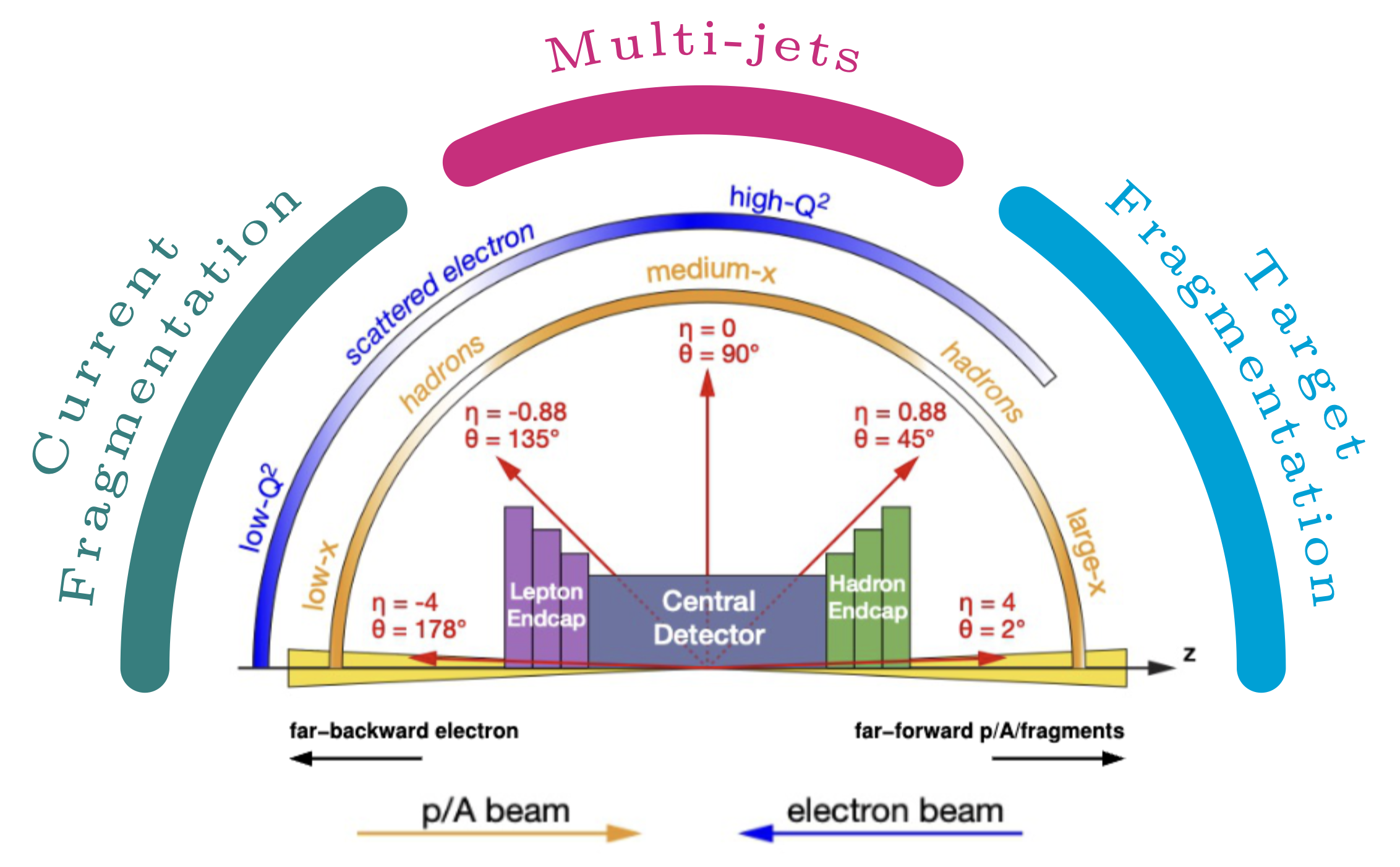}
    \caption{An illustration of the distinct kinematic regions in Deep Inelastic Scattering (DIS). The current fragmentation region involves the hadronization of the struck quark, while the target fragmentation region involves the hadronization of the nucleon remnant. Figure adapted from \cite{AbdulKhalek:2021gbh}.}
    \label{fig:TFRCFR}
\end{figure}

The study of DIS has also consistently pushed the boundaries of our theoretical understanding of QCD in Lorentzian spacetime. From the early development of the Light-Cone OPE~\cite{Wilson:1969zs,Christ:1972ms,Brandt:1970kg} used to analyze Bjorken scaling, to more recent theoretical advances connecting DIS physics to the analytic conformal bootstrap program~\cite{Komargodski:2012ek,Fitzpatrick:2012yx}, lepton-hadron scattering remains a vital arena for theoretical progress in QFT.

Past facilities like HERA provided a wealth of data on proton structure. Current experiments, such as those at the Continuous Electron Beam Accelerator Facility at Jefferson Lab, continue to explore nucleon structure with high precision. Looking towards the future, the Electron-Ion Collider (EIC)~\cite{Accardi:2012qut,AbdulKhalek:2021gbh}, planned to be built at Brookhaven National Laboratory, represents the next generation QCD machine dedicated to unlocking the secrets of nucleon and nuclear structure. Other future possibilities, such as the Electron-Ion Collider in China~\cite{Anderle:2021wcy}, are also being explored. The EIC promises unprecedented capabilities to address fundamental questions~\cite{Accardi:2012qut}, including:
\begin{itemize}
  \item How can we achieve a precise 3D tomography of the nucleon by mapping the spatial (impact parameter) and momentum distributions of its constituent sea quarks and gluons, revealing the fundamental structure underlying visible matter?
  \item How does the nucleon obtain its spin? What are the precise contributions from the intrinsic spins of quarks and gluons, and how can correlations between nucleon spin and parton transverse motion be probed?
  \item What constitutes unambiguous experimental evidence for the predicted universal regime of gluon saturation (CGC) governed by non-linear QCD dynamics, especially in nuclei? What are the emergent properties, collective behavior, and fundamental degrees of freedom of this dense gluon matter?
  \item How does the fundamental process of hadronization occur, transforming colored quarks and gluons into colorless hadrons? How is this process modified when occurring within the nuclear medium (``color filter"), and what can this teach us about the space-time dynamics of parton propagation, energy loss, and color neutralization?
\end{itemize}

\begin{figure}[t]
  \includegraphics[width=0.755\linewidth]{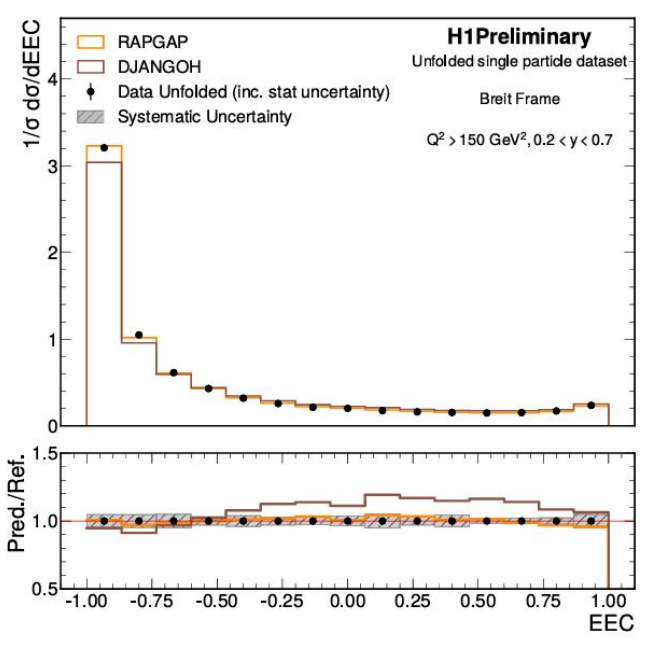}
  \caption{The first measurement of the transverse energy-energy correlator in electron-proton DIS, performed by the H1 collaboration in the Breit frame. This observable measures the transverse energy-weighted angular distribution of hadrons relative to the scattered electron. Figure from \cite{H1:2024dof}.}
  \label{fig:ep_measure}
  \end{figure}

DIS events offer a rich final state structure, typically divided into different kinematic regions as illustrated in \Fig{fig:TFRCFR}. The "current fragmentation" region is associated with the hadronization of the quark struck by the virtual photon. This region is analogous to the fragmentation process in $e^+e^-$ collisions and has been extensively studied.
The multi-jets region, where hard partonic scattering processes in DIS can lead to the production of two or more distinct jets. This region is valuable as it is often amenable to rigorous perturbative QCD calculations, offering opportunities to test QCD dynamics and probe the underlying partonic interactions with high precision. 
The "target fragmentation" region, involving the hadronization of the nucleon or nuclear remnant after the parton has been removed, offers complementary and unique insights into hadron structure, baryon number transport, and multi-parton correlations~\cite{Accardi:2012qut,AbdulKhalek:2021gbh}. Understanding this region requires advanced factorization frameworks~\cite{Trentadue:1993ka, Grazzini:1997ih}.

First step towards energy correlator measurement in $ep$ collisions has very recently been taken by the H1 collaboration. They performed a measurement of the transverse EEC in DIS using HERA data~\cite{H1:2024dof}, which is shown in \Fig{fig:ep_measure}. The observable is defined as:
\begin{align}
\frac{1}{\sigma} \frac{d\Sigma}{d\cos \theta} = \frac{1}{\sigma} \sum_{h} \int d\sigma_{ep \to e+h+X} \, z_h \delta(\cos \theta_{hP} - \cos \theta) \,,
\label{eq:ep_EEC_def}
\end{align}
with
\begin{equation}
  z_a \equiv \frac{P \cdot p_h}{P \cdot\left(\sum_i p_i\right)}
\end{equation}
where $p_h$ and $P$ are the momentum of the hadron $h$ and the incoming proton respectively, $\theta_{hP}$ is the angle between the scattered electron and the hadron, measured in the Breit frame, and $\sigma$ is the total DIS cross section in the measured kinematic region. In the back-to-back limit~($\cos\theta_{hP} \to -1$), a factorization theorem involving TMD PDFs, TMD FFs, and a soft function similar to the ones for EEC in $e^+e^-$ can be formulated~\cite{Li:2021txc}, see also Sec.~\ref{sec:CS_results}. This measurement thus provide an access to those quantities that are related to the 3D tomography of nucleon and hadronization. A detailed comparison between theory and experiments are therefore very welcome.

Besides the conventional TMD region~($\cos\theta_{hP} \to -1$), an exciting prospect for energy correlator studies lies in the target fragmentation region. It would be highly valuable to improve experimental resolution in order to probe the collinear region~($\cos\theta_{hP} \to 1$). This region is interesting as it provides insight into the dynamical transition from quarks and gluons to hadrons~\cite{Komiske:2022enw}, a phenomenon which has recently been observed in both $pp$~\cite{CMS:2024mlf,ALICE:2024dfl,ALICE:2025igw,talk_Hwang,Tamis:2023guc,STAR:2025jut,talk_Shen}, $pA$~\cite{talk_Anjali,talk_Anjali2}, $AA$~\cite{talk_Ananya,CMS-PAS-HIN-23-004,CMS:2025jam,CMS:2025ydi}, and $e^+e^-$~\cite{{Bossi:2024qeu,Bossi:2025xsi}} collision. While Pythia simulations suggest that a similar transition may occur in the target fragmentation region~\cite{Liu:2022wop}, this has yet to be confirmed experimentally. In the future, the EIC, with its collider kinematics and dedicated far-forward detectors integrated into the interaction region (including tracking, calorimetry, and particle identification capabilities extending to very small angles relative to the outgoing hadron beam)~\cite{AbdulKhalek:2021gbh}, will open up unprecedented opportunities to study target fragmentation products. 
We believe that energy correlators measurement in this region could open up new avenues for probing nucleon tomography and hadronization, which will be discussed further in \Sec{sec:nuclear}.

\begin{figure}
\includegraphics[width=0.8\linewidth]{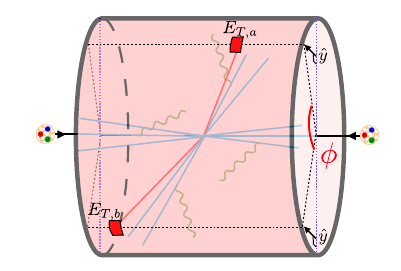}
\caption{The transverse energy-energy correlator (TEEC) at a proton-proton collider. Due to the presence of the beam, the TEEC is measured as a function of the azimuthal angle, $\phi$. In the collinear limit, it regains its universal source independent form, but in the back-to-back limit, it is sensitive to the nature of the colliding beams. Figure from \cite{Gao:2019ojf}.
}
\label{fig:TEEC_schematic}
\end{figure}

In summary, $ep$ and future $eA$ colliders provide a unique environment for studying hadron structure using energy correlators. Building on the historical legacy of DIS, these observables offer new ways to explore QCD dynamics, from the scaling behavior reminiscent of Bjorken scaling to detailed studies of fragmentation in both the current and target regions. Experimental exploration of such opportunities is still in the early stage. We look forward to more clues from data in the near future.

\subsection{$pp$ Colliders}

The highest energy colliders are hadron-hadron colliders, in particular the Large Hadron Collider (LHC) at CERN. The remarkably high collisions energies, $13$ TeV, opens up the opportunity to explore QCD in completely new regimes. In addition, the Relativistic Heavy Ion Collider (RHIC), while primarily focusing on heavy ion collisions, also provides valuable data on proton-proton collisions at lower energies. Hadron-Hadron collisions are much more complicated than their $e^+e^-$ or $ep$ counterparts. The state in which the energy correlators are measured cannot be viewed as one produced by a local operator acting on the vacuum.  Nevertheless, we will see that overcoming this complexity allows us to probe aspects of energy correlators in completely unexplored regimes. This has been one of the driving forces in reinvigorating the interest in energy correlator observables.

As compared to the case of $e^+e^-$ collisions, the initial state of hadron colliders motivates us to use hadron colliders to explore specific kinematic regimes of the energy correlators separately. While this may seem like a complication, we will see that for these specific kinematic regimes they are the best available colliders. 

\begin{figure}
\includegraphics[width=0.85\linewidth]{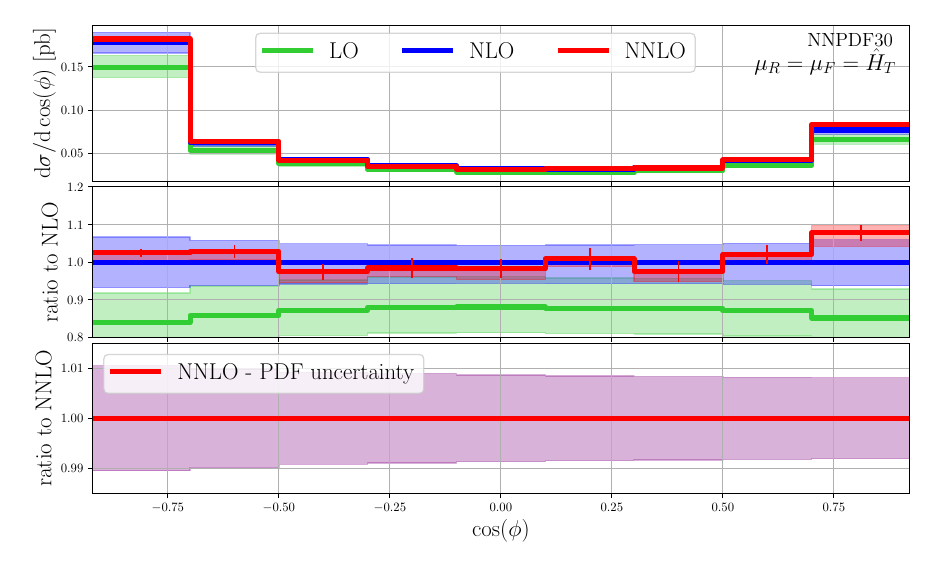}
\includegraphics[width=0.95\linewidth]{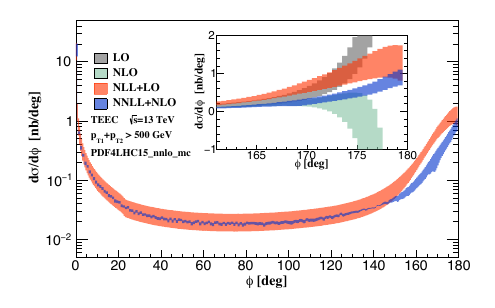}
\caption{Calculations of the TEEC at the LHC. In the top panel, the TEEC is computed on constituent jets at NNLO accuracy. Figure from \cite{Alvarez:2023fhi}. In the lower panel, the TEEC is computed on particles, and includes resummation in the back-to-back limit. Figure from \cite{Gao:2019ojf}.
}
\label{fig:TEEC_NNLO}
\end{figure}

In the case of $e^+e^-$ colliders, there were two kinematic limits where the dynamics of the energy correlator observables simplified greatly. We can understand which of these is modified by the presence of the initial state beams in the hadron collider context. The presence of the beams modifies the symmetries of the energy correlators, if one measures them at generic positions. As compared to the case of $e^+e^-$ collisions, where the local operator preserves an SO(3) symmetry, in the case of hadron-hadron collisions this is further broken down to an azimuthal symmetry about the beam. To achieve boost invariance, it is then standard to use hadron collider coordinates of rapidity, $\eta$ and azimuthal angle $\phi$.

There is another major difference between $e^+e^-$ colliders and hadron colliders, namely that due to the fact that the initial state is composed of complicated bound states of QCD, there is intrinsic interest by a large community in studying the initial state.   The choice of kinematic region that one is most interested in studying can therefore depend a lot on whether one is interested in studying the physics of the initial state, namely the nucleon structure, or the physics of the final state. We will highlight through the section how one can study both the physics of final state jets using the energy correlators, as well as the physics of the initial state.

\begin{figure}
  \includegraphics[width=0.9\linewidth]{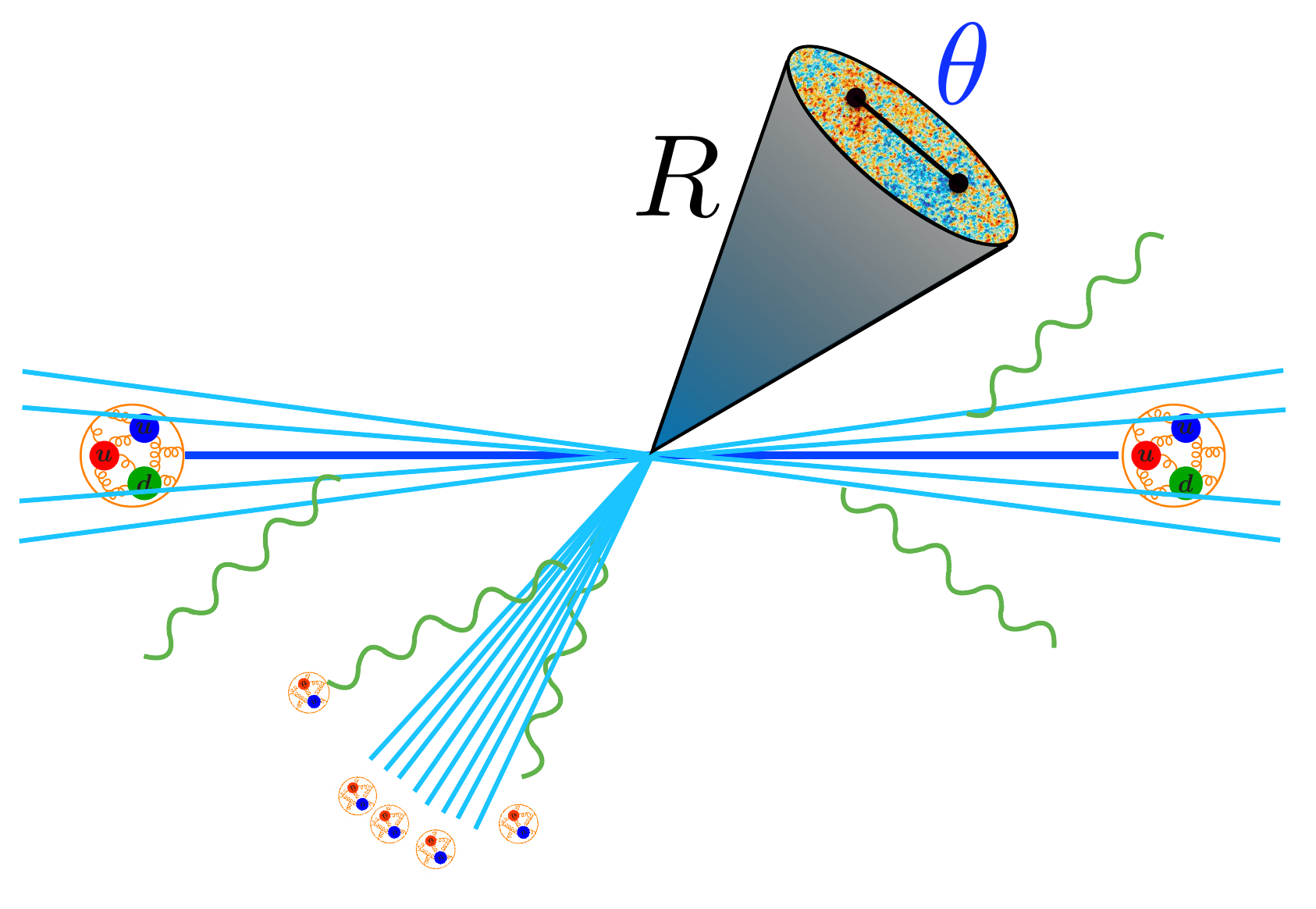}
  \caption{In hadronic events at the LHC, energy correlators are measured inside identified jets of radius $R$. Their universal scaling behavior is regained in the limit $\theta \ll R$. This approach has enabled measurements of the scaling behavior of the energy correlators at record precision energies. 
  }
  \label{fig:LHC_collinear_schematic}
  \end{figure}

For generic angles,  the appropriate generalization of the energy correlator observable at hadron colliders to suppress the contributions from the beam is
\cite{Basham:1978bw,Ali:1984yp}
\begin{align}\label{eq:TEEC_intro}
\frac{d\sigma}{d\cos \phi}=\sum\limits_{a,b} \int d\sigma_{pp\to abX} \frac{2 E_{T,a} E_{T,b}}{ |\sum_i E_{T,i}|^2 }   \delta(\cos\phi_{ab} - \cos\phi)\,.
\end{align}
This definition is illustrated in \Fig{fig:LHC_collinear_schematic}. As of the writing of this review, the full distribution in $\phi$ has not been measured at a hadron-hadron collider when it is calculated as correlations of particles. This is due to the experimental challenge of performing this measurement, due to the extremely large number of particles in the event. However, the two-point correlator has been measured on jets. In this case particles in the event are first clustered into jets, and these jets are correlated into pairs. This significantly complicates the theoretical description, and also degrades the angular resolution in the back-to-back and collinear limits. Nevertheless, it is an interesting test of perturbative QCD. This observable was calculated to a remarkable NNLO accuracy in \cite{Alvarez:2023fhi}, and is shown in \Fig{fig:TEEC_NNLO}.

Much like in the case of $e^+e^-$ collisions, the energy correlator in proton-proton collisions exhibits universal features in both the collinear and back-to-back limits. 
These probe different physics, the back-to-back limit is sensitive to the physics of the beams, i.e. to proton structure, while the collinear limit is sensitive to the physics of final state jet dynamics.

A factorization theorem describing the back-to-back limit of the energy correlator at hadron colliders was presented in \cite{Gao:2023ivm,Gao:2019ojf}. We will not derive the factorization theorem in any detail, but merely emphasize some of its key features. As compared to the case of $e^+e^-$, the dijets produced in the back-to-back limit, along with the beam, now form a plane, as illustrated in \Fig{fig:TEEC_schematic}. The angle $\tau = (1 - \cos\phi)/2$ between the detectors can be expressed in terms of transverse momenta of final state jet functions, a soft function, and TMD beam functions, as
\begin{align}
  \tau = \frac{\left(\frac{k_{3,y}}{\xi_3}+\frac{k_{4,y}}{\xi_4}+k_{1,y}+k_{2,y}-k_{s,y}\right)^2}{4 P_T^2}+\ldots \,.
\end{align}
This allows one to write down a TMD factorization formula for this observable \cite{Gao:2019ojf,Gao:2023ivm}
\begin{widetext}
  \begin{align}
\frac{d\sigma^{(0)}}{d\tau}
 =&\  \frac{1}{16 \pi s^2 (1 + \delta_{f_3 f_4}) \sqrt{\tau}}\sum\limits_{\text{channels}} \frac{1}{N_{\text{init}}}\int \frac{dy_3 dy_4 p_T dp_T^2}{\xi_1\xi_2} \int_{-\infty}^{\infty}\frac{db}{2\pi}e^{-2ib\sqrt{\tau} p_T} \mathrm{tr}\big[\mathbf{H}^{f_1 f_2 \to f_3 f_4}(p_T,y^*,\mu) \mathbf{S}(b, y^*, \mu,\nu) \big]\nn \\
&\ \cdot   B_{f_1/N_1}(b,\,\xi_1,\,\mu,\,\nu)\,B_{f_2/N_2}(b,\,\xi_2,\,\mu,\,\nu) J_{f_3}\left(b,\mu,\nu\right)
  J_{f_4}\left(b,\mu,\nu\right). 
\label{eq:b2b_fact_lhc}
\end{align}
\end{widetext}
As emphasized, in the back-to-back limit, we are sensitive to TMD beam functions. This observable is therefore of intrinsic interest for studying the structure of the proton. We will see that it is also interesting for studying the structure of large nuclei in p-A and A-A collisions in \Sec{sec:heavy_ion}. Similar formulas exist for color singlet plus jet production. It is important to emphasize that the factorization formula in Eq. \ref{eq:b2b_fact_lhc} does not incorporate possible factorization violating effects, and it is possible that it is modified at higher orders. It has been tested to N$^3$LL. This provides another motivation for more detailed studies of the TEEC in the back-to-back limit, where it is an interesting playground for studying factorization and its potential violation. This should be possible using the framework of \cite{Rothstein:2016bsq}.

In \Fig{fig:TEEC_NNLO}, we show the calculation of the TEEC on particles, including resummation in the back-to-back region. As compared to the TEEC on jets, the TEEC measured on particles allows a more detailed view of the kinematic limits. We hope that in the future there will be measurements of the TEEC on particles, in particular, in the back-to-back region, as well as an improved theoretical description.

The other interesting limit at hadron-hadron colliders is the collinear limit, where there is a plethora of new measurements, as well as rigorous theoretical calculations. This limit is largely insensitive to proton structure and allows the study of the final state physics of jets. For non experts in experimental aspects of hadron collider physics, we would like to briefly emphasize why measurements of energy correlators are performed in manner they are. As highlighted above, it is experimentally challenging to measure correlations on the entire event. Furthermore, proton-proton scattering is dominated by forward scattering, and therefore it is of practical importance to select events where there is a genuine hard scattering event. There are two ways of achieving this. The theoretically simplest is to use collisions which produce an energetic color singlet object, such as a $Z$ boson. This guarantees that there has been a hard scattering, and then one can compute the energy correlator on the radiation in the $pp\to Z+X$ collision. However, this is statistically challenging, and therefore does not achieve as high an energy reach as studying the energy correlators in pure QCD events. To study the energy correlators on pure QCD events, one can identify high energy jets, using a jet clustering algorithm. This ensures that there has been a hard scattering, and sets the energy scale for the event. One can then compute the energy correlator on the subset of particles inside the high energy jet. This is shown schematically in \Fig{fig:LHC_collinear_schematic}. As we will show shortly, this extra step does not modify the universal scaling behavior of the energy correlators in the small angle limit. However, it requires a slightly more involved perturbative calculation to accurately describe. Furthermore, it only allows a measurement of the energy correlator for angles smaller than the jet radius, $R$.  We therefore see that proton-proton colliders focus us on a particular region of the energy correlators, in particular their small angle limit. However, for studying this limit, the high energies of hadron colliders are unparalleled.

While it may seem like measuring the energy correlators inside an identified jet state significantly complicates the  underlying universal physics of the collinear limit, in the limit $\theta \ll R$, one can factorize the jet production from the energy correlator leading to the same universal scaling behavior as a function of $\theta$, as if the jet was not there. In the small angle limit, we can write a factorization formula \cite{Lee:2022ige,Lee:2024tzc,Lee:2024icn}
\begin{widetext}
\begin{align}
&\Sigma^{[N]}(x_L,z_J)=
&\int_{z_J}^1 \frac{dx}{x}~  \left[ H_i\left(x,\ln\frac{Q^2}{\mu^2},\mu \right)  \int_{0}^1 dy\, y^N\,\mathcal{J}_{ij}\left(\frac{z_J}{x},y,\ln\frac{x^2 Q^2 R^2}{4\mu^2},\mu \right) \right] J^{[N]}_{j}\left(\ln\frac{x_Ly^2 Q^2}{\mu^2},\mu\right)\,.\nn
\end{align}
\end{widetext}
Here $H_i$ is the inclusive single hadron hard function, much like in the factorization theorem for the small angle limit of the energy correlator in $e^+e^-$, and $\mathcal{J}$ describes the dynamics of the jet clustering algorithm \cite{Kang:2016ehg,Kang:2016mcy,Dai:2016hzf}. Their combination can be viewed as an effective hard function for the process. Importantly, the function $J^{[N]}_{j}$ is the same energy correlator jet function that appears in $e^+e^-$ colliders. This again makes sense from the perspective of the OPE.

An important aspect of this factorization theorem, is that it can be proven rigorously by extending the proofs of factorization for inclusive fragmentation in \cite{Collins:1981ta,Bodwin:1984hc,Collins:1985ue,Collins:1988ig,Collins:1989gx,Collins:2011zzd,Nayak:2005rt}. This is important if this factorization theorem is going to be used for precision measurements of the strong coupling constant, or the top quark mass.  Therefore, using the above factorization theorem, we are able to extend the nice properties of the energy correlators from $e^+e^-$ colliders to the complicated world of hadron colliders.

\begin{figure}
  \includegraphics[width=0.955\linewidth]{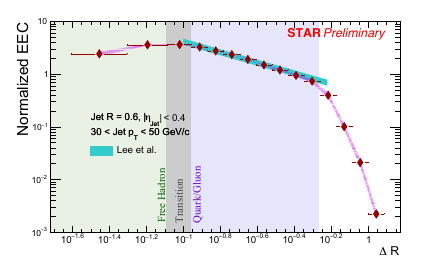}
  \caption{The measurement of the two-point energy correlator in the STAR experiment, compared with NLL perturbative predictions. Figure from \cite{Tamis:2023guc,STAR:2025jut}.
  }
  \label{fig:STAR_EEC}
  \end{figure}

  \begin{figure}
  \includegraphics[width=0.655\linewidth]{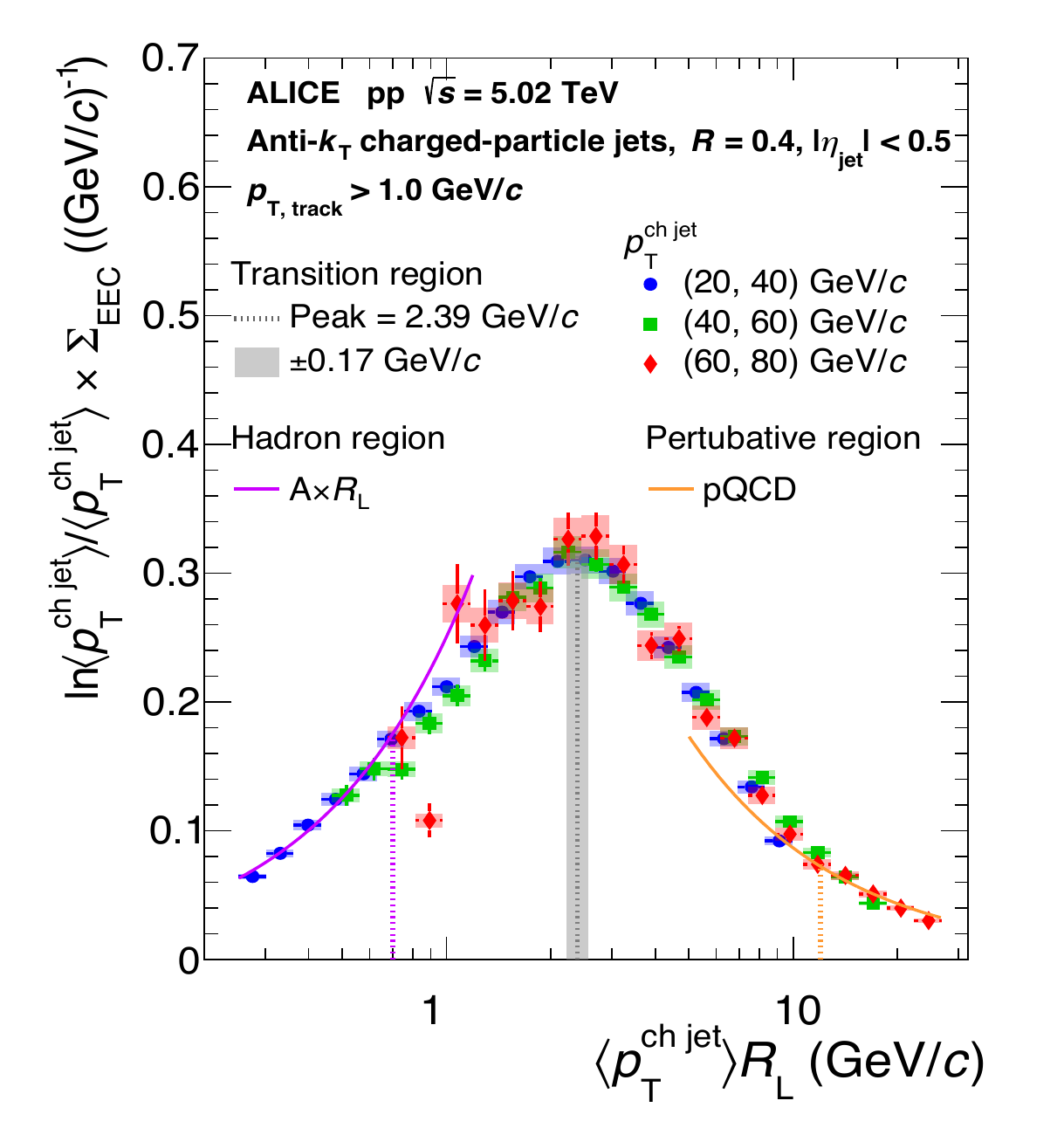}
  \includegraphics[width=0.955\linewidth]{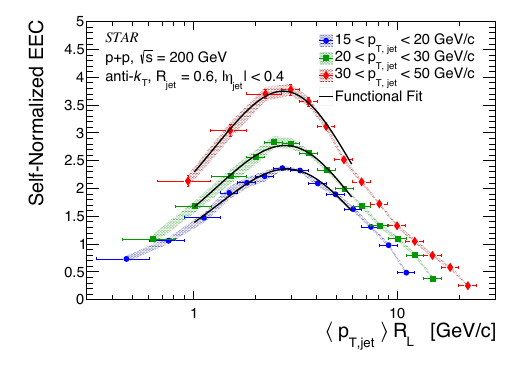}
  \caption{Measurements of the two-point energy correlator by ALICE (top) and STAR (bottom), emphasizing the transition from a perturbative scaling reflective of asymptotically free quarks and gluons, to an integer scaling reflective of freely interacting hadrons. Figures from \cite{ALICE:2024dfl} and \cite{STAR:2025jut}.
  }
  \label{fig:ALICE_EEC}
  \end{figure}

The simplest of the energy correlator observables is the two-point energy correlator. It is a function of a single variable, namely the angle between the two detectors. In the proton-proton collider context, the observable is expressed in terms of the boost invariant angle $\Delta R=\sqrt{\Delta \eta^2 + \Delta \phi^2}$, where $\eta$ is the rapidity, and $\phi$ is the azimuthal angle. In the small angle limit, and at central rapidity, this is essentially identical to the angle.

The first property we can probe is the small angle scaling of the two-point correlator. While this scaling is evident in $e^+e^-$ by the enhancement in the collinear limit, it is hard to study there due to large non-perturbative corrections. Recall that the OPE is an expansion in $\theta \ll 1$. However, in QCD, we want to have $\theta \gg \frac{\Lambda_{\text{QCD}}}{Q}$ to reduce non-perturbative corrections. To probe the scaling behavior, we therefore want to extend the range $\frac{\Lambda_{\text{QCD}}}{Q} \ll \theta \ll 1$. The energies of the LHC, where one has large samples of TeV scales jets makes this possible, and is a unique advantage of high energy hadron colliders.

Prior to recent studies, energy correlators had not been measured at hadron colliders. They have now been measured by two collaborations at the LHC (ALICE and CMS), as well as at RHIC (STAR). There are two  features of the energy correlator that one is interested in studying, and this motivates how they are plotted by the experimental collaborations. One is either interested in the scaling behavior in the asymptotically free regime, or in the transition between asymptotically free quarks and gluons, and hadrons, namely studying the confinement transition. 

\begin{figure*}
\includegraphics[width=0.755\linewidth]{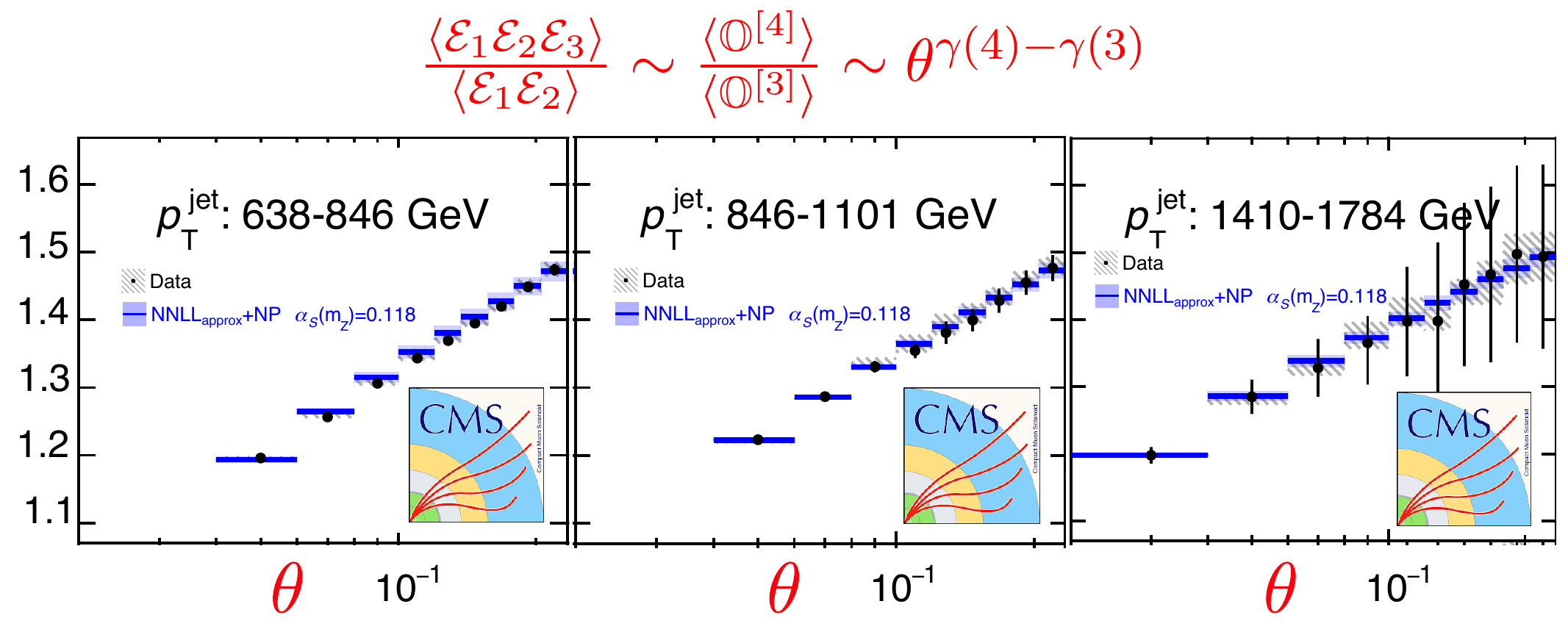}
\caption{CMS measurements of the ratio of the three-point to two-point projected energy correlators inside high energy jets at the LHC. This observable provides a new approach to precision measurements of the strong coupling constant. Figure adapted from \cite{CMS:2024mlf}.
}
\label{fig:CMS_projected}
\end{figure*}

We first consider studies emphasizing the scaling behavior. In this case, one typically uses log-log axes to emphasize the power law scaling. In \Fig{fig:CMS_scaling} we show a measurement of the two-point function function inside high energies jets at the LHC from the CMS experiment \cite{CMS:2024mlf}. Here a jet selection of 1410 GeV $\leq p_T \leq$ 1784 GeV was chosen. We can see by eye the perfect scaling over multiple orders of magnitude in angle. The ability to go to the tremendously high energies of the LHC has been transformative in revealing the  asymptotic limits of the energy correlator. The two-point correlator has also been measured in lower energy jets at proton-proton colliders by both the ALICE and STAR collaborations.  In \Fig{fig:STAR_EEC}, we show a measurement from STAR \cite{STAR:2025jut,Tamis:2023guc}, with jets in the range 30 GeV $\leq p_T \leq$ 50 GeV. We again see a clean scaling regime, although it is smaller due to the lower energy jets. The cutoff of the scaling behavior at larger angles corresponds to the correlators hitting the jet boundary.

One can also emphasize the transition between perturbative and non-perturbative physics. For this purpose the distribution is plotted in a log-linear fashion.  In \Fig{fig:ALICE_EEC} we show measurements from ALICE \cite{ALICE:2024dfl}, and STAR \cite{STAR:2025jut,Tamis:2023guc}. These measurements highlight the confinement transition: at larger angles (to the right of the peak), we observe a non-integer scaling behavior of quarks and gluons, and on the left, we observe a geometric scaling law associated with free hadrons. There is currently no known way to calculate the behavior in the transition region from first principles. However, the ability to precisely image the hadronization transition is quite remarkable. The plots in  \Fig{fig:ALICE_EEC}  are reminiscent of phase transitions, however, in this case no external parameter is being tuned. In this sense they are similar to dynamical quantum phase transitions \cite{Heyl:2018jzi}. It would be interesting to understand them better.

\begin{figure}
\includegraphics[width=0.955\linewidth]{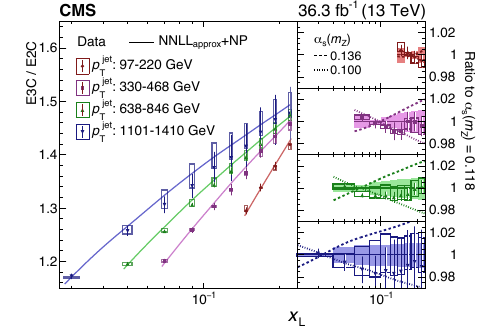}
\includegraphics[width=0.855\linewidth]{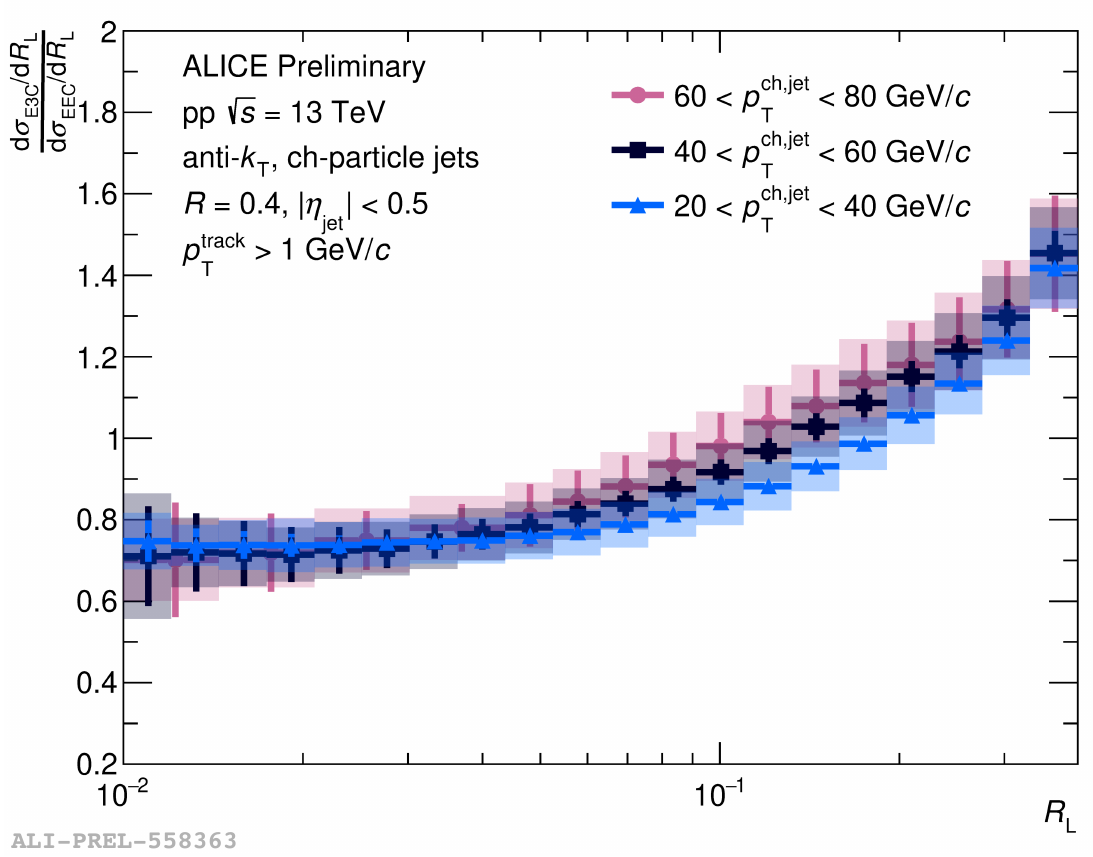}
\caption{Measurements of the ratio of the three-point to two-point projected correlator as a function of jet energy. Due to the running of the coupling in QCD, as well as the enhanced gluon fraction at lower jet energies, the scaling  increases at lower jet energies. We show results from both CMS at high energies, and ALICE at lower energies. Figures adapted from \cite{CMS:2024mlf} and \cite{talk_Ananya}.
}
\label{fig:CMS_alpha_run}
\end{figure}

We see that using high energy jets at the LHC allows us to explore the OPE limit of the energy correlators. Indeed, we can do much better, and in fact we can measure the anomalous dimensions of different operators that can appear in the OPE limit. The measurement of the complete shape of multi-point correlators begins to depend on many variables. One would therefore like to project out the information associated with the scaling behavior in a simple manner that is efficient experimentally. A method to do this was proposed in \cite{Chen:2020vvp}, and is referred to as projected energy correlators.

We define the projected $N$-point correlator as
\begin{align}
  \label{eq:projection}
\hspace{-0.5cm}  &\frac{d\sigma^{[N]}}{d x_L} \ = \int\! d\Omega_{\vec{n}_1} 
\! \int\! d\Omega_{\vec{n}_2} 
\delta (x_L - \frac{1 - \vec{n}_1 \cdot \vec{n}_2}{2} ) 
\prod_{k=3}^N \int \!  d\Omega_{\vec{n}_k}
\nn\\
&\ 
\Theta(\{\vec{n} \}) \int\! d^4 x\, \frac{e^{i q \cdot x}}{Q^N} \langle 0 | \cO^\dagger(x)   
{\cal E}(\vec{n}_1) {\cal E}(\vec{n}_2)  \ldots 
{\cal E}( \vec{n}_N)
\cO (0) | 0 \rangle \,,
\end{align}
where
\begin{equation}
\label{eq:areaint}
d \Omega_{\vec n} = \frac{1}{4 \pi} \sin\theta d \theta d \phi\,, 
\end{equation}
is the area element on the celestial sphere. The integration region for $d\Omega_{{\vec n}_k}$ is specified by
\begin{align}
  \label{eq:omega12k}
  \Theta(\{ \vec{n} \})  = \prod_{
\substack{1\leq i<j\leq N
\\
i+j > 3 
}
} \theta(|\vec{n}_1 - \vec{n}_2|  - |\vec{n}_i - \vec{n}_j|) \,.
\end{align}
In other words, we fix the largest angular distance between two detectors in the correlator to be $x_L = (1 - \cos\theta_{12})/2$, and integrate over all the additional shape information of the correlator. This observable is convenient, since it provides a function of a single variable that allows one to easily isolate the scaling exponents of operators on the leading Regge trajectories. The first illustration of the projected correlators was performed using CMS Open Data  presented \cite{Komiske:2022enw}, and was compared against NLL calculations in \cite{Lee:2022ige}. Precision calculations \cite{Chen:2023zlx} of this observable and its use for measurements of the strong coupling constant will be discussed in \Sec{sec:particle_alphas_pp}.

A useful feature of having a family of single variable correlator observables is that we can take the ratio \cite{Chen:2020vvp}. This has two beneficial features: first, it removes the classical scaling component;  Second, it cancels some of the non-perturbative corrections to the observables. The ratio of the three-point to two-point projected correlators as measured on both high energy jets in CMS \cite{CMS:2024mlf}, and lower energy jets in ALICE \cite{talk_Ananya}, is shown in \Fig{fig:CMS_projected}. We emphasize that the observed slope is a purely quantum mechanical effect arising from the anomalous dimensions of the twist-2 light-ray operators. The beauty of this is that it enables one to connect what is actually measured in hadron collider experiments with microscopic properties of the underlying theory, i.e. the spectrum of anomalous dimensions In this sense, we are able to measure the ``spectrum of a jet". It also provides an experimental test of the OPE of detector operators.

An interesting feature of the scaling is that, as compared to a conformal theory, it depends on the energy of the jet in which it is measured. In \Fig{fig:CMS_alpha_run}, we show a measurement of the ratio of the three-point to two-point projected energy correlators measured over a wide range of energies, from approximately 100 GeV to 1 TeV. It is observed that the slope is considerably larger at smaller energies. This arises due to two effects. On the one hand, the QCD coupling is larger at smaller energies, increasing the value of the anomalous dimensions. On the other hand, lower energy jets have a large gluon fraction, making the lower energy predictions more sensitive to the gluon anomalous dimension, as compared to the quark anomalous dimension.

\begin{figure}
\includegraphics[width=0.755\linewidth]{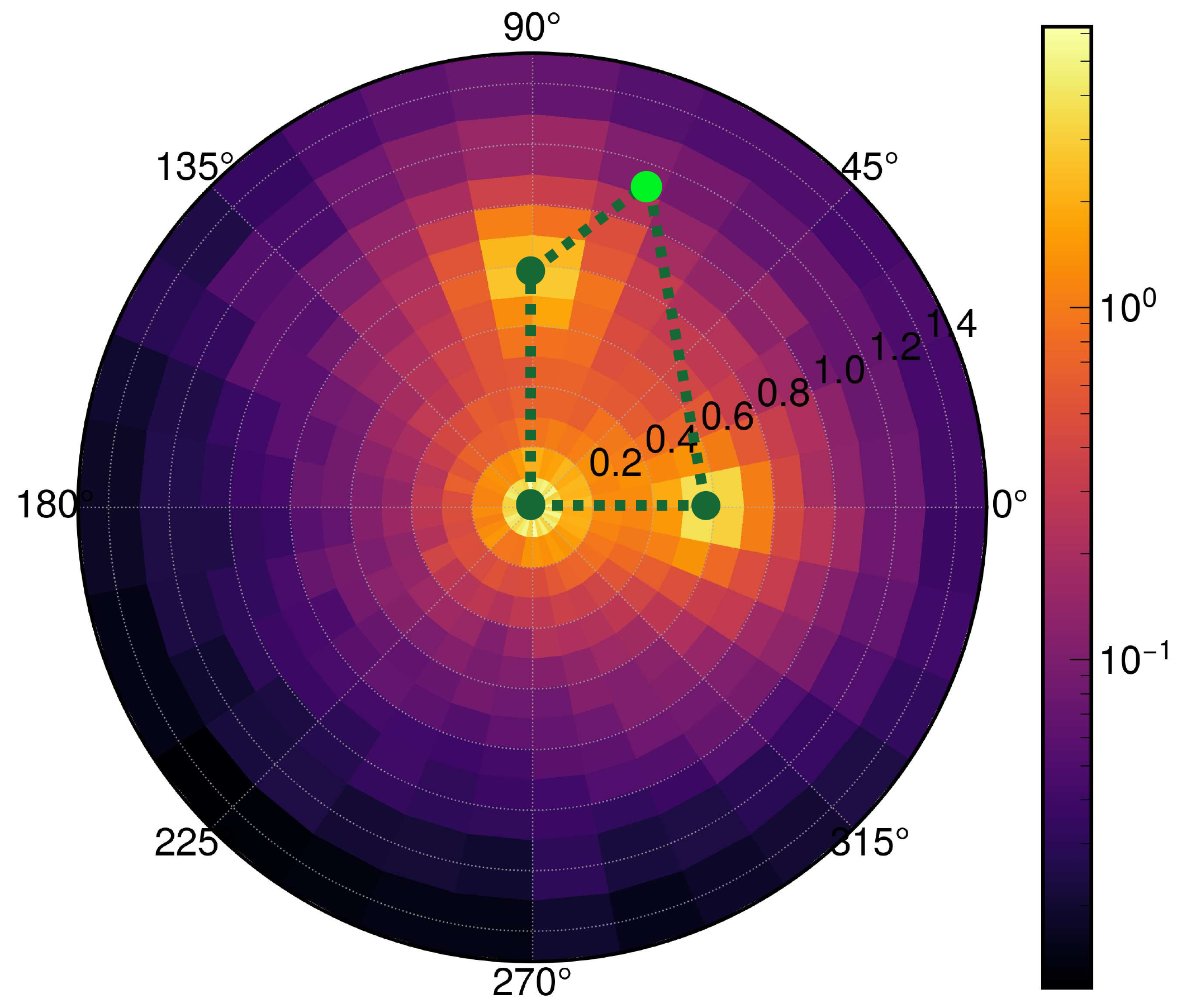}
\caption{A simulation of the shape dependent four-point correlator inside a high energy jet at the LHC. The dark green correlators are kept fixed, and the result is plotted as a function of the position of the light green detector. Such measurements of high point correlators are being pursued by the collaborations, providing a detailed view of the internal structure of jets.
}
\label{fig:4point_measure}
\end{figure}

Beyond just the scaling properties of the multi-point energy correlators, it is also possible to directly measure the shape dependence of higher point correlators in the collinear limit. These shapes become particularly interesting for probing other systems, such as the top quark, or the quark gluon plasma, as we will discuss in more detail. In \Fig{fig:decorated_opendata} we showed a measurement of the three-point correlator, and in \Fig{fig:4point_measure} we show a simulation of a soon to be released measurement of the four-point correlator.  One thing that is quite remarkable is that we are directly measuring the observable that one can compute, as was highlighted in \Sec{sec:multipoint}.

\begin{figure}
\includegraphics[width=0.955\linewidth]{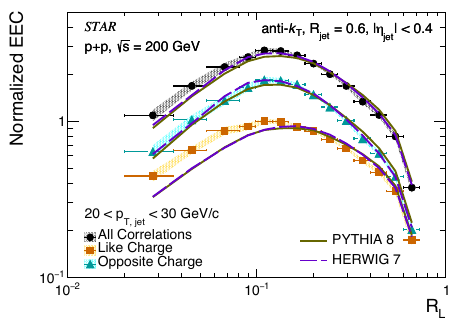}
\includegraphics[width=0.955\linewidth]{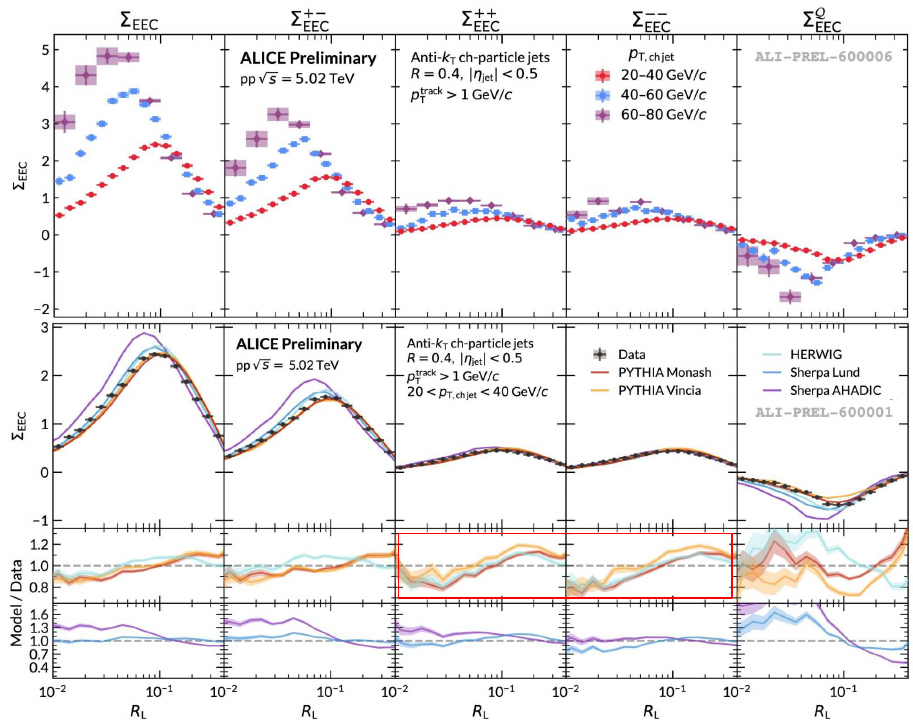}
\caption{Measurements of the $\langle \mathcal{E}_+  \mathcal{E}_+ \rangle $, and $\langle \mathcal{E}_-  \mathcal{E}_+ \rangle $ correlators by the STAR collaboration (upper panel) and ALICE collaboration (lower panel). Figures from \cite{STAR:2025jut} and \cite{talk_Hwang}.
}
\label{fig:fig_STAR_charged}
\end{figure}

In addition to measurements of the standard energy detector, it is also worth emphasizing that all modern particle detectors have particle ID, and can measure properties of the detected particles beyond just their energies. In particular, an example of a generalized detectors that can be measured in collider experiments is the energy flux on positive or negative particles, enabling the measurement of the correlation functions $\langle \mathcal{E}_+  \mathcal{E}_+ \rangle $, and $\langle \mathcal{E}_-  \mathcal{E}_+ \rangle $. Such correlators are interesting because they are sensitive to properties of hadronization. In particular, one expects enhanced correlations at small angles in the $\langle \mathcal{E}_-  \mathcal{E}_+ \rangle $ correlator in models of string fragmentation. The measurement of these correlators, was performed by both the STAR  \cite{STAR:2025jut} and ALICE \cite{talk_Hwang} collaborations, as shown in \Fig{fig:fig_STAR_charged}.  These measurements are not well produced by parton shower generators, and would be interesting to study further. It would also be interesting to make these measurements more differential, for example on strange particles, to further test hadronization models.

\begin{figure}
\includegraphics[width=0.55\linewidth]{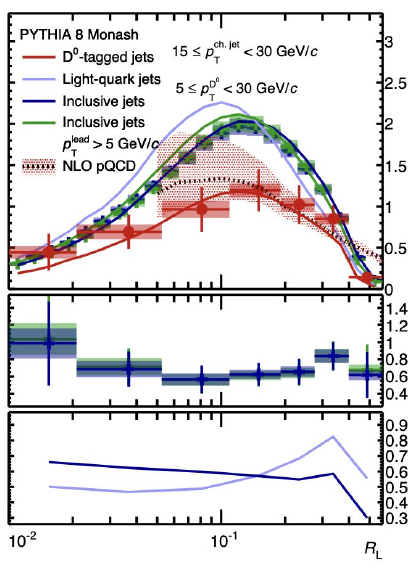}
\caption{A measurement by the ALICE collaboration of the two-point energy correlator on $D_0$ tagged jets, compared with inclusive jets. $D_0$ mesons are formed from b-quarks, whose mass provides an additional scale, $m_b \gg \Lambda_{\text{QCD}}$, at which the scaling behavior of the energy correlator stops. This results in a suppression of the correlator at smaller angles. Figure from \cite{ALICE:2025igw}.
}
\label{fig:bquark_measure}
\end{figure}

Another exciting opportunity associated with the identification of specific particle species is the opportunity to measure energy correlators on jets originating from massive quarks. One example where this can be achieved is the case of the b-quark, which hadronizes into a $D_0$ meson. The $D_0$ then undergoes a weak decay, but its momentum can be reconstructed from the momentum of the decay products, offering the opportunity to study energy correlators on $D_0$ jets, and test the intrinsic mass effects discussed theoretically in \Sec{sec:heavy_quarks}. A measurement of the energy correlator on $D_0$ jets was performed by the ALICE collaboration \cite{ALICE:2025igw}, and is shown in \Fig{fig:bquark_measure}. As compared with inclusive jets, the presence of the b-quark mass introduces an additional scale, $m_b \gg \Lambda_{\text{QCD}}$, which interrupts the otherwise nearly conformal scaling of massless QCD. Therefore one expects that the two-point correlator should turn over at a larger scale than for massless jets. This results in a suppression of radiation at small angles, and is  known as the dead cone effect. This is observed in \Fig{fig:bquark_measure}. Using current statistics, it is only possible to perform this measurement at relatively low energies where the perturbative scaling only lasts over a small region. With increased datasets it should be possible to measure this at higher energies, where there would be a larger separation between the perturbative and non-perturbative regimes. LHCb is a specialized detector for studying B-physics, and it would be particularly interesting to study energy correlators there.

\begin{figure}
\includegraphics[width=0.75\linewidth]{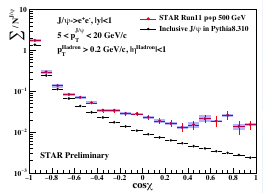}
\caption{A measurement of the J/Psi energy correlator by the STAR collaboration. A significant difference between data and parton shower simulations is observed, highlighting the importance of this observable for improving our description of charmonium formation. Figure from \cite{talk_Shen}.
}
\label{fig:jpsi_measure}
\end{figure}

Another class of interesting systems in QCD are bound states of heavy quarks, generically referred to as quarkonia, such as the J/Psi (bound state of two charm quarks), or bottomonium (bound state of two b-quarks). At modern colliders, these states can be produced with large transverse momentum. In \cite{Chen:2024nfl} a generalization of the energy correlator was introduced to study the energy distribution around quarkonia. A more detailed discussion of the physics of this observable is provided in \Sec{sec:quarkonia}. This observable was first measured by the STAR collaboration, and is shown in \Fig{fig:jpsi_measure}. As compared to the simpler cases of massless QCD, here we see a large disagreement between theoretical calculations and data. This illustrates the potential of using these measurements to improve our understanding of quarkonia systems, and the rich physics program that is possible at hadron colliders.

Another benefit of the high energies of modern proton-proton colliders is the opportunity to produce massive states, which decay hadronically. Precision measurements of energy correlators on their decay products enable a characterization of these states. Examples include bottom and top quarks, and  $W/Z/H$ bosons. We will discuss these more specialized cases in detail in \Sec{sec:particle}.

The Regge limit offers another important regime for study in high-energy jet production. Observables such as Mueller-Navelet jet production~\cite{Mueller:1986ey} are typically employed to probe the onset of BFKL dynamics~\cite{CMS:2016qng}. Very recently, it has been proposed that the Regge limit of QCD can also be investigated using a full-range EEC in proton-proton collisions~\cite{Chen:2025rjc}. For this purpose, it is necessary to use the full particle energies as weighting factors, rather than transverse energies, defining the correlator as:
\begin{align}
  \frac{d^2 \Sigma}{d \Omega_a d \Omega_b}=&\ \sum_{i, j} \int d \sigma_{p p \rightarrow i+j+X} E_i E_j \nonumber
  \\
  &\ \cdot \delta^{(2)}\left(\Omega_a-\Omega_{p_i}\right) \delta^{(2)}\left(\Omega_b-\Omega_{p_j}\right) \,,
\end{align}
where $\Omega_{a(b)}$ are are the orientation of the two detectors. 
LO calculation shows that at very large rapidity difference $\Delta Y$ the full-range EEC grows as $
\Delta Y e^{3 \Delta Y}$. This large $\Delta Y$ behavior can be reproduced using the LO Lipatov vertex~\cite{Lipatov:1976zz}, indicating its intimate connection with BFKL dynamics. It would be interesting to explore whether the full-range EEC can serve as a novel observable for probing BFKL dynamics at high energies.

Therefore, in summary, we see that proton-proton colliders are able to perform a variety of measurements of energy correlators, across a wide range of energies, and kinematic regimes. While the difficulties of the hadronic environment had previously prevented precision studies, advances in both theory and experiment have overcome this, opening the door to a precision program. We will survey applications of measurements of energy correlators in proton-proton colliders for both nuclear and particle physics in \Sec{sec:particle} and \Sec{sec:nuclear}.

\subsection{Nuclear Collisions}\label{sec:heavy_ion}

\begin{figure}
\includegraphics[width=0.85\linewidth]{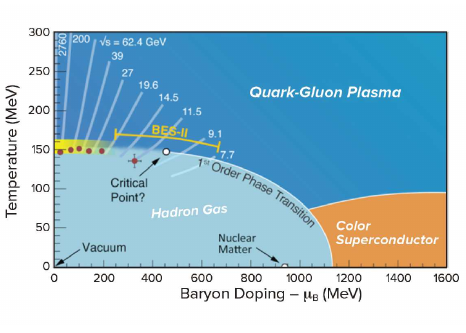}
\includegraphics[width=0.80\linewidth]{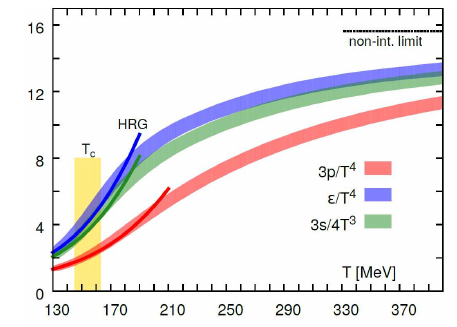}
\caption{Top Panel: The expected phase diagram of QCD. High energy collisions probe the region with small baryon density. Bottom Panel: Lattice calculations at $\mu_B=0$, where the transition is a cross-over. The horizontal line indicates the non-interacting limit which is approached for asymptotically free quarks and gluons.  Figures from \cite{Busza:2018rrf} and \cite{HotQCD:2014kol}.
}
\label{fig:phase_diagram}
\end{figure}

\begin{figure*}
\includegraphics[width=0.65\linewidth]{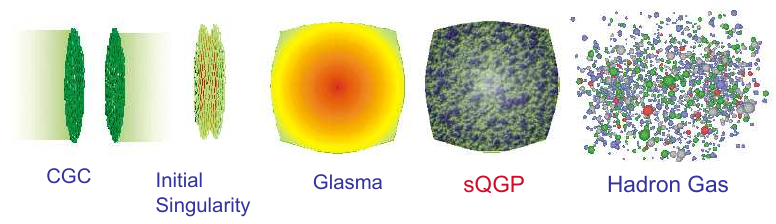}\qquad \qquad \qquad
\includegraphics[width=0.20\linewidth]{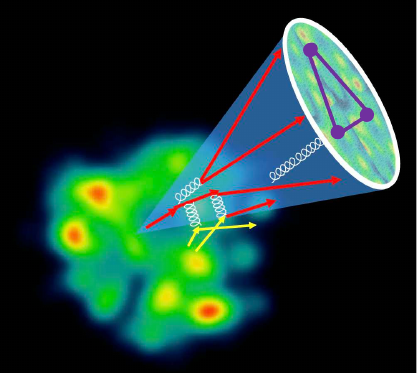}
\caption{Collisions of heavy nuclei provide the opportunity to study rich initial state dynamics, in the form of the color glass condensate, and final state dynamics, in the form of the QGP.  Figure from \cite{McLerran:2008uj}.  An energetic jet produced in a hard scattering propagates through this state. The detailed pattern of energy within the jet provides information about the microscopic dynamics of the QGP. Different kinematic regions of the energy correlator observables can be used to isolate and study the physics of these different phases of the collision. Figure from \cite{Yang:2023dwc}
}
\label{fig:qgp_visualize}
\end{figure*}

Beyond protons, hadron colliders also offer the possibility of colliding larger nuclei (which we will denote by $A$),  such as oxygen, gold and lead, in different, possibly asymmetric, combinations. This produces extremely violent collisions, whose representative event displays of collisions for different nuclei are shown in \Fig{fig:HIC_event_display}. The study of high multiplicity hadronic collisions has a remarkable history predating QCD itself. Indeed, it was the study of high multiplicity collisions by Hagedorn using the statistical bootstrap hypothesis that led to the prediction of an exponentially increasing spectrum of hadronic states, and a temperature, now referred to as the Hagedorn temperature  \cite{Hagedorn:1967tlw,Hagedorn:1965st}, at which hadronic matter boils.  This was later understood \cite{Polyakov:1978vu,Susskind:1979up,Cabibbo:1975ig} as a transition to a state where quarks and gluons are free, the so called ``quark gluon plasma" (QGP).

These early studies gave rise to the idea of studying extreme states of QCD matter, both at high temperature and high density in the laboratory, and presented many questions: What is the complete phase diagram of QCD?, What is the nature of the QGP? Can we reliably produce this state in the laboratory?  Understanding these questions is important for the study of early universe cosmology, and more generally, for studying the emergence of complex forms of matter, ``the condensed matter physics of QCD", from simply underlying laws.

Since this time, much has been learned both theoretically, and experimentally. The expected phase diagram of QCD is shown in \Fig{fig:phase_diagram}, along with markers indicating the regimes explored by different colliders. We will focus on high energy colliders, which explore the regime of $\mu_B\sim 0$. Here it has been understood through lattice calculations \cite{HotQCD:2014kol} that QCD exhibits a crossover transition to a phase in which quarks and gluons are free. 

Due to a heroic experimental program, it is possible to create the QGP in the laboratory. For interesting discussions of the history of the experimental heavy ion programs, we refer the reader to \cite{Busza:2025uid,Busza:2025gpg}. These experimental studies revealed a number of surprises. Instead of a plasma of weakly interacting quarks and gluons, the QGP is a strongly interacting fluid, with a near minimal shear viscocity to entropy density ($\eta/s$), sometimes referred to as the ``most perfect fluid".

The study of the QGP is a multi-disciplinary endeavor, which tries to understand both its bulk, and microscopic properties. For an excellent overview of the big questions of the field, we refer the readers to the excellent reviews  \cite{Busza:2018rrf,Harris:2023tti}. In this review, we will focus on one particular aspect of this, which is particularly suited to study using energy correlators and jet substructure. One of the remarkable features of the QGP is that at long distances it behaves as a strongly coupled liquid, however, when probed at high energies it is composed of asymptotically free quarks and gluons. One of the central problems is therefore to understand how asymptotically free quarks and gluons can form a strongly coupled liquid. 

The natural way to study this is similar to the original Geiger-Marsden experiment \cite{doi:10.1080/14786440408634197}, namely to shoot energetic quarks and gluons (jets) through the quark gluon plasma. Many valuable insights have been gained by examining the propagation of energetic partons through nuclear matter \cite{Busza:2018rrf,CMS:2024krd,ALICE:2022wpn,Cao:2020wlm,Apolinario:2022vzg,Cunqueiro:2021wls,Connors:2017ptx,Arratia:2019vju,Brewer:2021kiv,AbdulKhalek:2022hcn,Abir:2023fpo}. This approach allows us to probe the QGP at different energy scales, providing insight into its microscopic structure. This produces complicated patterns of energy flux, as illustrated in \Fig{fig:qgp_visualize}, which we can decode using energy correlators. Scales of the underlying QGP should imprint themselves into the angular scales of the energy correlator, making the energy correlator observables ideal for resolving the microscopic structure of the QGP. The study of jet substructure has had a significant impact on nuclear physics, for reviews, see \cite{Larkoski:2017jix,Asquith:2018igt,Marzani:2019hun,Cao:2020wlm,Apolinario:2022vzg, Cunqueiro:2021wls,Connors:2017ptx}.

From the perspective of the study of energy correlators, nuclear collisions are interesting for two reasons. On the one hand, the ability to produce QGP states in the lab opens up the opportunity to measure detector operators in non-trivial states, which is fascinating from a purely field theoretic perspective. On the other hand, we can use the theoretical control of energy correlator observables to learn about nuclear physics. There are two reasons why we believe that energy correlator observables are particularly well suited to improving our understanding of nuclear collisions. First, due to the complexity of nuclear collisions, it is often impossible to perform first principles calculations, therefore the interpretability of observables is extremely important. The beautiful feature of the energy correlators is that nuclear scales will imprint themselves as breaks in the otherwise smooth power laws of the energy correlators. The second reason is that the theoretical description of nuclear collisions necessarily involves the description of non-perturbative effects. The operator definition of the energy correlators opens the door to the characterization of these effects, and their understanding by symmetry principles.

The study of nuclear collisions is an immense, and highly developed field, involving many specialized techniques on both the theory and experiment sides.  For recent historical reviews, see \cite{Busza:2025uid,Busza:2025gpg}. We will discuss in more detail the physics cases for these different collisions in \Sec{sec:nuclear}, as well as some of the specific physics goals that measurements of the energy correlators will be useful in addressing. This section is intended to give an overview at a high level for those outside the nuclear physics community, highlighting how energy correlators can be measured in these environments, and how we expect modifications due to the nuclear collisions to manifest in the energy correlator observables themselves.

To understand measurements and uses of energy correlators in nuclear collisions, it is important to understand, at least at a high level, the different components of a nuclear collision. For concreteness, we consider the case of an A-A collision. The different components are shown schematically in \Fig{fig:qgp_visualize}. At a very elementary level, one has two large nuclei in the incoming state. In the limit of a large nuclei, it is possible to compute the gluon distribution using an effective field theory approach referred to as the color glass condensate \cite{Gelis:2010nm,Iancu:2003xm,McLerran:1993ni,McLerran:1993ka,McLerran:1994vd}. These nuclei collide, producing some extremely dense configuration of gluon fields, referred to as the Glasma \cite{Krasnitz:1999wc,Kovner:1995ts,Kovner:1995ja}. This Glasma then equilibrates into the quark gluon plasma (QGP), which then ultimately cools and hadronizes into the hadrons in the detector. Each aspect of this  collision is an entire field of study.

Nuclear collisions are therefore interesting to a broad range of communities interested in each of these different aspects of the collision, since they provide the opportunity to study them experimentally. One of the major difficulties in the study of  nuclear collisions is that most observables depend on all the different components of the collision, making it difficult to extract robust conclusions about any particular sub process.  At a very high-level, one of the primary goals is to have observables which isolate specific features of the collision. Namely one would like to be able to use energy correlator observables to study both the physics of the initial state, namely the color glass condensate, and the parton distribution functions of large nuclei, as well as the physics of the state produced by the collisions at early times, namely the Glasma and the QGP. Much like in the case of proton-proton collisions, different kinematic regimes of the energy correlators can be used to access these two types of physics. 

\begin{figure}
\includegraphics[width=0.75\linewidth]{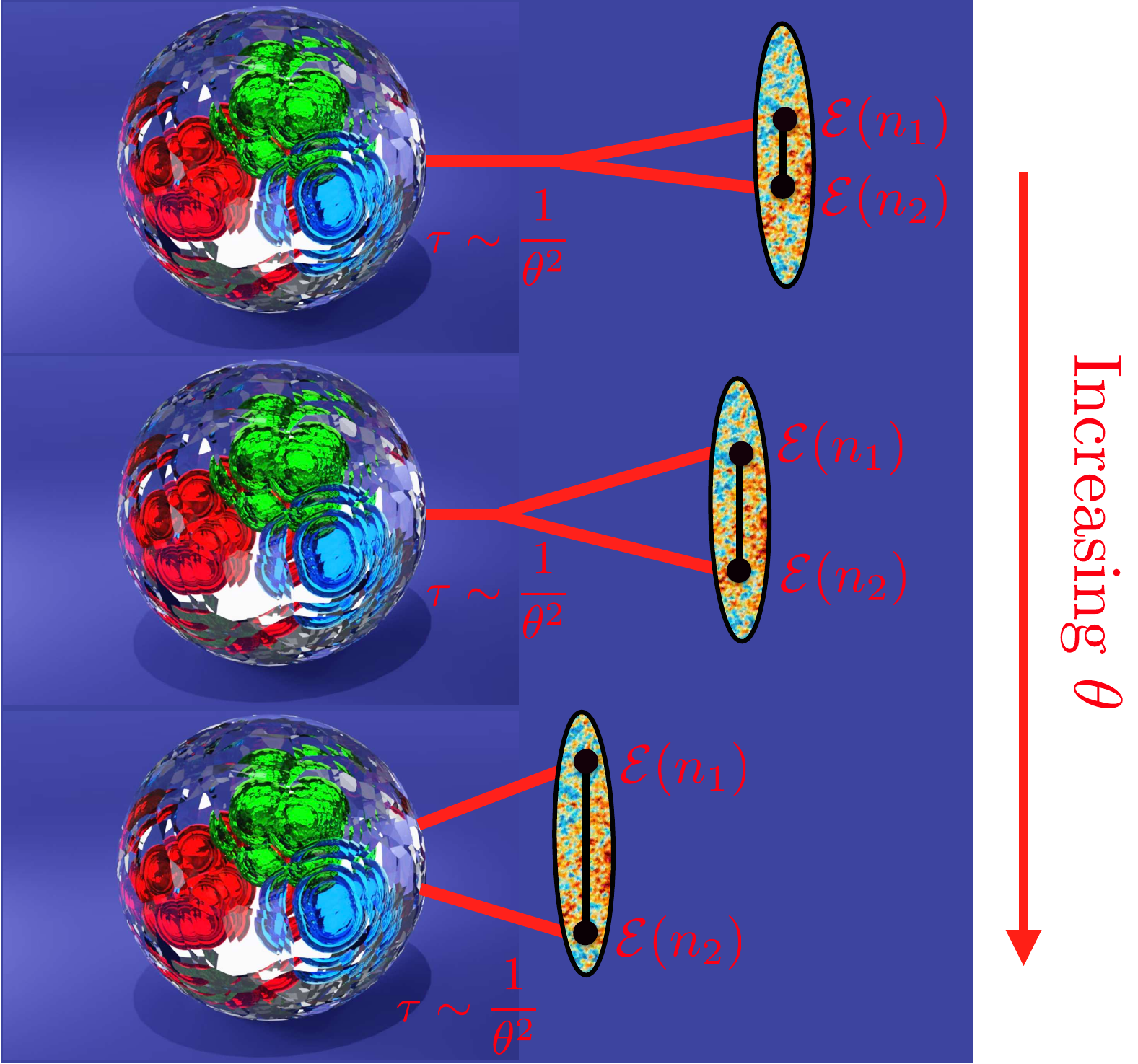}
\caption{The presence of a droplet of QGP imprints itself as a modification in the scaling behavior of the two-point correlator as one moves from small to larger angles. By identifying the modification in scaling, one can image a Fermi scale droplet using patterns in the asymptotic energy flux.
}
\label{fig:HIC_IRUV}
\end{figure}

In addition to the choice of observable, a second handle is the choice of nuclei. By changing the choice of nuclei one can enhance or suppress particular effects. In the case of p-A collisions, there are rigorous factorization theorems, where the nuclear modification is incorporated using higher twist operators \cite{Luo:1991bj,Qiu:1990xxa,Qiu:1990xy,Luo:1992fz}. Such collisions therefore provide an interesting case that is under better theoretical control. As one increases the size of the nuclei, at some point these factorization theorems break down, and one forms the QGP. When and where this occurs is an active area of research. Therefore, as compared to the case of p-p, where we are interested in precision physics, the goal of nuclear collisions, at the current stage, is to obtain a global picture, and develop observables that isolate particular aspects of the collision.

Energy correlators have only recently been measured in nuclear collisions for the first time, and have so far focused primarily on probing the structure of the final state. Much like in the case of proton-proton collisions, we study the energy correlators inside an identified high energy jet. In addition to the collision that forms the QGP, there are also additional hard scattering processes. These produce high energy partons which traverse the QGP. One therefore expects that any scales of the QGP will imprint themselves into the correlators, see \Fig{fig:qgp_visualize}, which will give access to the properties of the QGP.

It is easy to understand how the presence of a ball of QGP will modify the energy correlators using an intuitive argument. This also illustrates the UV/IR exchange for the light-ray operators. The ball of QGP has the size of a few Fermi. At very small angles, the particle splitting into the detectors is nearly on-shell, and propagates nearly to the detector before splitting, see \Fig{fig:HIC_IRUV}. As we increase the angular separation of the two detectors in the correlator, we push this splitting back into the QGP. At an angular scale corresponding to the size of the QGP, we expect to see a modification of the scaling behavior. As compared with hadronization, one can think of this as a UV modification.  We note that the energy correlators make this ideal, since any scale in the physical problem imprints itself as an angular scale in the energy correlators. Therefore even if we are not able to quantitatively compute the correlator in the QGP, we are able to understand its structure.

\begin{figure}
\includegraphics[width=0.555\linewidth]{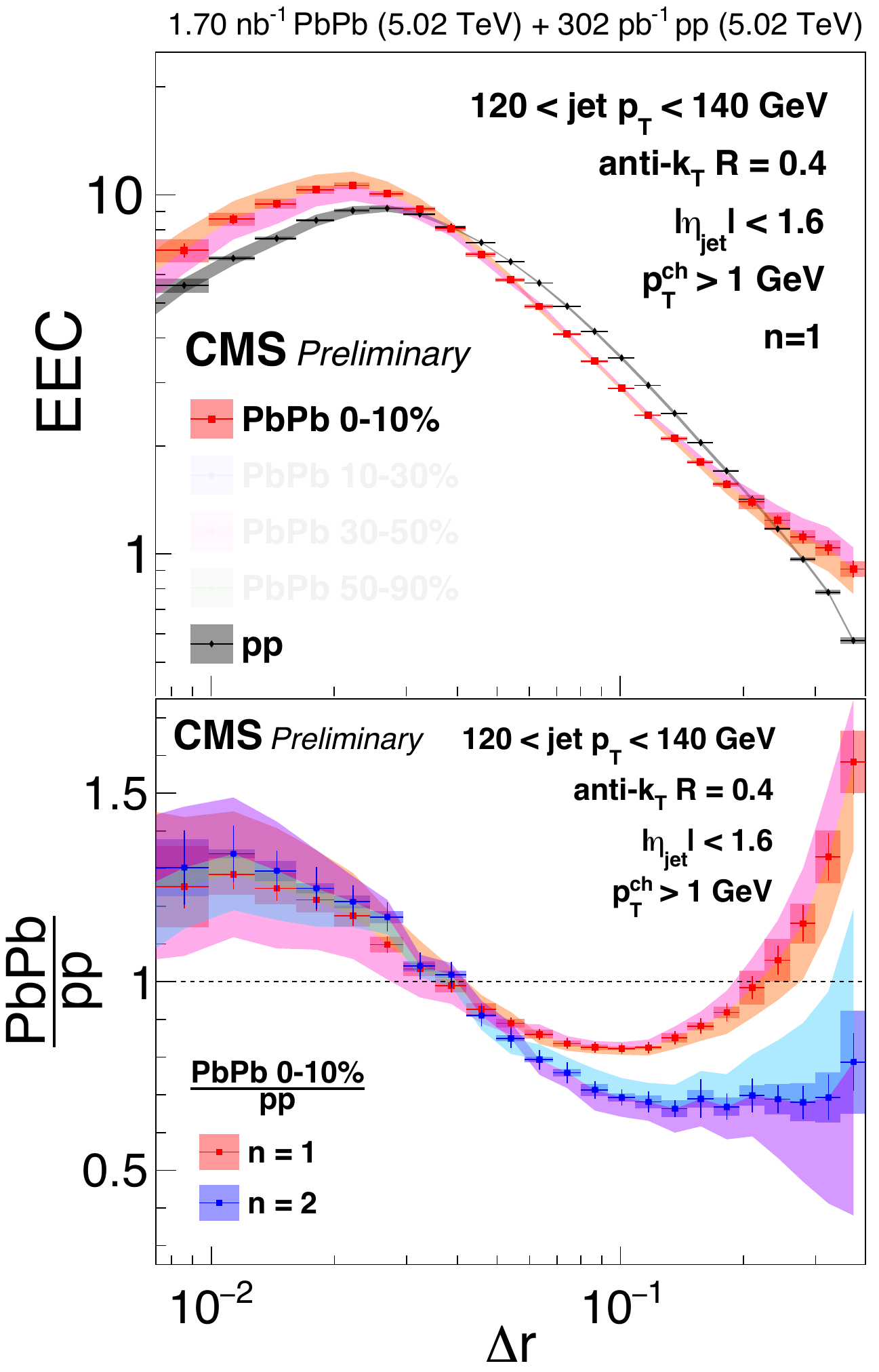}
\includegraphics[width=0.555\linewidth]{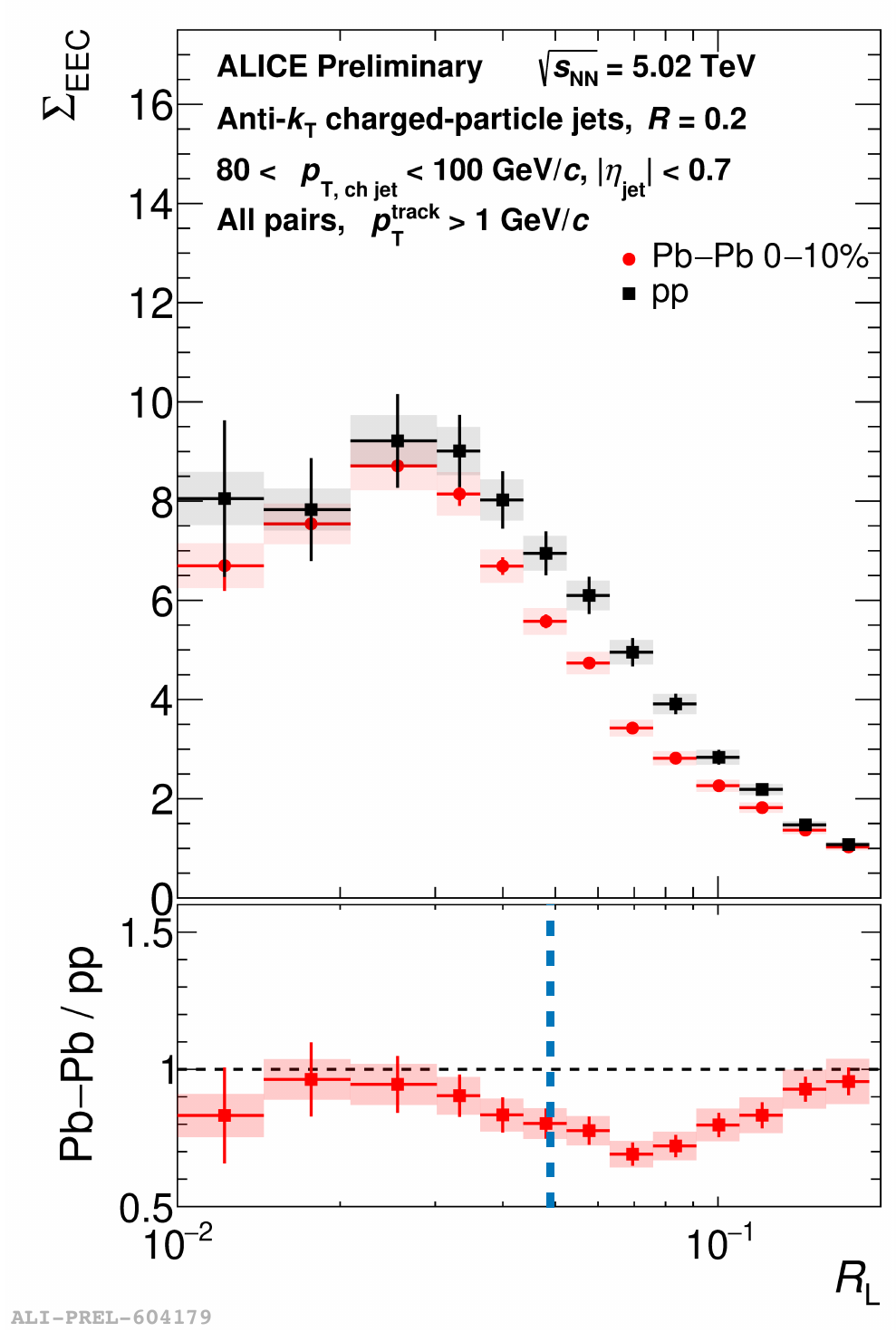}
\caption{Measurements of the two-point energy correlator in proton-proton and Pb-Pb collisions, and their ratios. In the upper two panels we show measurements by the CMS experiment, and in the lower panel by the ALICE experiment. Modification due to the presence of the QGP is observed at large angles. Figures from \cite{CMS-PAS-HIN-23-004,CMS:2025ydi} and \cite{talk_Ananya}.
}
\label{fig:CMS_raw_compare}
\end{figure}

The two-point correlator in Pb-Pb collisions has been measured by both the CMS \cite{CMS:2025ydi} and ALICE \cite{talk_Ananya} collaborations. The measurements are shown in \Fig{fig:CMS_raw_compare}. In both of these figures, we show the two-point correlator in both p-p and Pb-Pb collisions, as well as the ratio. There is a modification of the scaling behavior as we move from small angles towards large angles, which is particularly visible in the ratio. We find it quite remarkable that we can observe the presence of a ball of QGP that is only several Fermi in the patterns of asymptotic energy flux.

\begin{figure}
\includegraphics[width=0.855\linewidth]{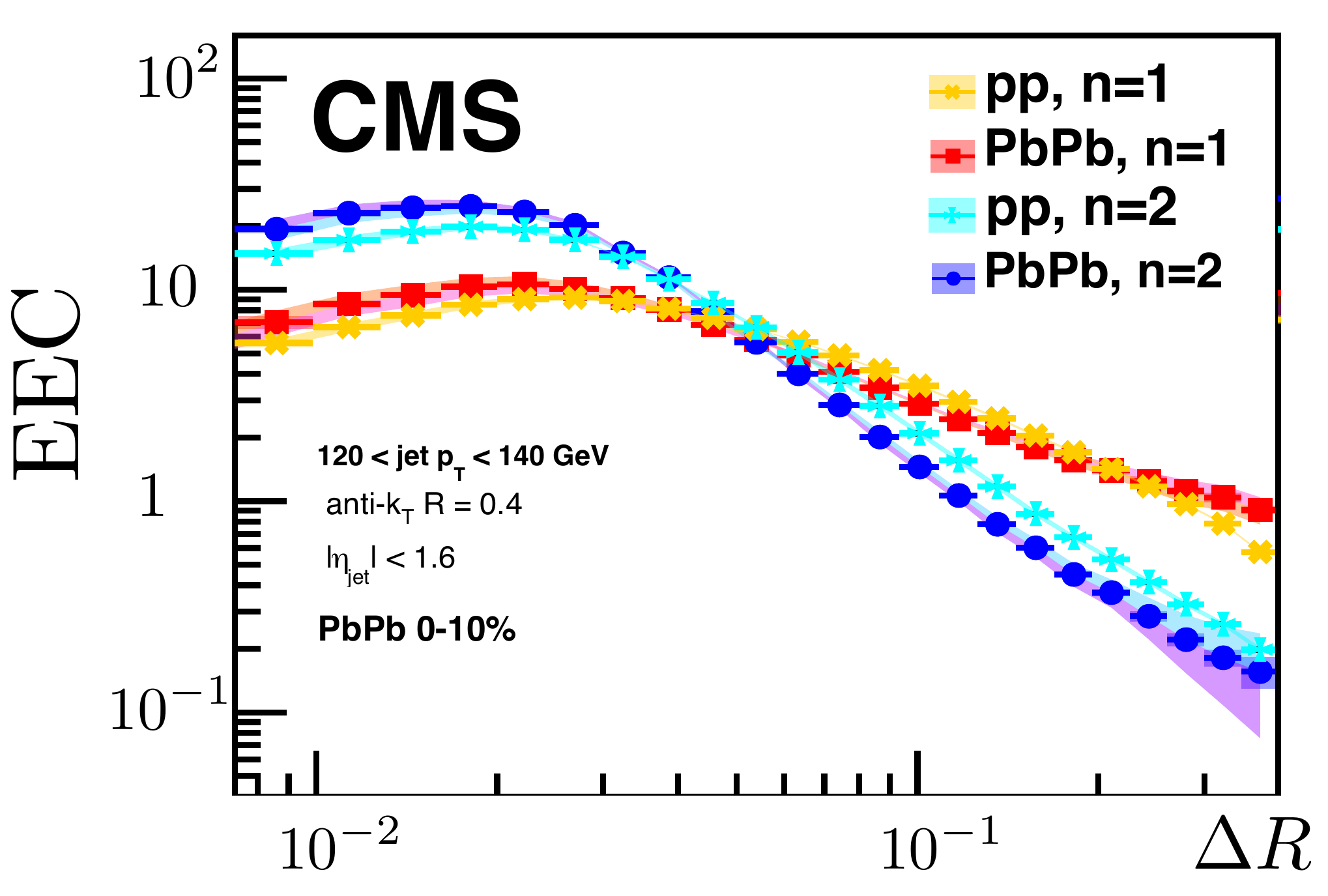}
\caption{Measurement of the two-point energy correlator $\langle \mathcal{E} \mathcal{E} \rangle$, and the detector correlator $\langle \mathcal{E}^2 \mathcal{E}^2 \rangle$, in both p-p and Pb-Pb collisions. Different scaling laws are observed for the two types of detectors. Figures from \cite{CMS-PAS-HIN-23-004,CMS:2025ydi}.
}
\label{fig:CMS_HIC_raw}
\end{figure}

The CMS collaboration has also performed a measurement of the $\langle \mathcal{E}^2 \mathcal{E}^2 \rangle$ in both p-p and Pb-Pb collisions, which is shown in \Fig{fig:CMS_HIC_raw}. This is interesting in the case of p-p collisions, since the use of different detector operators gives rise to a different scaling behavior, which is clearly visible in \Fig{fig:CMS_HIC_raw}. In the context of heavy ion collisions, it is interesting since it suppresses soft radiation.

One of the difficulties with the interpretation of heavy ion measurements, in particular in their comparison to measurements in proton-proton collisions, is that measurements are made as a function of the observed jet $p_T$. This is also modified due to the presence of the nuclear medium. Therefore, in a fixed $p_T$ bin, one can be comparing jets originating from underlying hard scattering processes with different energies, and different quark gluon fractions. This issue is referred to as bias. This can be seen in \Fig{fig:CMS_raw_compare} as a shift in the location of the hadronization peak. Although this makes the precise comparison interesting, it is in itself interesting, and in fact, the shift in the hadronization peak in the energy correlation  provides a definitive illustration of energy loss of jets in the QGP, as well as a measure of the magnitude of the energy loss.

\begin{figure}
\includegraphics[width=0.5755\linewidth]{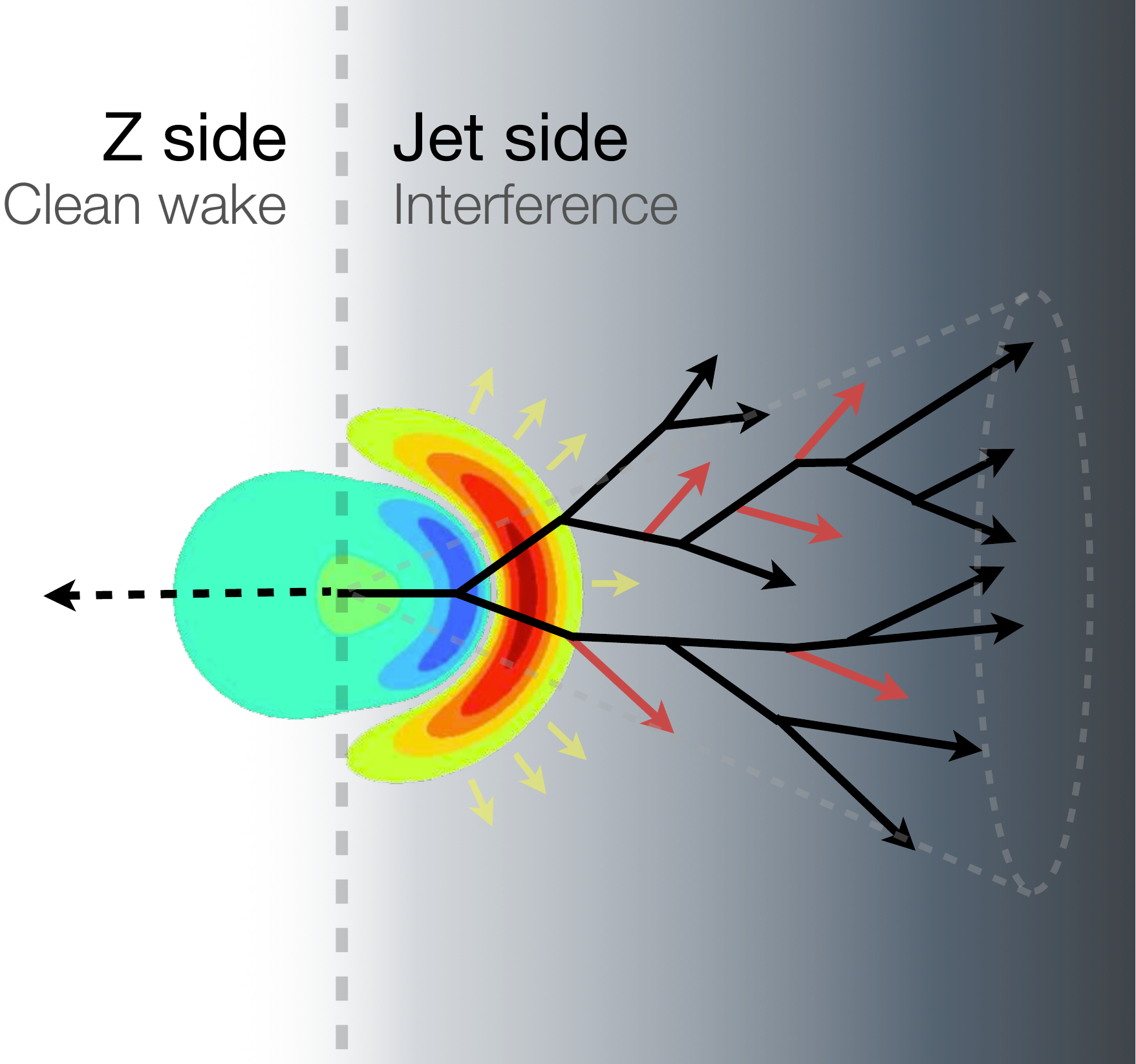}
\includegraphics[width=0.655\linewidth]{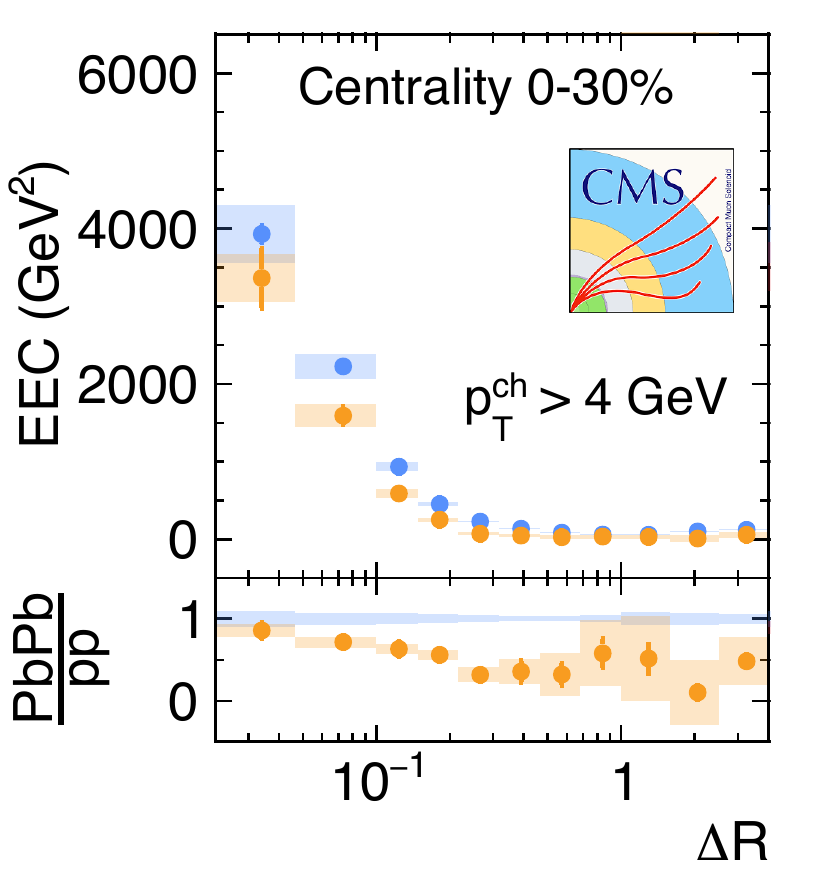}
\caption{A measurement of the two-point energy correlator in $Z+$ jet events in both p-p and Pb-Pb collisions. The use of a hard $Z$, here $p_T>40$ GeV for the Z boson, eliminates the need for a jet algorithm, allowing the measurement to be extended to larger angles. The ratio between p-p and Pb-Pb collisions is also shown, highlighting the modification due to the formation of the QGP. Figure from \cite{CMS:2025jam}.
}
\label{fig:Z_EEC_measure}
\end{figure}

 We will discuss some ways that it can be overcome for energy correlator measurements in the case of purely hadronic collisions in \Sec{sec:QGP_results}. However, one of the cleanest ways of overcoming bias is to use a color singlet Z-boson recoiling against a jet, as illustrated in \Fig{fig:Z_EEC_measure}. The first such measurement was recently performed by CMS \cite{CMS:2025jam}, in both proton-proton and Pb-Pb collisions. This is an extremely clean way of studying the energy correlators. Another excellent feature of this way of performing the measurement, is that it can be performed for all angles, unrestricted by the jet radius. In \Fig{fig:Z_EEC_measure} we show the first such measurement. In this case, the Z boson is required to have $p_T > 40$ GeV. The ratio is also shown in  \Fig{fig:Z_EEC_measure} suggesting an interesting pattern of modification due to the underlying QGP. As of the writing of this review, a theoretical understanding of this measurement has not been presented.

\begin{figure}
\includegraphics[width=0.75\linewidth]{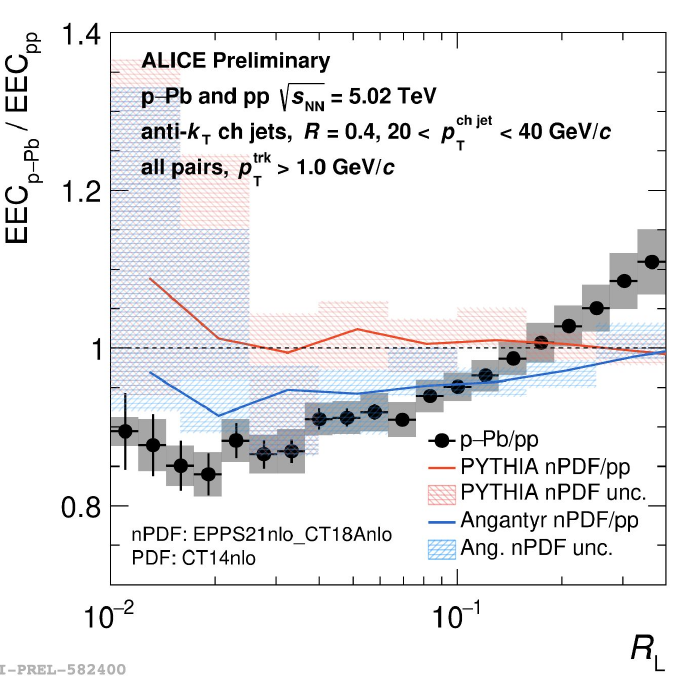}
\caption{Data from ALICE for the two-point energy correlator  in p-A collisions. Evidence for nuclear modification beyond nuclear PDFs is observed. Figure from \cite{talk_Anjali,talk_Anjali2}.
}
\label{fig:ALICE_pPb}
\end{figure}

In addition to Pb-Pb, energy correlators have also been measured in p-Pb collisions by the ALICE collaboration \cite{talk_Anjali,talk_Anjali2}. In this case, one no longer expects the formation of a QGP, nevertheless one expects nuclear modification as the jet passes through the nucleus. In \Fig{fig:ALICE_pPb} we show the ratio of the two-point correlator in p-Pb to that in p-p as measured by the ALICE collaboration. This is the first measurement of nuclear modification in p-Pb collisions using jet substructure.  Much as with the case of Pb-Pb collisions, there are numerous subtleties in taking this ratio. In particular, the quark-gluon fraction can change significantly between p-p and p-Pb led collisions due to the differing PDFs. This effect is more important to incorporate in the case of p-Pb collisions, since the overall nuclear modification is much smaller than for Pb-Pb. We will discuss several theoretical attempts to understand this measurement in \Sec{sec:QGP_results}.

These measurements have illustrated for the first time that it is possible to measure energy correlators in nuclear collisions. While so far the measurements have focused on the simplest case of the two-point energy correlator, much like in proton-proton collisions, it is also possible to measure higher point correlators, both shape dependent and projected, inside high energy jets produced in heavy ion collisions. While the two-point correlator is sensitive to the scale of the ball of QGP, higher point correlators are sensitive to the detailed underlying dynamics. Much like in cosmology, where the ``shapes" of non-gaussianities have been mapped out, we hope that it will be possible to map out the shapes of higher point correlators in different models, and that measurements of these correlators can improve our understanding of the dynamics of the QGP. Due to the complex environment of heavy ion collisions, one requires the ability to compute multi-point energy correlators extremely efficiently. There has recently been work in this direction \cite{Budhraja:2024xiq,Alipour-fard:2024szj}, which should prove useful in future studies of energy correlators in nuclear collisions..

So far energy correlators have been measured in both p-Pb and Pb-Pb collisions. It will be particularly interesting to study them on intermediate nuclei, such as oxygen-oxygen, which should be performed in the forthcoming oxygen runs at the LHC, whose physics program is reviewed in \cite{Brewer:2021kiv}. The future EIC will enable clean studies of electron-Ion collisions. These will also provide a useful benchmark for the interpretation of nuclear modification in a clean environment. Additionally, one can study energy correlators on other states produced in the QGP, such as the production of heavy quarks in the QGP. These will be studied using jet substructure with larger datasets. We will discuss these more detailed applications in \Sec{sec:nuclear}.

\begin{figure}
\includegraphics[width=0.755\linewidth]{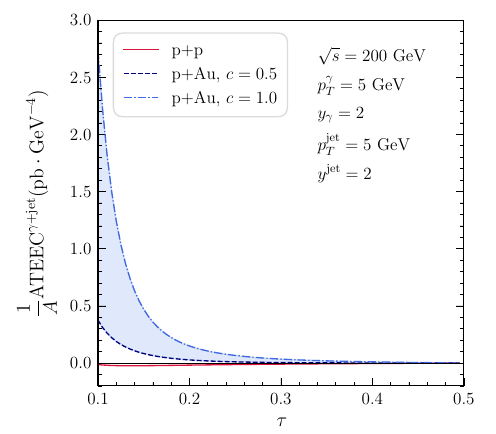}
\caption{The asymmetry of the TEEC as a probe gluon saturation. The modification occurs primarily in the back-to-back region (Here $\tau=(1+\cos(\phi))/2$.). Figure from \cite{Kang:2025vjk}.
}
\label{fig:gam_ATEEC}
\end{figure}

We also wish to emphasize that energy correlators can be used to study the initial state in heavy ion collisions. As in the case of proton-proton collisions, one way of doing this is to study energy correlators in the back-to-back limit. Studies of the energy correlators in the back-to-back limit in nuclear collisions include \cite{Kang:2025vjk,Kang:2024otf,Kang:2023oqj}, which have shown sensitivity to the color glass condensate and saturation. An illustration from \cite{Kang:2025vjk} is shown in \Fig{fig:gam_ATEEC}. Here the asymmetry of the TEEC is plotted as a function of $\tau=(1+\cos(\phi))/2$. Significant modification is observed in the back-to-back region. We believe that such measurements are extremely interesting, and should be pursued to complement the measurements of the energy correlator in the collinear limit.

\section{Energy Correlators in Particle Physics}\label{sec:particle}

In \Sec{sec:exp_opp} we discussed how different kinematic regions of energy correlator observables can be directly measured at collider experiments, allowing them to be used to study the Standard Model. The elegance of this approach is that it provides a direct relation between correlations in the macroscopic energy flux, which can be measured by experimentalists, and correlation functions in the underlying quantum field theory, such as the $\langle J T J \rangle $ correlators. Since these are the first measurements of the energy correlator observables in hadron colliders, it is natural to ask how they provide new ways of probing the Standard Model.

Jet substructure at the LHC has been successful in providing new ways to search for beyond the Standard Model physics, and to identify highly boosted Z/W/Higgs bosons. However, the theoretical complexity of jet substructure observables has made it difficult to achieve a precision jet substructure program enabling new ways of extracting Standard Model parameters. The simplicity of the energy correlators opens up this possibility. In particular, the operator product expansion allows one to reduce the description of the transverse structure of jets to a set of well defined scaling laws which can be computed with high precision. The philosophy of the energy correlator program is then to identify the simplest possible energy correlator observable that has sensitivity to the Standard Model parameters of interest. In this way we hope to reformulate collider physics measurements of Standard Model parameters as sharp questions about energy correlators. This is advantageous due to the fact that these simple observables can be computed to much higher perturbative accuracy than other more complicated jet substructure observables. Moreover, by having a rigorous operator definition of the observable, we are optimistic that future progress will enable a non-perturbative understanding of these observables. Here we highlight several cases where we believe that this reformulation will be most impactful.

\subsection{The Strong Coupling Constant}\label{sec:particle_alphas}

The strong coupling constant, $\alpha_s$, is one of the least precisely measured parameters of the Standard Model. Due to the importance of QCD corrections to all colliders physics observables, its knowledge is crucial for precision measurements of many processes, for example precision studies of the Higgs sector.

\begin{figure}
\includegraphics[width=0.855\linewidth]{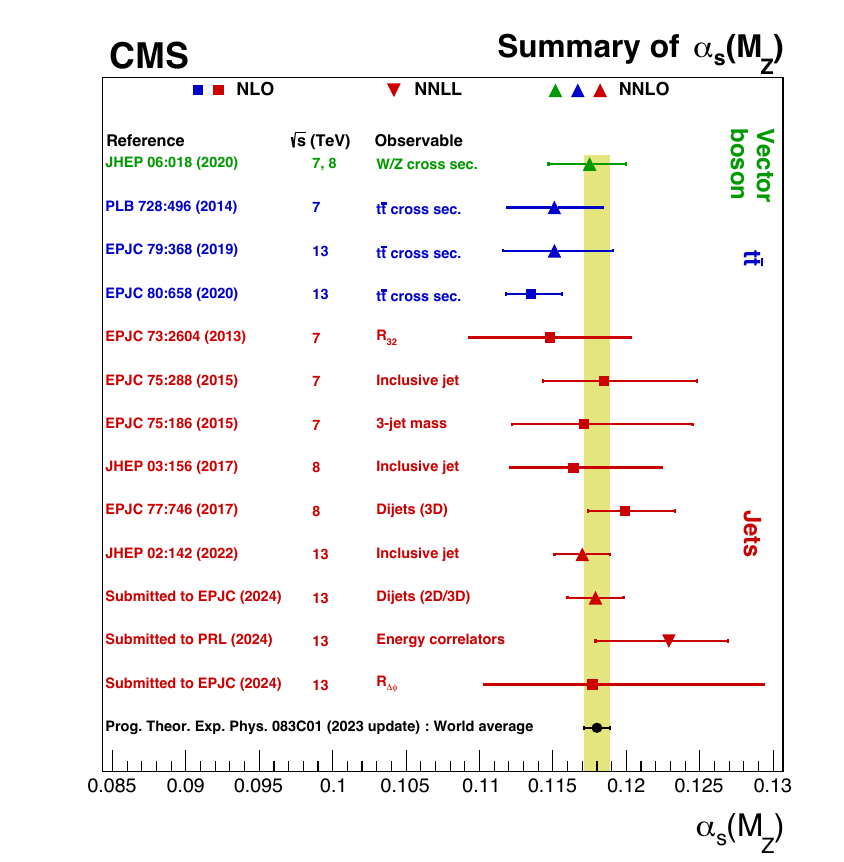}
\caption{A summary of measurements of $\alpha_s$ in hadron colliders. The recently introduced approach of measuring the scaling exponent of the energy correlators provides a competitive approach, whose theoretical and experimental uncertainties can be greatly improved. Figure from \cite{CMS:2024gzs}.
}
\label{fig:alphas_measure}
\end{figure}

While there are many approaches to measuring the strong coupling constant (For a review, see \cite{dEnterria:2022hzv,Huston:2023ofk}.), and we can not provide a review of them here, one way of measuring it is through precise calculations of event shapes. Factorization theorems developed within SCET, combined with progress in fixed order calculations have enabled precision calculations of event shapes, combining state of the art resummation, fixed order perturbation theory, and field theoretic definition of non-perturbative parameters. The most precise fits have been performed for the thrust observable \cite{Becher:2008cf,Abbate:2012jh,Abbate:2010xh}, the C-parameter \cite{Hoang:2014wka,Hoang:2015hka}, and the heavy jet mass \cite{Benitez:2025vsp}. For other recent discussions of these extractions and calculations, see \cite{Benitez:2024nav,Benitez-Rathgeb:2024ylc}. There is currently a discrepancy between these extractions from event shapes, and other extractions of the strong coupling constant. It is therefore highly desirable to resolve this discrepancy.

The observables used for extractions of the strong coupling constant are all Sudakov observables, and use the same treatment of non-perturbative corrections. It is therefore highly desirable to attempt to extract the strong coupling constant from asymptotic energy flux using observables that have different perturbative and non-perturbative structures, as well as using higher energy hadron collider data, which suppresses non-perturbative effects.  

Using the energy correlator observables, we are able to present two new approaches to measure the strong coupling constant: one at $e^+e^-$ colliders, and one exploiting the high energies of the LHC. We believe that these are excellent targets for both experimental measurements, and further theory development.

\subsubsection{Hadron Colliders}\label{sec:particle_alphas_pp}

Hadron colliders offer the advantage of measuring QCD processes at the highest possible energies, suppressing non-perturbative power corrections that scale like $\Lambda_{\text{QCD}}/Q$. The strong coupling constant can be measured from the precise calculation of many different QCD processes. For a summary from the CMS experiment, see \Fig{fig:alphas_measure}, and \cite{CMS:2024gzs}. For several recent measurements of the strong coupling constant using jet production processes, see e.g. \cite{Alvarez:2023fhi,Ahmadova:2024emn,ATLAS:2023tgo,CMS:2024hwr}.

\begin{figure}
\includegraphics[width=0.95\linewidth]{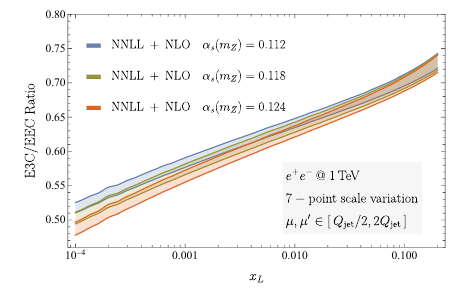}
\caption{Calculations of the three-point to two-point projected correlators at NNLL for different values of the strong coupling constant. These calculations were used to perform a precision extraction of the strong coupling constant. Figure from \cite{Chen:2023zlx}.
}
\label{fig:NNLL_projected}
\end{figure}

It has long been a goal of jet substructure to perform a precision measurement of the strong coupling constant. Until recently, this remained a dream, due to the complexity of precision calculations of jet substructure observables in the complicated environment of the LHC. The philosophy of the energy correlators is to identify a robust feature of the simplest possible observable that is sensitive to the strong coupling constant. We have seen that the most universal feature of the energy correlators, present in any state, is the scaling behavior in the collinear limit. As discussed, this scaling behavior can be cleanly identified in experiment by measuring ratios of the multi-point projected energy correlators \cite{Chen:2020vvp}. This provides a conceptually clean approach to extracting the strong coupling constant, since the scaling behavior, dictated by anomalous dimensions in QCD, are proportional to the strong coupling constant, and can be computed to high perturbative accuracy.

The ratio of three-point and two-point projected energy correlators was computed at NNLL accuracy in  \cite{Chen:2023zlx}, whose result is shown at two different energies in \Fig{fig:NNLL_projected}, along with its variation as a function of $\alpha_s$. The possibility of achieving a precision measurement using this scaling behavior was first realized  in \cite{CMS:2024mlf}, who extracted a value of the strong coupling constant
$\alpha_s(m_Z)=0.1229+0.0040-0.0050$, which is $\sim 4\%$ accuracy. This result broke a longstanding barrier in precision jet substructure. We also belief that it provides a remarkable story of the identification of universal features in energy flux observables in a toy model, impacting real world measurements at the LHC.

This measurement makes us optimistic about the possibility of a record precision measurement of the strong coupling constant using energy correlators at the LHC. It can be improved on the theory side in a number of ways. First, it will be important to include a proper field theoretic treatment of non-perturbative corrections. The formalism for this has been developed, and can be incorporated in future measurement \cite{Lee:2024esz,Chen:2024nyc}. Second, it is important to improve the perturbative accuracy of the hard function. This can be performed using the formalism of \cite{Generet:2025vth}. Furthermore, it would be interesting to measure higher point projected correlators, and perform a simultaneous fit to all of them. We believe that this can exploit the amazing high energies of the LHC to achieve a record precision measurement of $\alpha_s$.

\subsubsection{$e^+e^-$ Colliders}\label{sec:particle_alphas_ee}

In addition to providing a new approach to measuring the strong coupling constant at hadron colliders, it is also interesting to attempt a measurement of the strong coupling constant from a state-of-the-art calculation of the energy correlator in $e^+e^-$ colliders. This is appealing for two reasons. First, using the track function formalism, we can perform a re-analysis of the data, and perform the measurement on tracks, making the dataset independent of previous event shape measurements used for $\alpha_s$ extractions. Second, due to the simplicity of the energy correlator observable, we have remarkable theoretical control over its entire kinematic range. 

\begin{figure}
\includegraphics[width=0.95\linewidth]{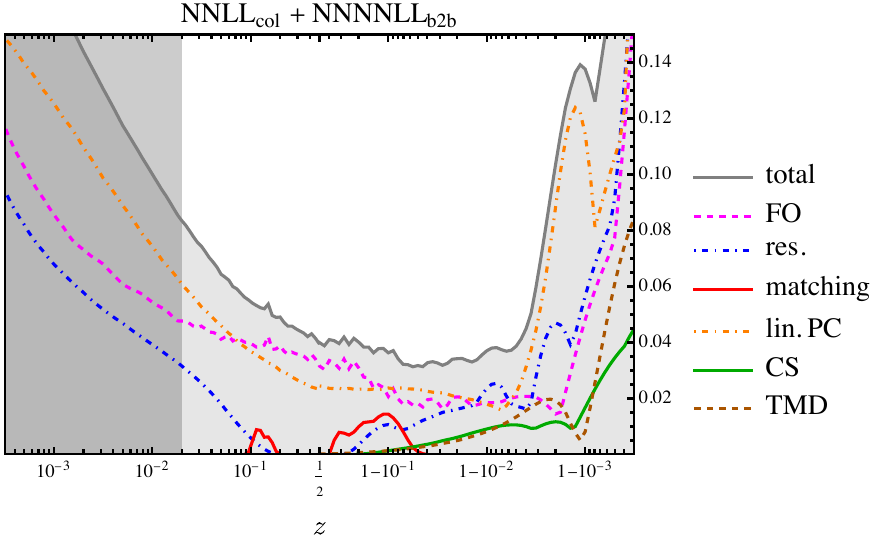}
\caption{Theoretical uncertainty estimates for the calculation of the EEC on tracks at LEP energies, using all state-of-the-art ingredients. Both perturbative and non-perturbative uncertainties are included. An accuracy of $\sim 4\%$ is achieved throughout the bulk of the distribution. Figure from \cite{talk_Max}.
}
\label{fig:ee_us_uncertainty}
\end{figure}

In \Fig{fig:ee_us_prediction} we show a state of the art prediction for the two-point energy correlator, computed on tracks. It includes NNLO perturbative calculations in the bulk of the distribution, N$^4$LL resummation in the back-to-back limit, combined with non-perturbative inputs from the lattice, and N$^2$LL resummation in the collinear limit \cite{talk_Max}. This provides the most accurate theoretical prediction ever for an event shape. In \Fig{fig:ee_us_uncertainty} we show a break down of the uncertainties in the theoretical calculation. In addition to further improving the perturbative description, we see that it will be important to improve our understanding of the non-perturbative power corrections to significantly reduce the uncertainties. We believe that this provides a promising observable for precision extractions of the strong coupling constant.

Another interesting approach to minimize the effect of non-perturbative power corrections is to the consider the asymmetry of the EEC, defined as
\begin{align}
\text{AEEC}(z)=\frac{1}{2} \left( \text{EEC}(z)-\text{EEC}(1-z) \right)\,,
\end{align}
 which was introduced already in the original paper \cite{Basham:1978bw}. This particular asymmetry was proposed to cancel non-perturbative corrections. 

 It is interesting to revisit this observable from a modern perspective, since we now have improved operator level understanding of the leading non-pertubative corrections to the energy correlator observables. In \Sec{sec:non_pert} we showed that the leading non-perturbative correction in the bulk of the distribution is given by 
\begin{align}
\frac{1}{\sigma} \frac{d\sigma}{dz}=\frac{1}{\sigma} \frac{d\hat \sigma}{dz}+\frac{1}{2}\frac{\bar \Omega_{1q}}{Q (z(1-z))^{3/2}}\,.
\end{align}
The contribution proportional to $\Omega_{1q}$ is indeed symmetric in $z$, and therefore completely cancels in the asymmetry. However, the perturbative structure of this observable is much more complicated, since it mixes collinear and Sudakov resummation. Nevertheless, now that we have both of these under control at high perturbative accuracies, we can precisely compute the EEC asymmetry. It will be interesting to pursue a precision program to compute this observable as precisely as possible. There are a number of concrete directions in which the calculation can be improved. On the perturbative side, it will be important to perform an analytic NNLO calculation of the energy-energy correlator, matching the perturbative accuracy in $\mathcal{N}=4$ sYM. It will also be important to extend the accuracy of the resummation, as well as to analytically compute higher twist corrections. On the non-perturbative side it will be important to increase the precision of lattice extractions of the Collins-Soper kernel, as well as of the linear power correction. Finally, it is interesting to explore its potential at future $e^+e^-$ colliders \cite{Lin:2024lsj}.   We are optimistic that there is significant progress to be made in these directions.

\subsection{The Top Quark Mass}\label{sec:particle_mt}

The top quark is the heaviest particle in the Standard Model, as such it plays an important role in electroweak vacuum stability \cite{Buttazzo:2013uya,Degrassi:2012ry,Elias-Miro:2011sqh,Andreassen:2014gha}, electroweak precision constraints \cite{Baak:2012kk,Baak:2014ora}, and in many extensions of the Standard Model. Remarkably the top quark mass is one of the worst known parameters of the Standard Model, and indeed, is the leading uncertainty in determining the stability/ metastability of the universe.  As with measurements of $\alpha_s$, precision measurements of the top quark is an immense subject to which we can not do justice in this short review. For reviews with complementary perspectives, we recommend \cite{Nason:2017cxd,Hoang:2020iah}.

\begin{figure}
\includegraphics[width=0.85\linewidth]{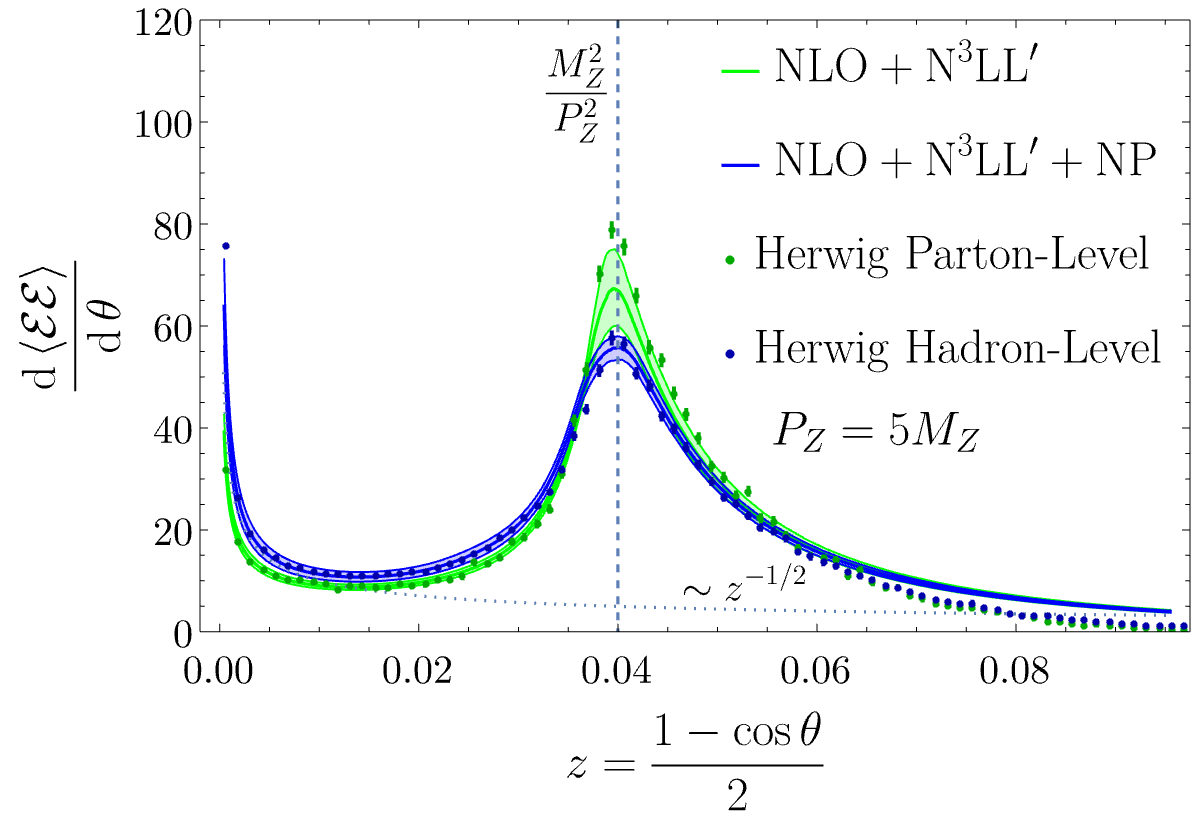}
\caption{The energy correlator on a jet from a decaying massive particle, in this case a $Z$ boson, has a peak at the characteristic angular scale set by the mass of the decaying particle.
}
\label{fig:Z_EEC}
\end{figure}

\begin{figure}
  \includegraphics[width=0.95\linewidth]{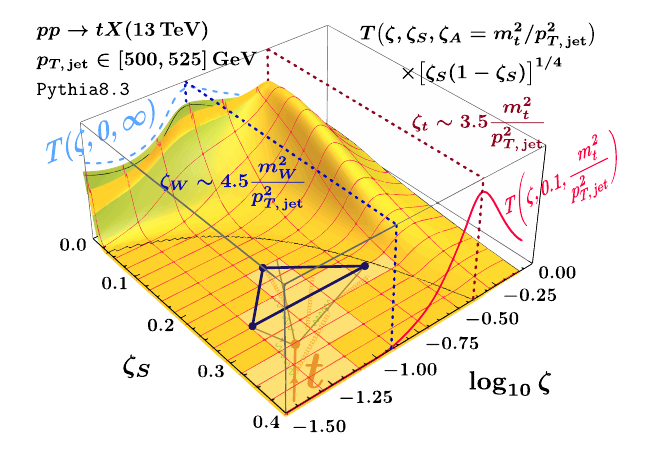}
    \includegraphics[width=0.95\linewidth]{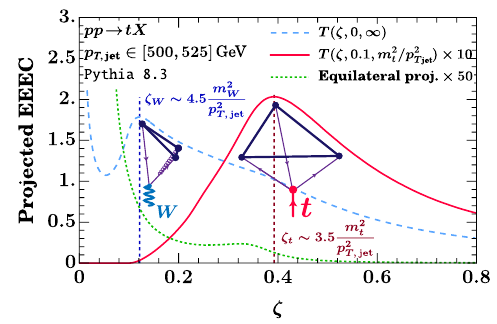}
  \caption{A plot of the three-point energy correlator on top quark decays. A detailed analysis of its shape,  enables the extraction of the top quark mass. Figures from \cite{Holguin:2023bjf}. }
  \label{fig:top_quark_1}
  \end{figure}

As compared to e.g. the Higgs boson, the complexity in precisely extracting the mass of the top quark is due to its strongly interacting nature. Observables must be formulated in terms of precise factorization theorems where the top mass appears in a specified scheme, requiring rigorous factorization theorems for the top quark mass sensitive observable. For a detailed discussion, see \cite{Hoang:2020iah}. The techniques to achieve such factorization theorems have been developed in the pioneering works of \cite{Fleming:2007qr,Fleming:2007xt}, along with the development of the appropriate mass schemes 
\cite{Hoang:2009yr,Hoang:2008yj,Hoang:2017suc}. For applications to the extraction the top quark mass using jet substructure, see \cite{Hoang:2017kmk,Bachu:2020nqn}.

In our opinion, the top quark is one of the most important targets for a precision measurement at the HL-LHC. We would therefore like to understand if it can be extracted from energy correlator observables. As in the case of the strong coupling constant, we would like to determine the simplest possible energy correlator observable that is sensitive to the top quark mass. We believe that then we will have the best chance of bringing it under theoretical control.

A program to extract the top quark mass using energy correlator measurements has been put forward in \cite{Holguin:2022epo,Holguin:2023bjf,Holguin:2024tkz}. It is based on the simple observation that if one measures the multi-point energy correlator on a decaying massive state, the mass of the state will imprint itself at a characteristic angle.

This can be illustrated in the case of the two-point correlator measured on a $Z$ boson decay, which is shown in \Fig{fig:Z_EEC}. As opposed to the simple power law which we observe in the case of massless jets, here we see at a peak at a characteristic angle. Quite remarkably, the detailed shape of this peak can be accurately computed, and is in fact just a boosted Sudakov. This can be understood due to the fact that in the case of the Z-boson, this distribution can be obtained by directly boosting the distribution computed in the rest frame. As discussed above, the energy correlator in the back-to-back limit is known with exceptional perturbative accuracy, enabling this peak to be computed accurately as well.

However, this simplicity comes with a cost, namely that we extract an angular scale, instead of a mass scale. This angular scale can only be converted to a mass scale if the energy of the $Z$-boson is precisely known. In general this is not the case at hadron colliders, since the energy must be reconstructed from the hadrons into which the $Z$ decays.

For the case of the top quark, an approach to overcome this has been proposed. In \Fig{fig:top_quark_1}, we show the three-point correlator measured on a top quark decay. The three-point correlator is chosen since the Born-level decay of the top quark proceeds into three-particles. The overall shape of this correlator is quite complicated, however, we can consider specific projections, which are shown in the lower panel. A lucky feature of the mechanism of the top quark decay is that it proceeds via a W-boson. By considering the squeezed limit of the correlator, we are able to access the mass of the W-boson as a peak in the energy correlator spectrum. We can therefore use the known W-boson mass as a standard candle to calibrate the angular scale, namely we can extract the top quark mass in units of the W-boson mass \cite{Holguin:2023bjf}. This provides the possibility of performing a theoretically clean measurement of the top quark mass using energy correlators.

Studies of the feasibility of this approach were performed in \cite{Holguin:2024tkz}, which studied in detail numerous sources of uncertainty. It was found under optimistic assumptions that an uncertainty of $\pm 0.3$ GeV can be obtained with the full high-luminosity LHC run. For a study of a related approach also using the energy correlators to extract the top quark mass, see \cite{Xiao:2024rol}. These studies illustrate the feasibility of this approach, and motivate the systematic theoretical study of energy correlators on top quark decays. We believe that this should be one of the primary targets for precision QCD measurement at the LHC.

\subsection{Quarkonium Dynamics}\label{sec:quarkonia}

\begin{figure}
\includegraphics[width=0.75\linewidth]{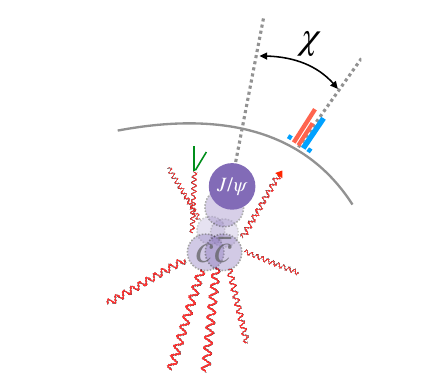}
\includegraphics[width=0.95\linewidth]{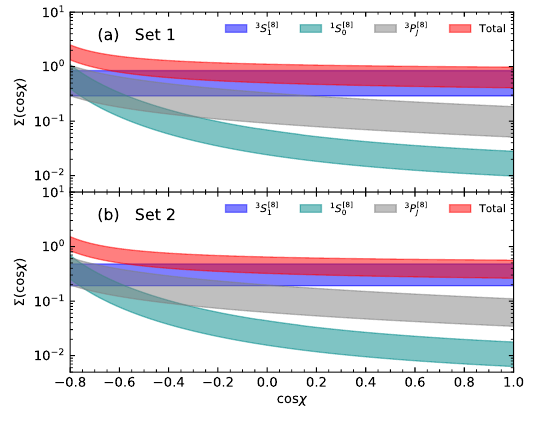}
\caption{Upper Panel: An illustration of the quarkonium energy correlator: a detector correlates radiation around the quarkonium as a function of angle. Lower Panel: The quarkonium energy correlator for two different sets of LDMEs. The contributions for LDMEs of different quantum numbers are shown as a function of angle. Figures from \cite{Chen:2024nfl}.
}
\label{fig:quarkonia_EEC}
\end{figure}

Quarkonia, non-relativistic bound states of heavy quarks, in particular charm or bottom quarks, provide interesting laboratories to study QCD, and played a seminal role in the development of QCD as a theory of the strong interactions \cite{Appelquist:1974zd}. Due to the non-relativistic nature of the bound states, and the presence of a heavy mass, $m_{c,b}$, these systems can be treated rigorously using effectively field theory techniques, in particular non-relativistic QCD  \cite{Bodwin:1994jh,Beneke:1997av,Brambilla:1999xf,Kramer:2001hh}. See  \cite{Brambilla:2004jw,Brambilla:2010cs} for reviews. Despite the foundational nature of these systems, there are in fact long standing discrepancies between theory and data for quarkonia production. 

With the advent of jet substructure at the LHC, it became possible to study energetic quarkonia produced inside jets, and there have been numerous approaches proposed to study quarkonia produced inside jets using jet substructure techniques. Examples include \cite{Baumgart:2014upa,Kang:2017yde,Zhang:2024owr,Bain:2017wvk,Bain:2016clc}. Nevertheless, this problem remains open.

Recently, the quarkonium energy correlator was introduced in \cite{Chen:2024nfl}. It is illustrated in \Fig{fig:quarkonia_EEC}, and is defined as
\begin{align}
\Sigma(\cos \chi) =\int d\sigma \sum_i \frac{E_i}{M} \delta (\cos \chi -\cos \theta_i)\,.
\end{align}
This observable measures the pattern of radiation around the charmonium. In particular, it measures the average energy emitted during hadronization. It was demonstrated that the charmonium correlator can probe different quarkonium production mechanisms, such as color-singlet and color-octet. In \Fig{fig:quarkonia_EEC} we also show a calculation of the charmonium energy correlator using different sets of long distance matrix elements (LDMEs), as well as the contributions as a function of angle from LDMEs of different quantum numbers.

The first measurement of the charmonium correlator was shown in \Fig{fig:jpsi_measure}, and exhibits a large disagreement between theory and data. This presents a significant opportunity to improve our understanding of quarkonium dynamics. Future measurements at higher transverse momentum, as well as for bottomonium, will provide a more complete picture, and will be particularly exciting.

\subsection{Beyond the Standard Model Searches}\label{sec:bsm}

\begin{figure}
\includegraphics[width=0.95\linewidth]{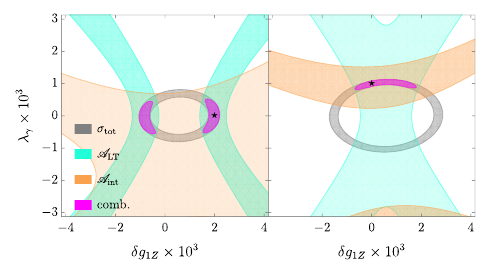}
\caption{In the left (right) plot, nomalous longitudinal (transverse) W boson production due to non-vanishing $\delta_{g_{1Z}} (\lambda_\gamma)$ is included. The constraints from both the energy correlator observable, as well as the measurement of the total cross section, are shown. Figure from \cite{Ricci:2022htc}.
}
\label{fig:BSM_polarized}
\end{figure}

Although we have focused primarily on applications of the energy correlators to precision measurements, it is also interesting to explore their application to beyond the Standard Model searches. Here we wish to highlight two particular applications, with the hope that as our understanding of these observables improve further, we will see many more.

In general, any precision measurement of QCD observables places strong constraints on any contributions that would modify the scaling behavior. This was originally emphasized in \cite{Kaplan:2008pt} who used measurements of the thrust observable at LEP to constrain light particles. This also applies to the remarkable new measurements of the energy correlators at the LHC, and it would be interesting to explore this in more detail and perform dedicated analyses.  This has been applied to the TEEC in \cite{Llorente:2024wpr}.

Since we have highlighted in this review that light-ray operators provide a connection between BFKL physics and DGLAP physics, we can also highlight a similar approach to the detection of new physics \cite{Kowalski:2012ur} from the Pomeron side. Instead of studying modifications to the scaling of energy correlator observables, this work suggests that new physics could modify the structure of discrete pomeron states. This idea has not received much attention, but would be interesting to explore using re-analyses of HERA data, or at the forthcoming electron-ion collider.

Another interesting special feature of the detector correlator observables, as compared to standard jet substructure observables, is that being functions of the kinematics on the celestial sphere, they preserve polarization information. This will be crucial to their application in nuclear physics, but it also has many applications in beyond the Standard Model searches, where new physics can modify the fraction of longitudinal and transversely polarized $W/Z$ bosons. Being able to detect this in hadronic, as compared to leptonic decays, is beneficial, and extends the jet substructure program to incorporating polarization information. This idea was applied in \cite{Ricci:2022htc}. They studied a simple model where the production of
 longitudinal modes is modified by a triple gauge coupling $\mathcal{L}\supset ig c_{\theta_W} \delta_{g_{1Z}} \left(W^+_{\mu \nu} W^-_\mu - W^-_{\mu \nu} W^+_\mu \right) Z_\nu$, while the production of transverse modes is controlled by $\mathcal{L} \supset i \frac{e}{m_W^2} \lambda_\gamma W^+_{\mu \nu} W^-_{\nu \rho}A_{\rho \mu}$. They could then study their ability to detect new physics in the 2d space of $(\delta_{g_{1Z}} , \lambda_\gamma)$, using energy correlator observables. An illustration of the potential of this technique is shown in \Fig{fig:BSM_polarized}, where in the left (right) plot, one has anomalous longitudinal (transverse) W boson production due to non-vanishing $\delta_{g_{1Z}} (\lambda_\gamma)$. The constraints from both the energy correlator observable, as well as the measurement of the total cross section, are shown. While the total cross section does not enable the determination of the polarization, the energy correlator does. It would be interesting to explore the feasibility of this approach in real experiments at the LHC.

We hope that in the future, precision measurements of the energy correlators can be used to place constraints on a wide range of proposals for BSM physics.

\section{Energy Correlators in Nuclear Physics}\label{sec:nuclear}

In this section, we present an overview of applications of energy correlator observables to nuclear physics. We have highlighted throughout this review that one of the primary features of the energy correlators is their ability to cleanly identify scales at which interesting dynamics occur. Such scales manifest in energy correlator observables as breaks in power law behavior, which can be searched for in experiment, even if a complete theoretical understanding of the dynamics occurring at this scale is not available. This feature is ideal for numerous problems in nuclear physics, due to the numerous scales present, such as the confinement scale, the scale of chiral symmetry breaking, the saturation scale, the Debye length, etc., many of which are non-perturbative in nature. In addition, as compared to many jet substructure observables, energy correlators preserve directional information, and are hence sensitive to spin, opening up many interesting opportunities in the study of energy flux.  Here we outline a broad program using energy correlators to address a number of the most important questions in nuclear physics, ranging from understanding nucleon structure, to studying the evolution of spin in the fragmentation process, to revealing the microscopic structure of the quark-gluon plasma.

\subsection{Energy Correlators for Nucleon Structure}

As explained in \Sec{sec:ep}, the existence of the highly energetic nucleon state in DIS represents an interesting challenge for understanding QCD, as well as an opportunity for using energy flow operators to explore the nucleon state through patterns in energy flux produced in the collisions.

Perhaps the most classic example of probing the partonic structure of nucleon states are the PDFs, which have a formal definition as a non-local light-like quark-antiquark correlator in a energetic nucleon state \cite{Collins:1981uw}
\begin{align}
  & f_{i / H}(x)= \int \frac{d y_{-}}{8\pi} e^{-i x P_{+} y_{-}/2 } \nonumber \\
  & \times\left\langle H(P) \right| \bar{\psi}_i \left(0, y_{-}, \boldsymbol{0}_T\right) \gamma_{+}\psi_i \left(0,0, \boldsymbol{0}_T\right)\left|H(P)\right\rangle \,, \label{eq:pdf}
  \end{align}
where $|H(P)\rangle$ is a state for nucleon $H$ with momentum $P$, $i$ is quark flavor label. Eq.~\eqref{eq:pdf} is defined in light-cone gauge. In a general covariant gauge, an additional light-like Wilson like connecting the two quark fields is needed.  The partonic structure of the nucleon is probed through the Fourier conjugate of the light-cone coordinate, which measures the fraction of longitudinal momentum carried by the struck parton. One can also probe more detailed structures of how the quarks are distributed in the nucleon, such as transverse-momentum distributions, Wigner distributions, or spin structures, by employing increasingly complicated correlators in the nucleon matrix element. 

\begin{figure}
\centering
\includegraphics[width=0.4\textwidth]{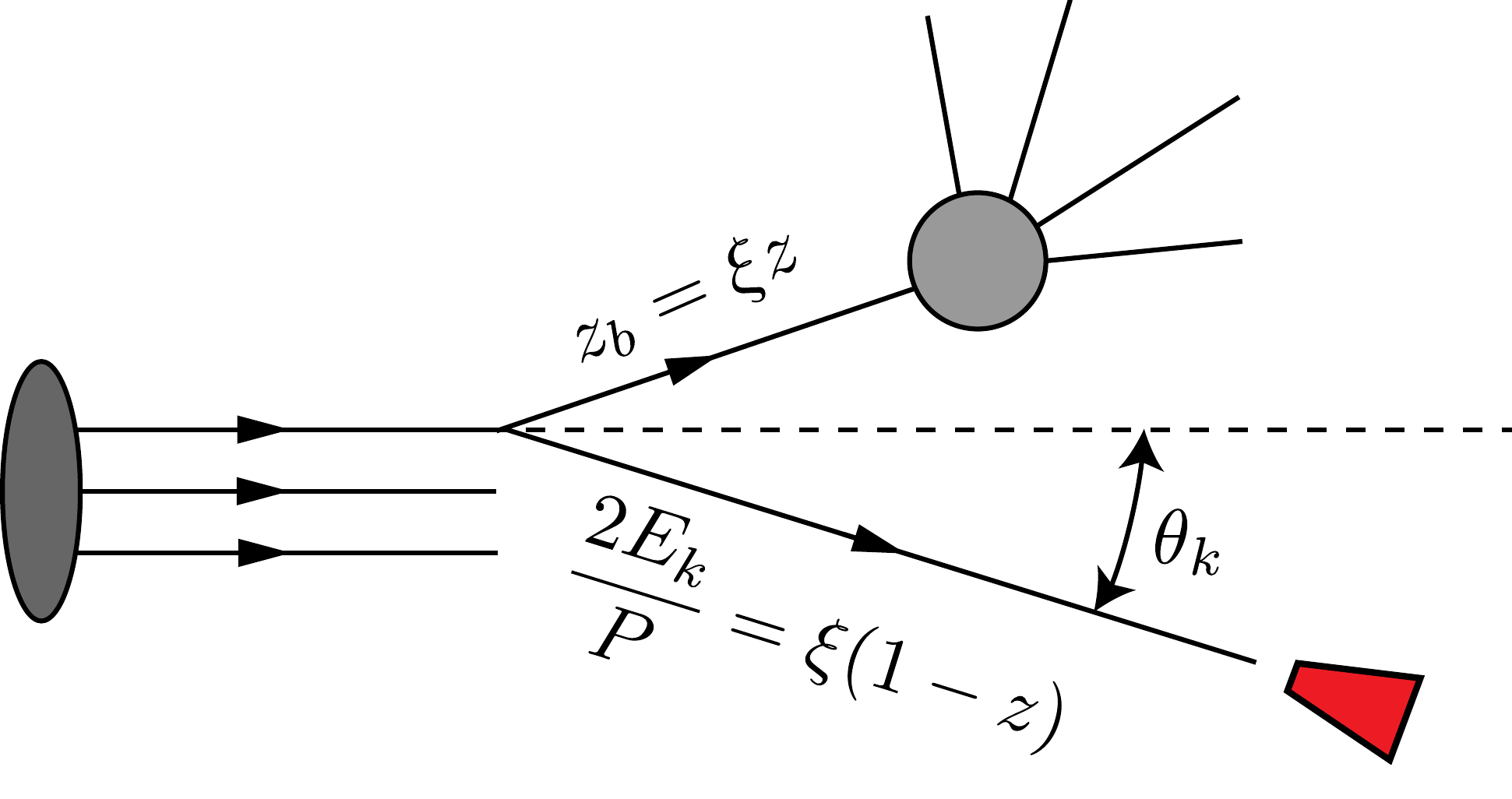}
\caption{A LO illustration for the measurement of NEC with one energy detector in the target fragmentation region. A parton with momentum fraction $z_b = \xi z$ participate in hard scattering, while a parton with energy $E_k$ from initial splitting is detected by a detector at angle $\theta_k$. Figure from \cite{Liu:2022wop}.}
\label{fig:nucleon_EEC}
\end{figure}

An intriguing question is whether insights into the nucleon can be gained by measuring the energy flow in the target fragmentation region. Such a measurement is formally defined in Eq.~\eqref{eq:nec_def}, and is referred to as the nucleon energy correlators (NECs)~\cite{Liu:2022wop,Cao:2023oef,Liu:2024kqt}: 
\begin{align}
  & f_{q, n}(z, \{\Omega_1, \ldots \Omega_n\})   =    \int \frac{d y_{-}}{8 \pi} e^{-i z P^{+} \frac{y^{-}}{2}} \nonumber \\
  \times &\langle P| \bar{\psi}_i  \left(0, y_{-}, \boldsymbol{0}_T\right)  \gamma_{+} \mathcal{E}(\Omega_1)  \ldots  \mathcal{E}(\Omega_n) \psi_i  \left(0, 0, \boldsymbol{0}_T\right)  |P\rangle \,.  \label{eq:nec_def}
\end{align}
For $n=1$, where only a single detector is present in the target fragmentation region, this setup is illustrated in Fig.~\ref{fig:nucleon_EEC}. Interestingly, a sum rule has been established that connects NECs at twist-2 and twist-3 levels to fracture functions~\cite{Chen:2024bpj}, which have been studied extensively in earlier works \cite{Trentadue:1993ka,Berera:1995fj}. Furthermore, it has been shown that Mellin moments of NECs are related to the moments of transverse-momentum-dependent parton distribution functions (TMD PDFs) \cite{Liu:2024kqt}, suggesting that NECs could serve as a powerful tool for nucleon tomography.

\begin{figure}[ht!]
  \centering
  \includegraphics[width=0.3\textwidth]{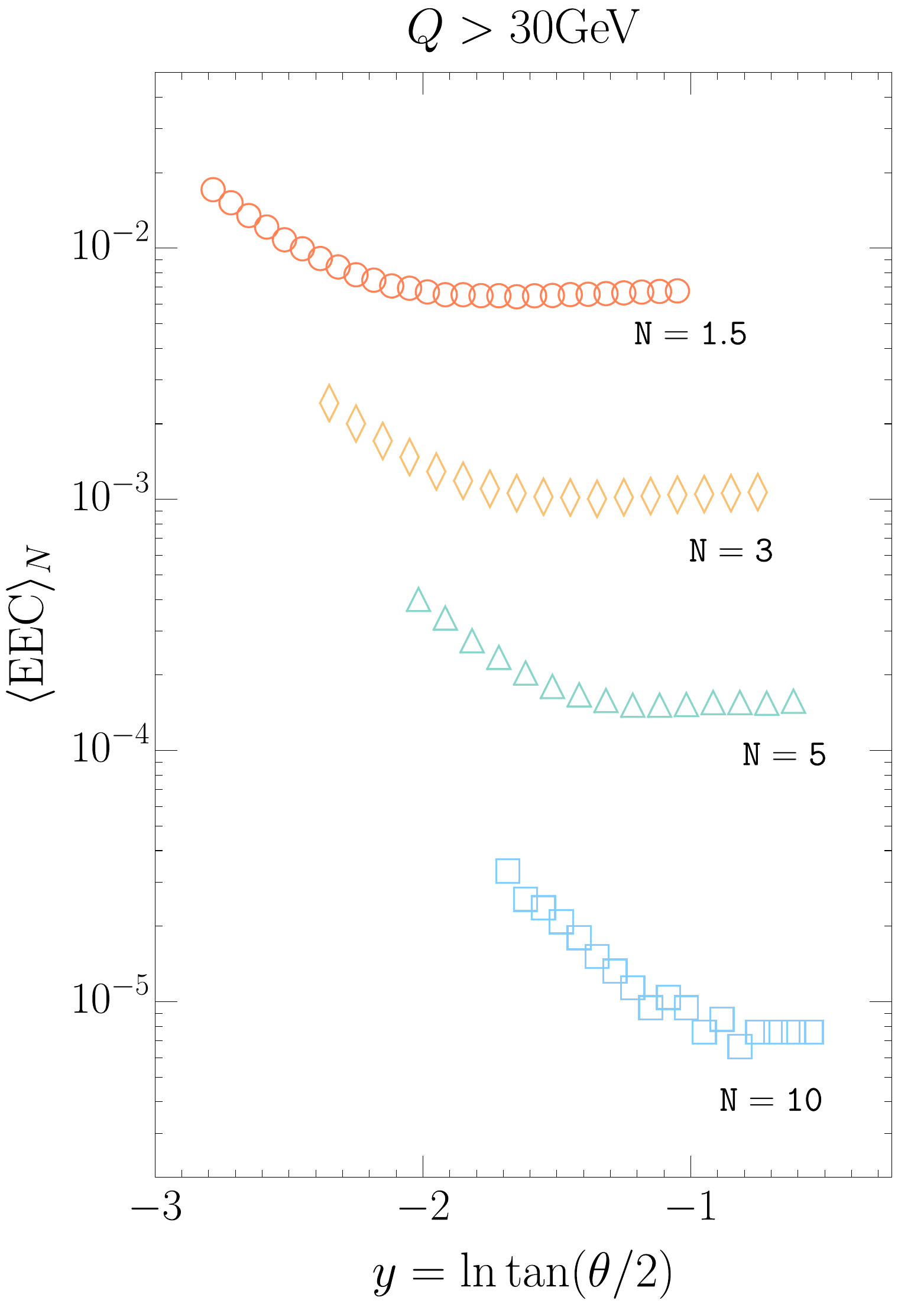}
  \caption{Scaling evolution as imprinted in the angular evolution for NECs. Shown here are different Mellin moments of  the NECs from a Pythia simulation. Figure from \cite{Liu:2022wop}.}
  \label{fig:nucleon_EEC-N-variation}
  \end{figure}

\begin{figure}[ht!]
  \centering
  \includegraphics[width=0.4\textwidth]{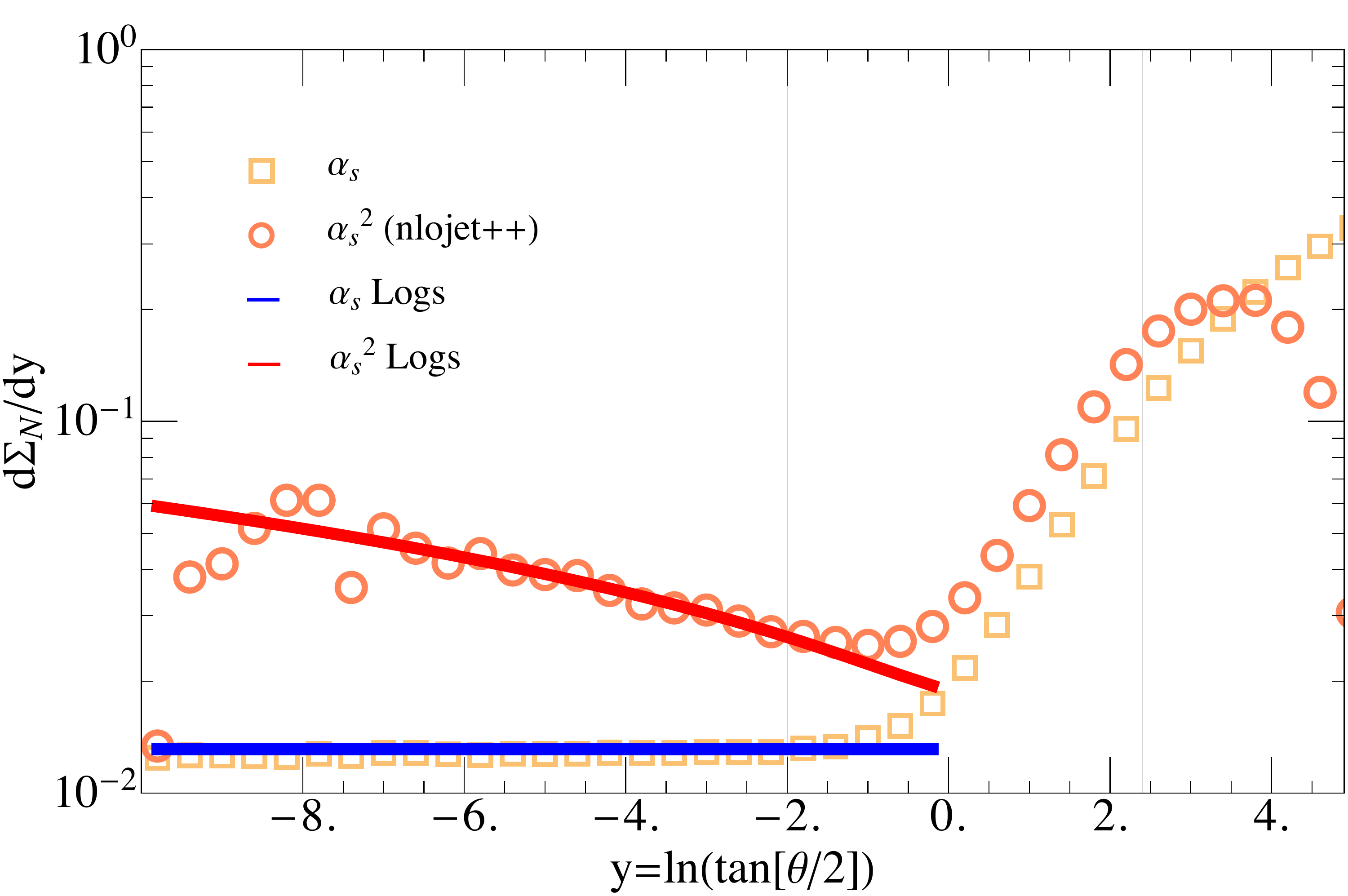}
  \caption{A comparison of the perturbative logarithmic terms in the matching coefficients for NECs,  between analytic predictions derived using a factorization formula and a numerical fixed-order calculation. Figure from \cite{Cao:2023oef}.}
  \label{fig:nucleon_EEC-singluar}
  \end{figure}

The NEC $f_{q, n}$ represents a non-perturbative object that encodes information about the nucleon's structure. In the perturbative regime ($\theta Q \gg \Lambda_{\rm QCD}$), it has been shown that the scaling evolution of $f_{q,1}$ in \eqref{eq:nec_def} is governed by the DGLAP kernel, similar to the evolution of standard PDFs. However, unlike traditional PDFs, the scaling evolution here manifests as an angular dependence, analogous to the behavior observed in EECs. Consequently, NECs can effectively probe the nucleon's structure at different scales by varying the angle $\theta$. In the perturbative region, we expect to observe a scaling behavior in $\theta$, where the scaling exponent is controlled by the DGLAP kernel. This behavior is demonstrated in Fig.~\ref{fig:nucleon_EEC-N-variation} using Pythia simulations, which show the angular scaling evolution for different Mellin moments of $f_{q,1}$. Here, the Mellin moments are taken with respect to the momentum fraction of the parton involved in the hard scattering. The scaling behavior has also been verified through numerical calculations at NLO using the program \texttt{nlojet++}~\cite{Nagy:2003tz}, as shown in Fig.~\ref{fig:nucleon_EEC-singluar} for $N=3$ at $Q^2 = 100\,\text{GeV}^2$.

In the remainder of this section, we highlight a few interesting applications of NECs.  One of the primary scientific goals of the EIC is to probe the scale of gluon saturation, where a CGC \cite{Gelis:2010nm,Iancu:2003xm,McLerran:1993ni,McLerran:1993ka,McLerran:1994vd} of gluons forms due to the highly nonlinear evolution of gluon densities at small $x$~\cite{Gelis:2010nm}. The NECs offer an intriguing possibility for observing gluon saturation and measuring the saturation scale~\cite{Liu:2023aqb}. The key idea is that the energy radiation pattern in the target fragmentation region depends on whether the gluon participating in the hard scattering originates from a free partonic state or a CGC state. Furthermore, since $\theta Q$ serves as a measure of the characteristic energy scale of the underlying process, varying the angle $\theta$ allows for a scan across different energy scales, providing a means to search for the saturation scale. 
  
\begin{figure}[t!]
  \centering
  \includegraphics[width=0.5\textwidth]{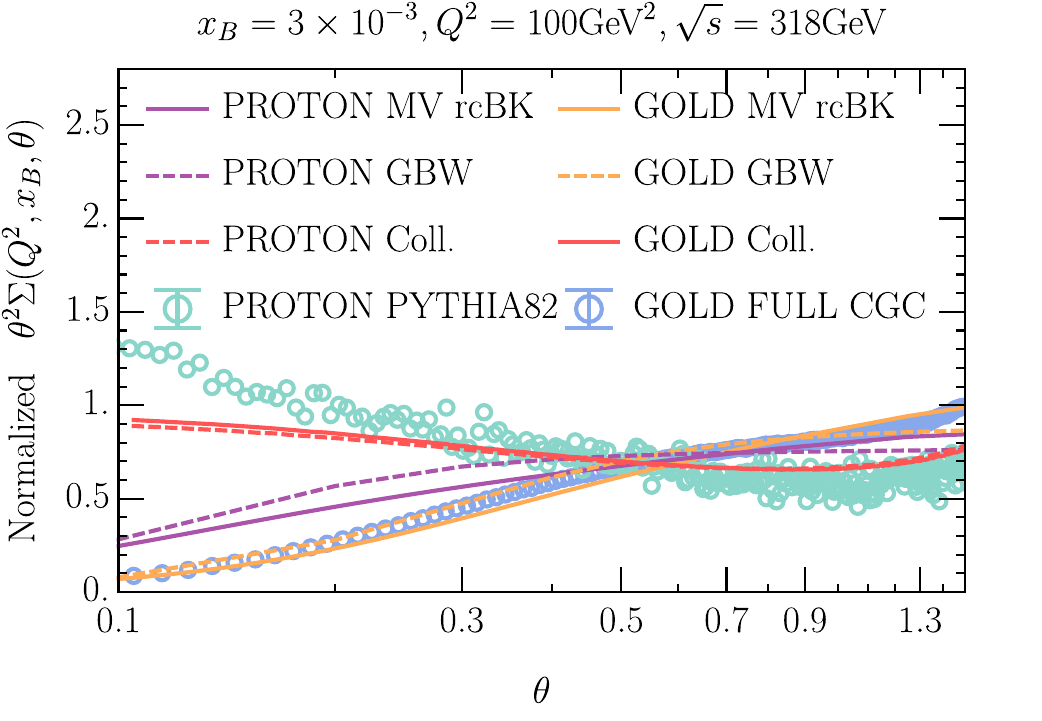}
  \caption{The normalized NEC for proton and Au at $\sqrt{s} = 318$ GeV. Figure from \cite{Liu:2023aqb}.}
  \label{fig:CGC_1}
  \end{figure}

Fig.~\ref{fig:CGC_1} presents predictions from \cite{Liu:2023aqb} for NECs at small angles using multiple approaches: the standard collinear approximation, Pythia simulations, models of gluon saturation, and full Color-Glass-Condensate calculations. Both the collinear approximation and Pythia simulations assume that the gluon is in a free partonic state and exhibit the typical scaling behavior expected for NECs. In contrast, calculations using models of gluon saturation and full CGC computations show suppression at small $\theta$, indicating that the gluons being probed are in the CGC state. A transition point from suppression to a plateau signifies the saturation scale. An experimental search for this transition point would therefore be of great interest.

\begin{figure}
  \includegraphics[width=0.95\linewidth]{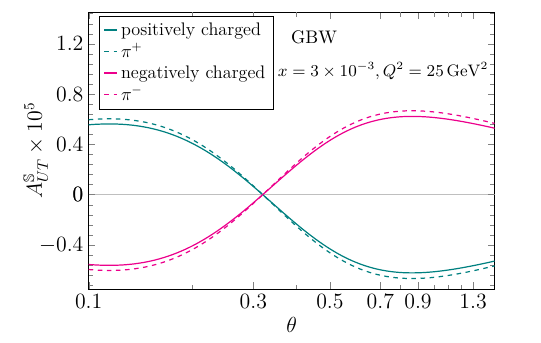}
  \caption{The SSA of track-based NEC as a function of polar angle $\theta$ with respect to beam axis in Breit frame for positively and negatively charged hadrons. Figure from \cite{Mantysaari:2025mht}.
  }
  \label{fig:odderon_EEC}
  \end{figure}

NECs have also been suggested as a tool to hunt for the spin dependent odderon in DIS~\cite{Mantysaari:2025mht}, see also \cite{Bhattacharya:2025bqa} for studies of the odderon with the TEEC. The odderon is a $C$-odd colorless gluonic bound state that has been predicted to exist in $t$-channel exchange in high-energy hadronic scatterings~\cite{Lukaszuk:1973nt}. It has received renewed interest recently due to experimental progress~\cite{D0:2020tig}. The spin-dependent odderon can generate a single-spin asymmetry~(SSA) in DIS at small-$x$~\cite{Zhou:2013gsa,Boer:2015pni} in the current fragmentation region. In the target fragmentation region, NECs, combined with the track functions to tag $C$-odd quantities, provide a novel probe of the spin-dependent odderon. Specifically, it has been suggested that the twist-3 component of NECs can be sensitive to the spin-dependent odderon and yield opposite-sign predictions for positively and negatively charged hadrons, as illustrated in Fig.~\ref{fig:odderon_EEC}. The IRC safety of NECs helps reduce the dependence on non-perturbative fragmentation functions, offering a cleaner probe for such phenomena at the EIC.
Besides the phenomenological applications discussed here, it would interesting to describe NECs in a more formal field theoretic language, much like the operator and correlation description for EECs. We refers to \cite{Caron-Huot:2022lff} for a related discussion.

\subsection{The Collins-Soper Kernel}\label{sec:CS_results}

\begin{figure}[t!]
  \centering
  \includegraphics[width=0.4\textwidth]{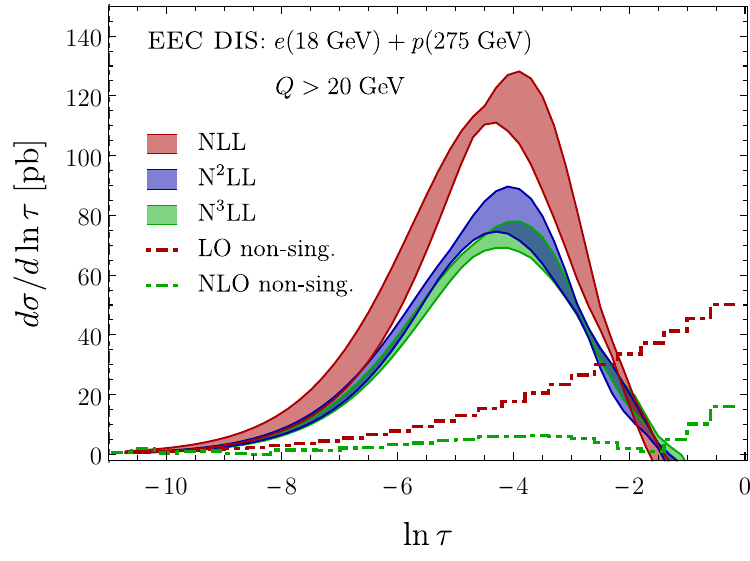}
  \caption{Theoretical prediction in the back-to-back limit for energy correlators in DIS with large logarithms resummed to N$^3$LL. Figure from \cite{Li:2021txc}.}
  \label{fig:res_log}
  \end{figure}

Energy correlators have also been defined and studied in the current fragmentation region for DIS~\cite{Meng:1991da,Li:2021txc,Li:2020bub}. In the Breit frame, one can define a back-to-back limit, where the polar angle $\theta$ between the energy detector and the beam axis approaches $\pi$. It has been shown that EECs in DIS provide a probe of TMD PDFs and TMD FFs in this back-to-back limit. A TMD-like factorization formula has been derived, establishing an interesting connection between energy correlators and TMD studies, which themselves constitute a significant area of research in nuclear physics~\cite{Li:2021txc}.

\begin{figure}[t!]
  \centering
  \includegraphics[width=0.4\textwidth]{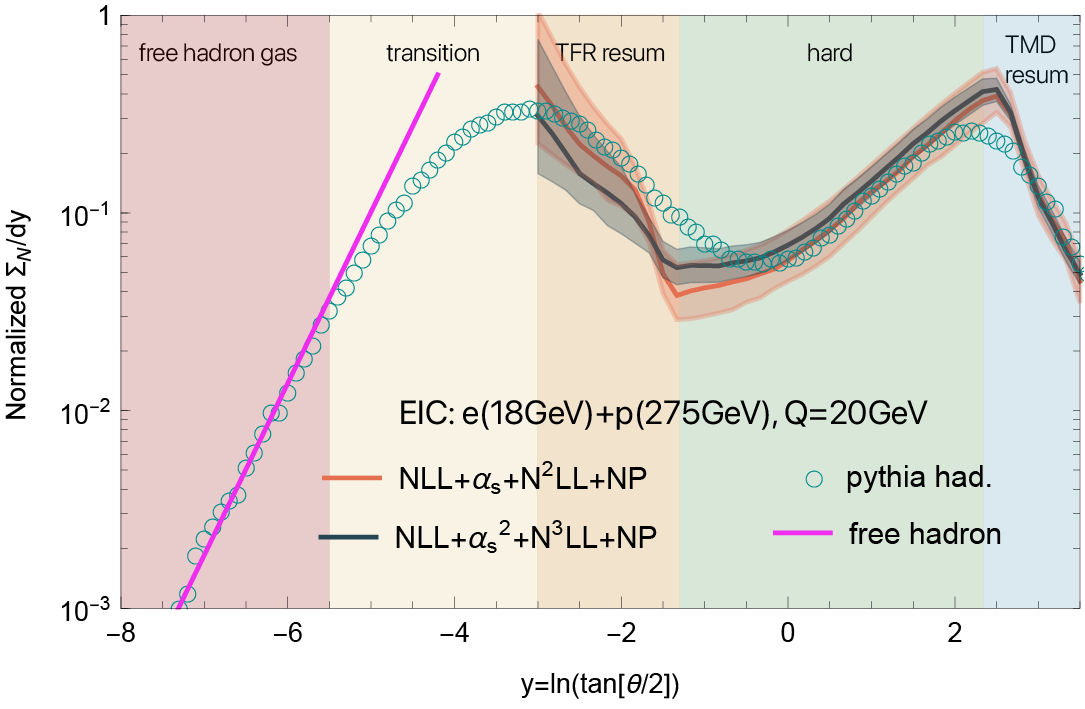}
  \caption{Energy correlators from the back-to-back limit in the current fragmentation region to the NECs in the target fragmentation region for EIC. Both predictions from RG-improved QCD and Pythia simulation are shown. Figure from \cite{Cao:2023qat}.}
  \label{fig:pythiahai}
  \end{figure}

Resummation in this limit has been carried out at an impressive N$^3$LL accuracy, with excellent convergence observed across different perturbative orders~\cite{Li:2021txc}. Recently, these calculations have been further extended to the full phase space, Fig.~\ref{fig:pythiahai}, bridging the back-to-back region in the current fragmentation region with the NECs in the target fragmentation region~\cite{Cao:2023qat}. It would be highly interesting to measure these observables in DIS experiments and compare them with theoretical predictions to extract quantities of interest in the nuclear community, such as the Collins-Soper kernel and various TMD functions.

\begin{figure}
\includegraphics[width=0.75\linewidth]{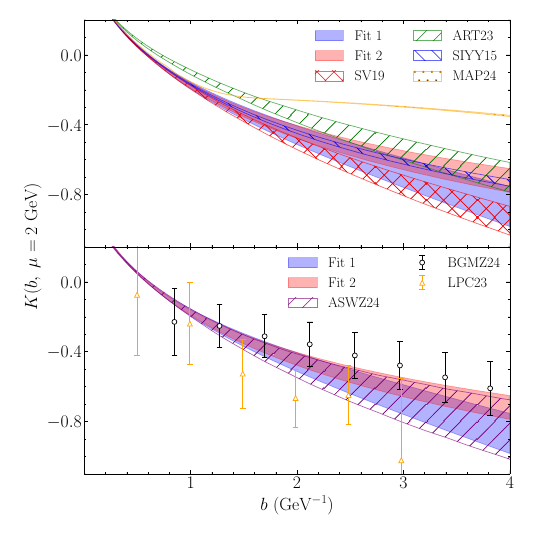}
\caption{An extraction of the Collins-Soper kernel from the back-to-back limit of the energy correlator observable. The results are compared with previous  extractions. Figure from \cite{Kang:2024dja}.
}
\label{fig:EEC_CSfit}
\end{figure}

 A glimpse of this promising approach was presented in \cite{Kang:2024dja} for $e^+e^-$, where the extracted values of the Collins-Soper kernel were compared with other extractions, as shown in \Fig{fig:EEC_CSfit}. The results are encouraging, underscoring the potential for further advancements using higher-order perturbative data and more precise measurements, such as those from track-based EECs as described in \Sec{sec:det_func}. 
  
Ultimately, these determinations can be cross-validated with high-precision lattice extractions~\cite{Avkhadiev:2024mgd,Avkhadiev:2023poz,Shanahan:2021tst,Shanahan:2020zxr,Shanahan:2019zcq}, or the lattice results can serve as input to enable precise extractions of other fundamental quantities, such as the strong coupling constant.

\subsection{Spin Physics with Energy Correlators}\label{sec:spin}

\emph{In Loving Memory of Fanyi Zhao}

\vspace{0.4cm}

Since its discovery, spin has played a central role in particle and nuclear physics, see e.g. \cite{Filippone:2001ux,Bass:2004xa,Aidala:2012mv,Ji:2020ena} for recent reviews. Spin plays an important role in both the physics of the colliding nuclei themselves, as well as in the final state. In the initial state, one of the key scientific goals of the EIC is to understand the origin of proton spin. In the final state, the study of spin in the fragmentation process provides access to chiral symmetry breaking, enabling it to be studied in asymptotic energy flux.

While the concept of the energy operator has been known for a long time, its application to studying parton and nucleon spin is relatively recent. Nevertheless, we believe that the energy operator opens new avenues for advancing spin physics. Traditional tools for probing spin include TMD PDFs, FFs, jets, and jet substructure. In comparison, energy correlators offer several advantageous features worth highlighting: energy flow observables are IRC safe; angular separation provides insights into parton-to-hadron evolution; and angles are unambiguously defined for individual particles.

\begin{figure}
\includegraphics[width=0.65\linewidth]{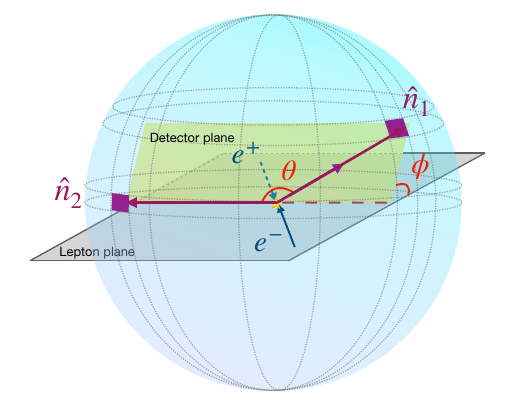}
\includegraphics[width=0.65\linewidth]{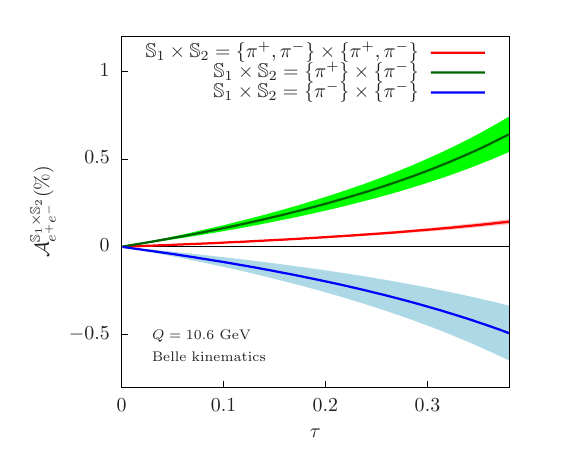}
\caption{Detailed studies of the angular dependence of energy correlator observables in $e^+e^-$ collisions enable measurement of non-perturbative functions, in this case the Collins function.  The magnitude of the contributions from the Collins effect are shown in the lower panel. Here $\tau\to 0$ corresponds to the back-to-back kinematics of the energy correlators.  Figure from \cite{Kang:2023big}.
}
\label{fig:ee_polarized}
\end{figure}

TMD PDFs and FFs have long served as powerful tools for unraveling the spin structure of the nucleon and studying parton-to-hadron fragmentation~\cite{Boussarie:2023izj}. However, their dependence on multiple variables, such as the momentum fraction and transverse momentum, introduces significant challenges in both theoretical calculations and experimental measurements. Fortunately, a fascinating opportunity arises from the close analogy between the factorization of EECs in the back-to-back limit and TMD factorization. This connection opens the door to exploring many polarized TMD functions using EECs, with the added advantage that EECs depend only on a single moment of the momentum fraction rather than a full non-perturbative function. Moreover, by forming ratio observables, it is possible to eliminate the non-perturbative normalization, simplifying the analysis.

This interesting avenue was first explored in \cite{Kang:2023big}, where azimuthal-dependent EECs were proposed for both $e^+e^-$ collisions and DIS. In the $e^+e^-$ case, the two detectors are placed in the back-to-back limit, and the azimuthal angle is defined relative to the detector plane and the lepton plane, as illustrated in Fig.~\ref{fig:ee_polarized}. Remarkably, the azimuthal correlation serves as a sensitive probe of the Collins effect \cite{Collins:1992kk}, which describes how a transversely polarized quark fragments into an unpolarized hadron. The calculations in \cite{Kang:2023big} reveal clear asymmetries in the azimuthal angle for different combinations of final states, as shown in the lower panel of Fig.~\ref{fig:ee_polarized}.  Building on this idea, \cite{Kang:2023big} also introduced a definition of EECs in the Breit frame of DIS, where the back-to-back limit provides access to the Sivers effect, which gives information on an unpolarized quark inside a transversely polarized nucleon \cite{Sivers:1989cc,Brodsky:2002cx,Collins:2002kn}. The resulting azimuthal asymmetries were computed in EIC kinematics, and shown to be observable.

The similarity between EEC factorization in the back-to-back limit and TMD factorization suggests that much of our understanding of TMDs can be seamlessly applied to EECs. Indeed, \cite{Kang:2023big} highlights two crucial examples: the Collins and Sivers functions, both of which provide deep insights into the chiral symmetry breaking of QCD. Clearly, this direction holds immense potential for future progress, including precision studies and further explorations of the rich interplay between spin and energy correlations.

High-energy jets at the LHC are typically unpolarized. However, effects due to gluon spin can be studied using multi-point energy correlators. This was first revealed by examining the analytic results for the EEEC inside a jet, where a small but non-zero azimuthal correlation emerges in the squeezed limit of the EEEC~\cite{Chen:2020adz}. This highlights the importance of having analytic results for theoretical predictions. In this limit, considering the azimuthal angle of the squeezed pair relative to the plane spanned by the energy detector at large angles (as shown in Fig.~\ref{fig:spin_shower}), a $\cos(2\phi)$ modulation appears at leading order (LO) in perturbation theory. This modulation arises from the interference of dynamically generated linearly polarized gluons in the intermediate state of splitting, analogous to the famous double-slit experiment but in spin space. Numerically, the effects are below the $5\%$ level due to cancellations between $g \to gg$ and $g \to q\bar{q}$ splittings, making them challenging to observe directly. Nevertheless, the spin modulation provides the first application of the light-ray operator product expansion (OPE), with the first all-order resummation formula for such effects derived in~\cite{Chen:2020adz}. It further inspired studies of azimuthal correlations in the EEEC beyond leading power~\cite{Chen:2021gdk}, leading to the discovery of collinear celestial blocks~\cite{Chen:2022jhb,Chang:2022ryc}. It will be interesting to extend these to higher point correlators. While the four-point energy correlator in the collinear limit has been computed in $\mathcal{N}=4$ sYM \cite{Chicherin:2024ifn}, it has not yet been computed in QCD. Due to its much more non-trivial structure, we expect it to contain a variety of interesting spin effects, which could potentially be studied at colliders.

\begin{figure}
  \includegraphics[width=0.65\linewidth]{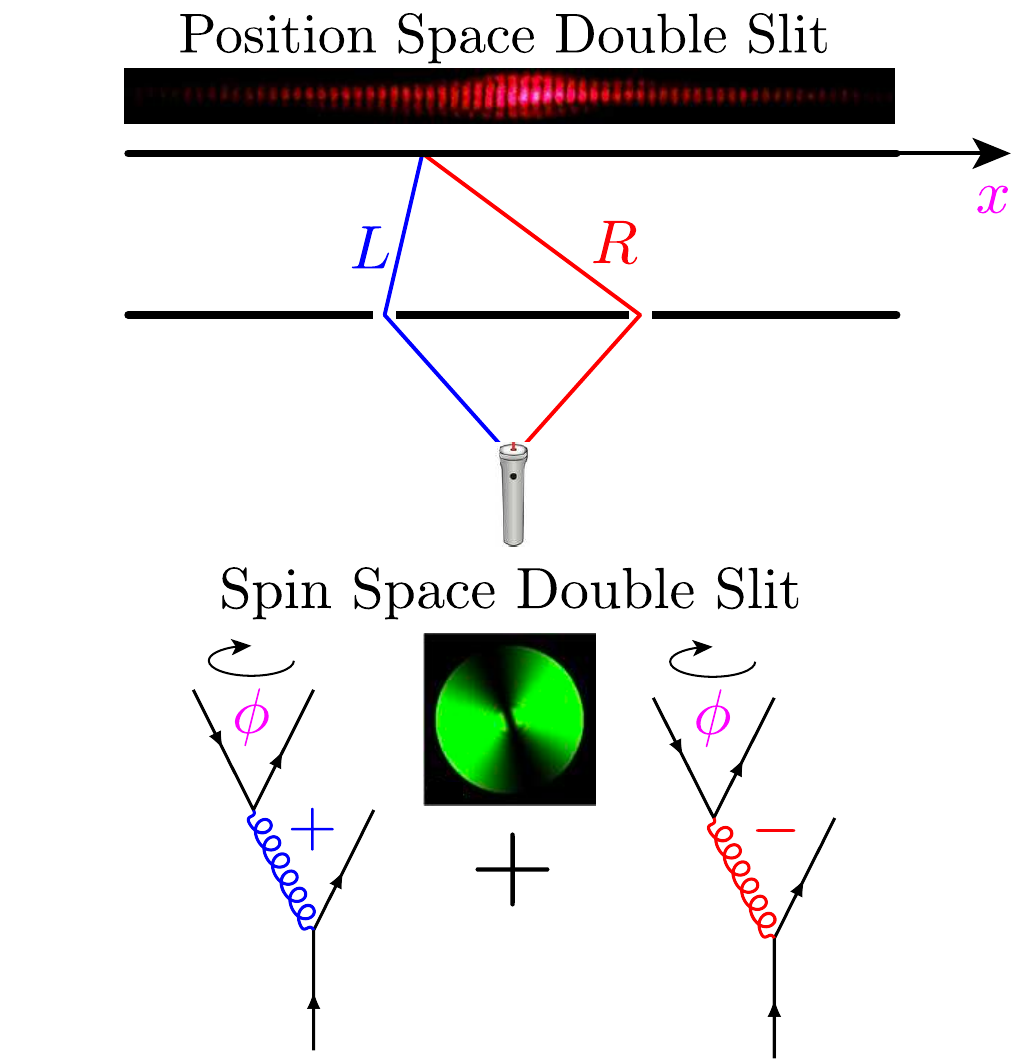}
  \caption{Interference pattern in azimuthal angle of EEEC in jet. Figure from \cite{Chen:2020adz}.
  }
  \label{fig:eeec_gluon_spin}
  \end{figure}

\begin{figure}[t!]
  \hspace{0.8cm}\includegraphics[width=0.35\textwidth]{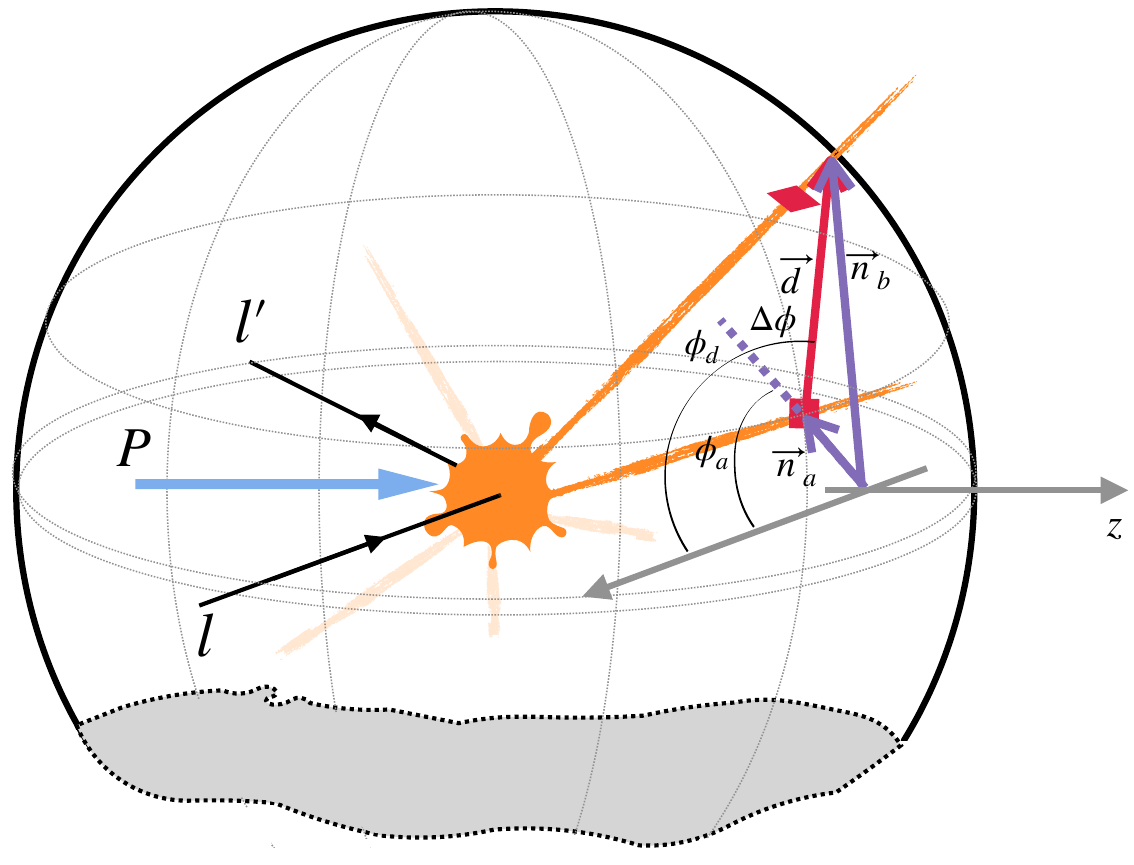}
   \includegraphics[width=0.45\textwidth]{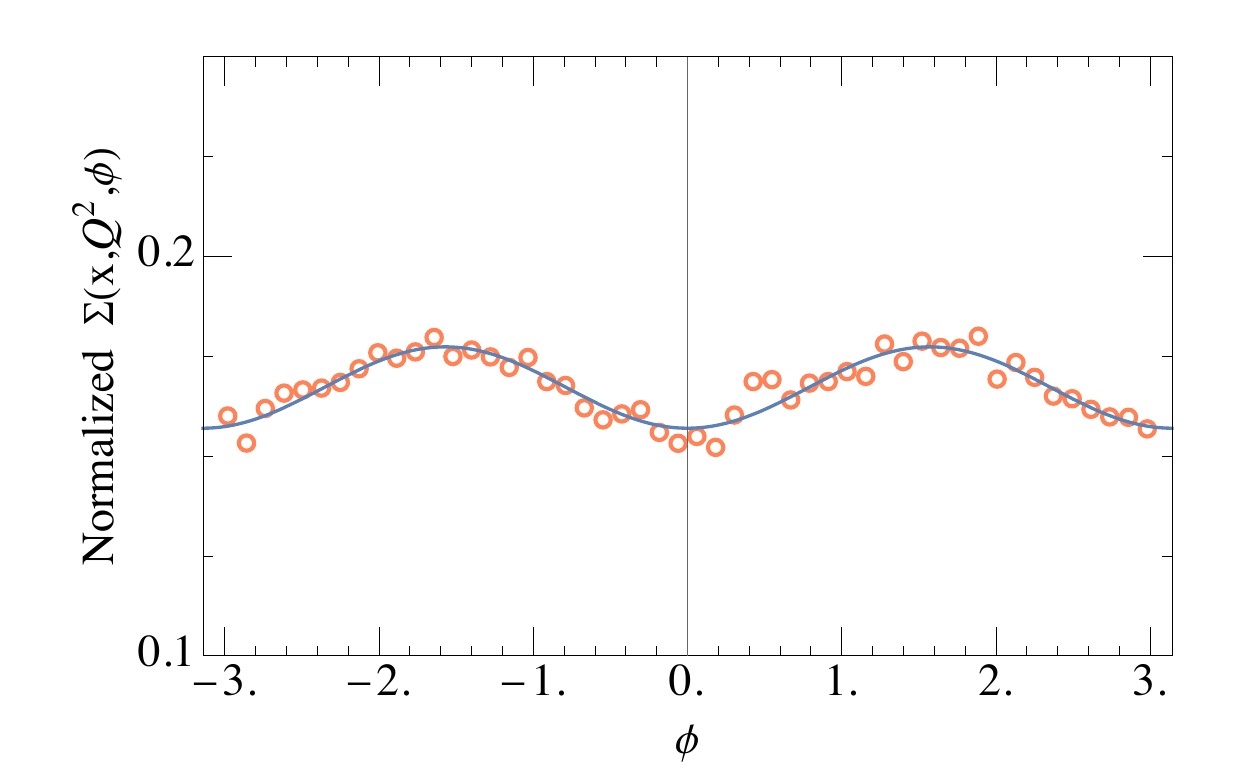}
  \caption{Top Panel: The detector configuration that measures the intrisic gluon polarization, where two energy detectors are positioned at $\vec{n}_a$ and $\vec{n}_b$. Bottom Panel: Azimuthal modulation due to linearly polarized gluon in the nucleon, as a function of the relative detector angle, $\phi$.  Figures from~\cite{Li:2023gkh}.}
  \label{fig:eec-solid-angle}
\end{figure}

The gluonic parton in nucleons contains both unpolarized and linearly polarized components~\cite{Mulders:2000sh,Catani:2010pd}. The origin of linearly polarized component is due to quantum interference effects. The linearly polarized gluon can be probed by the correlations of two energy detectors, and measures the modulation around the beam axis for one of the detector. This is similar to the interference pattern for EEEC in a jet, but with one of the detector crossed to the initial-state nucleon. An illustration of the detector setup is shown in Fig.~\ref{fig:eec-solid-angle}. 
Fig.~\ref{fig:eec-solid-angle} also shows a LO prediction for the $\phi_a$ modulation in the squeezed limit, where both detectors are close to the beam directions. In the case of EEEC, there is partial cancellation between $g\to gg$ and $g\to q\bar{q}$ splitting, reducing the sensitivity to gluon spin. In the DIS setup shown in Fig.~\ref{fig:eec-solid-angle}, there is no such cancellation at LO, as only $g\to q^*\bar{q}$ splitting is presented where the off-shell quark $q^*$ subsequently interact with the virtual photon. Closely related effects have also been proposed to measure Bell inequality violations in $pp \to Hjj$~\cite{Guo:2024jch}, and to provide new insights into the ridge effect in $pp$ collision~\cite{Guo:2024vpe}.

\subsection{Hot and Cold Nuclear Matter}\label{sec:QGP_results}

In this section we provide an overview of the study of energy correlators in hot and cold nuclear matter. As compared to the case of $e^+e^-$, $e^-p$, or $pp$ collisions, there do not exist completely rigorous frameworks for computing energy correlator observables in nuclear collisions from first principles. This makes this field both difficult, but also exciting. It also changes the nature of the questions one wishes to address, from precision calculations to robustly interpretable measurements.

Due to the complexity of nuclear collisions, their theoretical descriptions involves an incredibly broad range of physics, as illustrated in \Fig{fig:qgp_visualize}, ranging from hydrodynamics to perturbative QCD. Even within the context of jet substructure, there are a wide variety of approaches being developed, each focusing on different elements of the underlying physics, and taking those effects as primary. This section is not  intended as a review of theoretical approaches for the calculation of jet substructure observables in nuclear collisions. For excellent reviews on this topic, we refer the reader to \cite{Connors:2017ptx,Busza:2018rrf,Cunqueiro:2021wls,Apolinario:2022vzg}. Rather, the goal of this section is two fold: For those outside the field of nuclear collisions, we intend to give a flavor of the current level of theoretical understanding of energy correlator measurements in nuclear collisions, and highlight areas where we anticipate theoretical and experimental progress using energy correlators. For those familiar with nuclear collisions, we hope to emphasize the specific properties of the energy correlators that we believe make them promising for the study of nuclear collisions. 

Due to the lack of ability to precisely compute in the nuclear physics environment, we believe that the following two properties are essential:

\emph{Interpretability:} In most cases we will not be able to achieve a precision understanding. The ability to interpret observables, and establishing a clean mapping between the microscopic physics and the asymptotic observable then becomes key. Energy correlators are ideal in this respect, since scales in the microscopic description of the nuclear collision imprint themselves as breaks in power laws of the energy correlators. This is much in analogy with the case of the hadronization transition in the collinear limit: even though we do not understand its mechanism, we can identify it, and extract the scale at which it occurs using the energy correlators.

\emph{Theoretical Tractability:} The complexity of nuclear collisions implies that one will never be able to compute full distributions from first principles. One would like the ability to isolate particular features of an observable, for example scalings, that one can understand from first principles, without knowledge of the full complexities of the nuclear collisions. The light-ray OPE provides a systematic approach, reducing the complete description of the transverse structure of a jet, to the study of OPE coefficients which can capture the leading nuclear modifications.

\begin{figure}
\includegraphics[width=0.95\linewidth]{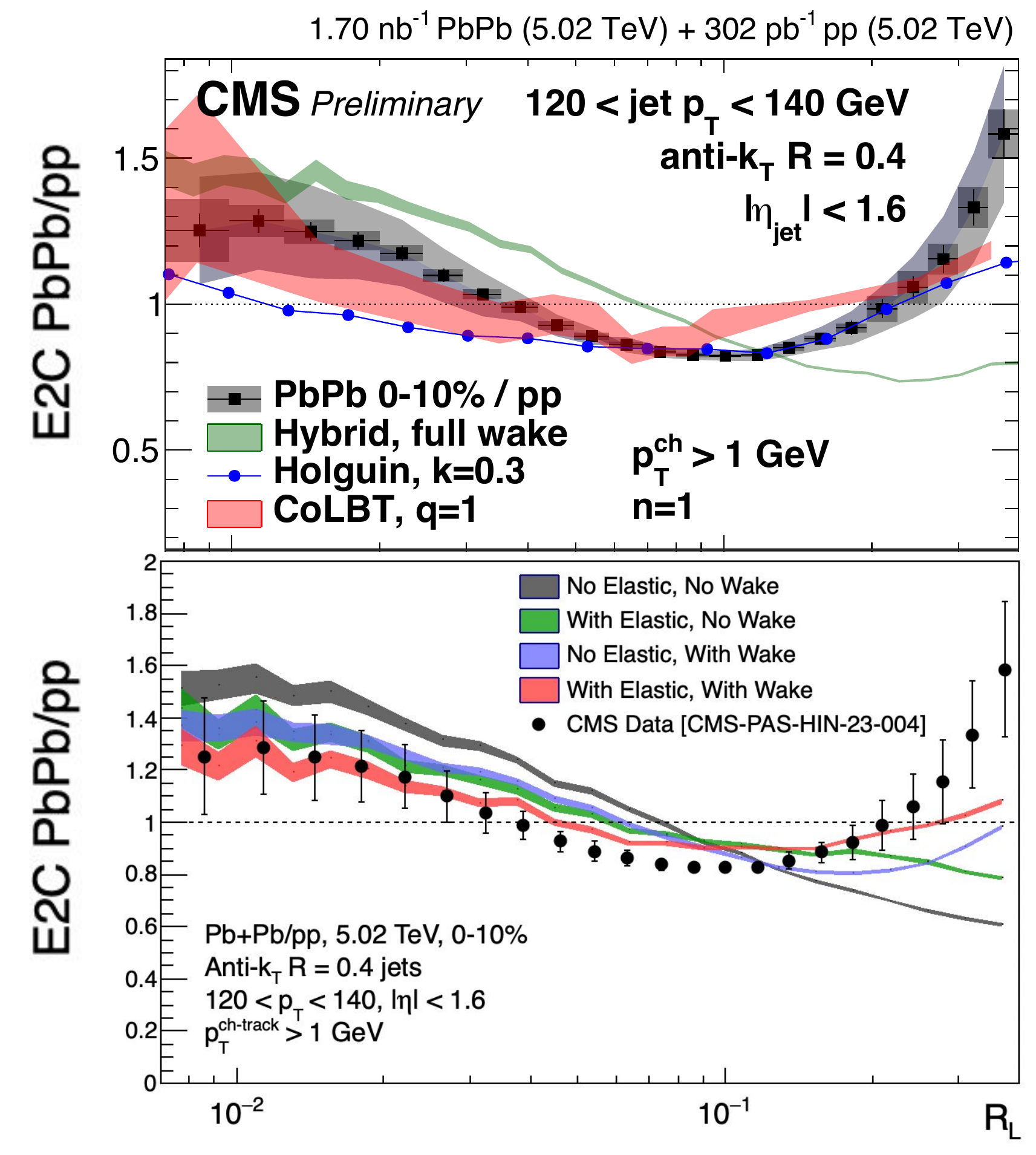}
\caption{A comparison of the CMS measurement of the ratio of the two-point energy correlator in Pb-Pb and p-p collisions with  different theoretical calculations. Details of the theoretical models are provided in the text. Figures from \cite{CMS-PAS-HIN-23-004} and \cite{talk_Arjun}.
}
\label{fig:CMS_raw_combo}
\end{figure}

Our goal in this section is to emphasize how energy correlators have already contributed in both these directions, and how they can continue to do so with both new experimental measurements, and new theoretical techniques. 

\begin{figure}
\includegraphics[width=0.755\linewidth]{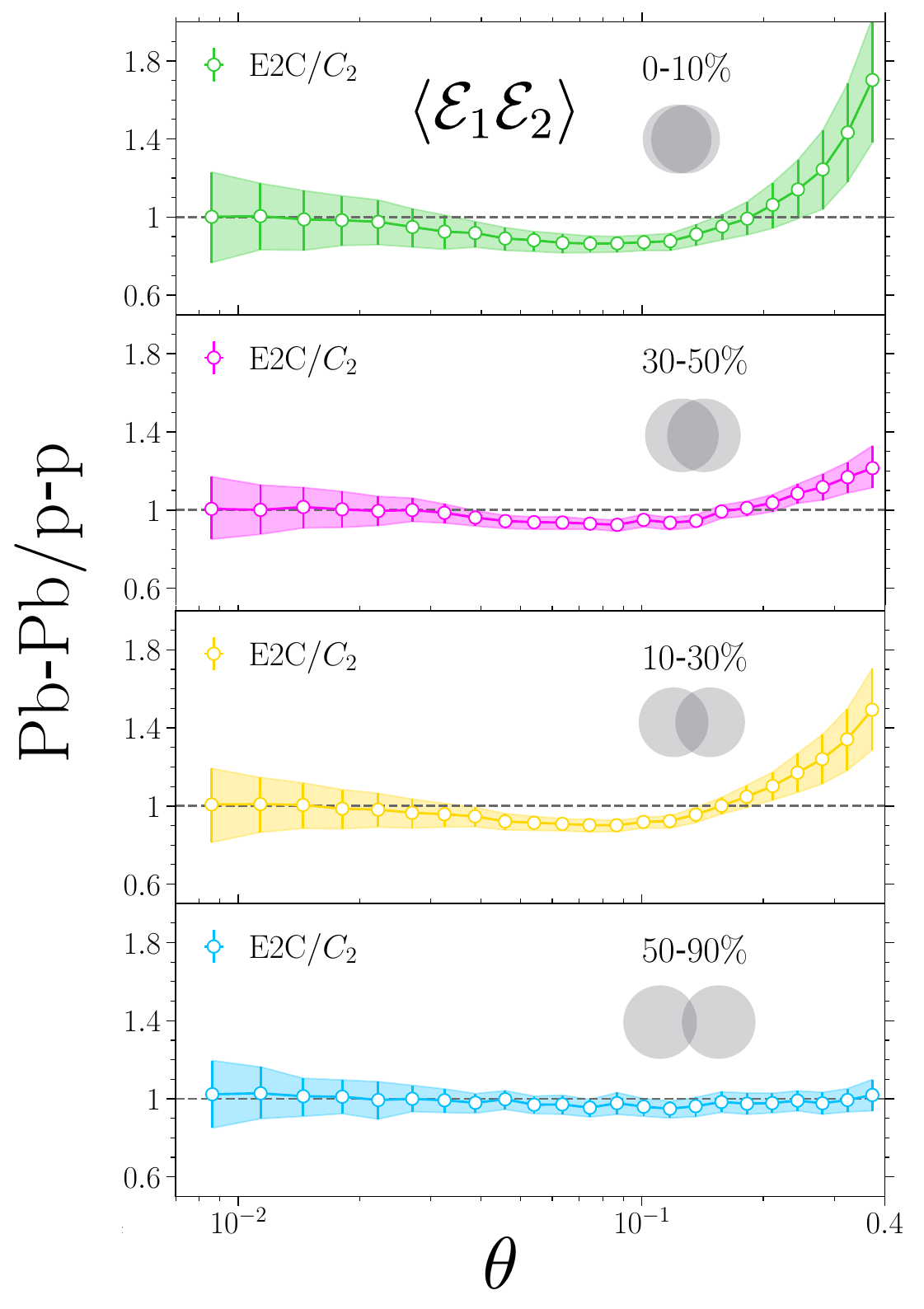}
\caption{The ratio of the two-point energy correlator in Pb-Pb and proton-proton collisions, after performing an unbiasing.  The results are shown as a function of centrality, which measures the overlap of the colliding ions (illustrated by the gray circles). The effect of the nuclear modification decreases as the ions collide in a more peripheral manner. Figures adapted from \cite{Andres:2024pyz}.
}
\label{fig:centrality_fixed}
\end{figure}

We begin with an overview of theoretical attempts to describe measurements of the energy correlator observables in heavy ion collisions. We then discuss future prospects both for improving our theoretical understanding of the current measurements, as well as for future measurements to further map out the structure of the QGP. We then consider the case of $p-A$ and $e-A$ collisions, highlighting attempts to provide a theoretical description of the measurement of the energy correlators in $p-A$, and future prospects for using this to obtain a coherent picture across collision systems.

The measurement of the two-point energy correlator in Pb-Pb collisions by the CMS and ALICE collaborations has prompted significant theoretical efforts to explain its structure.  A collection of references studying the two-point energy correlator in heavy ion collisions are \cite{Andres:2022ovj,Andres:2023xwr,Andres:2023ymw,Andres:2024ksi,Andres:2024pyz,Andres:2024hdd,Yang:2023dwc,Barata:2023bhh,Singh:2024vwb,Barata:2023zqg,Xing:2024yrb,Apolinario:2025vtx,Andres:2024xvk,Bossi:2024qho,Barata:2024ieg,Barata:2025fzd}. In \Fig{fig:CMS_raw_combo} the CMS measurement is compared with three different theoretical predictions. We first begin by highlighting a common issue in nuclear collisions, namely that the non-trivial shape of the measured result arises from a confluence of distinct effects: while we expected an enhancement of the energy correlator at large angles due to nuclear modification, we also observe an enhancement at small angles. This arises due to biases in the jet selection: jets are selected to have the same energy in both p-p and Pb-Pb. However, due to energy loss, this implies that the initiating hard parton had a higher energy in Pb-Pb collisions, modifying the two-point energy correlator in the hadronization transition region, as seen in \Fig{fig:CMS_raw_compare}. The ratio of Pb-Pb to p-p then takes a non-trivial shape due to an interplay of these two effects. Even though we wish to isolate only the nuclear modification at large scaling, its description requires also an understanding of jet quenching. We will discuss how this can be mediated shortly.

The three different theoretical predictions compared with data in \Fig{fig:CMS_raw_combo}  each incorporate different physics. The approach of Holguin et al. developed in \cite{Andres:2024ksi,Andres:2023ymw,Andres:2023xwr,Andres:2022ovj} is based on perturbative QCD, and uses a semi-hard implementation \cite{Dominguez:2019ges, Isaksen:2020npj} of the multiple scattering BDMPS-Z formalism \cite{Baier:1996kr,Baier:1996sk,Zakharov:1996fv,Zakharov:1997uu}. It focuses on the direct calculation of the energy correlator observable. The CoLBT approach uses a coupled Boltzmann transport \cite{Chen:2017zte,Chen:2020tbl,Zhao:2021vmu} (extending linear Boltzmann transport model \cite{Li:2010ts,He:2015pra,Cao:2016gvr,Luo:2023nsi}), the high-twist approach \cite{Guo:2000nz,Wang:2001ifa}, CLVisc hydrodynamics \cite{Pang:2012he,Pang:2014ipa,Pang:2018zzo}, and a hybrid hadronization model that combines hydrodynamics, quark coalescence for thermal and soft jet partons and fragmentation for hard partons~\cite{Zhao:2021vmu,Chen:2017zte,Chen:2020tbl,Yang:2021qtl,Yang:2022nei}. It attempts to provide a complete simulation of the entire nuclear collision, enabling the measurement of any observable, in particular the energy correlators. It was first applied to study the energy correlators in \cite{Yang:2023dwc}. Finally, the hybrid model \cite{Casalderrey-Solana:2016jvj,Casalderrey-Solana:2015vaa,Casalderrey-Solana:2014bpa} uses a description based off a combination of perturbative QCD for energetic partons, and a hydrodynamic model of the QGP inspired by $\mathcal{N}=4$ sYM. The hybrid model can be further improved by incorporating elastic scattering \cite{DEramo:2018eoy,Hulcher:2022kmn}, which is shown in the lower panel of \Fig{fig:CMS_raw_combo}. This is interesting for searching for quasi-particles in the QGP.

All these different descriptions provide a qualitative, although not exceptional, description of the data. However, this highlights the difficulties in understanding jet physics in heavy ion, namely there are many combined physical effects, and not sufficient precision in the theoretical description to distinguish between them. To proceed, we would therefore like to have more observables which can further distinguish different effects, approaches to isolate specific effects, and improved theoretical descriptions. We will discuss each of these in turn.

\begin{figure}
\includegraphics[width=0.65\linewidth]{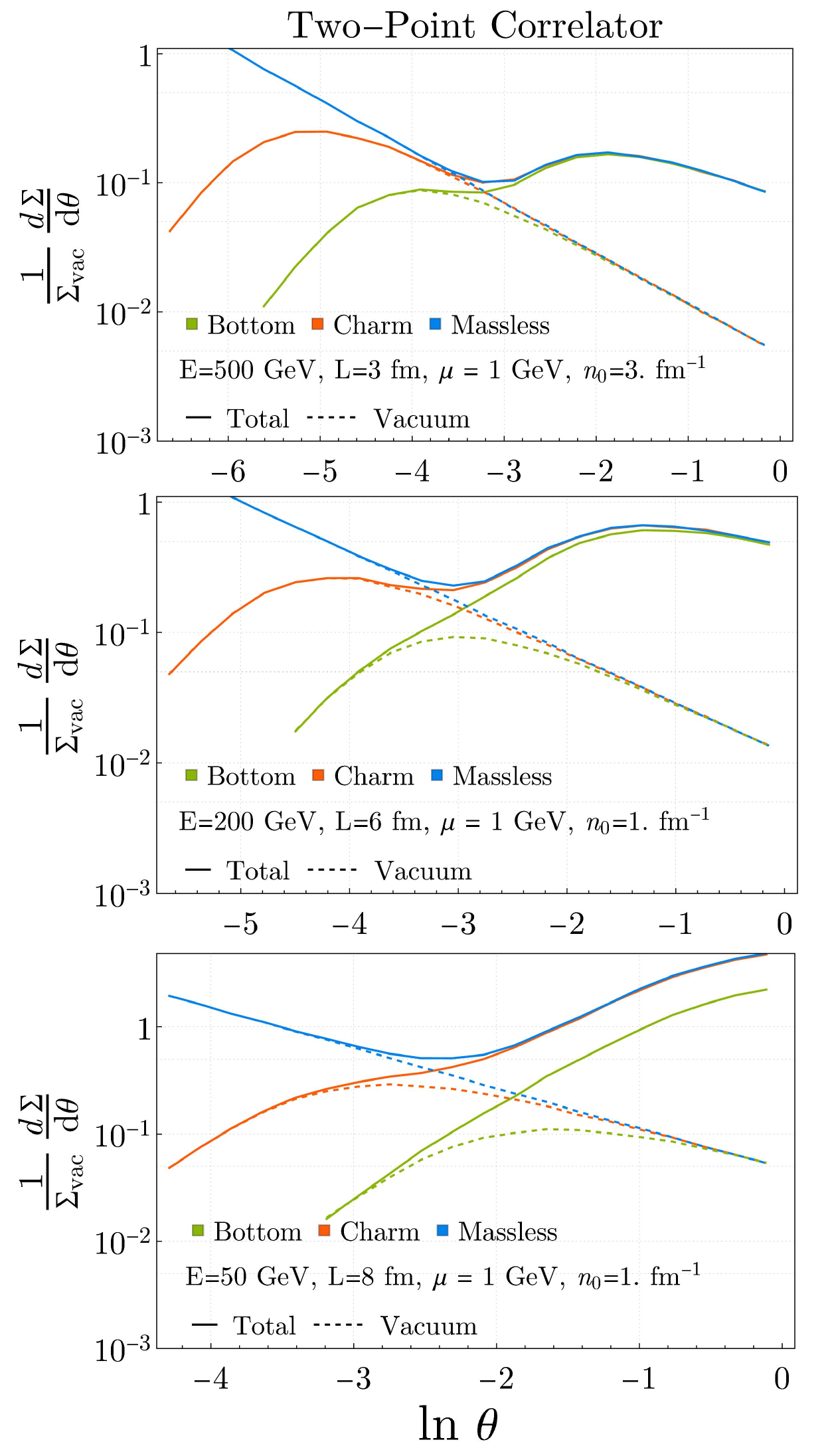}
\caption{The two-point energy correlator on heavy quark jets in nuclear collisions. At high energies, the deadcone is much smaller than the medium modification scale, and we clearly observe the two distinct scales. As we lower the jet energy the two scales approach each other, and we ``fill the deadcone". Figure from \cite{Andres:2023ymw}.
}
\label{fig:bquark_heavyion}
\end{figure}

We first discuss how the simple structure of the energy correlators can be used to isolate specific aspects of the underlying physics, even in the presence of jet selection bias. An important step in isolating the relevant physics was taken in \cite{Andres:2024pyz,Andres:2024hdd}. Using an understanding of the simple scaling behaviors of the energy correlators, Refs. \cite{Andres:2024pyz,Andres:2024hdd} introduced a transformation which removes the leading scaling contributions from jet selection bias. The distributions of the energy correlators after performing this transformation are shown in \Fig{fig:centrality_fixed}. The ratio of the two-point correlator in Pb-Pb and p-p collisions is shown for different centralities, a measure of the overlap of the colliding ions (illustrated by the grey circles). After performing the un-biasing, a much clearer picture of the underlying physics is revealed, emphasizing the enhancement due to nuclear modification as we move from small to larger angles. This modification also behaves as expected as the centrality is modified. Furthermore, after un-biasing, we are able to identify the angle, and therefore the scale, at which the nuclear modification arises. While highly successful, this approach does not remove all the potential biases, for example modifications in the quark gluon fraction. Another approach to obtaining unbiased measurements is to use a color neutral probe, such as a Z-boson. The first measurements of energy correlators recoiling against Z bosons was presented in \Sec{sec:heavy_ion}, and it will be important to pursue these measurements further, as well as to perform theoretical calculations for this process. 

\begin{figure}
\includegraphics[width=0.95\linewidth]{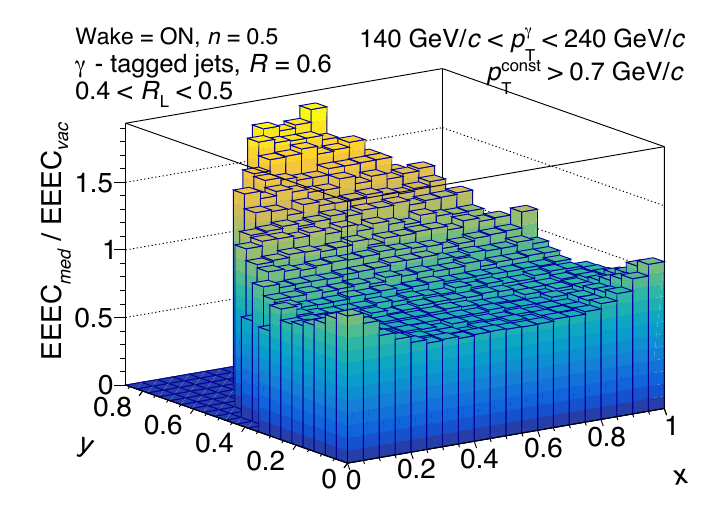}
\caption{The ratio of the three-point correlation function in QGP and vacuum, as computed using the hybrid model. Significant enhancement is seen, particularly in the ``equilateral" region. Figure from \cite{Bossi:2024qho}.
}
\label{fig:EEC_wake}
\end{figure}

To disentangle different effects and models, it will be important to measure a variety of different energy correlator observables, beyond the simplest two-point correlator on massless QCD jets. There are two directions which one can pursue: one is the measurement of the two-point correlator on jets produced by different flavor quarks, in particular heavy quarks, the other is the measurement of higher point correlators.

The study of energy correlators on heavy flavor jets in nuclear collisions was initiated in \cite{Andres:2023ymw}. As discussed in \Sec{sec:heavy_quarks}, an interesting feature of heavy quarks is that the intrinsic mass scale of the quark cuts off perturbative radiation at angles $\theta \leq m/p_T$. This is sometimes referred to as the dead cone \cite{Dokshitzer:1991fd}. It has long been realized that this provides an interesting opportunity in nuclear collisions \cite{Thomas:2004ie,Armesto:2003jh}, since medium induced radiation can fill the dead cone. The presence of radiation ``within" the deadcone provides an extremely sensitive probe of the properties of the medium.

Energy correlators are ideal for studying heavy quark jets in nuclear collisions, since they allow the distinct scales, namely the heavy quark mass scale, and the medium modification scale, to be cleanly resolved. Calculations of the two-point energy correlator on heavy quark jets propagating through a static block of QGP are shown in \Fig{fig:bquark_heavyion} for different values of the jet energy. At high energies, the deadcone is smaller than the medium modification scale, and we clearly observe the two distinct scales: one for the medium modification and one for the deadcone. As we lower the jet energy the two scales approach each other, and we ``fill the deadcone".  Other studies of energy correlators on heavy quark jets in nuclear collisions have been performed in \cite{Apolinario:2025vtx,Xing:2024yrb}. With the larger forthcoming datasets, it will be particularly interesting to have experimental measurements of heavy quark jets in nuclear collisions. The measurements of energy correlators on heavy quark jets in vacuum \cite{ALICE:2025igw} illustrate the experimental feasibility  of this approach.

\begin{figure}
\includegraphics[width=0.85\linewidth]{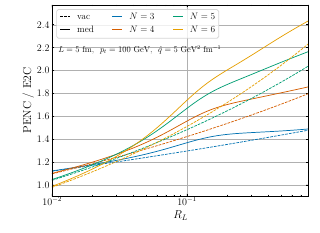}
\caption{Projected N-point energy correlators computed in vacuum and in medium. A clear modification in the scaling behavior is observed at large angles. Figure from \cite{Barata:2025fzd}.
}
\label{fig:PENC_heavyion}
\end{figure}

Beyond the two-point correlator, it will also be important to study higher point correlators in nuclear collisions. The two-point correlator can be thought of as identifying a scale, however, to probe the physics at that scale, one wants to measure higher-point correlators. This is much in analogy to the case of cosmology, where one would like to measure the three-point function (non-gaussianity) to identify the nature of inflation. Much like in cosmology, in nuclear collisions, we do not have a full understanding of the underlying microscopic description. An approach taken in cosmology is to categorize the shapes of higher point correlators predicted by different models. We believe that this should also be performed for different models of nuclear collisions. Measurements of higher point correlators can then distinguish between the different models.

A first study of higher point correlators in nuclear collisions was performed in \cite{Bossi:2024qho} using the hybrid model. In \Fig{fig:EEC_wake} we show the ratio of shape dependent three-point correlators between Pb-Pb and p-p collisions. We see an interesting modification, which peaks in the ``equilateral regime". It will be interesting to compute this ratio in other models of nuclear collisions. Refs. \cite{Bossi:2024qho,Barata:2025fzd} also studied the ratios of projected correlators in nuclear collisions. In \Fig{fig:PENC_heavyion}, we show ratios of the projected energy correlators in both vacuum and nuclear collisions. A clear modification is observed at large angles. Correctly predicting the modifications in the shape for different $N$-point correlators will provide important constraints on models. The medium modified triple collinear splitting functions in the opacity expansion have been computed \cite{Fickinger:2013xwa}, and it would be interesting to use them to compute the shape dependent three-point energy correlator. Extensions of the $\nu$ point correlators to nuclear collisions were also recently studied in \cite{Budhraja:2025ulx}.

\begin{figure}
\includegraphics[width=0.555\linewidth]{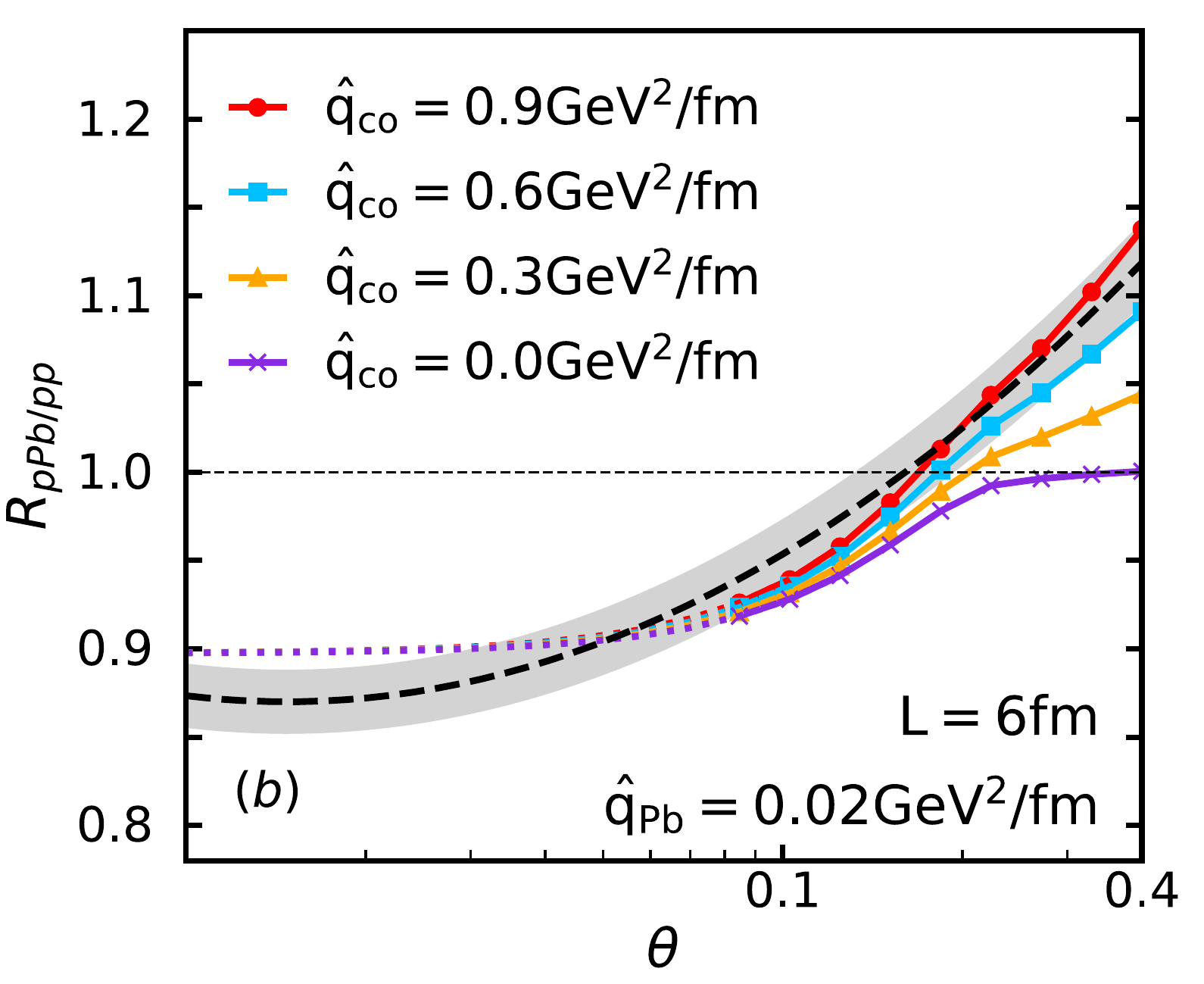}
\includegraphics[width=0.555\linewidth]{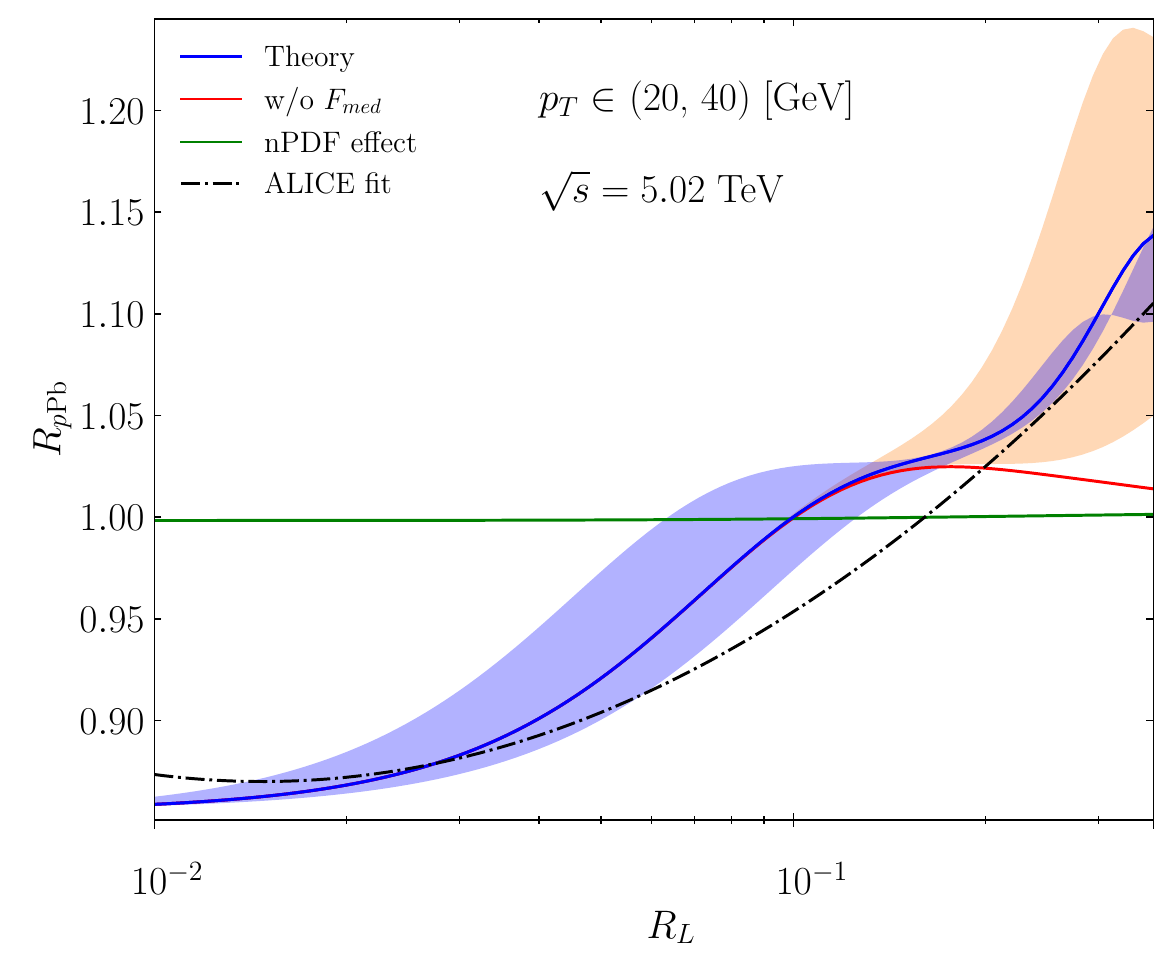}
\caption{A comparison of the ratio of two-point correlators in p-Pb and p-p with different theoretical predictions, as described in the text.  Figures from \cite{Fu:2024pic} and \cite{Barata:2024wsu}.
}
\label{fig:berndt_pA}
\end{figure}

Another direction which will provide significant insight, and aid to decouple different effects, is to study collisions of different nuclei. Currently energy correlators have been studied in p-A collisions, and they will hopefully also be studied in oxygen-oxygen collisions at the LHC. At the future EIC, they can be studied in e-A collisions. The particular appeal of e-A and p-A collisions is that they can be studied using rigorous factorization theorems \cite{Collins:1985ue,Collins:1989gx}, with nuclear modifications encoded in higher-twist operators \cite{Politzer:1980me,Ellis:1982cd,Ellis:1982wd,Jaffe:1983hp,Jaffe:1981td,Jaffe:1982pm,Qiu:1990xy,Qiu:1990xxa} whose matrix elements are enhanced by the size of the nucleus, $A^{1/3}$ \cite{Luo:1993ui,Luo:1994np,Luo:1992eq,Qiu:1991wg,Luo:1992fz,Luo:1991bj,Kastella:1989vd,Kastella:1989ux}. Therefore, while the effect of nuclear modification is smaller in these systems, there exist frameworks where it can be studied theoretically from first principles. In such collisions, one does not expect to produce a QGP, however, it allows one to isolate effects due to nuclear PDFs and final state interactions. These collisions therefore provide a crucial baseline on top of which to search for effects of the QGP. Achieving a consistent description of nuclear effects across collision systems is one of the primary goals of the nuclear program.

\begin{figure}
\includegraphics[width=0.955\linewidth]{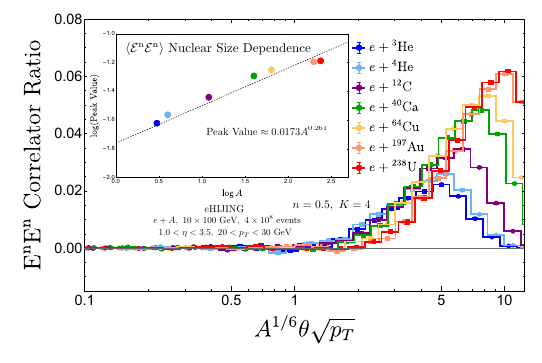}
\caption{Energy correlators in $eA$ collisions. The onset angle of nuclear modification is set by the nuclear size, $A^{1/6}$. Different sizes of nuclei can be identified from asymptotic energy flux.  Figure from \cite{Devereaux:2023vjz}.
}
\label{fig:cold_nuclear}
\end{figure}

Initial theoretical studies \cite{Fu:2024pic,Barata:2024wsu} of the ALICE p-Pb data are shown in \Fig{fig:berndt_pA} using different theoretical frameworks. Clear indication for nuclear modification is observed. This is the first measurement of nuclear modification in p-Pb using jet substructure, and illustrates the potential of the energy correlator observables. Energy correlators in final state jets in e-A collisions have also been studied in \cite{Devereaux:2023vjz,Fu:2024pic}. These will provide a particularly clean baseline. One interesting feature of the future electron ion collider will be the ability to study collisions on a wide variety of nuclei. In \Fig{fig:cold_nuclear}, we see that the nuclear size, $A^{1/6}$, is imprinted in the energy flux. We find it quite remarkable that we can see the sizes of different nuclei in the asymptotic energy flux.

\begin{figure}
\includegraphics[width=0.955\linewidth]{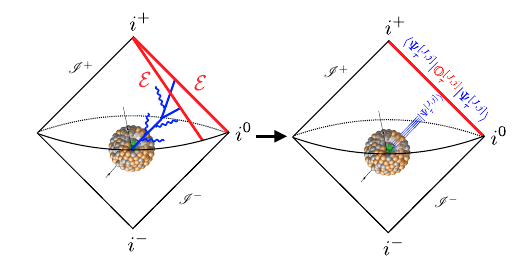}
\caption{A schematic of the operator product expansion in a large nuclear state. Figure from \cite{Andres:2024xvk}.
}
\label{fig:QGP_OPE_schem}
\end{figure}

Finally, we conclude this section with where we believe the energy correlator approach will be most powerful theoretically, which is through the full use of the operator product expansion, and symmetries. As highlighted in the discussion of the light-ray OPE in \Sec{sec:celestial_blocks}, the OPE reduces multi-point correlation functions to the study of (possibly higher twist) fragmentation, namely one point correlators of detector operators. Modifications from the nuclear state are then encoded in the relative sizes of the expectation values of these different light-ray operators. The OPE therefore acts to relate jet substructure observables, which benefit from the presence of a scale, and infrared safety, to the study of generalized fragmentation. This is illustrated schematically in \Fig{fig:QGP_OPE_schem}. Note that the nuclear collision modifies the state in which the energy correlators are computed, but not the theory.  Nuclear modification in single particle inclusive hadron production has been well studied, with nuclear modifications encoded in higher-twist operators \cite{Politzer:1980me,Ellis:1982cd,Ellis:1982wd,Jaffe:1983hp,Jaffe:1981td,Jaffe:1982pm,Qiu:1990xy,Qiu:1990xxa} whose matrix elements are enhanced by the size of the nucleus, $A^{1/3}$ \cite{Luo:1993ui,Luo:1994np,Luo:1992eq,Qiu:1991wg,Luo:1992fz,Luo:1991bj,Kastella:1989vd,Kastella:1989ux}. It has also been studied experimentally.

As in the case of perturbative QCD, ideally, we would be the specific operator that is being probed, and relate it to for example, transport coefficients of the underlying theory. While we are not yet at this stage, the light-ray OPE already tells us important details about the form of leading nuclear modification. To see this, recall the schematic form of the light-ray OPE truncated to twist-4 operators
\begin{align}
 \mathcal E(n_1)  \mathcal E(n_2) =\frac{1}{\theta^2}\mathbb{O}_{\tau=2}^{[J=3]}  +   \mathbb{O}_{\tau=4}^{[J=3]} +\cdots \,.
\end{align} 
We would now like to consider the evaluation of this in the expectation value of a nuclear state, and compare it to the same sum of light-ray operators evaluated in proton-proton collision. In the small angle limit, the twist-4 contributions are small in the case of proton-proton collisions. As was shown in the measurements of the energy correlators inside jets in proton proton collisions, and the extraction of $\alpha_s$, at small angles in the vacuum we can cleanly isolate the scaling of the twist-2 operator. Making this approximation, we therefore arrive at the following formula 
\begin{align}
   &\frac{  \langle \Psi_{QGP} | \mathcal E(n_1)  \mathcal E(n_2)| \Psi_{QGP} \rangle }{ \langle \Psi_{pp} | \mathcal E(n_1)  \mathcal E(n_2)| \Psi_{pp} \rangle} \\
   & =  \frac{  \langle \Psi_{QGP} |  \frac{1}{\theta^2}\mathbb{O}_{\tau=2}^{[j=3]}  +   \mathbb{O}_{\tau=4}^{[j=3]} +\cdots  | \Psi_{QGP} \rangle }{   \langle \Psi_{pp} |  \frac{1}{\theta^2}\mathbb{O}_{\tau=2}^{[j=3]}  +   \mathbb{O}_{\tau=4}^{[j=3]} +\cdots  | \Psi_{pp} \rangle    } \nn \\
   &   \sim  \frac{ \langle \Psi_{QGP} | \mathbb{O}_{\tau=2}^{[j=3]}  | \Psi_{QGP} \rangle   }{  \langle \Psi_{pp} |    \mathbb{O}_{\tau=2}^{[j=3]}    | \Psi_{pp} \rangle    }  +\theta^2   \frac{ \langle \Psi_{QGP} | \mathbb{O}_{\tau=4}^{[j=3]} | \Psi_{QGP} \rangle    }{  \langle \Psi_{pp} |    \mathbb{O}_{\tau=2}^{[j=3]}    | \Psi_{pp} \rangle    }    \,. \nn
\end{align}
This is quite a remarkable formula, since the $\theta$ dependence is predicted. The expectation values of light-ray operators depend on the details of the QGP, the kinematics of the collision etc, but not on $\theta$. Note that this is just using dimensional analysis on the celestial sphere. We note that in reality there are multiple twist-4 operators. Here we are only considering this expression at the classical level, so we can view this as their average value. 

Therefore, we see that the leading correction from medium modification can be captured by the $\theta^2$ scaling in the Pb-Pb to p-p ratio. This illustrates the power of the light-ray OPE. Much like how we were able to extract the strong coupling constant by identifying the scaling of the twist-2 operators, here we are able to identify the leading nuclear modification as the coefficient of a particular power law. From a computational perspective, it also reduces the problem from the calculation of a function of $\theta$ to the calculation of a number. 

\begin{figure}
\includegraphics[width=0.755\linewidth]{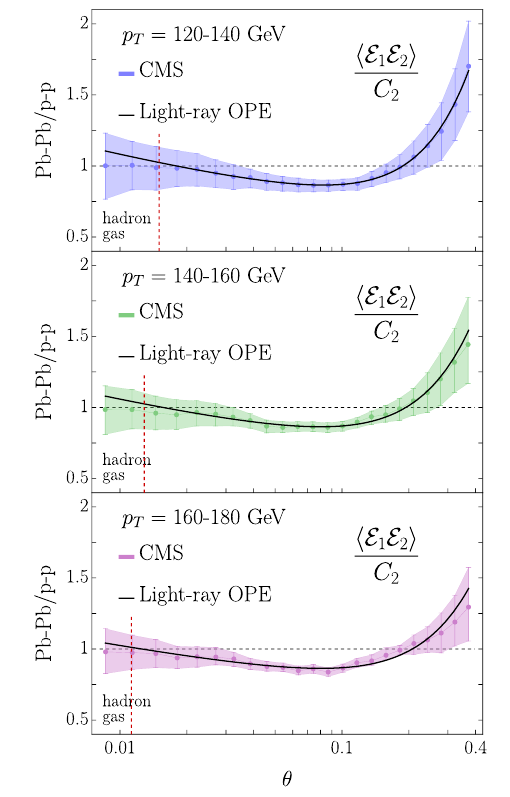}
\caption{A comparison of the unbiased CMS data for the ratio of the two-point correlator in Pb-Pb to p-p with predictions from the light-ray OPE. The light-ray OPE does not predict the magnitude of the enhancement, but it predicts its scaling as a function of $p_T$, which is well born out in data.   Figure from \cite{Andres:2024xvk}.
}
\label{fig:QGP_OPE}
\end{figure}

Ideally we could give an interpretation of this number in terms of transport coefficients or properties of the nuclear medium. This has not yet been achieved. Instead we can extract the value of this matrix element from a measurement at one value of $p_T$, and using our knowledge of its $p_T$ scaling, predict it at other $p_T$ values. This is shown in \Fig{fig:QGP_OPE}. This plot illustrates that the OPE predicts both the correct functional form, as well as the correct $p_T$ scaling. We find the fact that such simple arguments from symmetry principles can be used to understand the behavior of such complicated collisions quite remarkable. This approach was also used to study p-Pb collisions in \cite{Andres:2024xvk}.

\begin{figure}
\includegraphics[width=0.95\linewidth]{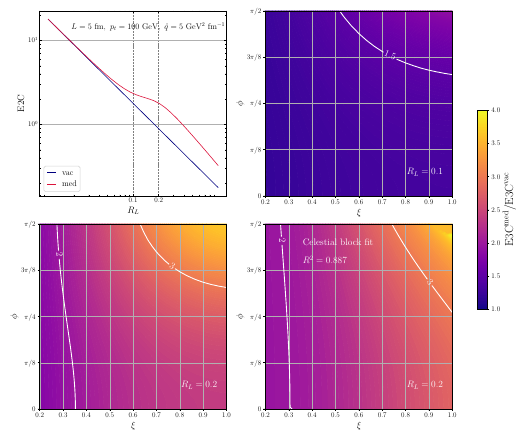}
\caption{The modification of energy correlators in nuclear collisions. Top Left: The scale of nuclear modification is identified by the two-point function. Top Right and Bottom Left: The three-point correlator is computed at the two angular scales identified in the two-point correlator. At larger angles it  shows  a large modification. Bottom Right: The three-point correlator is fit to a sum of celestial blocks, which indicate an enhancement in the OPE coefficients of the higher twist blocks.  Figure from \cite{Barata:2025fzd}.
}
\label{fig:threepoint_heavyion}
\end{figure}

Recently this approach was extended in \cite{Barata:2025fzd} to the case of the shape dependent three-point correlator. Although the full shape dependent three-point correlator cannot currently be computed exactly, it can be computed in a cascade approximation. In \Fig{fig:threepoint_heavyion} we first fix the overall scale at which we wish to look at the three-point correlator. The three-point correlator is then computed at two different scales as we pass this scale showing modification. Finally, we can fit this calculation to a sum over celestial blocks. We find that the twist-4 celestial blocks are enhanced in the medium modified calculation, as expected. Going forward, we anticipate that this can be a robust way of understanding multi-point correlators in complicated nuclear collisions.

The study of energy correlators in nuclear collisions is in its infancy. We believe that these observables have immense phenomenological and theoretical potential for transforming our understanding of these most complicated collisions. Nearly all of the measurements of the energy correlators in nuclear collisions have been performed within jets, in the small angle limit. It will be particularly interesting to extend these to measurements in the back-to-back limit \cite{Kang:2025vjk,Kang:2024otf,Kang:2023oqj} to improve our understanding of the initial state. It will also be important to extend the studies to incorporate anisotropies, enabling studies of the glasma initial state \cite{Avramescu:2024poa,Barata:2023zqg}. Finally, the picture will not be complete until there is a coherent picture across collision systems, and the study of energy correlators in neon-neon and oxygen-oxygen collisions will be crucial for this. Overall, we believe that this charts a rich experimental path for furthering our understanding of nuclear collisions using energy correlators. 

\section{Future Directions}\label{sec:open}

In this review, we have presented a broad overview of the physics of energy operators and their correlators, and their impact on nuclear and particle physics. Many of these observables have been measured in experiment only in the last year, and we have outlined throughout, a broad range of measurements that will further strengthen this connection between theory and experiment.

Throughout this review we have highlighted numerous directions in which the studies of energy correlators and detector operators can be extended. These include more detailed experimental measurements, explorations of detector operators in quantum gravity, an exploration of geometric and combinatorial structure for energy correlators, and an improved understanding of detector operators and the analytic properties of Regge trajectories in QCD, amongst many others. However, energy correlators also connect to a much broader range of theoretical topics, providing optimism that they can also have an impact on particle physics and nuclear phenomenology.  In this section we highlight a number of theoretical directions where we hope to see progress and connections built in the coming years, with the ultimate hope that they can drive further progress in connecting theory with real world experiments.

\subsection{Energy Correlators on the Lattice}\label{sec:open_lattice}

All current approaches to compute energy correlators are based on perturbative expansions in some parameter (coupling, charge), or take advantage of conformal symmetry. A primary feature of the energy correlator formulation of collider physics, is that it relates collider observables to correlation functions of local operators. While our ability to describe energy correlators in real world QCD has progressed tremendously due to progress in perturbative calculations, to go further will require non-perturbative techniques for calculating energy correlators in generic field theories.

\begin{figure}
\includegraphics[width=0.955\linewidth]{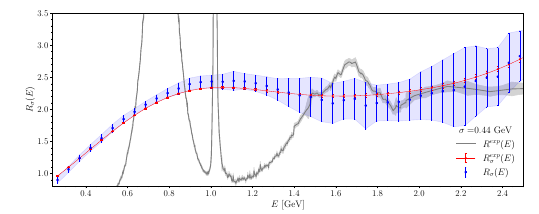}
\caption{The total cross section for $e^+e^- \to$ hadrons smeared over a bin width of $0.44$ GeV, computed using lattice QCD, and compared with smeared experimental data. Figure from \cite{ExtendedTwistedMassCollaborationETMC:2022sta}.
}
\label{fig:smeared_R_ratio}
\end{figure}

The most well established such technique is the lattice, which has had tremendous progress in the calculation of Euclidean quantities in QCD. There has recently been progress in computing the simplest observable, namely the inclusive cross section using lattice QCD. This was initiated in the works \cite{Bulava:2021fre,Hansen:2019idp,Hansen:2017mnd}. It has been applied to complete lattice simulations for both the smeared R-ratio \cite{ExtendedTwistedMass:2025tpc,ExtendedTwistedMassCollaborationETMC:2022sta}, and inclusive tau decays \cite{ExtendedTwistedMass:2024myu}. For recent discussions and theoretical progress, see \cite{Jay:2025dzl,Blum:2024hyr,Bergamaschi:2023xzx,Hashimoto:2025ohw,Bruno:2024fqc,Bulava:2023mjc,Rothkopf:2022fyo}.

In \Fig{fig:smeared_R_ratio} we show results of lattice calculations from the Extended Twisted Mass Collaboration  \cite{ExtendedTwistedMass:2025tpc,ExtendedTwistedMassCollaborationETMC:2022sta} for the total cross section $e^+e^- \to$ hadrons (more precisely the R-ratio) smeared over a bin width of $0.44$ GeV. This result is compared with experimental data, showing excellent agreement. This smearing is analogous to that proposed in the early days of QCD \cite{Poggio:1975af}, and washes out the resonance structure seen in the un-smeared data.

It will be important to generalize such techniques to the calculation of the $1$-point or 2-point energy correlators. While the available energies that can be computed on the lattice are limited by the lattice spacing to the few GeV range, this would be extremely interesting for comparison with measurements from low energy $e^+e^-$ colliders. 

Although much less developed than lattice QCD, quantum simulations hold great potential for the investigation of real time phenomenon in quantum field theories. Initial investigations of energy correlators using quantum computing were performed in \cite{Barata:2024apg,Lee:2024jnt}. It would be interesting to develop these further.

\subsection{Forward Scattering and the Total Cross-Section}\label{sec:open_forward}

\begin{figure}
\includegraphics[width=0.955\linewidth]{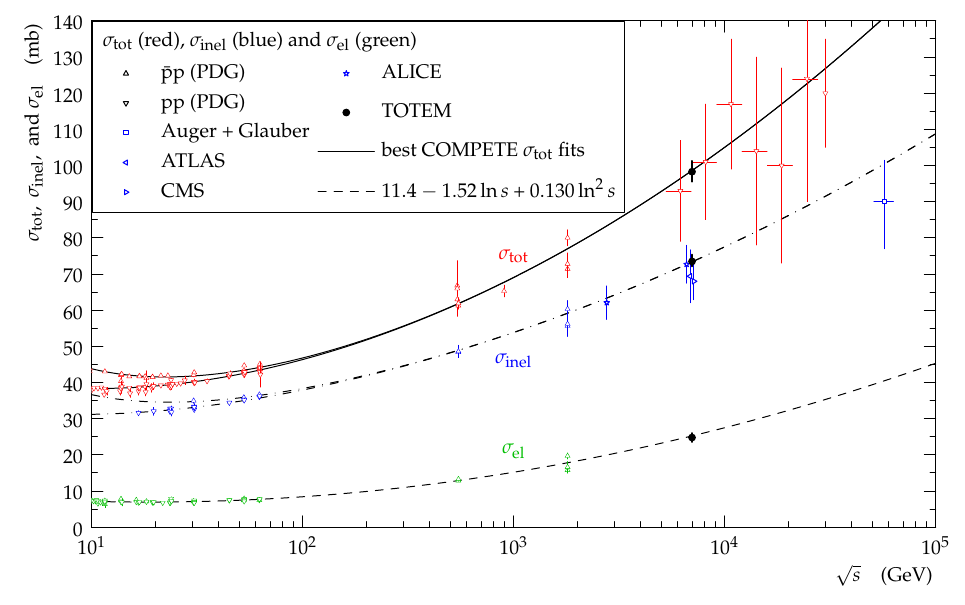}
\caption{Summary of measurements of the total cross section showing a rising cross section at large $s$. Figure from \cite{Antchev:2011vs}.
}
\label{fig:sigma_tot}
\end{figure}

In this review we have focused primarily on the connection to experiment via hard scattering. However, there is also a wealth of experimental data for very forward (Regge) scattering \cite{COMPETE:2002jcr,ArbiolVal:2024nrn,Antchev:2011vs,TOTEM:2012oyl,ATLAS:2014vxr}. See \cite{Amaldi:2015jhq} for an interesting overview of this program. This is shown in \Fig{fig:sigma_tot}, illustrating the growth of the total forward cross section in QCD. Much like the case of the energy correlators, this exhibits a beautiful nearly power law scaling. The quantitative understanding of this growth is a longstanding problem in QCD. While it is ultimately non-perturbative, we are optimistic that an improved understanding of the Regge limit in CFTs, combined with the S-matrix bootstrap, might shed some light on this longstanding problem.

In the Regge limit, we expect a CFT correlator to behave as $s^{\alpha(0)-1}$, where $\alpha(0)$ is the Regge intercept. In a unitarity CFT, $\alpha(0) \leq 1$ \cite{Caron-Huot:2017vep}, giving rise to the Regge boundedness of the correlator. One can have transient behavior, governed by $\alpha(0)^{\text{transient}}$, which satisifies $\alpha(0)^{\text{transient}} \leq 2$ in an unitary theory, by the bound on chaos \cite{Maldacena:2015waa}. For a nice summary, see \cite{Caron-Huot:2020ouj}.

In the perturbative regime, forward scattering in QCD is mediated by the BFKL Pomeron \cite{Lipatov:1985uk,Balitsky:1978ic,Kuraev:1977fs,Kuraev:1976ge,Lipatov:1976zz,Fadin:1975cb}, which predicts $\alpha(0)^{\text{transient}}=1+\mathcal{O}(\alpha_s) >1$. Evidence for the perturbative Pomeron has been seen in HERA data \cite{Ball:2017otu}. Fits to forward scattering data give  $\alpha(0)^{\text{transient}}-1\sim0.0808$ \cite{Donnachie:1992ny,Menon:2013vka}. This presents a beautiful target for the modern S-matrix bootstrap \cite{Paulos:2017fhb,Paulos:2016but,Paulos:2016fap}. Interesting developments include the identification of a ``Froissart amplitude" \cite{Paulos:2017fhb,Guerrieri:2021tak,Guerrieri:2023qbg,Bhat:2023puy} saturating bounds on the integrated cross section. It will be interesting to explore these directions further.

Additionally, we note that in a gapped theory, forward scattering satisfies the Froissart bound \cite{Froissart:1961ux,Martin:1965jj}
\begin{align}
\sigma^{\text{tot}}(s) \sim \frac{4\pi}{t_0} \log^2 \frac{s}{s_0}\,,
\end{align}
where $t_0$ is related to the lightest particle exchanged in the $t$-channel. While there has been some experimental evidence for Froissart like growth \cite{TOTEM:2017asr}, whether or not the Froissart bound is saturated in QCD, and the precise scaling of the cross section at asymptotic energies remain open, but fundamental questions in QCD.

There has been tremendous progress understanding the unitarization of forward scattering in QCD through multi-pomeron exchange. Indeed, this is where integrability in gauge theories was first found \cite{Lipatov:1993yb,Faddeev:1994zg}. In particular, the spectrum of states was solved in \cite{Derkachov:2001yn,Derkachov:2002wz,Korchemsky:2001nx,Korchemsky:1994um}. It would be interesting to revisit this from a modern perspective. 

A related approach is through the Reggeon field theory \cite{Gribov:1967vfb}, a $2+1$ dimensional field theory describing the dynamics in the transverse plane of the scattering. It would be interesting if one could bootstrap this field theory.  For early remarkable attempts in this direction, see \cite{Migdal:1973gz}.

We view this as an area with both data, and interesting open theoretical problems, where we hope to see progress in the near future.

\subsection{Correlator and S-matrix Bootstraps}\label{sec:open_bootstrap}

\begin{figure}
\includegraphics[width=0.755\linewidth]{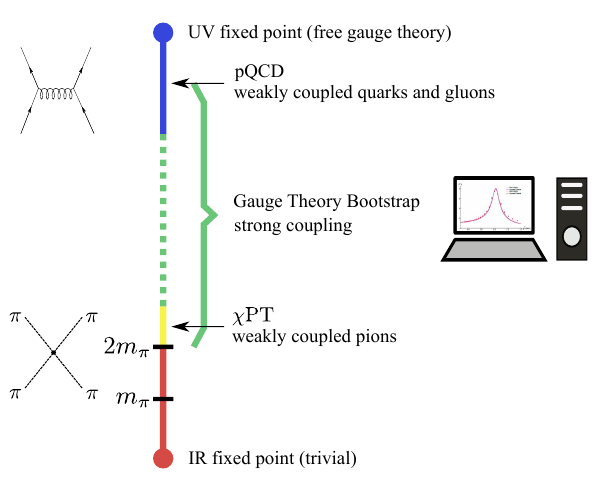}
\caption{The gauge theory bootstrap combines IR and UV information, with analyticity and unitarity, to make predictions about non-perturbative dynamics in strongly coupled gauge theories. Figure from \cite{He:2024nwd}.
}
\label{fig:gauge_theory_bootstrap}
\end{figure}

The last several years has seen the development of the modern non-perturbative S-matrix bootstrap \cite{Paulos:2017fhb,Paulos:2016but,Paulos:2016fap}. Much like the original S-matrix bootstrap \cite{Chew:1962eu,Smatrix:1,Smatrix:2}, this program aims to maximally utilize the constraints from analyticity and unitarity, to either directly constrain the space of scattering amplitudes, or to inject basic information, taken for example from experimental data or the lattice, to predict more complicated observables \cite{Karateev:2019ymz,Guerrieri:2024jkn,Guerrieri:2020bto,Guerrieri:2018uew,He:2024nwd,He:2023lyy,He:2025gws}. An illustration of this philosophy is shown in \Fig{fig:gauge_theory_bootstrap} for the particular case of the gauge theory bootstrap, which combines low energy information from pion scattering amplitudes, UV information from perturbative QCD,  with analyticity and unitarity, to make non-perturbative predictions about QCD. These approaches have enabled predictions about Lorentzian quantities in real world QCD. A striking example being a theoretical prediction of a new $2$ GeV tetraquark resonance \cite{Guerrieri:2024jkn}.

We believe it will be fruitful to have more interaction between the S-matrix bootstrap program, and the study of energy correlators: on the one hand, we believe that these approaches can be used to improve our understanding of the energy correlators in the non-perturbative regime, on the other hand, it will be interesting to explore how energy correlators observables can help the S-matrix bootstrap.

In the direction of the bootstrap approach helping the study of energy correlators,  we believe that the operator formulation of energy correlators makes them a particularly promising target. While so far the S-matrix bootstrap program has focused on S-matrix, form factors and spectral densities, it would be interesting to extend this to understand constraints on the three, or four point functions relevant for energy correlators. While the full analytic structure of three and four-point correlators in massive theories is not known, there has been significant progress working below certain mass thresholds in the study of light-by-light scattering in the context of precision calculations of the muon $g-2$ \cite{Ludtke:2024ase,Colangelo:2017fiz,Colangelo:2015ama,Colangelo:2014dfa,Colangelo:2018mtw}. It would be most exciting to see whether this could give constraints on the behavior of the energy correlators in the non-perturbative transition region in the collinear limit. Since this region of the energy correlators probes the transition from perturbative quarks and gluons to free pions, techniques that can consistently combine information from both regimes of QCD, as illustrated in \Fig{fig:gauge_theory_bootstrap}, seem particularly appealing.

In the direction of energy correlator observables helping the S-matrix bootstrap, while the S-matrix bootstrap has been successful at probing exclusive scattering amplitudes, it has primarily explored the universality class of amplitudes with minimal elasticity. Energy correlator observables, being inclusive in nature, may allow one to probe more inelastic processes, either enabling the construction of more realistic QCD amplitudes in a primal approach, or enabling one to rule out amplitudes in a dual approach. Another possible  point of contact is via the multi-particle amplitude bootstrap \cite{Guerrieri:2024ckc}, which has proven successful for massless $1+1$d theories. It would be very interesting to understand if this can be extended to higher dimensions, and/or incorporate masses. This could provide an interesting bridge between exclusive scattering amplitudes, and energy correlator observables.

\subsection{Bounding the Space of QFTs and Gravity}\label{sec:open_bound}

The last decade has seen tremendous progress in understanding the space of consistent QFTs and possible UV completions, using consistency relations and analytic properties of observables. These studies have primarily focused on correlation functions of local operators, and scattering amplitudes. Successes from the study of local correlators include the remarkable CFT bootstrap program \cite{Rattazzi:2008pe,El-Showk:2012cjh,El-Showk:2014dwa,Poland:2018epd}. Successes from the study of scattering amplitudes include understanding the uniqueness of string theory \cite{Guerrieri:2022sod,Guerrieri:2021ivu,Arkani-Hamed:2023jwn}, bounds on EFT Wilson coefficients  \cite{Arkani-Hamed:2018ign,Remmen:2019cyz,Arkani-Hamed:2020blm,Caron-Huot:2020cmc,Bellazzini:2020cot,Tolley:2020gtv,Bern:2021ppb,Caron-Huot:2021rmr,Bern:2022yes,Caron-Huot:2022jli}, and bounds on corrections to the graviton three-point vertex (CEMZ bound) \cite{Camanho:2014apa}. The CEMZ bound was also proven from the ANEC in \cite{Hartman:2022njz}.

\begin{figure}
\includegraphics[width=0.755\linewidth]{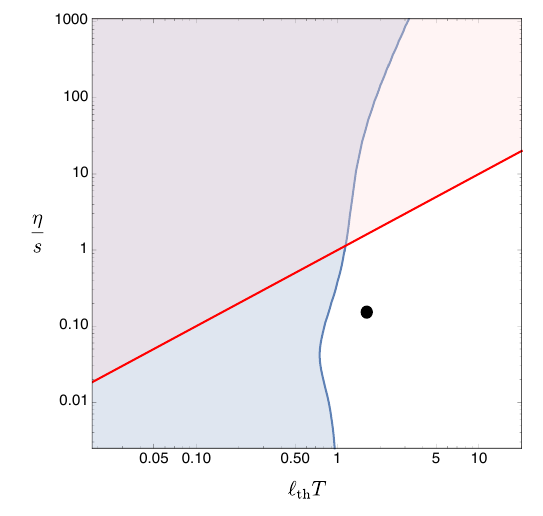}
\caption{An illustration of bounds on $\eta/s$ and $l_{\text{th}}$ (the thermalization length), obtained by combining the ANEC and eigenstate thermalization hypothesis. The dot corresponds to data for the QGP. Figure from \cite{Delacretaz:2018cfk}.
}
\label{fig:thermal_fig}
\end{figure}

Energy correlators (and more generally correlators of well defined detectors) provide examples of well defined observables in generic theories that do not admit an S-matrix, or correlation functions of local operators. Examples of such theories include gravity in $d\leq 4$, where there are infrared divergences (Although in certain cases this can be overcome \cite{Chang:2025cxc}.), or CFTs coupled to gravity. Even in the case of CFTs, where we have local operators, we have illustrated how certain questions constraints on field theories can be more easily seen through the Lorentzian lens of energy correlators. Despite this, the use of energy correlators to bound theories has focused on relatively special cases, in particular, one-point detector correlators in CFTs \cite{Hofman:2008ar}, and higher point correlators in the presence of a higher-spin symmetry  \cite{Maldacena:2011jn}, or extremal $a/c$ \cite{Zhiboedov:2013opa}. One interesting attempt to go beyond this was recently proposed in \cite{Riembau:2024tom}, which we hope can be further developed in the future.

It would be interesting to study the implications of higher point positivity of energy correlators more generally, or the implications of the crossing equations for multi-point detector correlators. This would be particularly interesting in the gravitational context, or for a CFT coupled to gravity.  Although the analytic structure of higher point correlators is not as well understood as the S-matrix, it would be interesting to develop this further, and to develop or axiomatize certain Regge boundedness or analyticity conditions. Since detector correlators are (to our knowledge) the only well defined observables in theories without S-matrices or local correlators, this provides an appealing direction for exploring the space of such theories.

Another interesting direction is to use energy correlators and the ANEC to bound thermal systems. The relation between positivity of energy flux and causality,  and its implications for thermal systems have been extensively explored,  both in the context of the AdS/CFT correspondence \cite{Brigante:2007nu,Brigante:2008gz,Buchel:2009tt,Hofman:2009ug,deBoer:2009pn,Camanho:2009vw,Camanho:2009hu,Buchel:2009sk,deBoer:2009gx,Myers:2010jv}, and from a field theoretic perspective \cite{Kulaxizi:2010jt}. In a recent example, \cite{Delacretaz:2018cfk}, ANEC positivity was combined with the eigenstate thermalization hypothesis to derive bounds on the shear viscosity to entropy ratio $\eta/s$, and the thermalization length $l_{\text{th}}$. These are shown in \cite{Delacretaz:2018cfk}, along with data for the QGP. It would be interesting to explore the implications of these results for understanding energy correlators in the QGP, particularly since they can now be directly measured at collider experiments

\subsection{Energy and Entanglement}\label{sec:open_QI}

The Bekenstein bound \cite{Bekenstein:1980jp,Casini:2008cr} implies deep connections between energy and entanglement. With the advent of the AdS/CFT correspondence, these have been significantly sharpened and refined \cite{Ryu:2006ef,Ryu:2006bv}, and are now central to our understanding of Quantum gravity

There now exist a variety of sharp statements, such as the Quantum Null Energy Condition \cite{Bousso:2015wca,Bousso:2015mna}, which generalize the ANEC. Although motivated by explorations in gravity, these hold as statements about QFT. The QNEC bounds the local energy by a particular second variation of the entropy
\begin{align}
\langle T_{kk}(p) \rangle \geq \frac{\hbar}{2\pi} \lim_{\mathcal{A}\to 0} S''_{\text{out}}/\mathcal{A}\,.
\end{align}
As compared to the ANEC, the QNEC is state dependent, and in particular,  the QNEC can be recovered from the ANEC in particular states \cite{Ceyhan:2018zfg}. 

It would be interesting to understand if these generalizations of the ANEC have any interesting relations for collider physics experiments. Asymptotic entropy bounds were studied in \cite{Bousso:2016vlt}, and it would be interesting to understand if these could place constraints on collider observables, or if measurements at collider experiments could be related to entropy. Another close relation between the two subjects arises from the fact that for certain null regions, the modular Hamiltonian is closely related to the ANEC, allowing entropy variations to be computed using the light-ray OPE, so that they are controlled by a similar physics to the collinear limit of energy correlator observables
\cite{Balakrishnan:2019gxl,Leichenauer:2018obf}. Could this enable studies of scaling behavior within high energy jets to be related to entanglement variations? We believe that such questions are deserving of more exploration, and more generally, we hope to see more exploration of techniques from quantum information in the study of energy correlators.

\subsection{Asymptotic Symmetries and Flat Space Holography}\label{sec:open_holography}

One of the primary goals of the study of detector operators is understanding how to characterize asymptotic fluxes in QFT and gravity. A major development in the energy correlator program has been the use of symmetries, in particular the action of the Lorentz group on the celestial sphere, to organize calculations.  This allows the identification of specific scaling laws, associated with quantum numbers of the Poincare group, as well as the decomposition into celestial blocks, which make these symmetries manifest. Remarkably, in general relativity, the asymptotic symmetries which preserve the boundary structure of asymptotically flat space times are not just the Lorentz group, but rather an infinite dimensional enhancement, referred to as the Bondi-van der Burg-Metzner Sachs (BMS) \cite{Bondi:1962px,Sachs:1962zza,Sachs:1962wk} group. This group includes supertranslations and superrotations \cite{Barnich:2011mi,Barnich:2010ojg,Barnich:2009se,Banks:2003vp}.  Much like for ordinary symmetries, these give rise to a rich set of conserved charges \cite{Ashtekar:1981bq,Dray:1984rfa,Wald:1999wa,Flanagan:2015pxa}. These relate quantities at $\mathcal{I}^+$ \cite{Dray:1984rfa,Ashtekar:1981bq,Wald:1999wa}, or between $\mathcal{I}^+$ and $\mathcal{I}^-$ \cite{Strominger:2013jfa,Pasterski:2015tva,Ashtekar:1979xeo}. For an overview of the beautiful mathematics and physics of this area, see \cite{Ashtekar:2014zsa}.

In the last decade, it has been realized that these symmetries have remarkable implications for scattering amplitudes. Starting with the realization that gravitational scattering exhibits BMS invariance \cite{Strominger:2013jfa}, there has been tremendous progress. This includes the discovery of new symmetries in QED \cite{He:2014cra,Kapec:2015ena} and Yang-Mills \cite{Strominger:2013lka}, and understanding that Weinberg's soft theorems \cite{Weinberg:1965nx} (and their subleading power generalizations \cite{Gell-Mann:1954wra,Low:1958sn,Low:1954kd,Burnett:1967km,DelDuca:1990gz}) are the consequence of asymptotic symmetries \cite{Lysov:2014csa,He:2014laa}, the discovery of new soft theorems \cite{Cachazo:2014fwa}, and an understanding of memory effects \cite{Strominger:2014pwa,Pasterski:2015tva,Pasterski:2015zua}. An understanding of these symmetries is also having an impact on practical calculations of gravitational waveforms for comparison with LIGO \cite{Elkhidir:2024izo,Bini:2024rsy}. For an overview of these developments, see \cite{Strominger:2017zoo}. 

\begin{figure}
\includegraphics[width=0.955\linewidth]{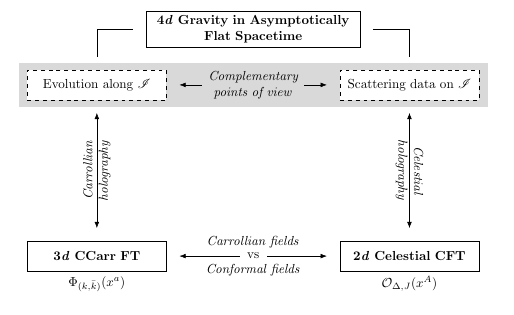}
\caption{A summary of different approaches to flat space holography. Light-ray/ detector operators naturally connect with both approaches, suggesting that they will play an important role in this story. Figure from \cite{Donnay:2022aba}.
}
\label{fig:celestial_fig}
\end{figure}

These advances raise many natural questions for the study of detector operators. What are the implications of asymptotic symmetries for detector correlators in gravity or Yang-Mills theories? It is known that light-ray operators satisfy the BMS algebra \cite{Cordova:2018ygx,Belin:2020lsr,Gonzo:2020xza}, and general light-rays \cite{Korchemsky:2021htm,Hu:2023geb,Hu:2022txx} satisfy infinite dimensional symmetry algebras similar to those observed in the study of asymptotic symmetries \cite{Strominger:2021mtt,Guevara:2021abz,Adamo:2021lrv}. How do these symmetries manifest for physical observables? What are the implications for local-in-angle conservation laws on asymptotic fluxes? For some interesting recent discussions of this topic, see \cite{Veneziano:2025ecv}. Color memory effects in Yang-Mills have been studied in \cite{Ball:2018prg,Pate:2017vwa}. Can these be observed in collider experiments? We believe that the study of detector operators in perturbative quantum gravity,  briefly discussed in \Sec{sec:QG}, is a particularly interesting playground for the further exploration of these topics.

Another primary motivation for understanding asymptotic symmetries in asymptotically flat space times comes from flat space holography. The general holographic principle \cite{tHooft:1993dmi,Susskind:1994vu,Bousso:2002ju} suggests that the dynamics of gravity in a particular spacetime region can be encoded on a lower dimensional boundary. While this has been extremely successful in AdS spacetimes with the celebrated AdS/CFT correspondence \cite{Maldacena:1997re,Witten:1998qj,Gubser:1998bc}, it has been much harder to realize in asymptotically flat space times (see e.g. \cite{Susskind:1998vk,Polchinski:1999ry,Giddings:1999jq,deBoer:2003vf,Arcioni:2003td,Arcioni:2003xx}). With recent improvements in our understanding of asymptotic symmetries, and their implications for scattering amplitudes, there has been greatly renewed interest in this topic, coming from two different directions, summarized in \Fig{fig:celestial_fig}. In one proposal, referred to as BMS/Carrollian holography \cite{Dappiaggi:2004kv,Dappiaggi:2005ci,Bagchi:2016bcd,Bagchi:2019clu,Bagchi:2019xfx,Laddha:2020kvp,Chen:2021xkw,Bagchi:2022owq,Duval:2014uva,Duval:2014lpa}, the result is a field theory on the 3d null boundary of spacetime. In a second approach, referred to as celestial conformal field theory \cite{Pasterski:2016qvg,Pasterski:2017ylz,Pasterski:2017kqt}, the holographic dual is a two-dimensional conformal field theory living on the celestial sphere.  For reviews of the celestial holography approach, see \cite{Strominger:2017zoo,Pasterski:2021rjz,Raclariu:2021zjz}. For reviews of the relation between these two approaches, see \cite{Donnay:2022wvx,Donnay:2022aba}.

It would be interesting to explore how light-ray/ detector operators fit into this picture. On the one hand, light-ray operators provide a natural operator living on the 3d null boundary, while on the other hand, they are categorized by quantum numbers on the celestial sphere, much like operators in the celestial CFT. They may therefore prove useful as a bridge between these approaches. At a more technical level, it would also be interesting to relate the light-ray OPE with the OPE of celestial currents \cite{Pate:2019lpp}. In a different direction, investigations of celestial holography in $(2,2)$ signature lead to Lorentzian dynamics on the ``celestial torus". In this case, light-ray operators in the celestial CFT may play an important role \cite{De:2022gjn,Hu:2022syq,Casali:2022fro,Jorge-Diaz:2022dmy}. In summary, we believe that there are many interesting open directions relating light-ray/ detector operators and flat space holography that are worth exploring.

\subsection{Energy Correlators in Table Top Experiments}\label{sec:open_table}

\begin{figure}
\includegraphics[width=0.955\linewidth]{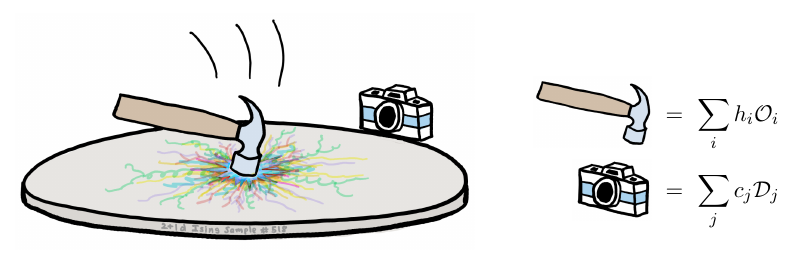}
\caption{An illustration of a realization of an energy correlator measurement in a condensed matter system. We hope that this can be made more concrete in the future. Figure from \cite{Caron-Huot:2022eqs}.
}
\label{fig:DSD_fig}
\end{figure}

We have highlighted throughout this review that detector operators and energy correlators provide an interesting class of observables for characterizing generic quantum systems. However, currently all the data we know of for energy correlator observables is from collider physics experiments, and probes the specific theory of QCD. It would be of particular interest to study energy correlator observables in other quantum systems, for example in condensed matter physics. A highly schematic illustration is shown in \Fig{fig:DSD_fig}. It is possible to make a variety of interesting condensed matter systems that behave as relativistic QFTs, such as graphene, or thin film superconductors, and it would be interesting to understand if such experiments are feasible in the laboratory. Another approach would be to realize such systems artificially, using either quantum computers or cold atoms.  In a more speculative direction, it would be interesting to further understand the relation between detector correlators in quantum gravity, and interferometer observables used experimentally to search for signals of the quantization of gravity \cite{Parikh:2020nrd,Parikh:2020fhy,Parikh:2020kfh,Banks:2021jwj,Zurek:2020ukz,Verlinde:2022hhs,Verlinde:2019ade,Verlinde:2019xfb,Carney:2024wnp}. See \cite{Sivaramakrishnan:2024ydy} for work in this direction.

We hope that this review, and the interesting recent progress in the study of energy correlators prompts their studies in more general systems, so that in a future review there is more to say on this topic.

\section{Conclusions and Outlook}\label{sec:conc}

50 years after the energy flux operator was introduced to characterize early QCD collider experiments, remarkable new measurements of energy correlators at modern colliders have revealed for the first time the behavior of these observables at extreme energies, in extreme kinematic limits, and in extreme states of matter.
Combined with tremendous theoretical developments, taking insight from wide ranging areas of theoretical physics including perturbative scattering amplitudes and the conformal bootstrap, this  transforms the possibility for interaction between collider experiment and formal theory.

In this review we have provided a broad overview of the physics of detector operators and energy correlator observables. 
We emphasized their connections to diverse areas in nuclear physics, particle phenomenology, and QFT, in an attempt to foster further interaction, and we highlighted the rich experimental opportunities these observables provide across collider systems.
This charts an experimental program with a broad range of physics goals, ranging from precision measurements of the top quark mass and strong coupling constant, to studies of spin physics, nuclear structure, and hot and cold nuclear matter.

Energy correlators are now, on the one hand, a practical tool for studying the Standard Model and nuclear physics, and on the other hand, a playground for understanding QFT and gravity. They provide a beautiful opportunity to connect deep theoretical ideas with the real world, and we look forward to their future exploration.

\section{Acknowledgements}

We are grateful to numerous colleagues who have significantly influenced the perspective presented here, and have contributed to the experimental and theoretical understanding of energy correlators. In particular, we would like to thank Lance Dixon, Kai Yan, Hao Chen, Yibei Li, Wouter Waalewijn, Jesse Thaler, Patrick Komiske, Juan Maldacena, David Simmons-Duffin, Murat Kologlu, Petr Kravchuk, Kyle Lee, Bianka Mecaj, Iain Stewart, Wenqing Fan, Jack Holguin, Xiaohui Liu, Carlota Andres, Fabio Dominguez, Cyrille Marquet, Xiaoyuan Zhang, Tong-Zhi Yang, Gherardo Vita, Gregory Korchemsky, Emery Sokatchev, Alexander Zhiboedov, Dimitry Chicherin, Anjie Gao, Aditya Pathak, Xin-Nian Wang, Tom Hartman, Steve Ellis, Massimiliano Procura, Shu-Heng Shao, Cyuan Han Chang, Hai Tao Li, Helen Caines, Ananya Rai, Andrew Tamis, Yenjie Lee, Yi Chen, Rithya Kunnawalkam Elayavalli, Evan Craft, Mark Gonzalez, Alexandre Homrich, Nima Arkani-Hamed, Joao Barata, Barbara Jacak, Csaba Csaki, Shounak De, George Sterman, Gregoire Mathys, Tom Appelquist, Hofie Hannesdottir, Sebastien Mizera, Giulio Salvatori, Alexandre Homrich, Krishna Rajagopal, Max Jaarsma, Hai Tao Li, Gabriel Cuomo, Matthew Walters, Enrico Herrmann, Julio Parra Martinez, Terry Generet, Rene Poncelet, Joao Barata, Youqi Song, Andrew Larkoski, Ben Nachman, Meng Xiao, Anjali Nambrath, Weiyao Ke, Enrico Herrmann, Riccardo Gonzo, Csaba Csaki, Fanyi Zhao, Austin Joyce, Jingjing Pan, Anastasia Volovich, Marcus Spradlin, Sabrina Pasterski, Berndt Mueller, Gabriel Cuomo, Matt Walters, Andrea Guerreri, Raju Venugopalan, Aida El-Khadra, Yue Zhou Li, Zhiquan Sun, Silviu Pufu, Pier Monni, Alfred Mueller, Xinan Zhou, Yang Zhang, Song He, Ming-xing Luo, Gang Yang, Feng Yuan, Werner Vogelsang, Duff Neill, Wenbiao Yan, Yuxiang Zhao, Ding Yu Shao, Zhen Xu, Ji-Chen Pan, Hao-Yu Liu, Yuxun Guo, Xiao Lin Li, Haotian Cao, Vladyslav Shtabovenko, Jun Gao, Zhongbo Kang, Jian-ping Ma, Yu Jia, Huajia Wang, Yuri Lensky, Sruthi Narayanan, Steve Sharpe, Max Hansen, Martin Kruczenski, Yifei He, Jim Brau, Michael Peskin, David Poland, Zahra Zahraee, Frank Coronado.

I.M. thanks the KITP Santa Barbara for hospitality while this review was written.
I.M. is supported by the DOE Early Career Award DE-SC0025581, and the Sloan Foundation. H.X.Z. is supported by the National Science Foundation of China under contract No.~12425505 and the Asian Young Scientist Fellowship.

\bibliography{EEC_ref.bib}{}
\bibliographystyle{apsrev4-1}
\newpage
\onecolumngrid
\newpage

\end{document}